\documentclass{aastex6}

\usepackage{epsfig}
\usepackage{hyperref}
\usepackage{graphicx}
\usepackage{epstopdf}
\usepackage{pslatex}
\usepackage{color}
\usepackage[caption=false]{subfig}
\usepackage{float}
\usepackage{subfloat}
\usepackage{cleveref}
 \usepackage{natbib}
 \usepackage{amsmath}

\usepackage{xcolor}

\newcounter{iso}
\newcommand{\rxn}{\refstepcounter{iso}%
                  (\oldstylenums{\theiso})}

\newcommand\osref[1]{(\oldstylenums{\ref{#1}})}

\defcitealias{2016ApJS..227...25R}{Paper I}
\defcitealias{2016ApJ...818...52T}{TTZ16}

\defcitealias{2012MNRAS.420.1019A}{AMI Consortium: Scaife et al. 2012}

\begin{document}

\title{Weak and Compact Radio Emission in Early High-Mass Star Forming Regions: II. The Nature of the Radio Sources} 
\bigskip\bigskip

\author{V. Rosero\altaffilmark{1,2,3}, P. Hofner\altaffilmark{3,10},  S. Kurtz\altaffilmark{4}, R. Cesaroni\altaffilmark{5}, C. Carrasco-Gonz\'alez\altaffilmark{4}, E. D. Araya\altaffilmark{6},  L. F. Rodr\'iguez\altaffilmark{4}, K. M. Menten\altaffilmark{7}, F. Wyrowski\altaffilmark{7}, L. Loinard\altaffilmark{4,7}, S. P. Ellingsen\altaffilmark{8}, \and S. Molinari\altaffilmark{9}}

\altaffiltext{1}{National Radio Astronomy Observatory, 1003 Lopezville Rd., Socorro, NM 87801, USA}
\altaffiltext{2}{Department of Astronomy, University of Virginia, Charlottesville, VA 22904, USA}
\altaffiltext{3}{Physics Department, New Mexico Tech, 801 Leroy Pl., Socorro, NM 87801, USA}
\altaffiltext{4}{Instituto de Radioastronom{\'\i}a y Astrof{\'\i}sica, 
Universidad Nacional Aut\'onoma de M\'exico, Morelia 58090, M\'exico}
\altaffiltext{5}{INAF, Osservatorio Astrofisico di Arcetri, Largo E. Fermi 5, 50125 Firenze, Italy}
\altaffiltext{6}{Physics Department, Western Illinois University, 1 University Circle, Macomb, IL 61455, USA}
\altaffiltext{7}{Max-Planck-Institute f\"{u}r Radioastronomie, Auf dem H\"{u}gel 69, 53121 Bonn, Germany}
\altaffiltext{8}{School of  Physical Sciences, University of Tasmania, Private Bag 37, Hobart, Tasmania 7001, Australia}
\altaffiltext{9}{INAF-Istituto di Astrofisica e Planetologia Spaziale, Via Fosso del Cavaliere 100, I-00133 Roma, Italy}

\altaffiltext{10}{Adjunct Astronomer at the National Radio Astronomy Observatory, 1003 Lopezville Road, Socorro, NM 87801, USA.}

\begin{abstract}
In this study we analyze 70 radio continuum sources associated with dust clumps and considered to be candidates for the earliest stages of high-mass star formation. The detection of these sources was reported by  \citet{2016ApJS..227...25R}, who found most of them to show weak (${\scriptstyle <}$1 mJy) and compact (${\scriptstyle <}\,$0\rlap.$^{\prime \prime}$6) radio emission.  Herein, we used the observed parameters of these sources to investigate the origin of the radio continuum emission.
We found that  at least $\sim 30\%$  of these radio detections are most likely ionized jets associated with high-mass protostars, but for the most compact sources we cannot discard the scenario that they represent pressure-confined H{\small II} regions.  
This result is highly relevant for recent theoretical models based on core accretion that predict the first stages of ionization from high-mass stars to be in the form of jets. Additionally, we found that properties such as the radio luminosity as a function of the bolometric luminosity of ionized jets from low and high-mass stars are extremely well-correlated.  Our data improve upon previous studies by providing further evidence of a common origin for jets independently of luminosity. 

\end{abstract}

\keywords{Ionized Jets  -- stars: formation -- techniques: high sensitivity -- techniques: interferometric}

\section{Introduction} \label{sec:intro}
Young high-mass stars (M $ \geq \,$ 8 M$_\odot$) probed in cm-wavelength interferometric studies typically appear as fairly bright (flux densities of $\sim$ few mJy to Jy)  regions of ionized gas  that are classified according to their size and emission measure, e.g., compact, ultracompact (UC), and hypercompact (HC) H{\small II} regions \citep[e.g.,][]{2005IAUS..227..111K}. It is generally thought that once  nuclear burning  has  begun the star  produces enough UV radiation to photoionize the surrounding gas. However, theories of the earliest stages remain poorly constrained by observations mainly due to the characteristics of the regions where they are born, which are highly dust-obscured, distant ($\gtrsim$ 1 kpc) regions that undergo rapid evolution, and they reach the zero-age main sequence (ZAMS) while still heavily accreting. In fact, an evolutionary sequence for high-mass stars has not yet been established \citep[e.g.,][]{2013A&A...550A..21S, 2014prpl.conf..149T}, although significant progress has been achieved on both observational and theoretical fronts \citep[e.g.,][]{2018ARA&A..56...41M}. The identification and study of
objects in the early stages of their evolution will help us to
  discriminate among proposed mechanisms for their formation; the two main
scenarios being core accretion (i.e., scaled-up version of low-mass
star formation) and competitive accretion (i.e., in which stars in a cluster attract each other while they accrete from a shared reservoir of gas; see \citealp{2014prpl.conf..149T}). The low-mass star formation process is modeled by
accretion via a circumstellar disk and a collimated jet/outflow that removes angular momentum and allows accretion to proceed \citep[e.g.,][]{1988ApJ...328L..19S}. The jet/outflow system is powered  magnetohydrodynamically by rotating magnetic fields coupled to either the disk (disk winds: e.g., \citealp{2000prpl.conf..759K}) and/or the protostar (X-winds: e.g., \citealp{1987ARA&A..25...23S}).  Additionally, protostellar collisions have been  proposed as an alternative mechanism for the formation of high-mass stars \citep{1998MNRAS.298...93B, 2005AJ....129.2281B}. \\

Massive molecular outflows are a common phenomenon in high-mass star forming regions (e.g., \citealp{1996ApJ...457..267S, 2002A&A...383..892B}); hence accretion disks and ionized jets similar to those found towards low-mass protostars are also expected. 
 In addition, several surveys toward high-mass star forming regions in the NIR spectral lines of H$_2$ have detected a large number of molecular jets \citep[e.g.,][]{2017ApJ...844...38W, 2015MNRAS.450.4364N}.
However, the current sample of known high-mass protostars associated with disks \citep[see review by][]{2016A&ARv..24....6B} and collimated jets \citep[e.g.,][]{1995ApJ...449..184M, 1998ApJ...502..337M, 2006ApJ...638..878C, 2008AJ....135.2370R} is inadequate to draw conclusions about the entire population.
The detection of sources at the onset of high-mass star formation and the measurement of their physical properties is essential  to test theoretical models of high-mass star formation \citep[e.g.,][]{2014prpl.conf..149T}. Furthermore, the  most sensitive instruments are necessary  to place significant constraints on the occurrence rate and parameters of these detections.\\

In  \citet[][hereafter Paper I]{2016ApJS..227...25R} we described our high sensitivity ($\sim$3 -- 10 $\mu$Jy beam$^{-1}$) continuum survey, which aimed to identify candidates in early evolutionary phases of high-mass star formation and to study their centimeter continuum emission. We observed  58 high-mass star forming region candidates using the Karl G. Jansky Very Large Array (VLA)\footnote{The National Radio Astronomy 
Observatory is a facility of the National Science Foundation operated under cooperative agreement by Associated Universities, Inc.} at 1.3 and 6 cm wavelengths at an angular resolution ${\scriptstyle <}\,$0\rlap.$^{\prime \prime}$6.  The 58 targets were grouped into three categories based on their mid and far-IR luminosity as well as the temperature of the cores: 25 hot molecular cores (HMCs), 15 cold molecular cores with mid-IR point source association (CMC--IRs), and 18 cold molecular cores (CMCs) devoid of IR point source associations. The cores in our sample cover a wide range of parameters such as bolometric luminosity and distance. They have similar masses and densities, however, the latter two types of cores---mainly found within infrared dark clouds (IRDCs)---have lower temperatures (T$\sim$ 10--20 K) than HMCs (T${\scriptstyle >}\,$50 K; depending on the probe and scale). In \citetalias{2016ApJS..227...25R} we reported  detection rates of 1/18 (6$\%$) CMCs, 8/15 (53$\%$) CMC-IRs and 25/25 (100$\%$) HMCs.
In several cases, we detected multiple sources within a region, which resulted in a total detection of 70 radio sources associated with 1.2 mm dust clumps.
The 100$\%$ detection rate of centimeter emission in the HMCs  is a higher fraction than previously reported. This suggests that radio continuum may be present, albeit weak, in {\it all} HMCs although in many cases it is only detectable with the superior sensitivity now available with the upgraded VLA.   Our results show further evidence for an evolutionary sequence in the formation of high-mass stars, from  very early stage cold cores (i.e., CMCs) to relatively more evolved ones (i.e., HMCs).\\

A number of physical processes can cause centimeter continuum emission associated with high-mass star forming regions (see \citealt{2012ApJ...755..152R} and \citealt{2013ApJ...766..114S} summaries of thermal and non-thermal emission detected at centimeter wavelengths from YSOs). Recently, \citet*[][hereafter TTZ16]{2016ApJ...818...52T}  developed a model to predict the radio emission from high-mass stars forming via core accretion. The \citetalias{2016ApJ...818...52T} model predicts that during the first stages of ionization the H{\small II} region is initially confined to the vertical (or outflow) axis and produces free-free emission with  similar features and parameters as observed towards ionized jets. Ionized jets are detected as weak and  compact centimeter continuum sources. At subarcsecond resolutions, they usually show a string-like morphology, often aligned  with a large-scale molecular outflow  of size up to a few parsecs. Ionized jets trace outflows on  smaller scales, providing the location of the driving protostar, that otherwise are deeply embedded in the natal clump and generally remain undetected at other wavelengths  due to the high extinction in the region \citep{1998AJ....116.2953A}.  However, less extincted sources may
have molecular jet counterparts visible in H$_2$ line emission from shocked gas.
These sources are also called `thermal radio jets' due to their characteristic  rising spectrum which is consistent with free-free radiation from ionized gas.  
The  ionization mechanism of these jets has been proposed to be \emph{shock-induced ionization} when the wind from the central protostar ionizes itself through  shocks due to variations in velocity of the flow or variations of the mass loss rate  (\citealt*{1987RMxAA..14..595C}; \citealt{1989ApL&C..27..299C}).  Unlike the simple model of a  uniform electron density H{\small II} region,  ionized jets and winds have a radial density gradient and thus  are partially optically thick.  \citet{1986ApJ...304..713R} discussed the behavior of  collimated jets and the dependency  of their physical parameters (such as temperature, velocity, density and ionization fraction) on  morphology, independently of the mechanism of ionization, and showed that the spectral index of a partially ionized jet  ranges  between $-0.1 \leq \alpha \leq 1.1$.\\

The detection of ionized jets toward  high-mass stars at their early stages, as predicted by \citetalias{2016ApJ...818...52T}, can help to distinguish between accretion scenarios (highly organized outflows are expected from core accretion but not from competitive accretion scenarios; \citealt{2016ApJ...821L...3T}),  and ultimately will give us insight about accretion disks around high-mass stars. Several systematic studies searching for ionized jets have been reported in the literature. \citet{2012ApJ...753...51G},  from a sample of 33 IR luminous objects, detected 2 ionized jets using the Australia Telescope Compact Array (ATCA) with a 4$\sigma$ detection limit and an image rms ($\sigma$) of $\sim$0.1-0.2 mJy beam$^{-1}$  at 4.8 and 8.6 GHz.  \citet{2016A&A...585A..71M} observed  11 high-mass YSOs using the Jansky VLA and detected 5 collimated ionized jets and 6 ionized wind candidates with a 3$\sigma$ detection limit and an rms $\sim$ 11 $\mu$Jy beam$^{-1}$ at $\sim$ 6.2 GHz. \citet{2016MNRAS.460.1039P} observed 49 high-mass YSOs using the ATCA and detected 16 ionized jets and 12 jet candidates  with a 3$\sigma$ detection limit and an rms $\sim$ 17 $\mu$Jy beam$^{-1}$ at $\sim$ 5.5 GHz. Additionally, the Protostellar Outflow at the EarliesT Stage (POETS) survey is undertaking a search of radio-jets  using the VLA with an angular resolution of $\sim$0\rlap.$^{\prime \prime}$1 and an image rms of $\sim$10 $\mu$Jy beam$^{-1}$ \citep{2018A&A...619A.107S, 2019A&A...623L...3S}. In our radio continuum Jansky VLA survey we observed 58 high-mass star forming regions and detected 70 radio sources with a 
5$\sigma$ detection limit and an rms $\sim$ 5 $\mu$Jy beam$^{-1}$ at $\sim$ 6 GHz \citep{2016ApJS..227...25R}. \citealt*{2018A&ARv..26....3A} is a recent comprehensive review of ionized jets in star forming regions. \\

The main goal of this  paper is to investigate the nature of the 70 detected radio sources 
reported in \citetalias{2016ApJS..227...25R}. The observations along with the complete list of targets, coordinates, radio  detections and derived observational parameters are presented in \citetalias{2016ApJS..227...25R}. In Section \ref{sec:models} we examine several scenarios to explain the origin of the ionized gas emission and we study the physical properties of the detected sources.  Section \ref{discussion_paperII}  contains a discussion of the viability of the different scenarios. In Section \ref{conclusions_paperII}
we summarize our findings. Additionally, Appendix \ref{app:lum} shows the bolometric luminosity estimates for these high-mass star forming regions using {\it Herschel}/Hi--GAL data and Appendix \ref{mom_rate_appe} shows a study of the momentum rate of ionized jets.

\section[Models  for the Radio Emission]{ Models Considered for the Radio Emission} \label{sec:models}

\subsection{Low-mass Young Stellar Objects}\label{yso}

The main goal of our high sensitivity continuum survey presented in  \citetalias{2016ApJS..227...25R} 
was to detect radio emission from high-mass protostars. However, there exists a variety of sources that could also appear as radio detections in our images.
In \citetalias{2016ApJS..227...25R}  we considered  contamination by extragalactic radio
sources, and found that only a small number of extragalactic sources are expected to be observed within
the typical dust clump size of $\sim$30$^{\prime \prime}$ (8 and 2 sources in the 6 and $1.3\,$cm bands, respectively for the entire sample). 
A more likely source of contamination would be the presence of low-mass YSOs which are expected
to be present in regions of high-mass star formation  \citep[e.g.,][]{2013A&A...554A..48R}. We are thus interested in identifying possible low-mass class 0 -- class III YSOs  that could have 
been detected in our survey toward high-mass star forming regions. 

\begin{figure}[!h]%
    \centering
    \includegraphics[width=0.7\linewidth, clip=True]{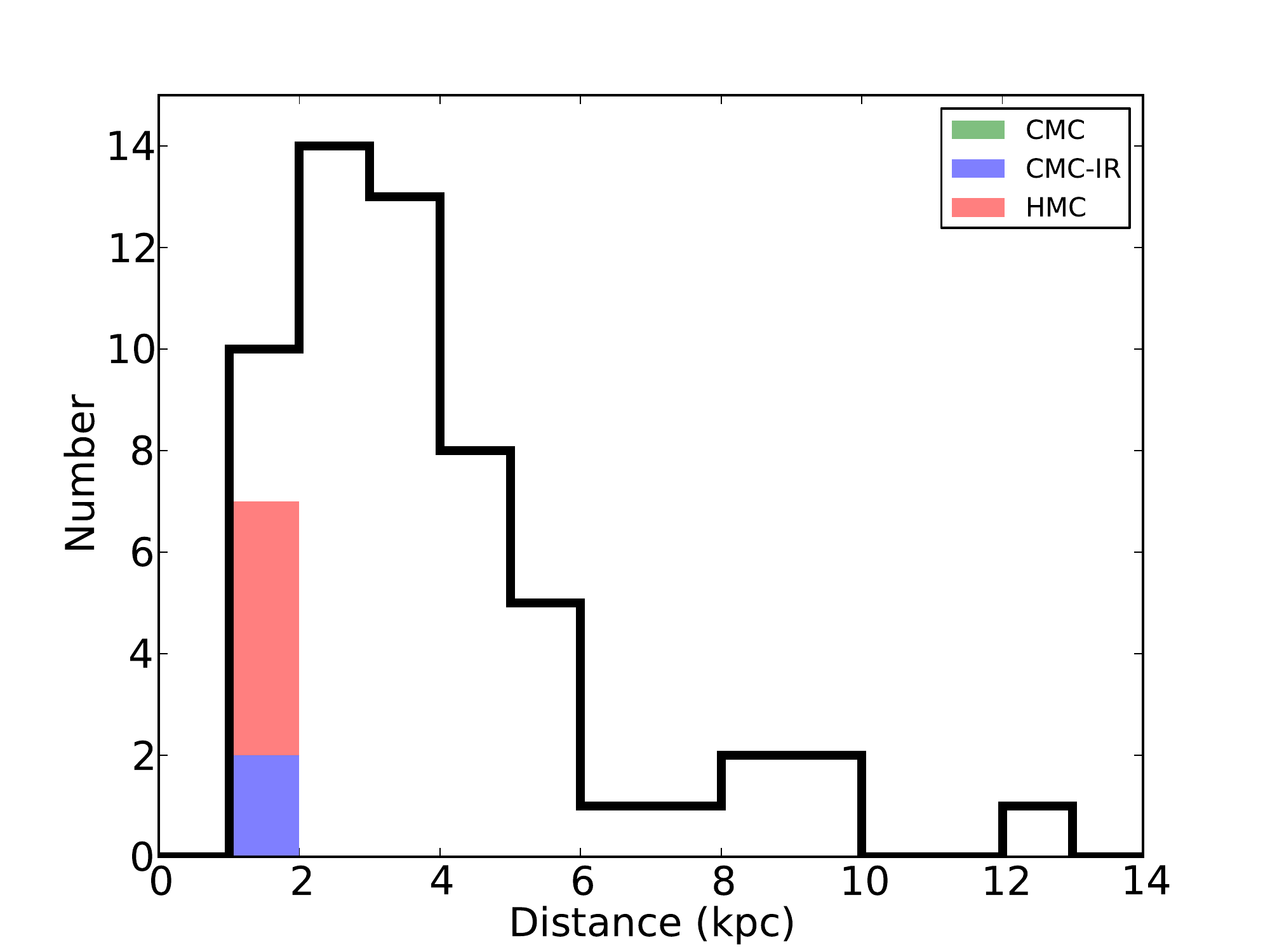}%
     \caption[The distance distribution of our targets is shown as a black line]{\small{The distance distribution of our targets is shown as a black line. The color histogram shows the number of targets where we expect the detection of T-Tauri 
     stars which are as bright as J162749.85--242540.5 in Ophiuchus. This is the case for 5 HMCs and 2 CMC-IRs.}
}%
    \label{T_Tauri_hist}%
\end{figure}

A large sample of low-mass YSOs has been  observed with the VLA at 4.5 and 7.5 GHz as part of  the Gould Belt survey (i.e.,  Ophiuchus at a distance of 120 pc:  
\citealp{2013ApJ...775...63D}; Orion at 414 pc: \citealp{2014ApJ...790...49K}; Serpens at 415 pc: \citealp{2015ApJ...805....9O}; Taurus-Auriga at 140 pc: \citealp{2015ApJ...801...91D}, and
Perseus at 235 pc: \citealp{2016ApJ...818..116P}). 
The brightest low-mass YSO in the entire Gould Belt survey (excluding the Orion region) was found in the Ophiuchus region (source J162749.85--242540.5,
a class III YSO, i.e., weak-lined T-Tauri star, with S$_{7.5\,GHz} = 8.51\,$mJy, \citealt{2013ApJ...775...63D}).  To determine whether such an object would have been detected in our survey,
we scaled its flux density to the assumed distance\footnote{Distances were taken from the literature, and are listed in Table \ref{SED_Parameters}. 
Most distances are kinematic; only a few regions have trigonometric parallax measurements.}
of each of our targets, and compared its scaled flux density to our adopted detection limit of  $\geq$ 5 times the image rms
at 7.4 GHz for each of our regions. We found that such a YSO would not be detected in any of our targets located at distances beyond $2\,$kpc.
Since the majority
of our targets exceed this distance (see Figure \ref{T_Tauri_hist}), we conclude that for most of our observed regions the detected radio sources are not low-mass YSOs.\\

There are 10 regions in our survey that are located at distances $\leq$ 2 kpc. However, given the 7.4 GHz image rms for these regions, only in 7 of them would we have detected the brightest low-mass YSO of Ophiuchus.
These regions are  five  HMCs: 18517$+$0437, 20126$+$4104, 20293$+$3952, 20343$+$4129, G34.43$+$00.24mm1, and  two CMC--IRs: LDN1657A$-$3  and
UYSO1. In these 7 regions we detected a total of 13 radio sources within the FWHM of the mm clumps: 10 towards HMCs and 3 towards CMC--IRs. That some of these 
sources are  possibly low-mass YSOs can be seen in the case of  IRAS 20126$+$4104: Besides the well-studied high-mass protostars associated
with radio sources 20126$+$4104 A and 20126$+$4104 B, the radio source G$78.121+3.632$ in this region (see  \citetalias{2016ApJS..227...25R} , Table 4) corresponds to the source I20var, which was
 discussed by \citet{2007A&A...465..197H}. This is a highly 
variable radio source and  has observational properties consistent with a flaring T-Tauri star. In the same region, we have also detected a new object of similar characteristics. Radio source 20126$+$4104 C 
was detected for the first time in our survey although several high sensitivity observations of this region have been made in the past \citep[][]{2007A&A...465..197H}.

Hence, 20126$+$4104 C is clearly 
variable in the radio regime, and is a candidate for a low-mass pre-main sequence star.
Additionally, the radio source LDN1657A$-$3 A, which has a negative spectral index ($\alpha=-1.2$), is also a candidate for a variable radio source, where the emission is probably
caused by non-thermal processes on the surface of a T-Tauri star.  While the observational properties of these sources are consistent with low mass YSOs, we note that alternative explanations are possible \citep[e.g.,][]{2018A&A...612A.103C}.

In summary, while some degree of contamination by low-mass YSOs probably exists in our survey for the nearest sources,
for the majority of our targets the detected radio sources are 
very likely not contaminated by emission from low-mass YSOs.

\subsection{H{\small II} Regions}
In \citetalias{2016ApJS..227...25R}  we reported the detection of 70 radio continuum sources associated with three different types of mm clumps and we calculated their 5--25 GHz spectral index ($\alpha$) using power-law fits of the form $S_{\nu}$ $\propto$ $\nu^{\alpha}$.  The spectral index values and the fits to the data for all the radio detections are reported in \citetalias{2016ApJS..227...25R} in Table 4 (electronic version) and in Figure 4, respectively. The range of spectral indices found was $-$1.2 to 1.8 (see Figure 5 in \citetalias{2016ApJS..227...25R}). Based on their radio spectra, we classify  these sources as flat spectral index ($-$0.25$<\alpha<$0.2),  positive spectral index ($\alpha \geq$ 0.2), and negative spectral index ($\alpha <$-0.25). Thus, we have 10 sources with flat, 44 sources with positive, and 9 sources with negative spectral index. For the remaining 7, there is not a clear estimate of the spectral index.

The radio sources have a variety of morphologies. Excluding the sources without spectral index information, there are 6 extended sources, 8 sources with elongated structures,
and the majority of sources (49) are compact with respect to our synthesized beam. In this section we consider whether a family of H{\small II} region models could  explain sources with flat and positive spectral index.

\subsubsection{Extended Sources}\label{ext_sources}

Among the sources detected in our survey associated with $1.2\,$mm dust emission, there are six sources that are clearly extended at cm wavelengths with
respect to the $\sim$0\rlap.$^{\prime \prime}$4 resolution of the maps, hence they are candidates for  H{\small II} regions,
i.e. photoionized gas.
 These sources are relatively bright (S$_{25.5\,GHz} \approx$ 1\,mJy), and are found mostly toward HMCs.
Moreover, they generally show a flat spectral index, indicative of optically thin free-free emission. 
For five of these sources, we calculate the physical properties from the $25.5\,$GHz continuum flux using the formulae from \citet{1994ApJS...91..659K},
which assume spherical symmetry, and optically thin emission from a uniform density plasma with T$_{e}= 10^{4}$\,K. The results are listed in Table \ref{HII_Parameters}, where column 
1 is the region name, column 2 is the 
specific radio source, and columns 3 and 4 are the frequency ($\nu$) and radio flux (S$_{\nu}$), respectively.  Column 5 is the observed linear size (diameter)
of the radio source ($\Delta s$) at 3$\sigma$ rms level in the image, column 6 is the emission measure (EM), column 7 is the electron density (n$_e$),
column 8 is the 
excitation parameter (U) and column 9 is the logarithm of the Lyman continuum  flux (N$^{\prime}_{Ly}$)  required for ionization.   We use log\,N$^{\prime}_{Ly}$ to 
estimate 
the spectral type of the ionizing star (listed in column 10) using the tabulation in \citet{1973AJ.....78..929P}, further assuming that a single ZAMS star is photoionizing the nebula and 
producing the 
Lyman continuum flux. The distances used for these calculations are listed in Table \ref{SED_Parameters}, and the near kinematic distance is adopted  when the region has a distance ambiguity.

\begin{deluxetable}{l c c c c c c c c c}
\tabletypesize{\scriptsize}
\tablecaption{Extended Sources:  Parameters  from Radio Continuum  \label{HII_Parameters}}
\tablewidth{0pt}
\tablehead{
\colhead{Region}                  & 
\colhead{Radio }  &
\colhead{$\nu$}   &
\colhead{S$_{\nu}$ }               & 
\colhead{$\Delta s$}  & 
\colhead{EM/$10^{5}$ }             & 
\colhead{n$_{e}/10^{3}$ } & 
\colhead{U } & 
\colhead{log\,N$^{\prime}_{Ly}$ } &
\colhead{Spectral } 
 \\[2pt]
\colhead{}                            & 
\colhead{Source}                      & 
\colhead{(GHz)}                      & 
\colhead{($\mu$Jy)}                      & 
\colhead{(pc)}                      & 
\colhead{(pc\,cm$^{-6}$)}                      &  
\colhead{(cm$^{-3}$)}                      & 
\colhead{(pc\,cm$^{-2}$)}                     & 
\colhead{(s$^{-2}$)}  &
\colhead{Type\tablenotemark{a}}                         \\[-20pt]\\}
\startdata
18470$-$0044        & C      & 25.5    &       3490   & 0.085   &   8.5  &   3.2  &   9.2  &  46.4  &  B0.5   \\[3pt]
18521+0134          & B      & 25.5    &        378   & 0.057   &   2.5  &   2.1  &   4.7  &  45.5  &  B1   \\[3pt]
19035+0641          & B      & 25.5    &       2270   & 0.009   &  39.6  &  21.1  &   3.4  &  45.1  &  B1   \\[3pt]
20293+3952          & C\tablenotemark{b}      & 25.5    &       1560   & 0.019   &   1.9  &   3.2  &   2.1  &  44.4  & B2   \\[3pt]
20343+4129          & A      & 25.5    &        881   & 0.005   &  17.5  &  18.5  &   1.8  &  44.3  &  B2   \\[3pt]

\enddata
\tablenotetext{\text{a}}{Using the tabulation in  \citet{1973AJ.....78..929P}.}
\tablenotetext{\text{b}}{Includes radio source 20293$+$3952 B (see Figure 2 in \citetalias{2016ApJS..227...25R}).}
\end{deluxetable}

The measured sizes of the five sources listed in Table \ref{HII_Parameters} are all below $0.1\,$pc, which according to  \citet{2005IAUS..227..111K} would suggest a classification as ultra (UC)- or hypercompact (HC)
H{\small II} regions. However, the calculated emission measures and electron densities are an order of magnitude or more  smaller than  typical values for such H{\small II} regions. The typical values of the emission measure and electron density for UCH{\small II} regions are EM$\gtrsim 10^{7}$ pc cm$^{-6}$ and n$_{e}\gtrsim 10^{4}$ cm$^{-3}$ and for HCH{\small II} regions are EM$\gtrsim 10^{10}$ pc cm$^{-6}$ and n$_{e}\gtrsim 10^{6}$ cm$^{-3}$ \citep{2005IAUS..227..111K}.
We note that four of these resolved extended radio sources (18470--0044 A, 18521$+$0134 B, 19035$+$0641 B  and 20293$+$3952 C) are offset $\sim$ 2/3 of the radius from the center of the mm clumps,
and are thus located at their outskirts. A plausible explanation for the lower electron densities and emission measures is that early B-type stars have
formed near the edge of the dust clumps where the density of the surrounding medium is much lower than in the center.

We will comment on two additional resolved and extended radio sources detected in this survey. The first one is  18470--0044 A, which was previously observed by \citet{2011ApJ...739L..17H} at 25.5 GHz using the VLA in the C-configuration.  This radio source has an offset of 7\rlap.$^{\prime \prime}$2 with respect to the peak of the mm clump associated with IRAS 18470--0044  reported by \citet{2002ApJ...566..945B}. Our image towards this region is affected by sidelobes due to the large flux and extended emission of a 
nearby radio source, and we were not able to accurately measure the radio flux at 1.3 cm. This source is among the detections without a clear estimate of its spectral index. However, based on the flux 
reported by \citet{2011ApJ...739L..17H},  and our measured flux at 6 cm (see \citetalias{2016ApJS..227...25R}), this radio source has a flat spectrum, and it is likely an H{\small II} region ionized by a 
B2 ZAMS star. 

The second bright, and resolved, radio source is G34.43$+$00.24mm2 A, which is the only extended source detected towards an IRDC clump in our survey. Interestingly, this radio source 
has a spectral index of $\alpha= -0.5$ (see \citetalias{2016ApJS..227...25R}) and is associated with a 24 $\mu$m point source, and  at least two molecular outflows \citep{2007ApJ...669..464S}. This radio 
source was originally detected at 6 cm by \citet[labeled by them as Mol 74]{1998A&A...336..339M}  and \citet{2004ApJ...602..850S} using the VLA (FWHM $\sim$6\rlap.$^{\prime \prime}$$\times
$3\rlap.$^{\prime \prime}$). The  flux densities at 6 cm reported in those studies are consistent with our data.  This radio source has an offset of 7\rlap.$^{\prime \prime}$8 with respect to the 
center of the $1.3\,$mm clump G34.43$+$00.24mm2  detected by \citet{2006ApJ...641..389R}.\\

We found that at least 36$\%$ of the HMCs have an extended radio source within the 1\rlap.$^{\prime}$8 FWHM primary beam at 25.5 GHz. However, most of them are located slightly outside the FWHM of the mm clump, and are not discussed in this study. These extended radio sources are: 18089$-$1732   G12.890$+$0.495, 18182$-$1433   G16.584$-$0.053, 18470$-$0044  G32.113$+$0.097, 18521$+$0134  G34.749$+$0.021, 19012$+$0536  
G39.389$-$0.143 and  19266$+$1745 G53.037$+$0.115.    All these sources and their radio continuum parameters are reported in \citetalias{2016ApJS..227...25R}.

\subsubsection{Compact Sources}\label{compact_sources_sect}

In \citetalias{2016ApJS..227...25R} we characterized the change in flux density with frequency using a power law. 
We reported the detection of  36 compact radio sources  (51$\%$ of total detections) with a rising spectrum ($\alpha >$0.2), of which 
7 were not detected at 6 cm, thus a lower limit for their spectral index was estimated. We also detected 5 compact radio sources with a flat cm spectrum.
In this section we investigate whether a uniform density UC/HC H{\small II} region can explain the observed fluxes and spectral indices for compact sources with rising spectra.

While optically thin free-free emission results in a flat spectral index ($\alpha = -0.1$), a rising spectrum implies appreciable optical depth in the emitting gas.
To fit our spectra, we thus will require a turnover, (i.e. $\tau_{\nu} \sim 1$) near the intermediate frequency of our observing bands, around $\nu_{t}=$14.7 GHz,
which in turn requires an emission measure near 9$\times$10$^{8}$ pc\,cm$^{-6}$. For fitting the data we use a uniform density, spherical H{\small II} region model
with electron temperature of $T_{e}=$ 10$^{4}$\,K
as shown in equation (11) from \citet{1975A&A....39..217O}. We find that 
for all compact, rising spectrum sources detected in our survey, within the given uncertainties, a uniform density H{\small II} region spectrum can be reasonably fit to the radio continuum data.
Examples of the fits are shown by the continuous blue line in Figure \ref{HII_fit_examples}. 
 The fits for all the 36 compact radio sources with rising spectral index is shown in Appendix \ref{all_HII_fit} Figure \ref{HII_fit_app}.


\begin{figure}[!h]%
    \centering
  \hspace*{-0.5cm} 
    \includegraphics[width=0.35\linewidth, clip]{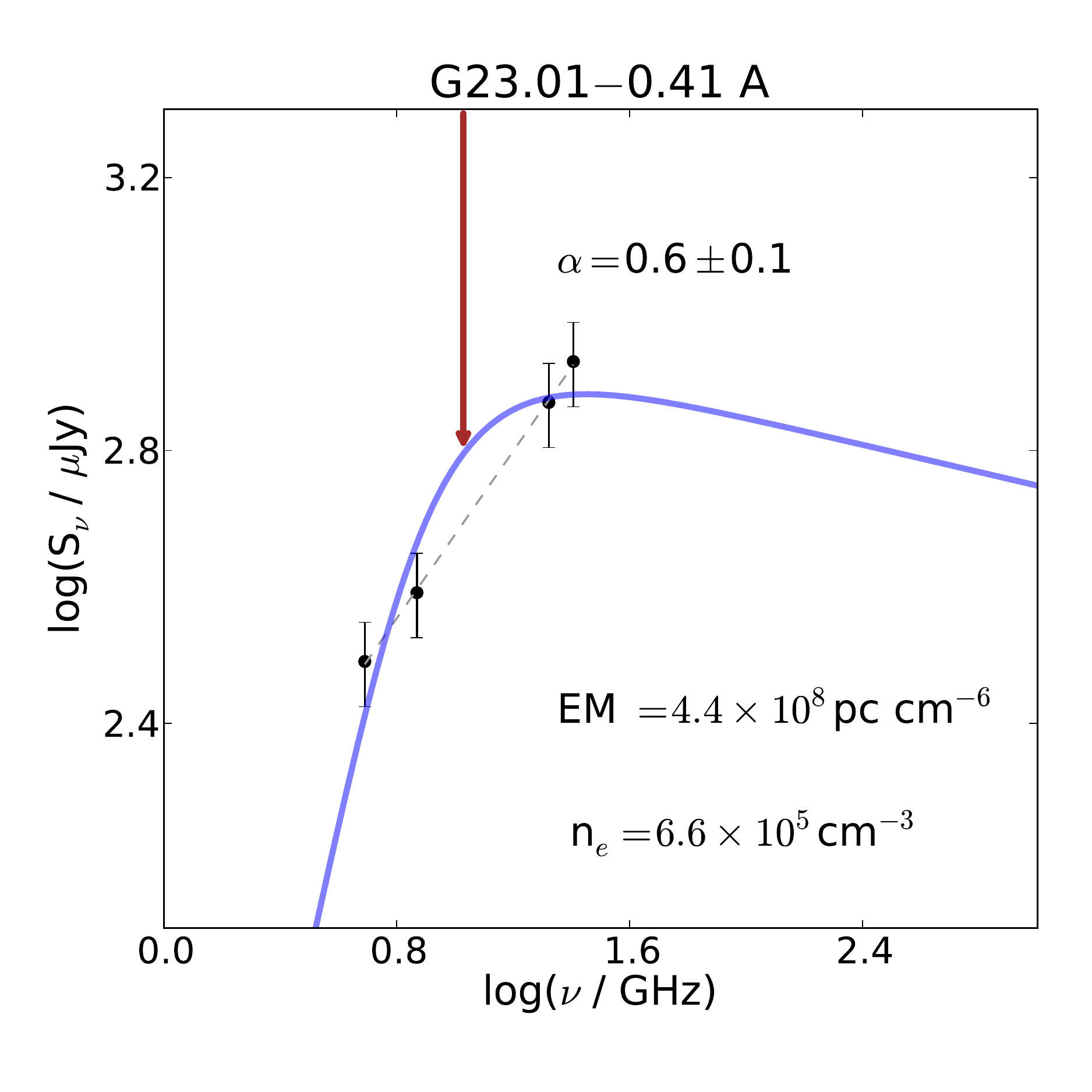}%
        \includegraphics[width=0.35\linewidth, clip]{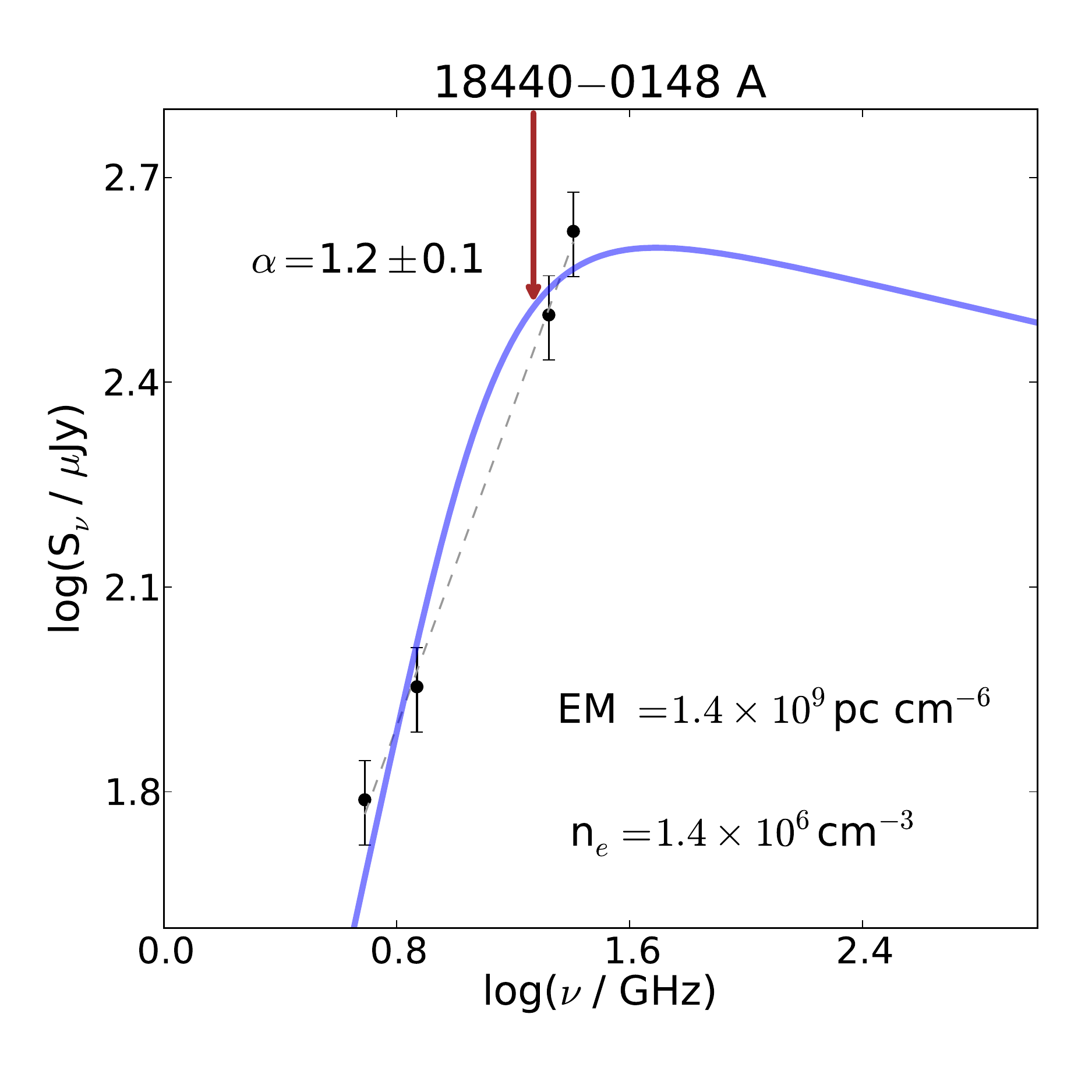}%
        \includegraphics[width=0.35\linewidth, clip]{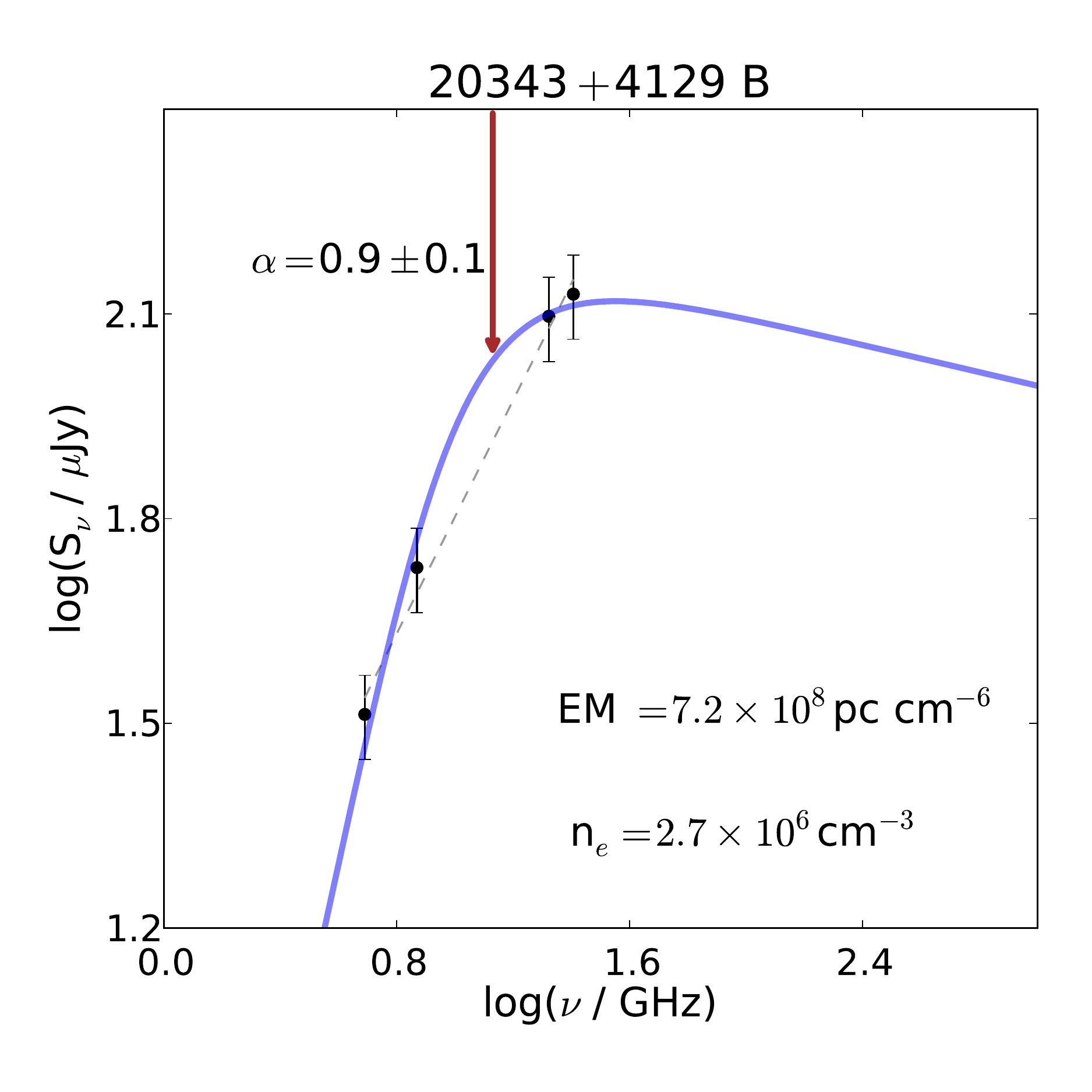}%
     \caption[Spectra of the compact radio sources G23.01--0.41 A  (left), 18440--0148 A (center) and 20343$+$4129 B  (right)]{\small{Spectra of the compact radio sources G23.01--0.41 A  (left), 18440--0148 A (center) and 20343$+$4129 B  (right). Error bars are an assumed uncertainty of 10$\%$ from the flux densities added in quadrature with an assumed 10$\%$ error in calibration. The  continuous blue line is the H{\small II} region fit using a spherical, constant density model. The red arrow indicates the frequency where $\tau_{\nu}=$1 in each model fit. The dashed line is the best fit to the data from a power-law of the form $S_{\nu}$ $\propto$ $\nu^{\alpha}$.}}\label{HII_fit_examples}%
\end{figure}

Our fitting results show that the generally quite weak emission from these very compact, rising spectrum sources implies a very small size
for the emitting regions. The sizes are much smaller than our angular resolution, and are on the order of the initial Str\"{o}mgren sphere radius.
The initial Str\"{o}mgren sphere radius ($R_{s}$) depends on the Lyman continuum (N$_{Ly}$) flux and the ambient molecular density (n$_{H_{2}}$) as
stated in equation 1 from \citet{1996ApJ...473L.131X}:

\begin{equation}
R_{s}= 4104.7 \left( \frac{N_{Ly}}{10^{49}s^{-1}}\right)^{1/3}\left(\frac{n_{H_{2}}}{10^{5}cm^{-3}}\right)^{-2/3}  \text{[au]}  .
\end{equation}

\begin{figure}[!h]%
    \vspace{1.2cm}
    \centering
    \includegraphics[width=0.5\linewidth, clip]{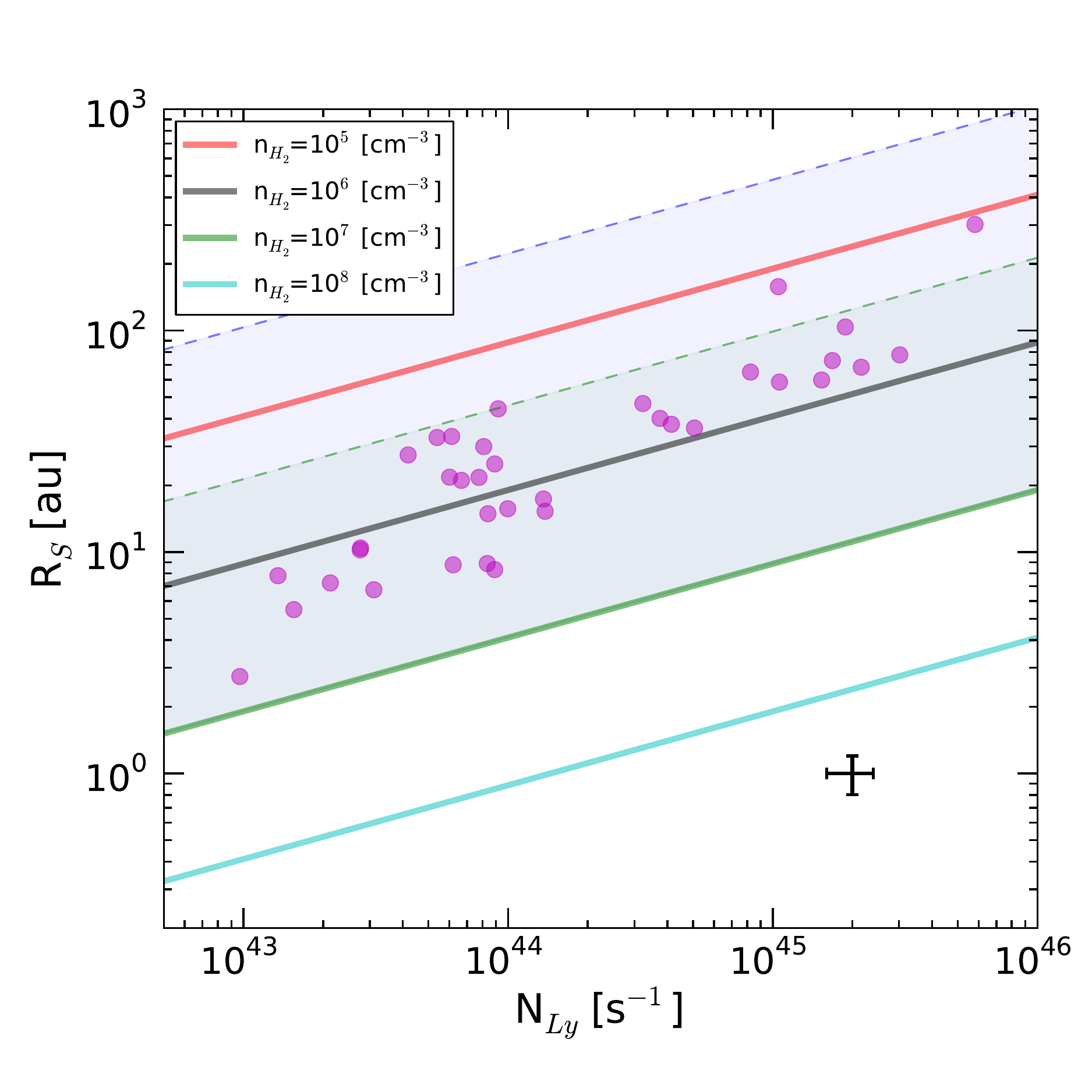}%
     \caption{\small{Initial Str\"{o}mgren sphere  radius as a function of the Lyman continuum for compact sources with rising spectra. The solid lines represent the  ambient molecular density for n$_{H_{2}}=$ 10$^{5}$ --  10$^{8}$ cm$^{-3}$. The solid purple dots represent the radius of the
     H{\small II} regions ($\Delta\,s/2$) as implied by the spherical, constant density H{\small II} region model fits.  The dashed green line represents the lower limit of  $R_{Turb}$ (for $\xi =$1) and the  dashed blue line presents $R_{Th}$ if the initial ambient molecular density for both cases is n$_{H_{2}}=$ 10$^{7}$\,cm$^{-3}$. Shaded areas  represent the path from their initial  radius up to their final Str\"{o}mgren sphere radius if the sources were born at a density of n$_{H_{2}}=$ 10$^{7}$\,cm$^{-3}$ (green and blue shaded areas for turbulence and thermal pressure confinement, respectively).
The error bars in the bottom right corner correspond to a 20$\%$ calibration uncertainty. 
}}%
    \label{Stromgren_sphere}%
\end{figure}

We show the relation R$_{s}$ versus N$_{Ly}$, represented by the solid lines, for n$_{H_{2}}=$ 10$^{5}$, 10$^{6}$,  10$^{7}$ and 10$^{8}$ cm$^{-3}$ in Figure 
\ref{Stromgren_sphere}. To place our sources in this diagram, we estimated the Lyman continuum flux (N$_{Ly}$) from our $25.5\,$GHz flux density, with
the formulae of \citet{1994ApJS...91..659K}, and use the radii ($\Delta\,s/2$) derived from the spectral fitting.  These data are listed in Table~\ref{tab:fig3}.
Our sources, represented as solid purple dots
in Figure \ref{Stromgren_sphere}, cluster around the expected Str\"{o}mgren radius for an initial density of $10^{6}$\,cm$^{-3}$. 

When nuclear burning begins and a high-mass star produces enough UV photons to photo-ionize the surrounding material, the initial Str\"{o}mgren sphere radius is reached (to within a few percent) in a recombination timescale, $t_{r}= $(n$_{H_{2}}$$\beta_{2}$)$^{-1}$[s], where $\beta_{2}$= 2.6$\times$10$^{-13}$\,cm$^{3}$s$^{-1}$ is the recombination coefficient \citep{1980pim..book.....D}. For any reasonable initial density this time scale is extremely short ($< 1\,$yr), and the initial
Str\"{o}mgren sphere radius is reached almost instantaneously. The highly over-pressured ionized region will then begin to expand, and hence the initial
Str\"{o}mgren sphere is a very short-lived configuration, and therefore it is unlikely that the large number of sources detected represent this evolutionary stage.

After  formation of the initial Str\"{o}mgren sphere around a star, the UC H{\small II} region is highly overpressured,  and as a result, it expands approximately at the sound speed until approaching pressure equilibrium with the ambient medium. 
\citet{1995RMxAA..31...39D} and \citet{1996ApJ...473L.131X} have studied  the confinement of UC H{\small II} regions in a molecular core by thermal, and thermal plus turbulent pressure, respectively. 
Assuming pressure equilibrium between ionized and surrounding molecular gas, \citet{1996ApJ...473L.131X}  gives the final radius of the ionized region (R) as :

\begin{equation}
R = R_{s} \left(   \frac{2k\xi T_{H^{+}}}{m_{H_{2}}\sigma_{v}^{2} + kT_{k}}\right)^{2/3}  ,
\end{equation}

\noindent
where $T_{H^{+}}$ is the temperature of the ionized region, $\xi$ is a turbulence factor ($>$ 1) that takes into account the pressure due to stellar winds and turbulence in the ionized gas, $\sigma_{v}$ is the velocity dispersion produced by turbulence and $T_{k}$ is the kinetic temperature of the surrounding molecular gas. 
Using typical values for the physical conditions in regions where high-mass stars form, we can  test whether the sources discussed in this section could be ionized regions in pressure equilibrium with the surrounding molecular gas.
While molecular line observations with single dish instruments indicate average densities of n$_{H_{2}}=$ 10$^{5}$\,cm$^{-3}$ over the $1\,$pc clump sizes \citep[e.g.,][]{2000ApJ...536..393H}, interferometric measurements of high-mass star forming cores
have revealed central densities of n$_{H_{2}}= 10^7 - 10^{10} $\,cm$^{-3}$ on scales $< 0.1\,$pc \citep[e.g.,][]{1990ApJ...362..191G, 2015A&A...573A.108G}. Following \citet{1996ApJ...473L.131X}, we adopt values of $T_{H^{+}}=$ 10$^{4}$\,K, $\sigma_{v}=$ 2 km\,s
$^{-1}$ (FWHM$\,\sim$\,5 km\,s$^{-1}$) and $T_{k}=$ 100 K. Assuming $\xi = 1$, and evaluating the above equation with these numbers we get $R_{turb} = 11.2 R_{s} $ for the case of thermal plus turbulent pressure, and
$R_{th} = 54.3 R_{s}$ for thermal pressure only. These relations are shown in Figure \ref{Stromgren_sphere} for n$_{H_{2}}= 10^7$\,cm$^{-3}$ as green and blue dashed lines, respectively. 
Considering  the location of our data points in Figure \ref{Stromgren_sphere}, we can exclude the extremely high densities of $10^{10} $\,cm$^{-3}$ as found by \citet{2015A&A...573A.108G}, which would predict
much smaller source sizes. On the other hand, our data points are located within the  shaded areas that represent the path from their initial Str\"omgren radius up to their final radius in pressure equilibrium,
if the sources were born at a density of n$_{H_{2}}=$ 10$^{7}$\,cm$^{-3}$ (green and blue shaded areas for turbulence and thermal pressure confinement, respectively).

An estimate of the expansion time $\tau_{expansion}$ for an ionized region can be obtained assuming that it  expands at its sound speed ($C_{s} \sim$ 10 km\,s$^{-1}$). To expand to $\sim$\,200 au,
then $\tau_{expansion} \sim R/C_{s} \sim $ 100 yr. Thus, an initial Str\"omgren sphere will expand fairly quickly and, as suggested by \citet{1995RMxAA..31...39D} and \citet{1996ApJ...473L.131X}, 
the ionized regions can remain compact as long as the molecular core provides the outside pressure. Observations of  UCH{\small II} regions and HMCs suggest that this time is on order $10^5\,$yr
\citep{1989ApJ...340..265W,2001ApJ...550L..81W}, and it hence appears that  our sources could be ionized regions around newly formed stars in pressure equilibrium in their
molecular cores. While our calculations have not been fitted to a particular source, the observed scatter in Figure \ref{Stromgren_sphere} can be accounted for with a varying amount of turbulence in the molecular gas,
i.e. if the molecular line FWHM varies between $\sim$ 7 -- 20 km\,s$^{-1}$ for the case of an  ambient molecular density of n$_{H_{2}}=$ 10$^{7}$\,cm$^{-3}$.

It is  interesting to note that we found that the radius of the extended sources discussed above are within the pressure equilibrium zone for an initial density of n$_{H_{2}}=$ 10$^{5}$\,cm$^{-3}$. 
These sources are located on the outskirts of the mm core, and one might ask whether they have migrated out of the molecular core center, or if they were born in their current location. Assuming stellar velocities between 2 and 12 km\,s$^{-1}$ 
\citep{2007ApJ...660.1296F} in order for them to travel to the half power point of the cores (FWHM median angular size for HMC $=$18\rlap.$^{\prime \prime}$ at a distance of 4 kpc), times between
around 10$^{5}$ yr and 10$^{4}$ yr, respectively, are needed. While migration toward lower density regions is thus possible, we note that we do not find any strong evidence for cometary regions which would be predicted due to
bow shocks between molecular and ionized gas \citep{1990ApJ...353..570V}.

\subsubsection{Lyman Continuum}

\begin{figure}[!h]
    
    \centering
    \includegraphics[width=0.5\linewidth, clip]{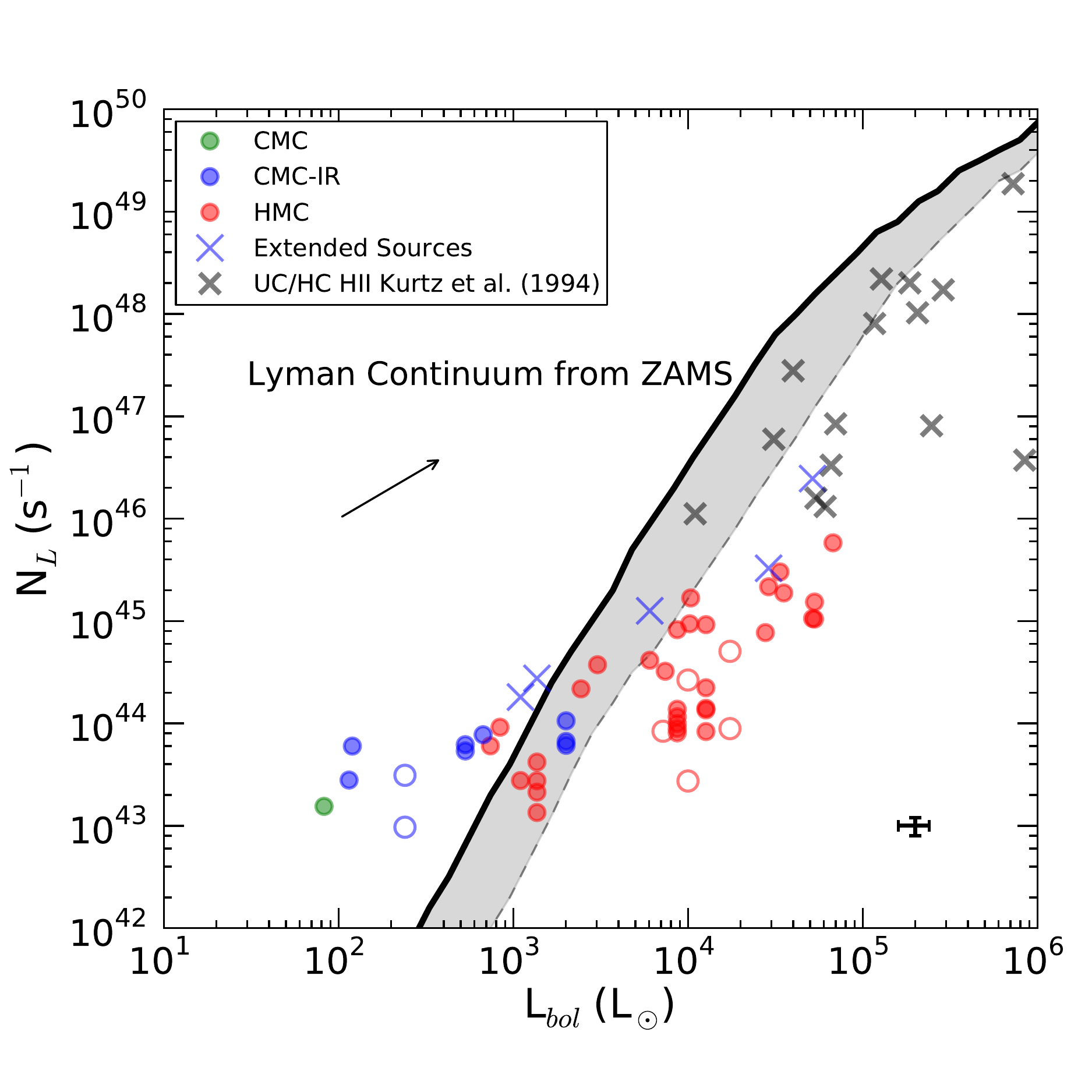}
     \caption{\small{Lyman continuum measured at 25.5 GHz as a 
     function of the bolometric luminosity for all detected sources with flat or rising spectra in our sample. The bolometric luminosity is mainly estimated from  {\it Herschel}/Hi--GAL data (except for the open circles for
     whose the bolometric luminosity  is from the literature). The circles  represent the compact sources with flat or rising spectra, while the blue  $\color{blue} \times $ symbol represents the flat spectrum extended 
     sources from \S \ref{ext_sources}. UC H{\small II} regions from \citet{1994ApJS...91..659K} are represented by the gray $\color{gray} \times $ symbol. The continuous black line is the expected Lyman continuum photon rate of a 
     single  ZAMS star at a given luminosity, and the shaded area gives these quantities for the case of a cluster \citep{2013A&A...550A..21S}. The arrow indicates how much a point would move if the distance were increased by 
     a factor of 2. 
The error bars in the bottom right corner correspond to a 20$\%$ calibration uncertainty. 
}}
    \label{Lyman_cont_plot}

\end{figure}

An additional point to consider to understand the nature of our detections is the Lyman continuum photon rate  as a function of the bolometric luminosity. We analyze this relation for all the sources 
with a flat or a rising spectrum (including extended, elongated structure, as well as compact morphology) as shown in Figure \ref{Lyman_cont_plot}. The Lyman continuum photon rate is estimated from the radio 
continuum flux at 25.5 GHz and the bolometric luminosities for our regions are estimated  from  {\it Herschel}/Hi--GAL fluxes, and from ancillary data (see \S \ref{app:lum}).  We list these data in Table~\ref{tab:fig4}.
For data taken from the literature, care was taken that the  Lyman continuum flux and bolometric luminosities refer to the same distance. For sources with distance ambiguity, we use the near kinematic distance.
In Figure \ref{Lyman_cont_plot}, compact and elongated sources are represented by filled circles if the bolometric luminosities are estimated in this work  (see \S \ref{app:lum}), or open circles if the luminosity
is taken from the literature. The extended sources from \S \ref{ext_sources} are represented by the  $\color{blue} \times $  symbol. 
The continuous black line is the expected Lyman continuum photon rate from a single zero-age main-sequence (ZAMS) star at a given luminosity,  and the shaded area
bounded by the solid black line shows the expected Lyman continuum from a stellar cluster of the same N$_{Ly}$. For more details on these curves see \citet{2013A&A...550A..21S}.
Thus, H{\small II} regions ionized by stellar UV photons from a single early-type star are expected to lie on the black line. If,  on the other hand, the Lyman continuum comes from a cluster of stars (a likely scenario for high-mass 
stars) rather than from a single ZAMS star, the expected  N$_{Ly}$ is lower, and should be located within the shaded area \citep{2015A&A...579A..71C}.

 As seen in Figure \ref{Lyman_cont_plot}, only a small fraction of our sources fall in the shaded area of the plot  indicating direct stellar photoionization. Most of the HMC sources (red open/filled circles) lie
 below the curve of the expected Lyman continuum flux, and hence are underluminous at radio wavelengths.
 On the other hand, the majority of the sources detected towards CMCs and CMC--IRs are located in the so-called ``forbidden area'' above the Lyman continuum line, showing  an excess of Lyman continuum compared 
 to the expected value based 
on their luminosities. This is true even if the sources are corrected by the distance, i.e., when there is ambiguity in the kinematic distance of the source or the value of the distance is incorrect. For reference, the arrow in the plot indicates the amount 
that a point will move if the distance increases by a factor of 2. If the distance changes by any other factor the point will move parallel to the arrow. Additionally, there is  a possibility that some  bolometric luminosities  are underestimated (see \S \ref{app:lum}), however if this is the case we believe that the luminosities will shift to the right by less than 0.5 dex.

\citet{2013A&A...550A..21S} have reported Lyman continuum excess for several sources in an 18 and 22.8 GHz survey of high-mass star forming regions with the Australian Telescope Compact Array (ATCA).
Interestingly, $\sim$70$\%$ of their H{\small II} regions with Lyman excess are associated with molecular clumps belonging to two types of sources that are in the earliest evolutionary stages of high-mass stars based on their classification (equivalent to our CMCs and CMC-IRs clumps).  
Additionally, \citet{2015A&A...579A..71C}  found  Lyman  continuum excess for about 1/3 of their  sample of 200 compact and UC H{\small II} regions selected from the CORNISH survey \citep{2013ApJS..205....1P}.  Their sources with Lyman continuum excess are also in an earlier evolutionary phase within their sample. Both studies argued that the Lyman excess is not easily justified, leaving room for two possible scenarios, invoking additional
sources of  UV photons from an  ionized jet, or from an accretion shock in the protostar/disk system. \citet{2016A&A...588L...5C} suggested that the Lyman excess is  produced by  accretion shocks, based on outflow (SiO) and infall (HCO$^{+}$) tracer observations towards the 200 H{\small II} regions studied in \citet{2015A&A...579A..71C}. 

It is important to mention that due to our selection criteria (see \citetalias{2016ApJS..227...25R}) the sources studied by \citet{2013A&A...550A..21S} and \citet{2015A&A...579A..71C} are much brighter at radio wavelengths than the ones from our work, with radio luminosities at 5 GHz of $\sim$ 10$^{2}$--10$^{6}$ mJy kpc$^{2}$ vs 10$^{-2}$--10 mJy kpc$^{2}$ in our sample. In Figure \ref{Lyman_cont_plot} we also show several UC H{\small II} regions from  \citet{1994ApJS...91..659K}, denoted by the $\color{gray} \times $ symbol. These sources seem to be produced by higher free-free emission compared with our sample, suggesting that our sources represent a different population of radio sources. \citet{1999RMxAA..35...97C} based on selection criteria similar to ours,  detected sources with low radio luminosities like the ones in this work. Furthermore, these low radio luminosities are typical of thermal jets, with UV photons that are produced by shocks from collimated winds from the protostar  with the surrounding material \citep[e.g.,][]{1996ASPC...93....3A}. Thus, while the above analysis of the cm SEDs suggests a model of pressure confined H{\small II} regions for our compact sources, the Lyman continuum photon rate  as a function of the bolometric luminosity
shown in Figure \ref{Lyman_cont_plot} does not lend strong support to this model. A further possible explanation for the compact sources with rising spectra, as well as for several elongated sources
detected in our survey is that they arise from thermal jets. We explore this scenario in the following section.   

\subsection{Ionized Jets}\label{ionized_jet_sect}

\begin{figure}[htbp]
\centering
\begin{tabular}{c}

\includegraphics[width=0.7\textwidth,clip=true, trim = 10 240 10 80, clip, angle = 0]{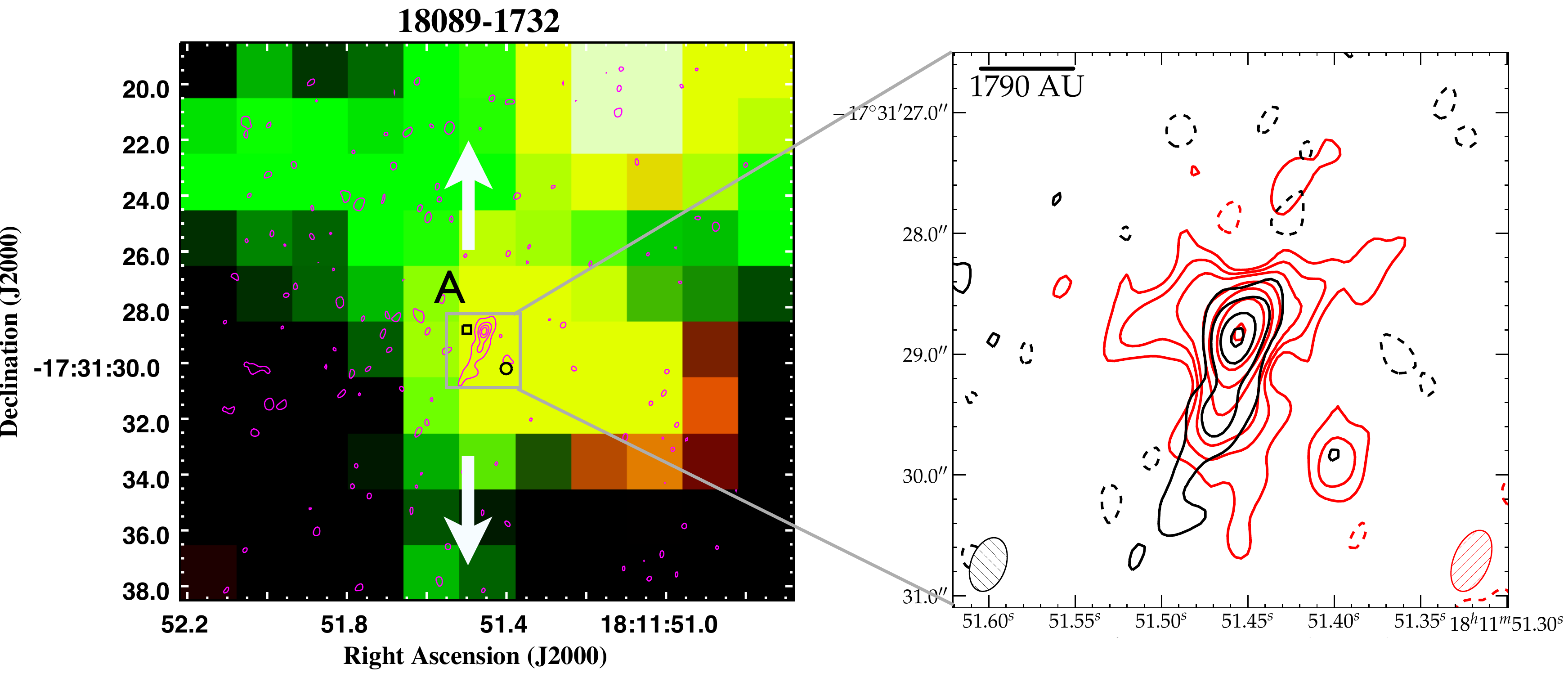}  \\ 
\includegraphics[width=0.68\textwidth,clip=true, trim = 10 240 10 60, clip, angle = 0]{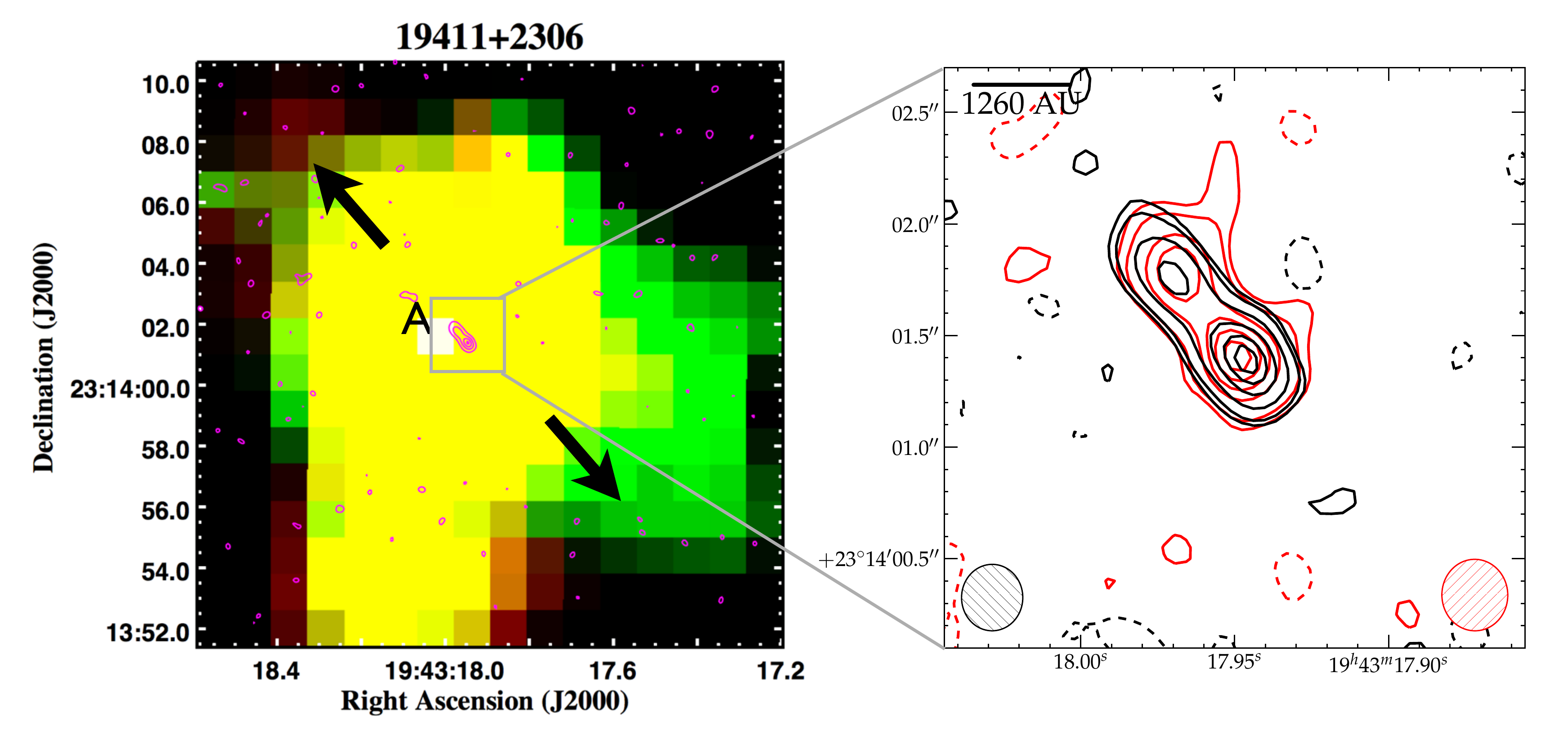}    
\end{tabular}
\caption{\small{\emph{Spitzer} IRAC GLIMPSE three-color (3.6$\mu m$-blue, 4.5$\mu m$-green and 8.0$\mu m$-red )
images of two ionized jet candidates, overlayed with VLA 6 cm continuum emission contours. 
Note that both regions show 4.5 $\mu$m excess emission. In the right panel we show
an enlarged version of the radio continuum from \citetalias{2016ApJS..227...25R}.
{\bf Top: 18089-1732 A:}  The arrows represent the direction of a the north-south bipolar SiO outflow detected by  \citet{2004ApJ...616L..23B, 2010ApJ...724L.113B}. The black circle and the square are the 6.7 GHz CH$_{3}$OH and H$_{2}$O masers reported in \citep{2002A&A...390..289B}, respectively.
VLA 6 cm contour levels are ($-$2.0, 3.0, 10.0, 25.0, 40.0) $\times $6 $\mu$Jy beam$^{-1}$, and 1.3 cm contour levels ($-$1.5, 3.0, 5.0, 7.5, 15.0, 25.0, 45.0, 95.0) $\times$  10 $\mu$Jy beam$^{-1}$.
{\bf Bottom: 19411+2306 A:} The arrows represent the direction of the detected CO  outflow by \citet{2002A&A...383..892B}. \citet{2002ApJ...566..931S} reported that 6.7 GHz CH$_{3}$OH and H$_{2}$O
 masers were not detected for this source in their survey.  VLA 6 cm contour levels are ($-$2.0, 2.0, 3.0, 6.0, 10.0, 13.0, 15.0) $\times$ 5.5 $\mu$Jy beam$^{-1}$, and 1.3 cm contour levels ($-$2.0, 2.0, 3.0, 6.0, 8.0, 10.0, 12.0) $\times$  8 $\mu$Jy beam$^{-1}$.  }}
 \label{fig:spitzer_examples}
\end{figure}

\begin{figure}[htbp]
\centering
\begin{tabular}{c}
\includegraphics[width=0.7\textwidth,clip=true, trim = 10 200 10 80, clip, angle = 0]{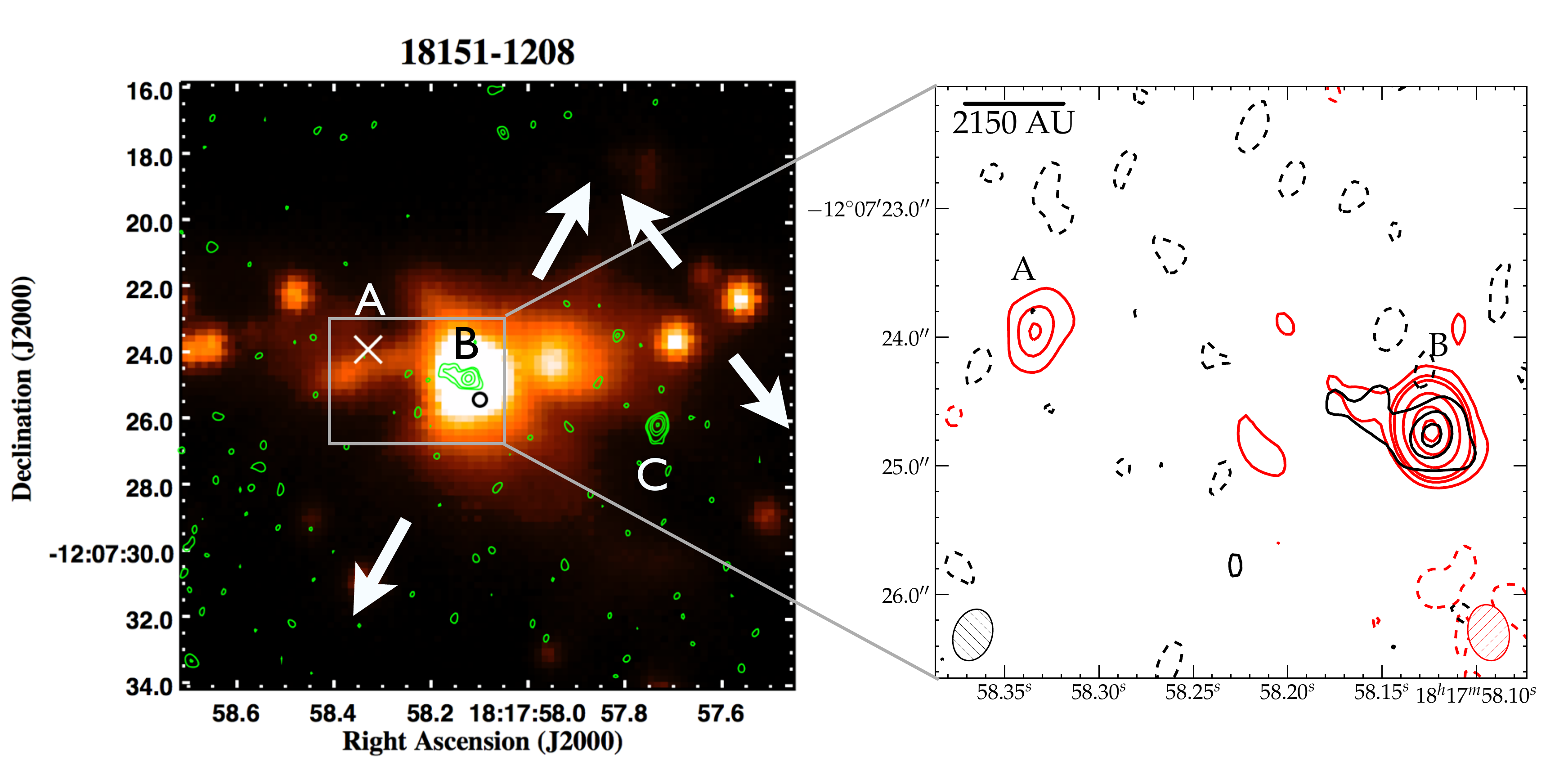} \\ 
\includegraphics[width=0.68\textwidth, clip=true, trim = 10 150 10 80, clip, angle = 0]{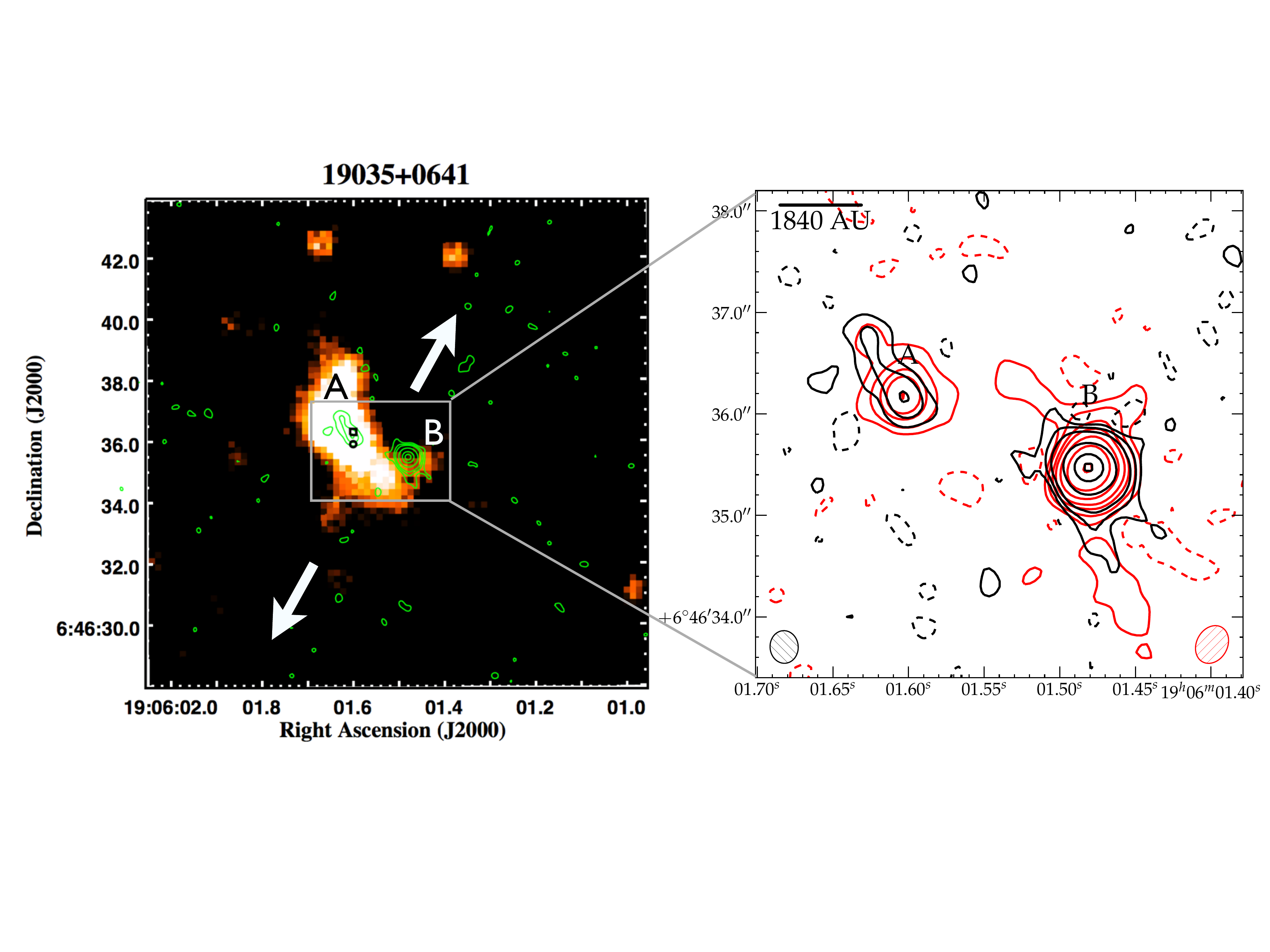}   
\end{tabular}
\caption{\small{UKIDSS {\emph K}-band images of two ionized jet candidates, overlayed with VLA 6 cm continuum emission contours. In the right panel we show
an enlarged version of the radio continuum from \citetalias{2016ApJS..227...25R}. 
{\bf Top: 18151$-$1208 B:} The arrows represent the direction of the two nearly perpendicular  CO  outflows detected by \citet{2011ApJ...729...66F}. A blue-shifted component of a molecular outflow going in the direction
of 18151$-$1208 B is seen in Figure 4 of \citet{2011ApJ...729...66F} but it is not discussed by the authors. The black circle  is the 6.7 GHz CH$_{3}$OH  maser from \citet{2002A&A...390..289B}. The x symbol represents the position of an 
additional radio source detected at 1.3 cm reported in \citet{2016ApJS..227...25R}. VLA 6 cm contour levels are ($-$2.0, 3.0, 9.0, 15.0) $\times$ 6 $\mu$Jy beam$^{-1}$, and 1.3 cm contour levels ($-$2.0, 3.0, 6.0, 8.5, 20.0, 40.0, 60.0)   $\times$  8 $\mu$Jy beam$^{-1}$.
{\bf Bottom: 19035+0641 A:} The arrows represent the direction of the detected CO and HCO$^{+}$ outflows \citep{2002A&A...383..892B, 2010A&A...517A..66L}. The black circle  and the square are the 6.7 GHz CH$_{3}$OH   and  
H$_{2}$O masers from \citet{2002A&A...390..289B}, respectively. VLA 6 cm contour levels are ($-$2.5, 3.0, 10.0, 25.0, 120.0, 280.0, 380.0) $\times$  4 $\mu$Jy beam$^{-1}$, and 1.3 cm contour levels ($-$2.0, 5.0, 10.0, 20.0, 30.0, 50.0, 90.0, 170.0) $\times$  8 $\mu$Jy beam$^{-1}$.}}
 \label{fig:UKIDSS_examples}
\end{figure}

Based on the low radio luminosities (S$_{5\,GHz}\,$d$^{2}$ $\sim$ 10$^{-2}$ -- 10 mJy kpc$^{2}$) of our detected sources, we need to consider the possibility that the source of ionization is not a ZAMS star, but rather 
that their nature is that of a thermal, ionized jet produced by shock ionization as described in \S \ref{sec:intro}.
Support for this hypothesis comes from a subset of resolved sources from  our survey. We have characterized 12 jet candidates based on their elongated, or string-like morphology in conjunction with an
association with a molecular outflow. These sources are listed in Table \ref{jet_cand_list}, where column 1 is the name of the region, column 2 are the radio sources that are thought to be part of the ionized jet, and 
column 3 lists the approximate direction of the ionized jet. Column 4  shows the approximate direction of the molecular outflows associated with the centimeter continuum emission as found in the literature. Column 5 
indicates if the centimeter continuum emission is a new detection or if it has been  detected in previous studies.  Column 6 lists the references for the molecular outflow detections and previous centimeter continuum 
detections if any.  Examples for these sources are shown in Figures \ref{fig:spitzer_examples}a, \ref{fig:spitzer_examples}b, and \ref{fig:UKIDSS_examples}a, \ref{fig:UKIDSS_examples}b. To our knowledge, 6 of these ionized jet candidates are new detections. In the cases of previous detections of centimeter continuum emission towards the listed regions, our high-sensitivity observations described in Paper I
generally show the elongation or string-like morphology of a jet for the first time \citep[e.g.,][]{2017ApJ...843...99H}. Furthermore, this subset of resolved jet candidates have the expected spectral index ( $0.2 \leq \alpha \leq 1.2$) for ionized jets,  several of them are associated with  6.7 GHz CH$_{3}$OH masers and H$_{2}$O 
masers, and they have excess emission at 4.5 $\mu$m,  which may trace shocked gas via H$_{2}$ emission in outflows or  scattered continuum from an outflow cavity \citep[e.g.,][]{2011ApJ...729..124C, 
2013ApJS..208...23L}. In some cases, like towards the ionized jet in 18182$-$1433, some of the radio sources  have  negative spectral indices, consistent with non-thermal lobes, since it is thought that when very strong shock waves from a fast jet move through a magnetized medium, some of the electrons are accelerated to relativistic velocities producing synchrotron emission \citep{2003ApJ...587..739G, 2010Sci...330.1209C}. \citet{2016MNRAS.460.1039P} and \citep{2018A&A...619A.107S, 2019A&A...623L...3S} also reported the detection of  ionized jets with non-thermal lobes (see also review by \citealt{2018A&ARv..26....3A}).

\begin{deluxetable}{l c c c c c c}
\tabletypesize{\scriptsize}
 \renewcommand*{\arraystretch}{1.5}
\tablecaption{Ionized Jets  \label{jet_cand_list}}
\tablewidth{0pt}
\tablehead{
\colhead{Region}                  & 
\colhead{Radio Source}        &
\colhead{Jet Direction}          &
\colhead{Outflow Direction}   &
\colhead{ H$_{2}-$Jet Direction}   &
\colhead{New Detection}             &
 \colhead{Reference}      \\[-20pt]\\}
\startdata
\setcounter{iso}{0}	
G11.11$-$0.12P1     &    A, C, D  &   NE$-$SW   &   E$-$W, NE$-$SW\tablenotemark{a}   &  E$-$W   &    y   & \rxn \label{rxn:Wang2014}  \rxn \label{rxn:Rosero2014} \rxn \label{rxn:Lee2013}  \\
18089$-$1732         &    A           &   N$-$S        &    N$-$S                 &   no/very weak\tablenotemark{b}     & n          & \rxn \label{rxn:Beuther2004} \rxn \label{rxn:Beuther2010} \rxn \label{rxn:Zapata2006}      \\
18151$-$1208         &    B          &   NE$-$SW   &   NW$-$SE\tablenotemark{c}    &  NW-SE  &    n     &  \rxn \label{rxn:Fallscheer2011} \rxn \label{rxn:Hofner2011}  \rxn \label{rxn:Davis2004}   \rxn \label{rxn:Varricat2010} \\
18182$-$1433         &    A$-$C\tablenotemark{d}   &   E$-$W        &   NE$-$SW, NW$-$SE     &   E$-$W   &    n       &  \rxn \label{rxn:Beuther2006}   \rxn \label{rxn:Moscadelli2013}  \rxn \label{rxn:Lee2012}        \\
IRDC18223$-$3      &    A$-$B\tablenotemark{e}       &    NE$-$SW  &  NW$-$SE\tablenotemark{f}      &  SE-NW  &  y     &  \rxn \label{rxn:Fallscheer2009}  \rxn \label{rxn:Beuther2005}   \\   
 G23.01$-$0.41      &    A    &    NE$-$SW         &  NE$-$SW  &  non-detection   & n & \rxn \label{rxn:Sanna2016} \rxn \label{rxn:Araya2008} \rxn \label{rxn:Sanna2018} \osref{rxn:Lee2013}\\
18440$-$0148         &    A          &    NW-SE       &  \nodata\tablenotemark{g}        &  non-detection  &    y    &     \rxn \label{rxn:Navarete2015}      \\
18566$+$0408        &    A$-$D\tablenotemark{h}  &    E$-$W        &   NW$-$SE   &  non-detection  &    n    &    \rxn \label{rxn:Zhang2007}\rxn \label{rxn:Araya2007}  \rxn \label{rxn:Hofner2017} \osref{rxn:Lee2013}                                   \\
19035$+$0641        &    A          &    NE$-$SW   &   NW$-$SE  &  no/very weak\tablenotemark{b}    &  y   &    \rxn \label{rxn:Lopez_Sep2010}       \\
19411$+$2306        &    A          &    NE$-$SW   &   NE$-$SW  &   detection\tablenotemark{b}    &  y   &    \rxn \label{rxn:Beuther2002bb}            \\
20126$+$4104        &   A$-$B   &    NW$-$SE   &   NW$-$SE, S$-$N   &   NW$-$SE  &   n  &  \rxn \label{rxn:Su2007}  \rxn \label{rxn:Shepherd2000} \rxn \label{rxn:Hofner2007} \rxn \label{rxn:Cesaroni1999}  \rxn \label{rxn:Cesaroni2013}  \\
20216$+$4107        &    A          &    NE$-$SW   &  NE$-$SW   &   NE$-$SW   &    y   &   \osref{rxn:Lopez_Sep2010}  \osref{rxn:Navarete2015}     \\
\enddata
\tablenotetext{\text{a}}{ALMA unpublished data (Rosero et al. in prep).}
\tablenotetext{\text{b}}{ T. Stanke and H. Beuther (private communication).}
\tablenotetext{\text{c}}{A blue-shifted component of a molecular outflow going in the direction of 18151$-$1208 B is seen in Figure 4 of \citet{2011ApJ...729...66F} but it is not discussed by the authors.}
\tablenotetext{\text{d}}{Radio source B has a negative spectral index and radio source A has an upper limit value in its spectral index. Their fluxes are not included in 
 Figures \ref{fig:rad_bol_lum} and \ref{fig:Tanaka_tracks} (right panel). 
}
\tablenotetext{\text{e}}{Radio source B has an upper limit value for the flux at 4.9 GHz and its value is not included in 
 Figures \ref{fig:rad_bol_lum} and \ref{fig:Tanaka_tracks} (right panel). 
}
\tablenotetext{\text{f}}{A blue-shifted component of a molecular outflow going in the direction of IRDC18223$-$3  is seen in Figure 5 of \citet{2011ApJ...729...66F} but it is not discussed by the authors.}

\tablenotetext{\text{g}}{\citet{2002ApJ...566..931S} report the presence of CO (2--1) wings towards this region, but contour maps of the molecular outflow are not available.}
\tablenotetext{\text{h}}{Radio sources C and D have  upper limit spectral indices that are consistent with being negative and their fluxes were not included in 
 Figures \ref{fig:rad_bol_lum} and \ref{fig:Tanaka_tracks} (right panel). 
}
\tablecomments{Generally there are multiple molecular outflows in each of these high-mass star forming region. We reference the ones that are located closest to the centimeter continuum emission. The  `y' and `n'  indicates if the centimeter radio continuum detection  is new or if it has been detected in a previous study, respectively.\\
 \osref{rxn:Wang2014}  \citet{2014MNRAS.439.3275W};  \osref{rxn:Rosero2014}  \citet{2014ApJ...796..130R}; \osref{rxn:Lee2013} \citet{2013ApJS..208...23L}; \osref{rxn:Beuther2004} \citet{2004ApJ...616L..23B}; \osref{rxn:Beuther2010} \citet{2010ApJ...724L.113B};   \osref{rxn:Zapata2006} \citet{2006AJ....131..939Z}; \osref{rxn:Fallscheer2011}  \citet{2011ApJ...729...66F}; \osref{rxn:Hofner2011}  \citet{2011ApJ...739L..17H}; \osref{rxn:Davis2004}  \citet{2004A\string&A...425..981D}; \osref{rxn:Varricat2010}  \citet{2010MNRAS.404..661V}; \osref{rxn:Beuther2006}  \citet{2006A\string&A...454..221B};  \osref{rxn:Moscadelli2013}  \citet{2013A\string&A...558A.145M};  \osref{rxn:Lee2012}  \citet{2012ApJS..200....2L}; \osref{rxn:Fallscheer2009} \citet{2009A\string&A...504..127F}; \osref{rxn:Beuther2005} \citet{2005ApJ...634L.185B}; \osref{rxn:Sanna2016} \citet{2016A\string&A...596L...2S}; \osref{rxn:Araya2008} \citet{2008ApJS..178..330A}; ; \osref{rxn:Sanna2018} \citet{2019A\string&A...623A..77S}; \osref{rxn:Navarete2015}  \citet{2015MNRAS.450.4364N}; \osref{rxn:Zhang2007}  \citet{2007A\string&A...470..269Z}; \osref{rxn:Araya2007}  \citet{2007ApJ...669.1050A}; \osref{rxn:Hofner2017} \citet{2017ApJ...843...99H}; \osref{rxn:Lopez_Sep2010}  \citet{2010A\string&A...517A..66L}; \osref{rxn:Beuther2002bb}  \citet{2002A\string&A...383..892B}; \osref{rxn:Su2007} \citet{2007ApJ...671..571S}; \osref{rxn:Shepherd2000}  \citet{2000ApJ...535..833S}; \osref{rxn:Hofner2007}  \citet{2007A\string&A...465..197H}; \osref{rxn:Cesaroni1999} \citet{1999A\string&A...345..949C};  \osref{rxn:Cesaroni2013} \citet{2013A\string&A...549A.146C}
 }
\end{deluxetable}

As listed in Table~\ref{jet_cand_list}, at least 5 of the ionized jet candidates are aligned in the same direction as a large scale molecular outflow (see Figures \ref{fig:spitzer_examples} for examples). The other 
ionized jet candidates appear to be associated with molecular outflows where the directions are approximately perpendicular. In Figure \ref{fig:UKIDSS_examples} we present the examples of 18151$-$1208 B and 19035+0641 
A where we show VLA~6$\,$cm continuum emission contours overlayed on a UKIDSS\footnote{United Kingdom Infrared Telescope (UKIRT) Infrared Deep Sky Survey (UKIDSS) Galactic Plane Survey \citep{2007MNRAS.379.1599L}.} {\emph K}-band (2.2 $\mu m$) image. It is interesting to note that the putative ionized jets and the 
UKIDSS {\emph K}-band emission in both cases are elongated in the same direction. This together with the fact that the ionized jets are located nearly at the peak of the UKIDSS {\emph K}-band emission could indicate 
that the latter is tracing scattered light from the central protostar that is escaping from an outflow cavity \citep{2013ApJS..208...23L}. The observed misalignment between cm continuum emission and  the dominating molecular flow in the region could be explained by the existence of two flows, where the molecular outflow associated
with the jet is weaker, and hence undetected. This could in fact be the case for 18151$-
$1208 B, where a blue-shifted component of a CO molecular outflow observed with the Submillimeter Array appears to be aligned in the direction of the ionized jet
(see Figure 4 of \citealt{2011ApJ...729...66F}), although this outflow component is not discussed by the authors. Another possible explanation for the misalignment in the directions of the ionized jet and the molecular outflow is that they are subjected to precession, where 
the flow axis changes from the small to the large scale as suggested by e.g., \citet{2000ApJ...535..833S} and \citet{2005A&A...434.1039C} for the case of 20126$+$4104,  \citet{2013A&A...558A.145M} to explain the 
case of  18182$-$1433 and \citet{2007ApJ...669.1050A} for 18566$+$0408. 

\begin{figure}[!h]
\centering
\begin{tabular}{cc}
\includegraphics[width=0.34\textwidth,  clip, angle = 0]{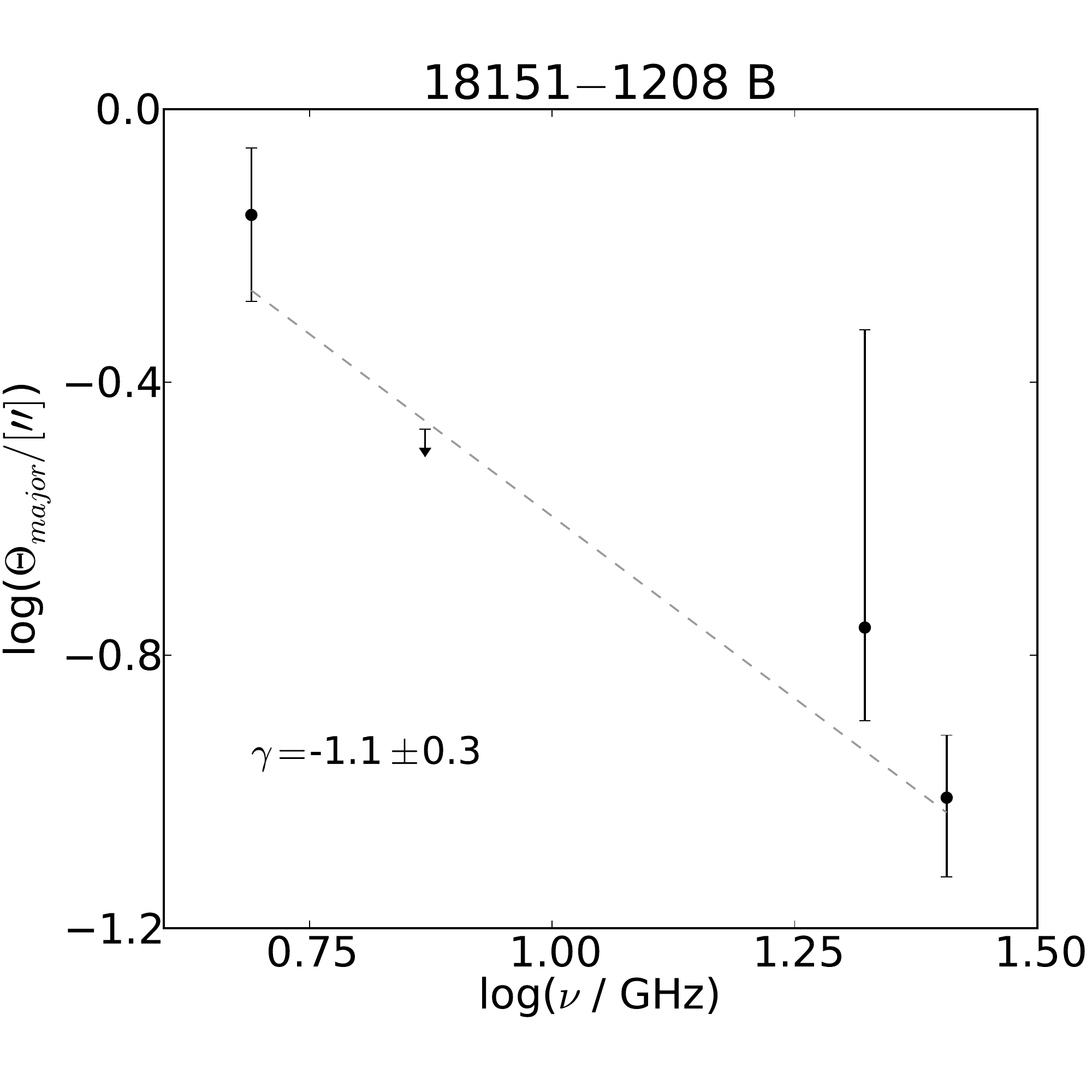}  & 
\includegraphics[width=0.34\textwidth,  clip, angle = 0]{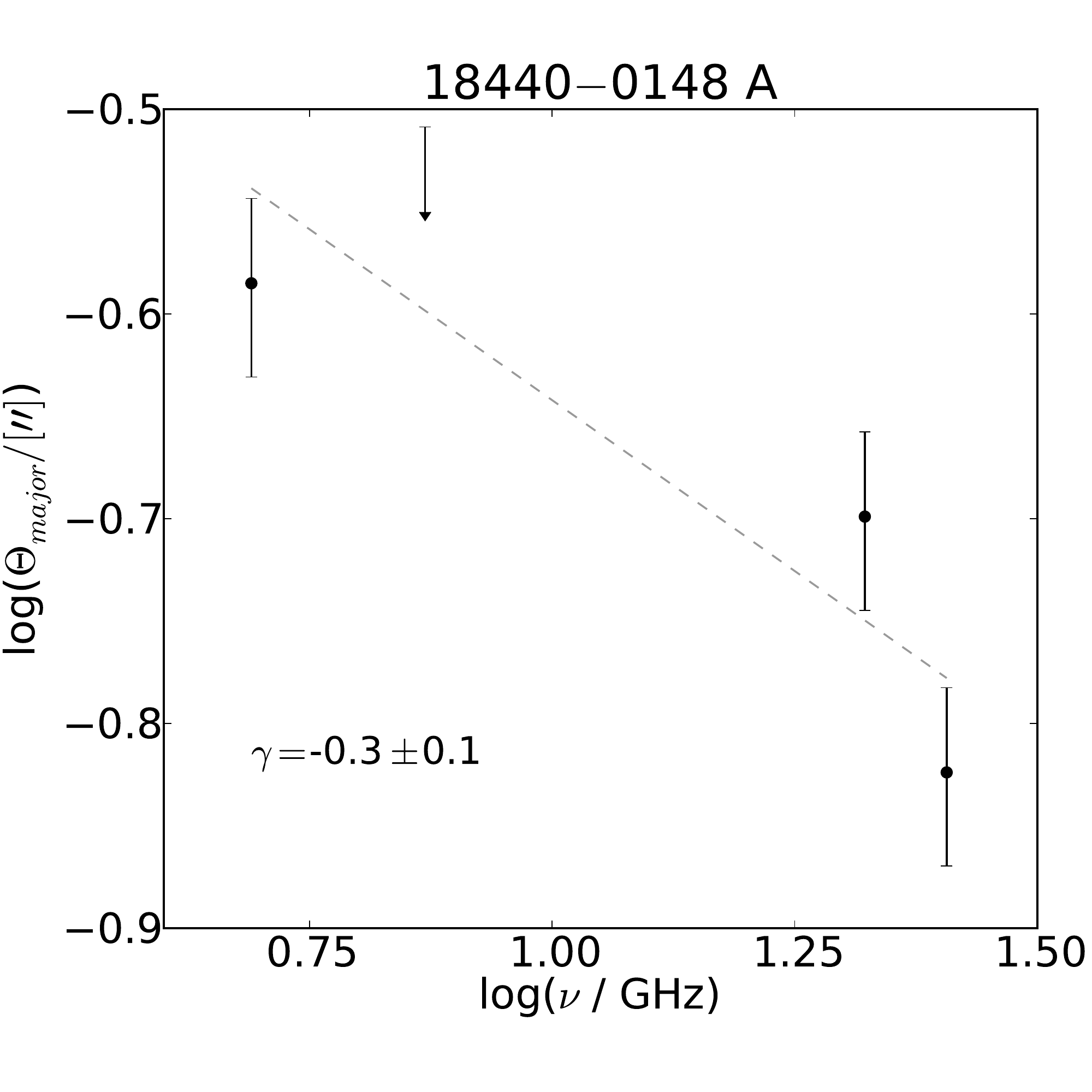}  \\
 \vspace{-1.cm} 
\end{tabular}
\caption{\small{Deconvolved major angular axis as a function of frequency for the ionized jet candidates 18151$-$1208 B, 18440$-$0148 A and 18566$+$0408 B. The arrows represent the size limit value from the synthesized beam of the map at the given frequency. The dashed line is the power law fit of the form $\theta_{maj} \propto \nu^{\gamma}$.}}
 \label{fig:size_freq}
\end{figure}

For a further test of their jet nature, we have also attempted to estimate the deconvolved sizes of the central jet components using the CASA task  {\tt imfit}. This was possible for 3 sources within the subsample of jet candidates listed in
Table~\ref{jet_cand_list}. Figure \ref{fig:size_freq} shows the deconvolved major axis as a function of frequency for 18151$-$1208 B and 18440$-$0148 A (the case of 18566$+$0408 B is reported in \citealt{2017ApJ...843...99H}). Within the uncertainties these radio sources follow the relation $\theta_{maj} \propto \nu^{\gamma}$, where a major axis index of $\gamma = -0.7$ is expected for a biconical ionized wind or jet \citep{1986ApJ...304..713R}. Therefore, at least in these 3 cases, we have further evidence for the jet nature of these specific radio sources. 

In summary, for the subsample of elongated continuum sources listed in Table~\ref{jet_cand_list} it is very likely that the nature of these sources are ionized jets at the base of a molecular outflow.



\defcitealias{2011MNRAS.415..893A}{AMI Consortium: Scaife et al. (2011}
\defcitealias{2012MNRAS.420.1019A}{2012)}

As mentioned above, most of our detected radio sources with a rising spectrum are compact, i.e., spatially unresolved, or marginally resolved. Several of these sources are associated with molecular 
outflows, and 6.7 GHz CH$_{3}$OH  and 22 GHz H$_{2}$O masers as found in the literature. More precisely, only   6 of the 25 regions where we detected a radio source with a rising spectrum, are not 
associated with molecular outflows, or adequate data that would trace such outflows do not seem to exist.
Furthermore, after taking into account the shape of the synthesized beam of our VLA observations  some of 
these radio sources appear slightly elongated in a certain direction. Examples are 18264$-$1152 F and G53.25$+$00.04mm2 A (see \citetalias{2016ApJS..227...25R}, Figure~2). Therefore, we now investigate the 
possibility that  the compact radio continuum sources with a rising spectrum represent ionized jets. 

\begin{figure}[h]
\centering
\begin{tabular}{ccc}
\hspace*{\fill}%
\includegraphics[width=0.5\textwidth, clip=true, angle = 0]{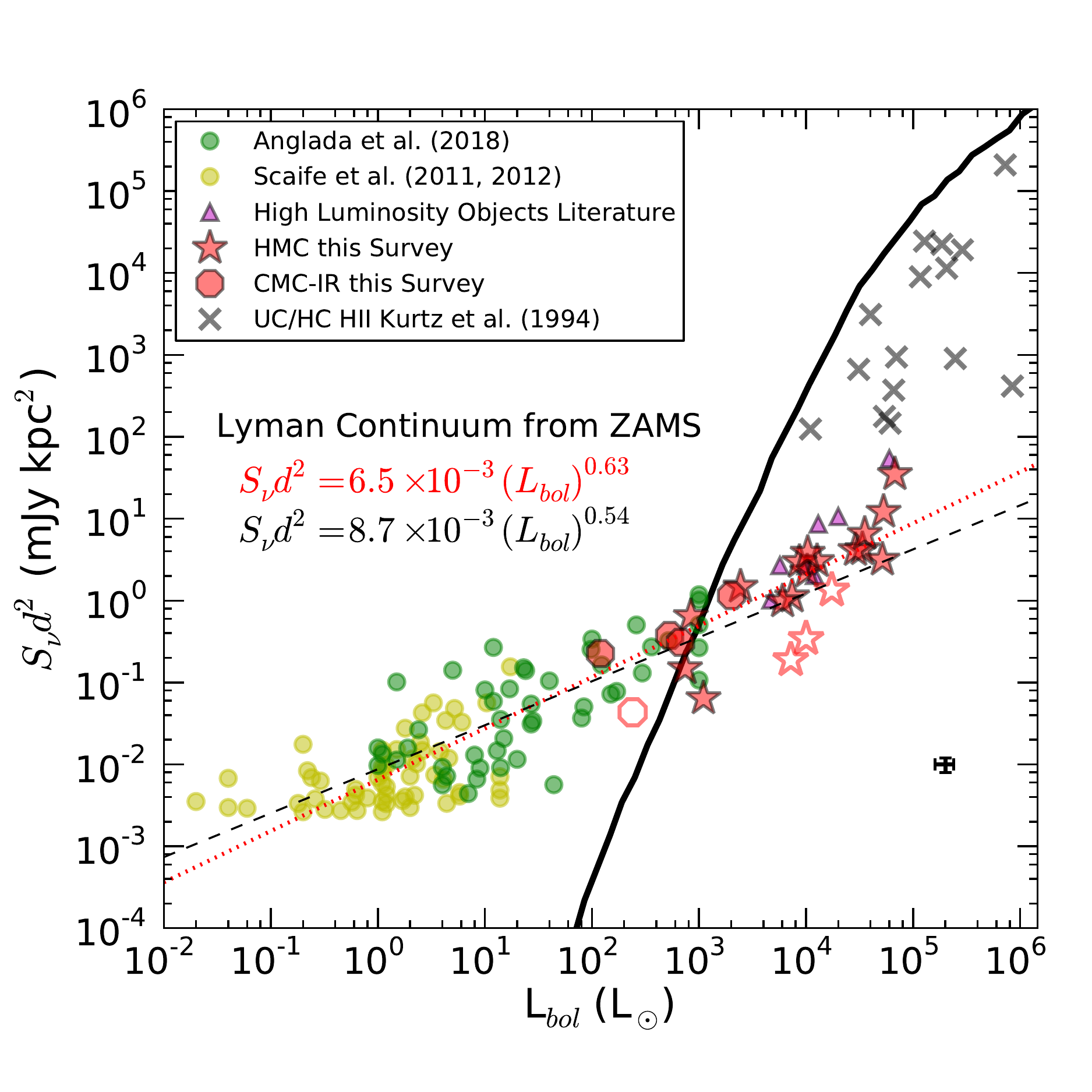}
\end{tabular}
\caption{\small{Radio luminosity at 4.9 GHz  as a function of the bolometric luminosity. The red stars and octagons are our ionized jet and jet candidates listed in Table \ref{jet_cand_list} and \ref{candidates}  towards HMCs and CMC-IRs, respectively.  The bolometric luminosity for the red symbols is mainly estimated from our {\it Herschel}/Hi--GAL data (except for the open symbols whose bolometric luminosity information is from the literature).  The green circles represent ionized jets associated with low-mass protostars  ($1\,L_{\odot} \leq L_{bol} \leq 1000\,L_{\odot}$)   from \citet{2018A&ARv..26....3A} and the yellow circles are the very low luminosity objects (VeLLOs) and  low-mass protostars from \citetalias{2011MNRAS.415..893A, 2012MNRAS.420.1019A}. 
The purple triangles represent ionized jets from high-mass stars as found in the literature, 
from \citet{2008AJ....135.2370R} and \citet{2016A&A...585A..71M}. The  $\times$ symbols are UC and HC H{\small II} regions from \citet{1994ApJS...91..659K}. 
The dashed line relation shows the positive correlation found by \citet{2015aska.confE.121A} derived for jets from low-mass stars. The red dotted line is our best fit to the data including ionized jets from low, intermediate and high-mass YSOs, but excluding the sources from \citetalias{2011MNRAS.415..893A, 2012MNRAS.420.1019A}.
The error bars in the bottom right corner correspond to a 20$\%$ calibration uncertainty.
}}
 \label{fig:rad_bol_lum}
\end{figure}

A  statistical way of investigating the nature of our compact sources is to study the energy contained in the ionized gas. In Figure \ref{fig:rad_bol_lum} we show the 
radio luminosity S$_\nu\,$d$^2$ of all the components of the ionized jet (or the jet candidate) as a function of the bolometric luminosity of the region. As in Figure \ref{Lyman_cont_plot} above, the black line is the radio luminosity expected from the Lyman continuum
flux at a given bolometric luminosity if it arises from photoionization of a single ZAMS star. In addition to the compact, rising spectra sources from our survey we also show in Figure \ref{fig:rad_bol_lum} as green circles
the radio luminosity from low-mass stars ($1\,L_{\odot} \leq L_{bol} \leq 1000\,L_{\odot}$) associated with ionized jets from \citet{2018A&ARv..26....3A} and as yellow circles the radio luminosity of very low luminosity objects (VeLLOs) and  low-mass protostars detected at 1.8 cm, and reported in \citetalias{2011MNRAS.415..893A, 2012MNRAS.420.1019A}. In order to compare the sources from \citetalias{2011MNRAS.415..893A, 2012MNRAS.420.1019A} and \citet{2018A&ARv..26....3A} with our $4.9\,$GHz data, we scaled their
fluxes using a factor of 0.48 and 0.76, respectively, assuming that those sources have a spectral index $\alpha=0.6$, which is the canonical value of ionized jets. 
 The scaling factors are calculated using 
$\frac{S_{\lambda_{1}}}{S_{\lambda_{2}}}= \left(\frac{\lambda_{2}}{\lambda_{1}} \right)^{\alpha}$. 

It is well known that for low mass YSOs the radio luminosities are correlated with the bolometric luminosity, and we show the correlation
$\frac{S_{\nu}d^{2}}{\text{mJy kpc}^{2}}= 8.7 \times 10^{-3} \left( \frac{L_{\text{bol}}}{L_{\odot}} \right)^{0.54}$ first found by \citet{1995RMxAC...1...67A} and recently updated by \citet{2018A&ARv..26....3A}. The black dashed line is the best fit to the green circles in Figure \ref{fig:rad_bol_lum}, which are  the low-mass ionized jets presented by \citet{2018A&ARv..26....3A}.
It is clear from Figure \ref{fig:rad_bol_lum} that the sources from \citetalias{2011MNRAS.415..893A, 2012MNRAS.420.1019A} also follow this relation, although their data were observed at low resolution ($\sim$ 30\rlap.$^{\prime \prime}$) and there is not enough information that proves that they correspond to ionized jets. 
\citet{1995RMxAC...1...67A} used this observed correlation to explain the apparent excess ionization levels from low mass YSOs by shock induced ionization from jets,
as modeled by \citet*{1987RMxAA..14..595C} and \citet{1989ApL&C..27..299C}. 
A handful of detections of ionized jets towards high-mass stars  in recent years suggested that this correlation appears to also hold for stars with luminosities up to $\sim$ 10$^{5}$ L$_{\odot}$ (see 
\citealt{2016MNRAS.460.1039P}). We have added these objects from \citet{2008AJ....135.2370R} and  \citet{2016A&A...585A..71M} as purple triangles in Figure \ref{fig:rad_bol_lum} and the data has been properly scaled to our frequency of 4.9~GHz assuming $\alpha=0.6$. The data from our survey (\citetalias{2016ApJS..227...25R}) in conjunction with improved estimates of the 
luminosities based on Herschel data (see \S \ref{app:lum}) allow us to further populate this plot and test if a correlation exists. In Figure \ref{fig:rad_bol_lum} the red stars and octagons are our radio sources with rising spectrum detected
towards HMCs and CMC-IRs, respectively, and we see that most sources are located very close to the relation found by \citet{1995RMxAC...1...67A} up to luminosities of $\sim$ 10$^{5}$ L$_{\odot}$. In fact, a fit of the data 
including low, intermediate and high-mass YSOs is shown as a red dotted line and the result is similar to what was found by  \citet{1995RMxAC...1...67A}. We excluded the sources from \citetalias{2011MNRAS.415..893A, 2012MNRAS.420.1019A} from our fit since it is unclear if those source are indeed  ionized jets.
 We take this result as a strong indication that the weak, and compact radio sources which we found in our survey are caused by the same mechanism which causes the radio emission the low mass YSOs, namely it is caused by ionized jets.  We also note that of the 6 compact radio sources where currently no observational association with molecular flows is known,
 5 match our fit (red dotted line in Fig 8) of the $S_\nu\,d^2$ vs $L_{bol}$ relationship.
 
 
\defcitealias{2011MNRAS.415..893A}{AMI Consortium: Scaife et al. (2011}
\defcitealias{2012MNRAS.420.1019A}{2012)} 
 
\begin{figure}[h]
\centering
\includegraphics[width=0.5\textwidth, clip=true, angle = 0]{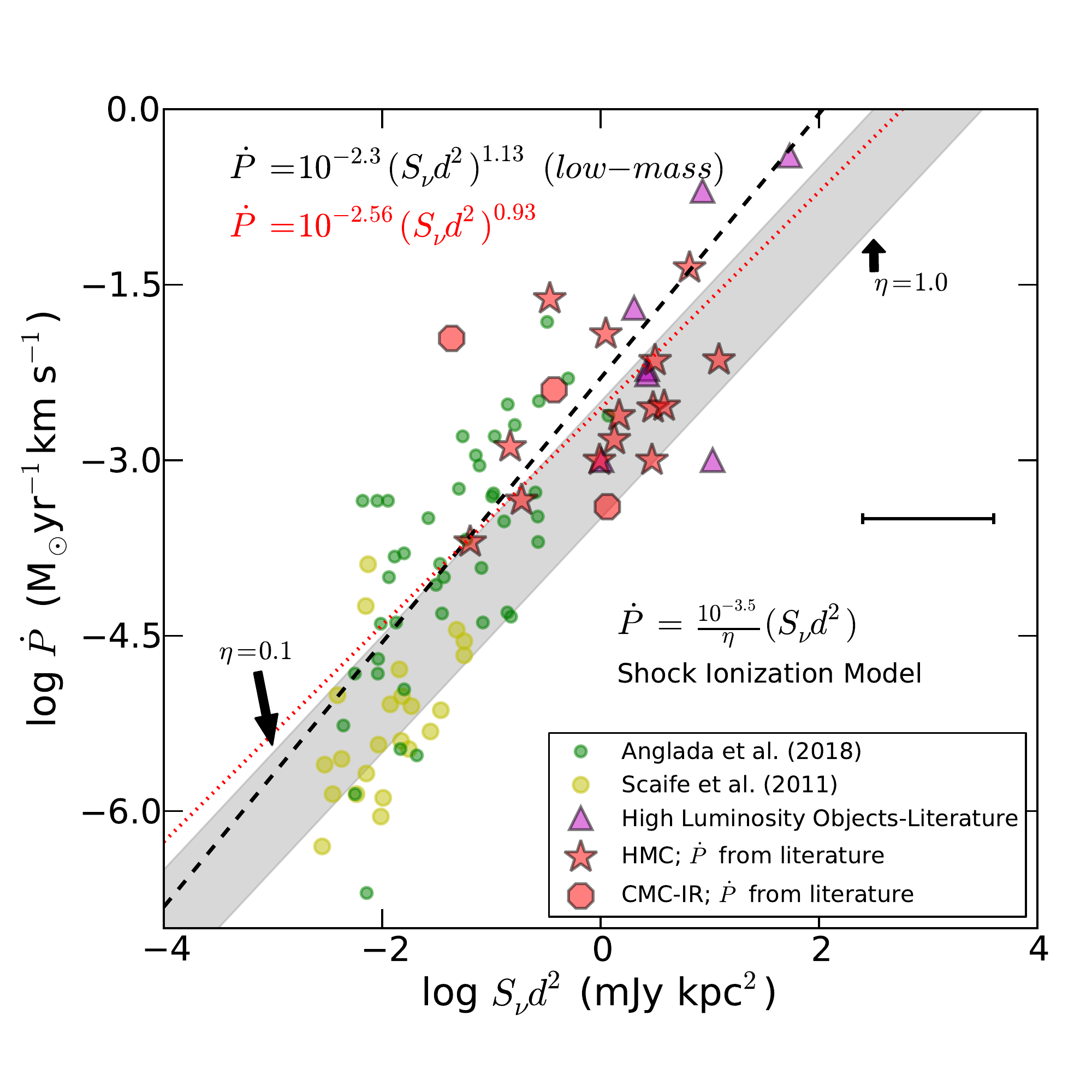}
\caption{\small{Momentum rate of the molecular outflow  as a function of the radio luminosity at 4.9 GHz. The red stars and octagons are our ionized jets and jet candidates listed in Table \ref{jet_cand_list} and \ref{candidates}  towards HMCs and CMC-IRs, respectively, symbols for which there is information of the momentum rate.  The momentum rate values of the molecular outflow for all the sources including our data are collected from the literature.  The green circles represent ionized jets associated with low-mass protostars  ($1\,L_{\odot} \leq L_{bol} \leq 1000\,L_{\odot}$)   from \citet{2018A&ARv..26....3A}  and the yellow circles are the very low luminosity objects (VeLLOs) and  low-mass protostars from \citetalias{2011MNRAS.415..893A, 2012MNRAS.420.1019A}. 
The purple triangles represent ionized jets from high-mass stars as found in the literature, 
from \citet{2008AJ....135.2370R} and \citet{2016A&A...585A..71M}. The  $\times$ symbols are UC and HC H{\small II} regions from \citet{1994ApJS...91..659K}. 
The dashed line relation shows the positive correlation found by \citet{1995RMxAC...1...67A} derived for jets from low-mass stars. The red dotted line is our best fit to the data including ionized jets from low, intermediate and high-mass YSOs, but excluding the sources from \citetalias{2011MNRAS.415..893A, 2012MNRAS.420.1019A}. The gray shaded area corresponds to the momentum rate as predicted by the shock ionization model from \citet*{1987RMxAA..14..595C} for values of the shock efficiency fraction of $\eta = 0.1$ and $\eta = 1.0$. 
 The error bar in the middle right  corresponds to a 20$\%$ calibration uncertainty. The error in $\dot{P}$  is not represented in the figure because it 
varies widely, and depends strongly on how different authors have gathered the data.
}}
 \label{fig:Curiel_plot}
\end{figure}

Furthermore, in Figure \ref{fig:Curiel_plot} we show the momentum rate ($\dot{P}$) of the molecular outflow as a function of the radio luminosity (S$_\nu\,$d$^2$) of the ionized jet estimated from our flux values at 4.9 GHz (symbols and colors are the same as  in Figure  \ref{fig:rad_bol_lum}). The momentum rate of the molecular outflows comes from information from the literature for our ionized jets (and jet candidates), if available, and the values are in most cases from single dish observations.  For consistency, we have scaled the physical values, so that they are based on the same distance. However,
many uncertainties remain due to the inhomogeneity of the data set. In particular, the values for the momentum rate come from observations taken
by different authors, using different spectral lines, as well as different telescopes. Hence, the large scatter in Figure \ref{fig:Curiel_plot} is not unexpected,
and the creation of a homogenous data set for the $\dot{P}$ versus $S_\nu\,d^2$ relation will be an important future task.

In spite of the large scatter, the correlation seen in  Figure \ref{fig:Curiel_plot} indicates that the more radio luminous the protostar are the more powerful they are in pushing outflowing material. This correlation, which has been studied by several authors (e.g., \citealt{1995RMxAC...1...67A}, \citealt{2008AJ....135.2370R}, \citetalias{2011MNRAS.415..893A, 2012MNRAS.420.1019A}, \citealt{2018A&ARv..26....3A}), follows the shocked-induced ionization model introduced by \citet{1987RMxAA..14..595C, 1989ApL&C..27..299C}, suggesting that the ionization of thermal jets is due to shocks. The  shocked-induced ionization model implies $\left(\frac{S_{\nu}d^{2}}{mJy\,kpc^{2}}\right) = 10^{3.5} \eta \left(\frac{\dot{P}}{M_{\odot}\,yr^{-1}\,km\,s^{-1}}\right)$ at $\nu=5$ GHz where $\eta$ is the shock efficiency fraction or the fraction of material that gets ionized by the shocks, which for low-mass protostars has been observationally found to be around 10$\%$ (or $\eta=0.1$).  \citet{2018A&ARv..26....3A} suggested that the ionization fraction of jets in general is low ($\sim 1 - 10\%$). With the current data, and due to the large scatter seen in the correlation of  $\dot{P}$ vs S$_\nu\,$d$^2$, we cannot yet properly quantify how the efficiency fraction changes with the luminosity of the protostar (e.g., if the ionization in thermal jets associated with high-mass protostars is higher than for low-mass protostars). Therefore, a uniform survey to measure the momentum rate of the molecular outflows associated with  ionized jets (ideally with  comparable resolutions) will be fundamental to further constrain this model. \citet{2018A&ARv..26....3A} discussed both  correlations shown in Figure~\ref{fig:rad_bol_lum} and Figure~\ref{fig:Curiel_plot}  in great detail  and they interpreted them as an indication that the mechanism of ionization, accretion and ejection of outflows associated with protostars do not depend on their luminosities.

\section{Discussion}\label{discussion_paperII}
Results reported in \citetalias{2016ApJS..227...25R} of detection rates  of CMC (6$\%$), CMC-IR (53$\%$) and HMCs (100$\%$) provide further evidence for an evolutionary sequence in the formation of high-mass stars, from a very early stage type of cores (i.e., CMCs) to relatively more evolved ones (i.e., HMCs). The fraction of centimeter wavelength sources detected towards HMCs is  higher  than previously expected towards this type of cores and suggests that radio continuum may be detectable at weak levels in all HMCs. The lack of radio detections for some objects in the sample (including most CMCs) provides interesting constraints and are ideal follow up candidates for studies of the earliest stages of high-mass stars.
It is important to note that it is likely that the ionized material from  jets or  HC H{\small II} regions associated with these type of cores remains undetected at our sensitivity, thus in order to rule out these regions as pre-stellar cores deeper observations are required or alternative tracers for ongoing star formation in these cores need to be identified.

Here we consider some constraints on the nature of the  centimeter continuum emission detected in these cores and clumps towards high-mass star forming regions. As described in \S \ref{yso}, most of our radio detections arise from high-mass YSOs and at least for 7 regions some of the radio detections could potentially arise from (solar-like mass) T-Tauri stars. Also, we detected at least 10 radio sources associated with the mm cores/clumps with a flat spectral index, most of them resolved sources, which are most likely UC H{\small II} regions.  Understanding the nature of the rising spectral index sources has proven to be more challenging. These compact radio sources appear to be well fitted (within the uncertainties) when using either a homogeneous H{\small II} region or a power-law fit  (as shown in Figure \ref{HII_fit_examples}).  
Therefore, we will discuss two plausible scenarios, that the radio sources are either  UC/HC H{\small II} regions or that the emission arises from shock ionized jets.

\subsection{H{\small II} regions}

For the first scenario, when fitting the sources in terms of a homogeneous H{\small II} region, the solutions required a significantly smaller size (several of them an order of magnitude smaller) for the H{\small II} region than the upper limit given by the FWHM synthesized beam. However, since these calculations assumed a pure hydrogen nebula, we must consider whether internal dust absorption can make the regions as small as the H{\small II} region model is predicting.   \citet{1989ApJS...69..831W} suggested that, even if the dust absorbs 90$\%$ of the UV photons, the radius of the H{\small II} region is reduced by only a factor of 0.46. Thus, dust absorption alone appears insufficient to explain the small region sizes predicted by the H{\small II} region model used to fit our data. As shown in Figure \ref{Stromgren_sphere}, these sources could be explained as turbulence-pressured confined  H{\small II} regions if they are born in a clump with density of n$_{H_{2}}=$ 10$^{7}$\,cm$^{-3}$ and assuming  velocity dispersions of $\sigma \sim$ 3--8 km~s$^{-1}$ (FWHM$\sim$7--20 km~s$^{-1}$). However, it is not clear if high velocity dispersions are common towards  the dense clumps harboring high-mass stars, since   $\sigma \sim$ 2 km s$^{-1}$ seems to be more typical. Measuring the line width of an optically thin tracer on $\sim$100 au scale would provide a decisive constraint on the velocity dispersion. We also found that the sources could be consistent with having been born in a denser environment of $n_{H_{2}}\approx$ 10$^{8}$\,cm$^{-3}$. Arguably, \citetalias{2016ApJ...818...52T}  predicts  that such a density for the ionized region is already too high for a protostar of  8 M$_{\odot}$ to 24 M$_{\odot}$.  Recently,  \citet{2016A&A...585A..71M} detected  compact radio sources towards high-mass YSOs with similar physical characteristics to the ones found in this survey, where the Lyman continuum derived from the bolometric luminosities always exceeds the one obtained from the radio luminosities (as seen in Figure \ref{fig:rad_bol_lum}). From their analysis they conclude that those sources cannot be HC or UC H{\small II} regions, unless the ionized gas has a density gradient (e.g., model IV of \citealt{1975A&A....39..217O}).

Additionally, for the extended UC H{\small II} regions detected at the outskirts of the mm cores there are two scenarios: either they were born in a low density clump of $n_{H_{2}}\approx$ 10$^{5}$\,cm$^{-3}$ or  they were born at a higher density and have migrated out of the center potential.  The latter scenario requires large stellar dispersion velocities ($\gtrsim$10 km~s$^{-1}$), which are not typical unless the source is a runaway OB star that has been dynamically ejected.   Observed stellar dispersion velocities, for instance for Orion's brightest  population is  only $\lesssim$3 km~s$^{-1}$  \citep[e.g.,][]{2005AJ....129..363S,2009ApJ...697.1103T}, which makes the former scenario more plausible. However, \citet{2007ApJ...660.1296F} predicted that stellar velocities up to $\lesssim$13 km~s$^{-1}$ are likely for core densities of 10$^{7}$\,cm$^{-3}$, and  that these high stellar velocities carry the star to lower density regions of the core/clump, where the H{\small II} region is free to expand. We are leaning to favor the scenario  where the sources have migrated since it allows 
to explain the occurrence of both compact and extended emission in the same protocluster (e.g., \citealt{2007prpl.conf..181H}). 
\newpage         
\subsection{Radio Jets}
         
Now we discuss the second scenario where the radio emission of the radio compact sources with rising spectrum is due to shock ionization. 
The observable properties of several of our radio detections indicate that they likely have a jet nature: one can argue that the low centimeter emission from the majority of the sources detected in this survey, their free-free
spectral index being in the range $0.2 \leq \alpha \leq 1.8$  and their association with molecular outflows indicate that even those sources without an elongated radio morphology are also ionized jets or stellar winds that are conical, accelerating and/or recombining.  From the analysis in \S\ref{compact_sources_sect} and \S \ref{ionized_jet_sect} we inferred that from the 44 sources with rising spectral index, approximately 12 of them are ionized jets (see Table \ref{jet_cand_list}) and 13 are jet/wind candidates (see Table \ref{candidates}). 
In fact, half of the jet candidates in Table \ref{candidates} have a spectral index 
$\alpha \approx 0.6$ and all but two of them (UYSO1 A and 18521$+$0134 A) have a spectral index $\alpha \leq 1.0$, which is consistent with the expected value of a spherical, isothermal and constant velocity ionized wind \citep[e.g.,][]{1975A&A....39....1P}. As stated before, the deviation from the value $\alpha = 0.6$ could be due to acceleration or recombination within the flow.

\begin{deluxetable}{l c c c c}
\tabletypesize{\scriptsize}
 \renewcommand*{\arraystretch}{1.5}
\tablecaption{Ionized Jet/Wind Candidates  \label{candidates}}
\tablewidth{0pt}
\tablehead{
\colhead{Region}                  & 
\colhead{Radio Source}        &
\colhead{Outflow Direction}   &
\colhead{ H$_{2}-$Jet Direction}   &
 \colhead{Reference}      \\}
\startdata
\setcounter{iso}{0}	
UYSO1                    &    A   &   NW$-$SE    &   \nodata  & \rxn \label{rxn:Forbrich04}   \\
18264$-$1152         &    F  &   NW$-$SE     &  E$-$W  & \rxn \label{rxn:Sanchez-Monge2013} \rxn \label{rxn:Navarete2015}    \\
18345$-$0641        &    A   &    NW$-$SE    &   very weak &\rxn \label{rxn:Beuther2002} \rxn \label{rxn:Varricat2013}  \rxn \label{rxn:Varricat2010}  \\
18470$-$0044        &    B   &   E$-$W          &   no/very weak\tablenotemark{a}      &  \osref{rxn:Beuther2002}   \\
18517$+$0437       &    A    &   N$-$S          &   very weak   & \rxn \label{rxn:Lopez_Sep10} \osref{rxn:Varricat2010} \\
18521$+$0134       &    A   &    \nodata\tablenotemark{b}     &     non-detection       &   \rxn \label{rxn:Cooper13}   \\
 G35.39$-$00.33mm2        &    A   &  \nodata            &  \nodata   &    \nodata   \\
18553$+$0414       &    A   &    \nodata\tablenotemark{c}     &   non-detection         &    \osref{rxn:Navarete2015}   \\
19012$+$0536       &    A   &    NE$-$SW      &   non-detection   &   \osref{rxn:Beuther2002} \osref{rxn:Navarete2015}   \\
G53.25$+$00.04mm2        &    A   & \nodata    &     \nodata       &    \nodata   \\
 19413$+$2332      &    A   &      \nodata\tablenotemark{d}     &     \nodata      &   \osref{rxn:Beuther2002}  \\
20293$+$3952      &    E\tablenotemark{e}    &     NE$-$SW    &     detection    &  \rxn \label{rxn:Beuther04} \rxn \label{rxn:Palau07_a} \osref{rxn:Varricat2010} \\        
 20343$+$4129      &    B   &     E$-$W      &   non-detection     &  \rxn \label{rxn:Palau07}  \osref{rxn:Cooper13}\\        
\enddata
\tablenotetext{\text{a}}{ T. Stanke and H. Beuther (private communication).}
\tablenotetext{\text{b}}{\citet{2002ApJ...566..931S} reports non-detection of CO (2--1) wings towards this region, although an outflow  could be present at an inclination angle of $< 10^{\circ}$ to the plane of the sky.}
\tablenotetext{\text{c}}{\citet{2002ApJ...566..931S} reports the presence of CO (2--1) wings towards this region, but contour maps of the molecular outflow are not available.}
\tablenotetext{\text{d}}{CO outflow is detected in the region, but the data does not show a clear bipolar structure.}

\tablenotetext{\text{e}}{Radio source E has an upper limit value for the flux at 4.9 GHz and its value is not included in Figures \ref{fig:rad_bol_lum} and \ref{fig:Tanaka_tracks} (right panel).}
\tablecomments{The dots indicates that there is not enough information available about observations of the molecular outflow in the literature.\\
\osref{rxn:Forbrich04} \citet{2004ApJ...602..843F}; \osref{rxn:Sanchez-Monge2013}  \citet{2013A\string&A...557A..94S};  \osref{rxn:Navarete2015}  \citet{2015MNRAS.450.4364N}; \osref{rxn:Beuther2002}  \citet{2002A\string&A...383..892B};  \osref{rxn:Varricat2013}  \citet{2013A\string&A...554A...9V}; \osref{rxn:Varricat2010}  \citet{2010MNRAS.404..661V}; \osref{rxn:Lopez_Sep10} \citet{2010A\string&A...517A..66L}; \osref{rxn:Cooper13} \citet{2013MNRAS.430.1125C}, \osref{rxn:Beuther04} \citet{2004ApJ...608..330B}; \osref{rxn:Palau07_a} \citet{2007A\string&A...465..219P}; \osref{rxn:Palau07} \citet{2007A\string&A...474..911P}.}

\end{deluxetable}

\subsection{H{\small II} regions vs Radio Jets}

In Figure \ref{fig:rad_bol_lum} we compared the radio luminosity with the bolometric luminosity using the radio flux at 4.9 GHz.  When fitting the ionized jets (and jet candidates) from low, intermediate and high-mass protostars using a power-law (represented by the dotted red line) we find an index of 0.63$\pm$0.04 with a correlation coefficient  of $r=$0.89 which yields the relation $S_{\nu}d^{2}$ [mJy kpc$^{-2}$]=$ 6.5 \times 10^{-3}$ (L$_{bol}$/L$_{\odot}$)$^{0.63}$.  This result is comparable with the index found by \citet{2016MNRAS.460.1039P} of 0.64$\pm$0.04 for jets spanning luminosities from $\sim 10^{-1}$ to $10^{5}$ L$_{\odot}$, although their fit has a lower correlation coefficient ($r=$0.73). Their estimates for bolometric luminosities, which include {\it Herschel}/Hi--GAL data for most of their sources, are similar to ours. Therefore, the scatter in their data may come from the radio fluxes. \citet{2016MNRAS.460.1039P} have stated that some of their jets have high flux densities probably because those objects represent a transition  between a jet and H{\small II} region stages. 
 Further, it is important to note that a similar relation between the bolometric luminosity and the luminosity of shocked H$_2$ emission from molecular jets
has been reported by \citet{2015A&A...573A..82C} for sources with a wide range of bolometric luminosities. These studies together with our
refined relation point to a common flow mechanism from YSOs of any luminosity.

\begin{figure}[htbp]
\centering
\begin{tabular}{cc}
\vspace{-0.7cm}
\includegraphics[width=0.46\textwidth, clip=true, angle = 0]{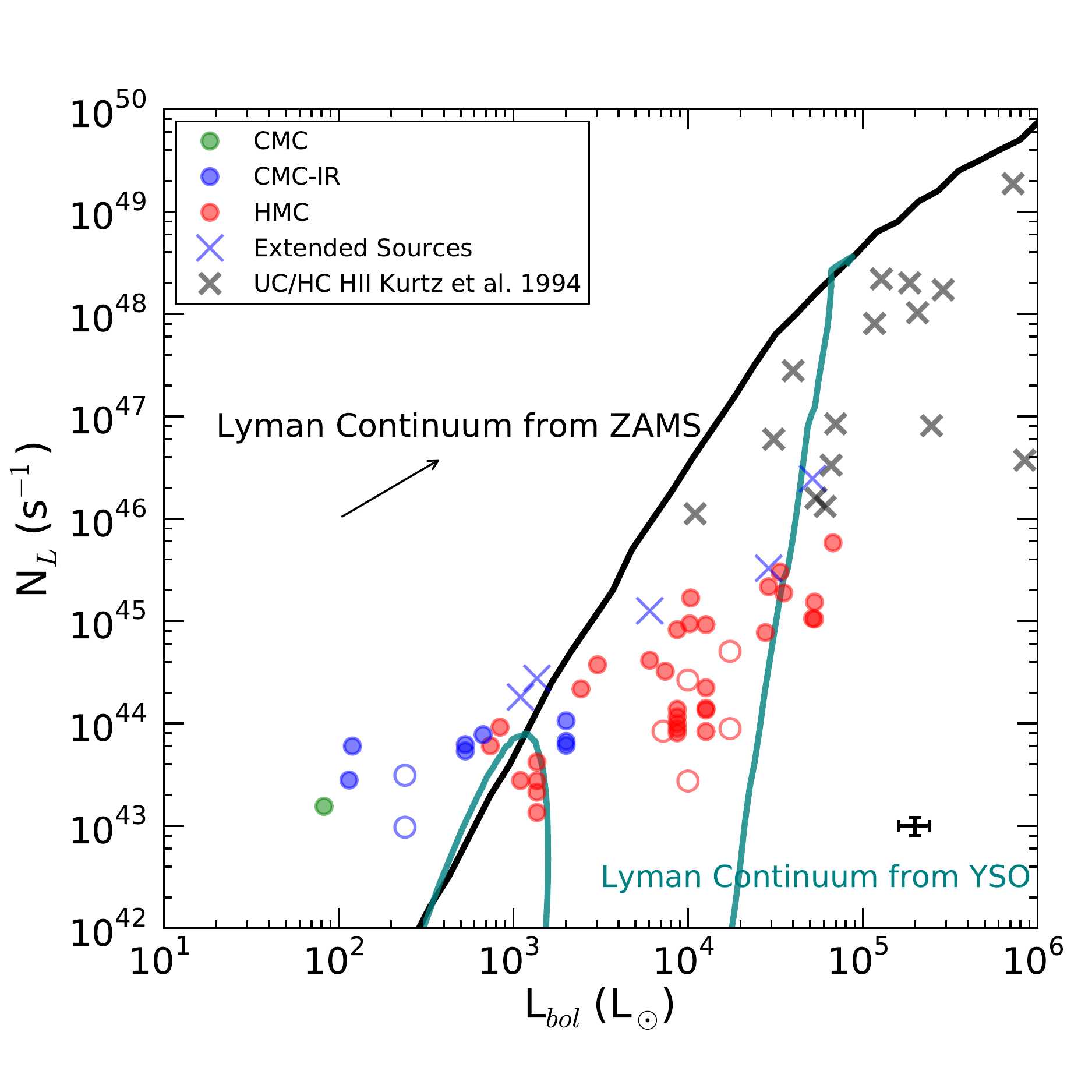} &\hspace{-1.6em} 
\includegraphics[width=0.46\textwidth,  clip=true, angle = 0]{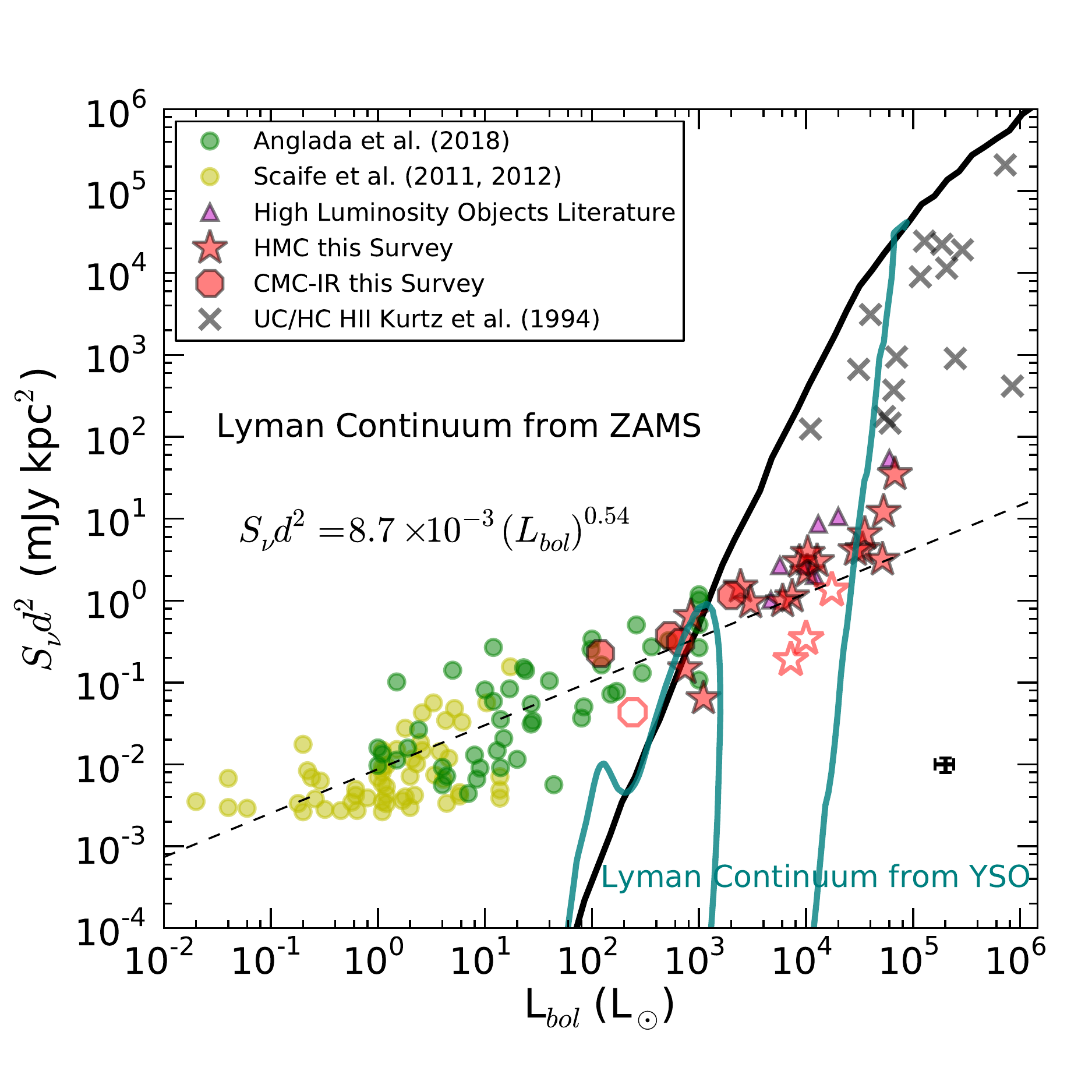}   
\end{tabular}

\caption{\small{Lyman continuum (left) and radio luminosity (right) as a function of the bolometric luminosity. Symbols and colors are the same as used in Figures \ref{Lyman_cont_plot} and \ref{fig:rad_bol_lum}, except that now we also show the estimated Lyman continuum from the \citetalias{2016ApJ...818...52T} model for an optically thin H{\small II} region based on the ionization of a protostar (cyan continuous line). The stellar model evolution  starts with a  core mass of M$_{c}=$ 60 M$_{\odot}$ and a mass surface density of ambient clump of $\Sigma_{cl}=$~1~g~cm$^{-2}$. The black continuous line is the Lyman continuum from a ZAMS star. 
 The error bars in the bottom right corner correspond to a 20$\%$ calibration uncertainty. 
}}
 \label{fig:Tanaka_tracks}
\end{figure}

Until very recently, the stellar evolutionary models that have been used to analyze this type of sources  correspond to more evolved objects (i.e., ZAMS star). However, the recently introduced \citetalias{2016ApJ...818...52T} model predicts the ionizing luminosity of a protostar which will allow us to compare our data with a more appropriate part of the evolutionary track.  These evolutionary stellar models mainly depend on the accretion history, this is the mass of the core (M$_{c}$) and the mass surface density of the ambient clump ($\Sigma_{cl}$). Figure \ref{fig:Tanaka_tracks} shows the same relations as those in Figs. \ref{Lyman_cont_plot} and \ref{fig:rad_bol_lum}, but now we also consider an evolutionary track for a YSO which is represented by the cyan continuous line for an initial core mass of M$_{c}=$ 60 M$_{\odot}$ and a mass surface density of ambient clump of $\Sigma_{cl}=$~1~g~cm$^{-2}$ (\citetalias{2016ApJ...818...52T} fiducial case). This cyan track shows the evolutionary sequence of the ionizing photon luminosity as a function of the protostellar luminosity. Its shape shows each of the physical stages in the evolution of the protostar: accretion stage, swelling stage (as seen with the decrease in the ionizing luminosity as the temperature decreases), contraction stage (increase of the ionizing luminosity as the temperature also increases) and nuclear burning stage when the protostar reaches the ZAMS (represented by the black continuous line; for more discussion on this evolutionary track see \citealt{2014ApJ...788..166Z}; \citetalias{2016ApJ...818...52T}).  The left panel of Figure \ref{fig:Tanaka_tracks} shows that
 for the majority of the radio sources detected  towards CMCs and CMC--IRs, the  Lyman continuum excess (for the fiducial case L$_{bol} \sim$10$^{2}$--10$^{3}$) is still evident and it is not likely due to photoionization. Additionally, the evolutionary  track for a YSO shows how the ionizing luminosity decreases as the protostar swells while accreting its mass and before it enters the Kelvin-Helmholz contraction (for the fiducial case L$_{bol} \sim$10$^{3}$--10$^{4}$). This further indicates that the measured radio flux for most of our radio sources detected towards HMCs are also very unlikely to be photoionized by the central object.
 
Model calculations presented by \citet{2002ApJ...568..754K,2003ApJ...599.1196K,2007ApJ...666..976K}  predict that high accretion rates on the order of  $10^{-4} - 10^{-3}$ M$_{\odot}$yr$^{-1}$ can choke off the H{\small II} region to very small sizes producing very low radio continuum; see also Section 5 of \citet{1995RMxAC...1..137W}. This might be a possible scenario for some of our more compact sources, but additional evidence is necessary such as high-resolution mm observations of infall tracers to determine mass infall rates for these sources. Based on the analysis and discussion presented above, we are inclined to favor the scenario that most of our compact sources (see Table \ref{candidates}) with rising spectrum are ionized jets.  However, the confirmation of these radio sources as shocked ionized gas requires further observational and theoretical work. Additional  observations and tests are necessary in  order to have conclusive information of the nature of these detections. Higher resolution data ($\lesssim$ 0\rlap.$^{\prime \prime}$1) of the radio continuum is required to resolve the ionized jets and estimate their degree of  collimation. Additionally, high resolution millimeter data will help us to disentangle multiple outflows, possibly being driven by protostellar clusters as expected toward high-mass star forming regions and to study the  kinematics of the outflow material. Masers arise from the hot core regions and their  association with ionized material is very important. They indicate the evolutionary stage \citep[e.g.,][]{2018A&A...619A.107S} of the exciting object and allow detailed studies of the kinematics at a smaller scale, very close to the powering high-mass YSO and the disk/jet interface.   For resolved sources, long-term monitoring of the ionized jet is necessary in order to estimate proper motions,
velocities of the radio jets and evolution in the morphology of the jet.

\section{Summary and Conclusions}\label{conclusions_paperII}
In this work we investigate the nature of the 70  radio sources
reported in \citetalias{2016ApJS..227...25R}.  These radio sources were observed using the VLA at 6 and 1.3 cm towards a sample of high-mass star forming region candidates having either no previous radio continuum detection or a relatively weak detection at the 1 mJy level. We have explored several scenarios such as pressure confined H{\small II} regions and ionized jets to explain the origin of the ionized gas emission and we have studied the physical properties of the detected sources. Based on our results we favor the scenario that $\sim 30 - 50 \%$ of our radio detections are ionized jets and/or jet knots. These sources, listed in Tables \ref{jet_cand_list} and \ref{candidates}, have observational properties that are not expected towards regular H{\small II} regions such as the correlation of their radio luminosity and bolometric luminosity and the correlation of the momentum rate of the molecular outflow with the radio luminosity of the ionized jet. Such correlations have been found observationally towards ionized jets associated with high-mass protostars of different luminosities and are also predicted in recent theoretical models such as the \citetalias{2016ApJ...818...52T} model. However, for the most compact radio continuum detections we cannot rule out the scenario that they correspond to pressure confined H{\small II} regions.  Our main results from this survey are summarized below: 

\begin{itemize}
\item We detected centimeter wavelength sources in 100$\%$ of our HMCs, which is a higher fraction than previously expected and suggests that radio continuum may be detectable at weak levels in all HMCs.  The lack of radio detections for some objects in the sample (including most CMCs) contributes  evidence that these clumps are in an earlier evolutionary stage than HMCs, providing interesting constraints and ideal follow up candidates for studies of the earliest stages of high-mass stars.

\item At least 10$\%$ of our detected radio sources are consistent with non-thermal emission and likely due to either  active magnetospheres in T-Tauri stars (possibly for the few regions located at a distance $<$ 2 kpc) or synchrotron emission from fast shocks in disks or jets.

\item For the most compact radio detections, the sources are consistent with being small pressure confined H{\small II} regions. Also, we cannot completely exclude the possibility that these sources are gravitationally trapped H{\small II} regions. 

\item The majority of our detected radio continuum sources ($\sim$80$\%$) have spectral indices ($-$0.1$<\alpha<$2) that are consistent with thermal (free-free) emission from ionized gas. 

\item Most of the radio sources with a rising spectrum detected towards clumps at an earlier evolutionary stage (i.e., CMCs and CMC--IRs) show Lyman continuum excess, consistent with previous results.  This can be explained either by  UV photons from shocks producing an ionized jet or shocks in an accretion flow onto the disk.

\item For most of the radio sources with a rising spectrum detected towards HMCs, the estimated Lyman continuum is lower than  expected if the radio flux comes from a single ZAMS star. This could indicate that the origin of the measured radio flux is not from HC/UC H{\small II} regions but  shock ionized jets.

\item We detected at least 12 ionized jets (6 of them are new detections to the
knowledge of the authors) based on their spectral index, morphology and molecular outflow associations. For several of the previously detected jets, we detected additional knots or lobes that are part of the collimated structure. Additionally, we detected at least 13 jet/wind candidates. 

\item We found that ionized jets from low and high-mass stars are very well correlated. This is consistent with previous studies and is further evidence of a common origin for jets of any luminosity.

\end{itemize}

\acknowledgments
 We thank the anonymous referee, whose comments improved this manuscript.
Support for this work was provided by the NSF through the Grote Reber Fellowship Program administered by Associated Universities, Inc./National Radio Astronomy Observatory.  P. H. acknowledges support from NSF grant AST--1814011. C. C-G. acknowledges support from UNAM DGAPA--PAPIIT grant number IA102816, IN10818. E. D. A. is partially supported by NSF grant AST--1814063. We thank K. E. I. Tanaka for providing the stellar model evolutionary tracks predicted by  the \citetalias{2016ApJ...818...52T} model. We thank K. E. I. Tanaka, K. Johnston and J. Marvil for  useful discussions. Herschel is an ESA space observatory with science instruments provided by European-led Principal Investigator consortia and with important participation from NASA. This work is based in part on observations made with the Spitzer Space telescope, which is operated by the Jet Propulsion Laboratory, California Institute of Technology under a contract with NASA. P.H. acknowledges support from NSF grant AST$-$0908901 for this project.  Some of the data reported here were obtained as part of the UKIRT Service Program. The United Kingdom Infrared Telescope is operated by the Joint Astronomy Centre on behalf of the UK Particle Physics and Astronomy Research Council.
This research made use of APLpy, an open-source plotting package for Python hosted at http://aplpy.github.com.

\vspace{5mm}

\software{CASA \citep{2007ASPC..376..127M}, APLpy \citep{2012ascl.soft08017R}.}

\appendix

\section{{\it Herschel}/Hi--GAL  Luminosity Estimates}\label{app:lum}
We obtained far--IR images  from the {\it Herschel}  Space Observatory \citep{2010A&A...518L...1P} infrared Galactic Plane Survey (Hi--GAL) project \citep{2010PASP..122..314M} toward 52 of our regions. Hi--GAL mapped 2$^{\circ}$  wide strips on the
sky in the galactic longitude range $|${\it l}$|$ $<$ 60$^{\circ}$,  in the five wavelength bands, 70\,$\mu$m, 160\,$\mu$m, 250\,$\mu$m, 350\,$\mu$m and 500\,$\mu$m,
with angular resolutions of 5\rlap.$^{\prime \prime}$, 13\rlap.$^{\prime \prime}$, 18\rlap.$^{\prime \prime}$, 25\rlap.$^{\prime \prime}$, 36\rlap.$^{\prime \prime}$, respectively.
These data provide an improvement of a factor of $\sim$10 in spatial resolution compared to IRAS, thus allowing a better estimate 
of luminosities by lowering the contribution from unrelated nearby sources. In Figure \ref{fig:Higal_images} we show the Hi--GAL images for our targets, except for the regions
LDN1657A--3, UYSO1, 18151$-$1208, 18517$+$0437 and 20126$+$4104 which are located outside the region covered by Hi--GAL.

We measured the flux densities using an algorithm written in GILDAS\footnote{http://www.iram.fr/IRAMFR/GILDAS}, where we use a suitable polygon to enclose the source at each wavelength and integrate
the flux over this area. Table \ref{SED_Parameters} shows the name of the region in column 1 and the values of the flux densities are given in columns 2 through 9. 
As the angular resolution worsens with increasing wavelength, there are a number of regions where the emission appears very extended and becomes blended with nearby cores.
Specifically, IRDCs  G25.04$-$00.20, G28.23$-$00.19, G28.53$-$00.25 and G30.97$-$00.14 have millimeter cores that are highly blended; for these sources we only report an upper limit for the luminosities. 
Also, for some regions we can only estimate lower, or upper limits for the flux densities at 24\,$\mu$m and 70\,$\mu$m, due to absorption, or non-detection. In the latter case, we do not
include these data points when estimating the luminosity.  The specific cases are identified in Table \ref{SED_Parameters} column 12 or are pointed out with footnotes. Additionally, for a handful of sources there was not enough ancillary data available and therefore their  bolometric luminosity  might be underestimated (e.g., 20216$+$4107, 20293$+$3952 and 20343$+$4129).

We also gathered images at 24\,$\mu$m from {\it Spitzer}/MIPS \citep{2004ApJS..154...25R} for the construction of the spectral energy distributions (SEDs) for
our targets.  The Spitzer/MIPS data are from the Multiband Imaging Photometer for Spitzer Galactic Plane Survey (MIPSGAL; \citealt{2005AAS...207.6333C}).
Sub-mm fluxes obtained from the APEX Telescope Large Survey of the Galaxy (ATLASGAL\footnote{The ATLASGAL project is a collaboration between the Max-Planck-Gesellschaft, the European Southern Observatory (ESO), and the 
Universidad de Chile.}) \citep{2014A&A...568A..41U}   and Bolocam Galactic Plane Survey (BGPS) v2 \citep{2015ApJ...799...29E} taken at 870 $\mu$m and 1.1 mm, respectively,  are also used for  estimating the luminosities.
Figures \ref{SED_CMC}, \ref{SED_CMC_IR} and \ref{SED_HMC} show the SEDs of the CMCs, CMC--IRs and HMCs of 52 out of the 57 regions observed with the VLA. We estimated the luminosity of each 
region integrating the continuum spectra by linearly interpolating the flux densities in the SED. The distance used to calculate the luminosities  and the estimated luminosities are given in Table \ref{SED_Parameters} 
columns 10 and 11, respectively.

In Figure \ref{IRAS_Herschel} we compare the luminosities derived in this work with values
from the literature. For the case of HMCs (red points) the literature
values are mostly based on far-infrared data taken with the Infrared Astronomical Satellite (IRAS) database \citep{2002ApJ...566..931S}.
In almost all cases our luminosities are smaller than the IRAS luminosities.
This is likely due to the inclusion of extended emission or unrelated sources
in the large IRAS beams. However the difference is usually $<$ 0.3 dex.

For the CMCs and CMC-IRs (green, and blue points), the literature values
for the luminosities were mostly taken from \citet{2010ApJ...715..310R} who
constructed SEDs using a combination
of Spitzer and ground based telescope data. For such sources, the scatter in Figure  \ref{IRAS_Herschel} is somewhat
larger than for HMCs. This could be due to a variety of factors; however most
sources have a consistent luminosity within a factor of $<$ 0.5 dex.

For the analysis performed in this work we have used the luminosities which
we have derived as described above, with exception of sources where we only
derived limits (G25.04--00.20mm1, G25.04--00.20mm3, G28.23-00.19mm1, G28.23-00.19mm3, G28.53--00.25mm1, G28.53--00.25mm2, G28.53--00.25mm4,  G28.53--00.25mm6, G30.97--00.14mm1 and G30.97--00.14mm2) or those sources that do not have HI--GAL data (see above). In both  cases, if available,  we used the literature values instead.


\clearpage

\begin{figure}[htbp]
\centering
\begin{tabular}{cc}
\hspace*{\fill}%
\includegraphics[width=0.52\textwidth,  trim = 20 20 20 15, clip, angle = 0]{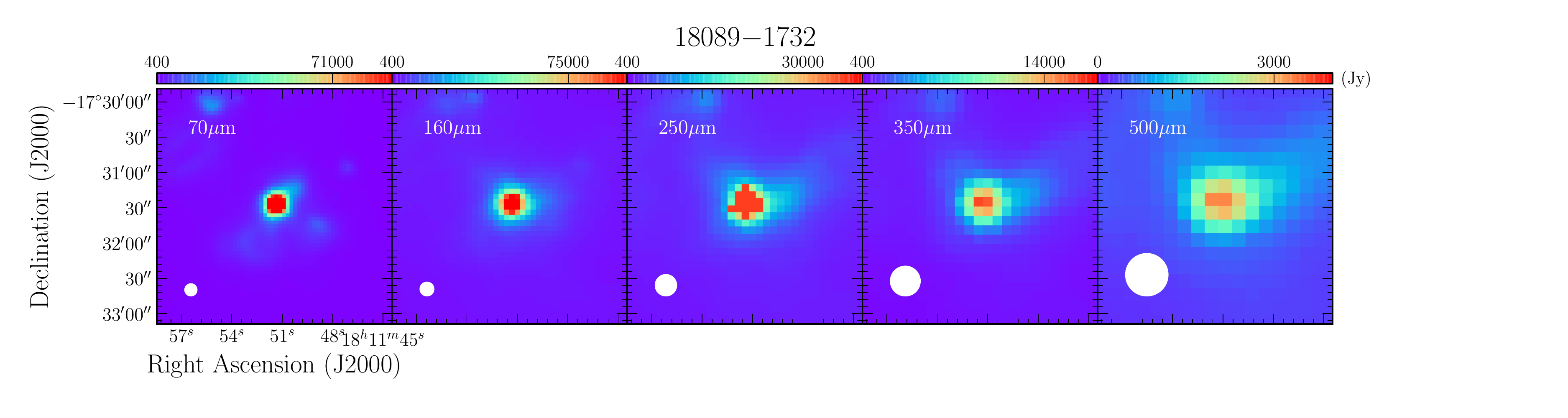}\hspace{-0.8cm} &
\includegraphics[width=0.52\textwidth,  trim = 20 20 20 15, clip, angle = 0]{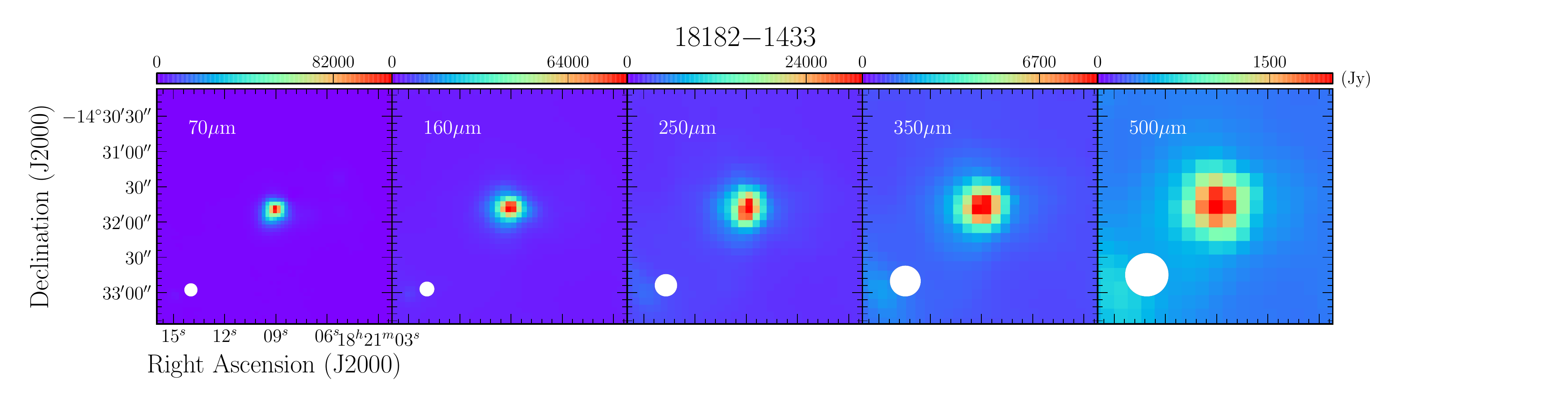}\\
\includegraphics[width=0.52\textwidth,  trim = 20 20 20 15, clip, angle = 0]{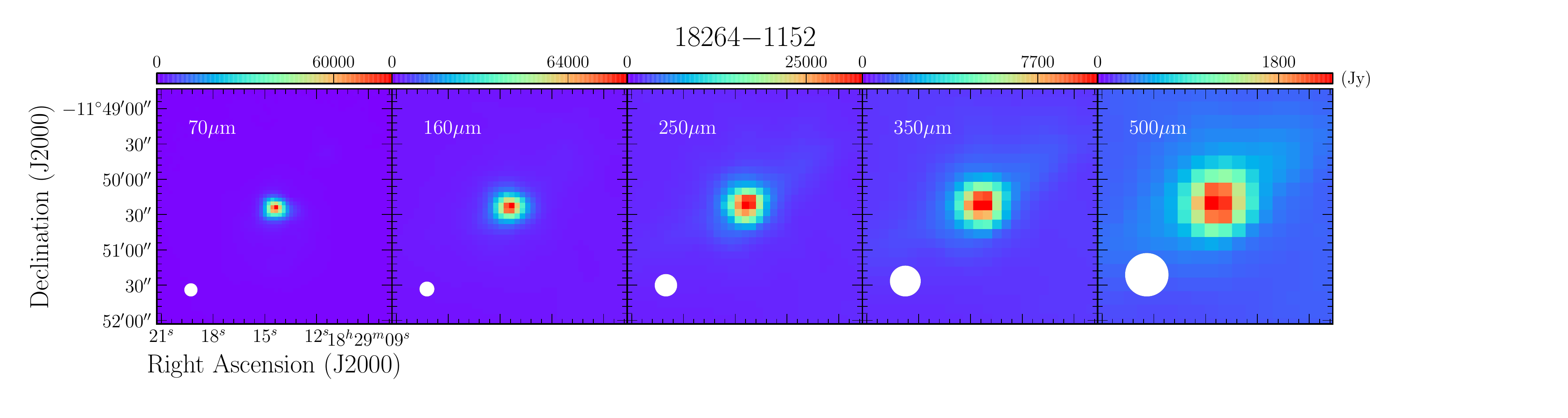}\hspace{-0.8cm} &
\includegraphics[width=0.52\textwidth,  trim = 20 20 20 15, clip, angle = 0]{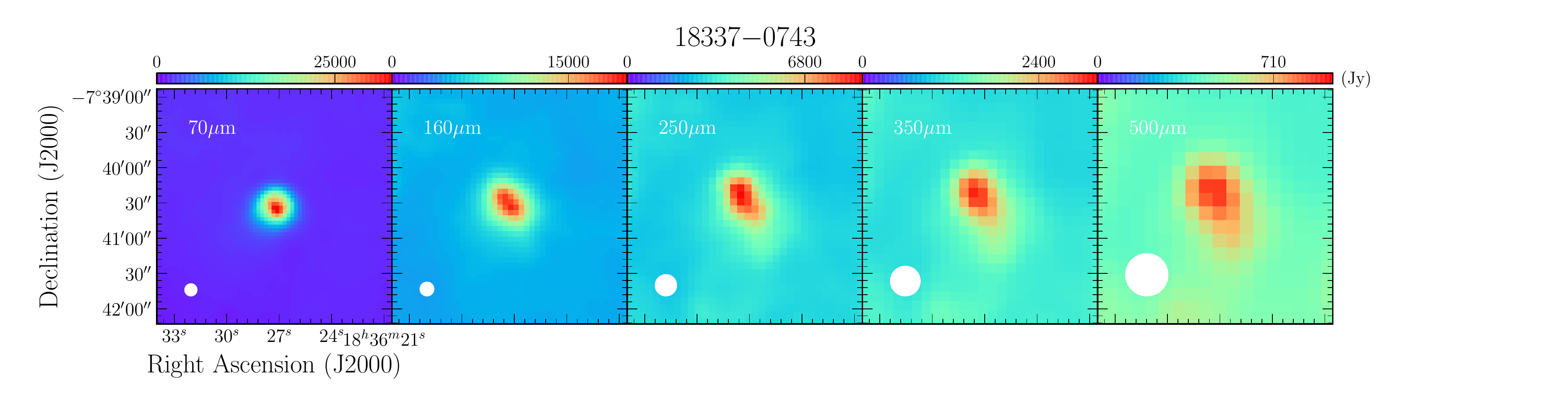} \\
\includegraphics[width=0.52\textwidth,  trim = 20 20 20 15, clip, angle = 0]{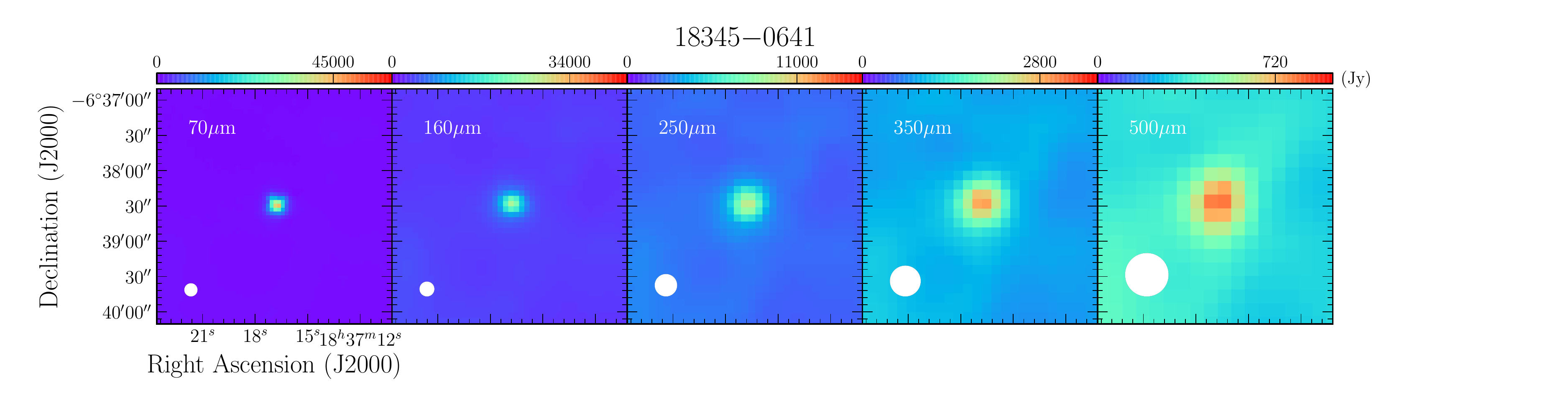}\hspace{-0.8cm} &
\includegraphics[width=0.52\textwidth,  trim = 20 20 20 15, clip, angle = 0]{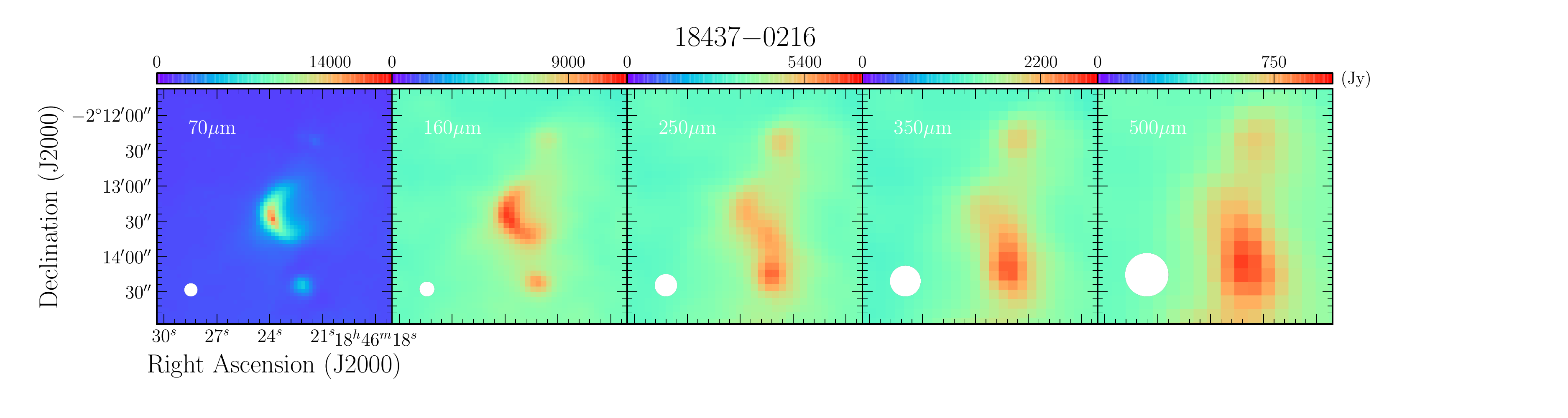}\\
 \includegraphics[width=0.52\textwidth,  trim = 20 20 20 15, clip, angle = 0]{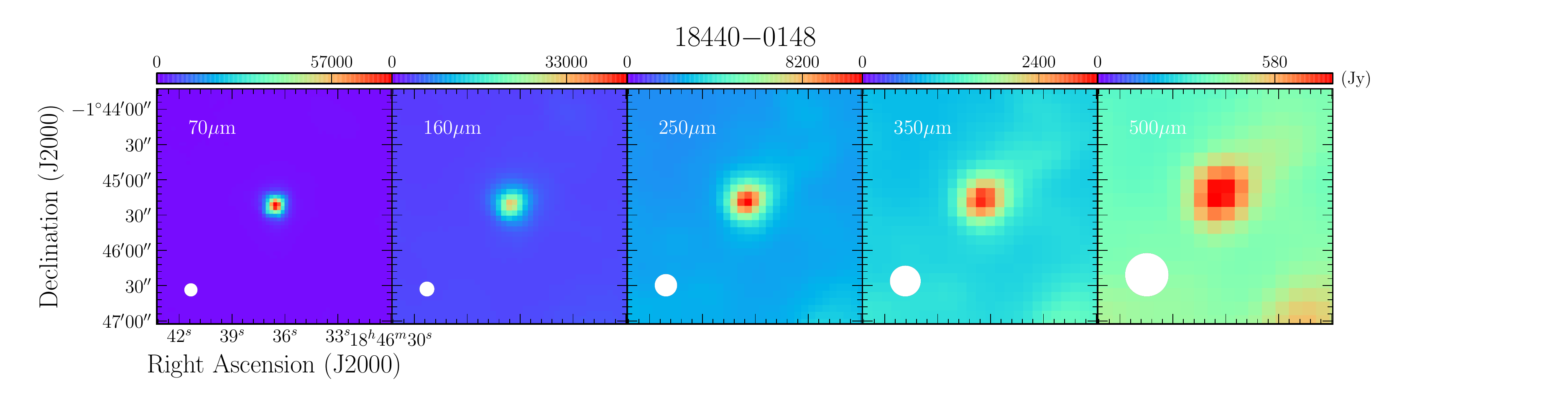}\hspace{-0.8cm} &
 \includegraphics[width=0.52\textwidth,  trim = 20 20 20 15, clip, angle = 0]{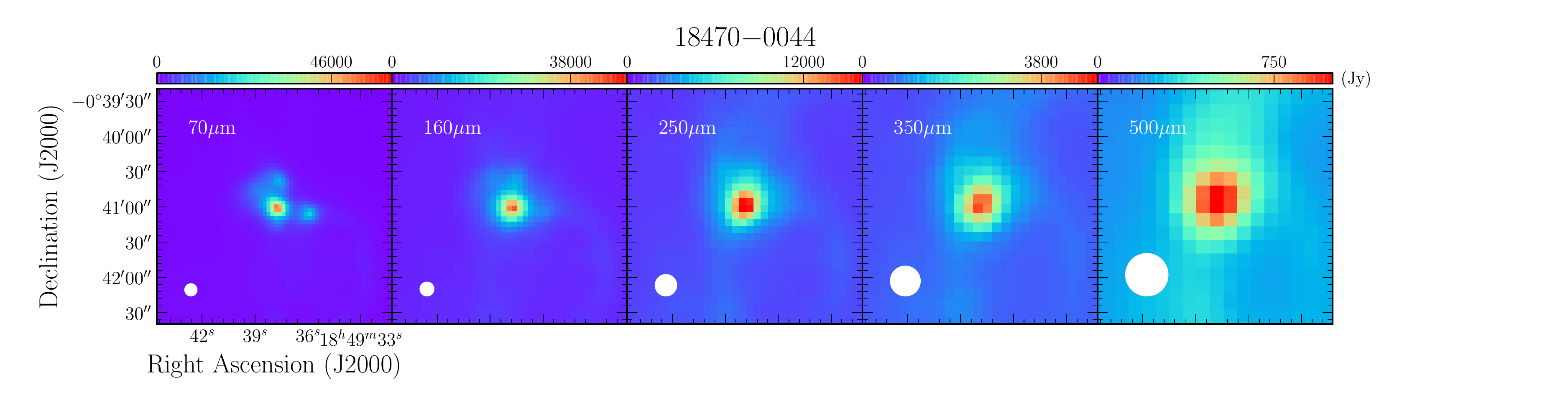}\\
 \includegraphics[width=0.52\textwidth,  trim = 20 20 20 15, clip, angle = 0]{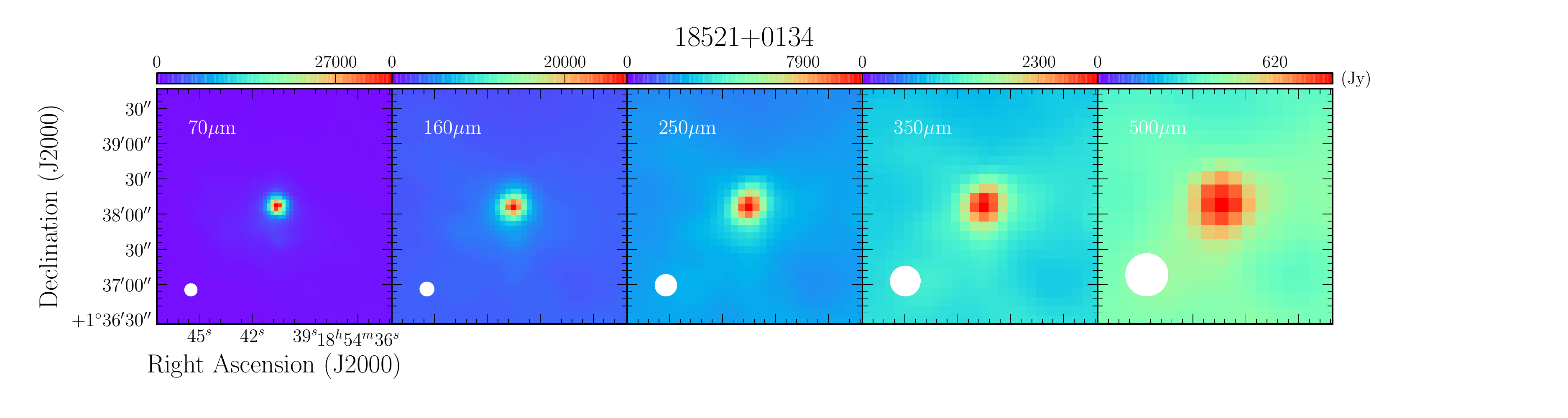}\hspace{-0.8cm} &
 \includegraphics[width=0.52\textwidth,  trim = 20 20 20 15, clip, angle = 0]{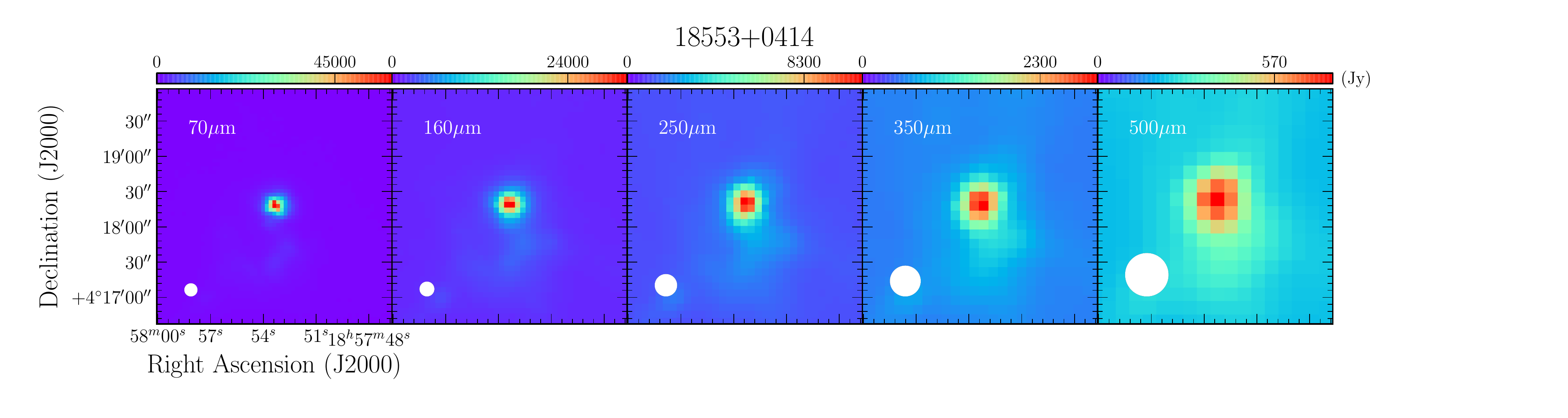}\\
 \includegraphics[width=0.52\textwidth,  trim = 20 20 20 15, clip, angle = 0]{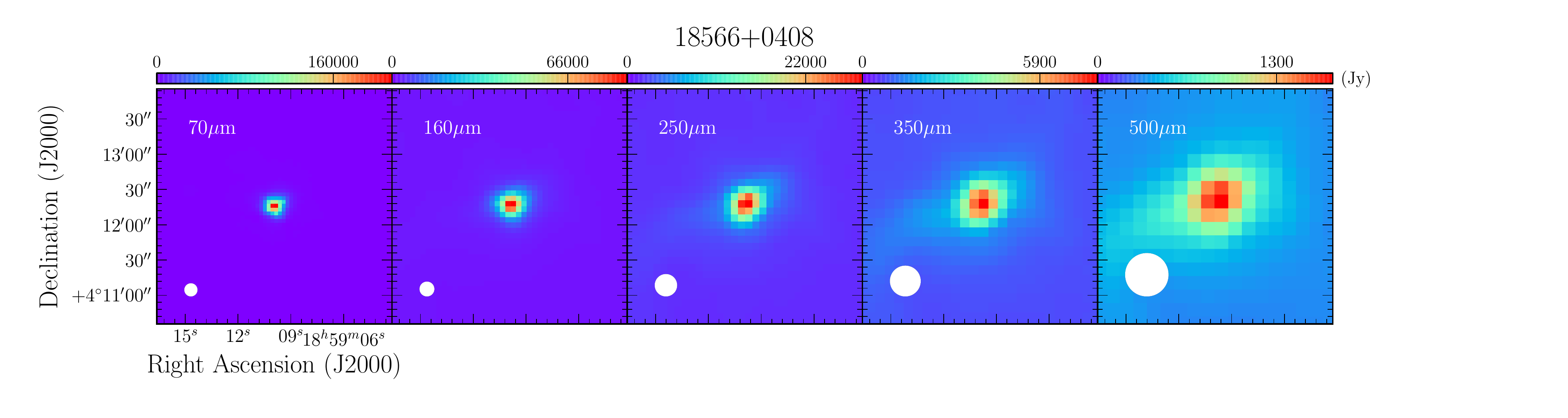}\hspace{-0.8cm} &
 \includegraphics[width=0.52\textwidth,  trim = 20 20 20 15, clip, angle = 0]{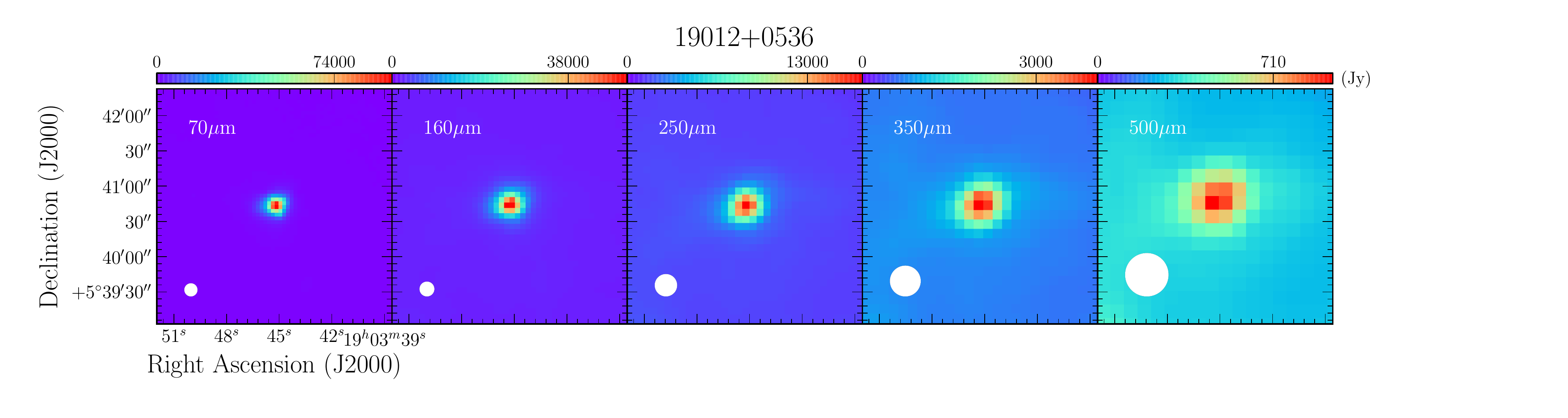}\\
 \includegraphics[width=0.52\textwidth,  trim = 20 20 20 15, clip, angle = 0]{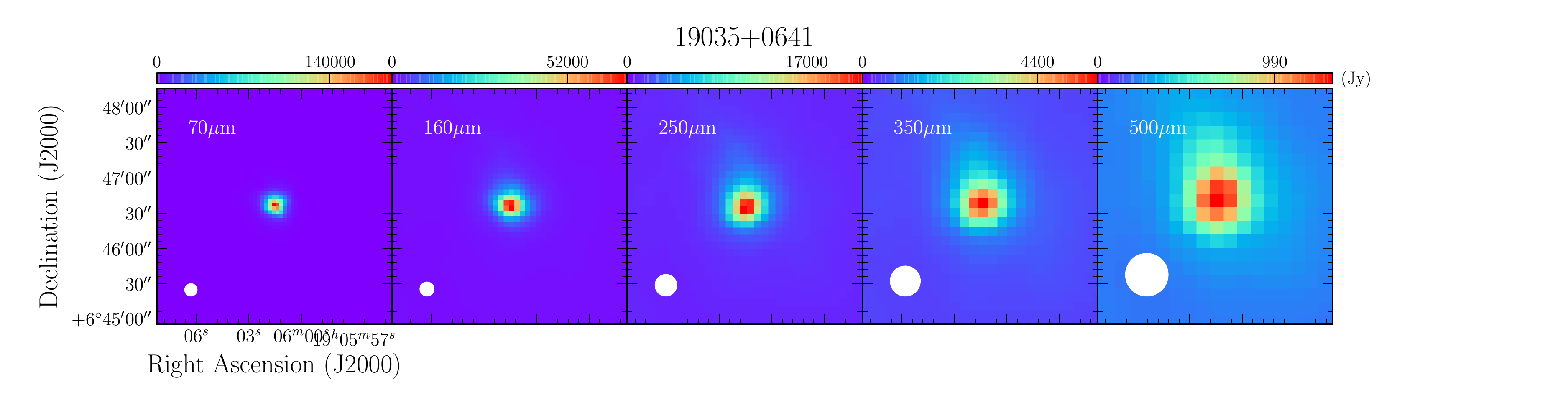}\hspace{-0.8cm} &
 \includegraphics[width=0.52\textwidth,  trim = 20 20 20 15, clip, angle = 0]{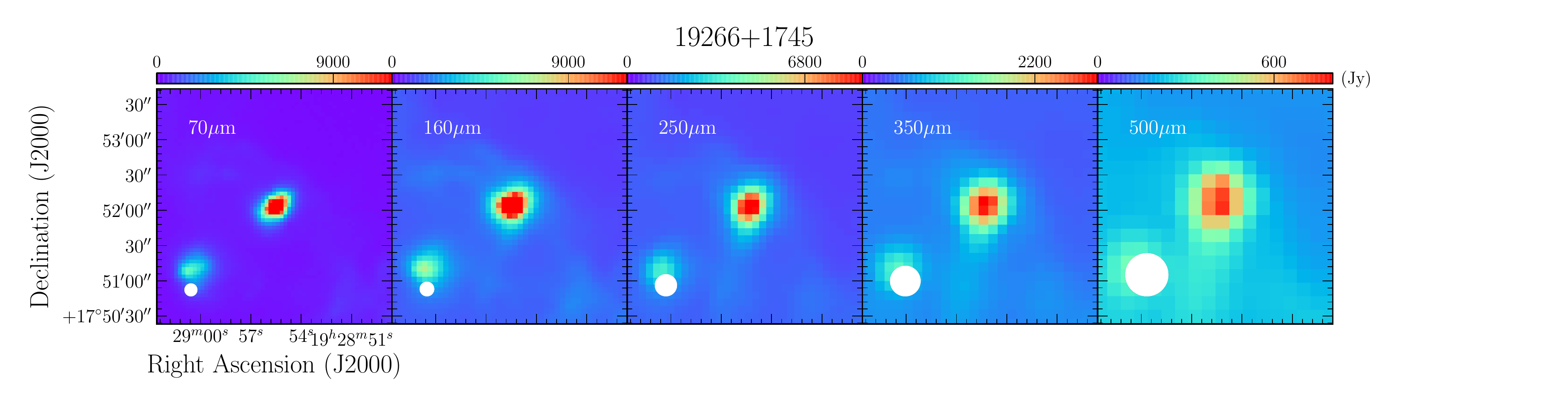}\\
 \includegraphics[width=0.52\textwidth,  trim = 20 20 20 15, clip, angle = 0]{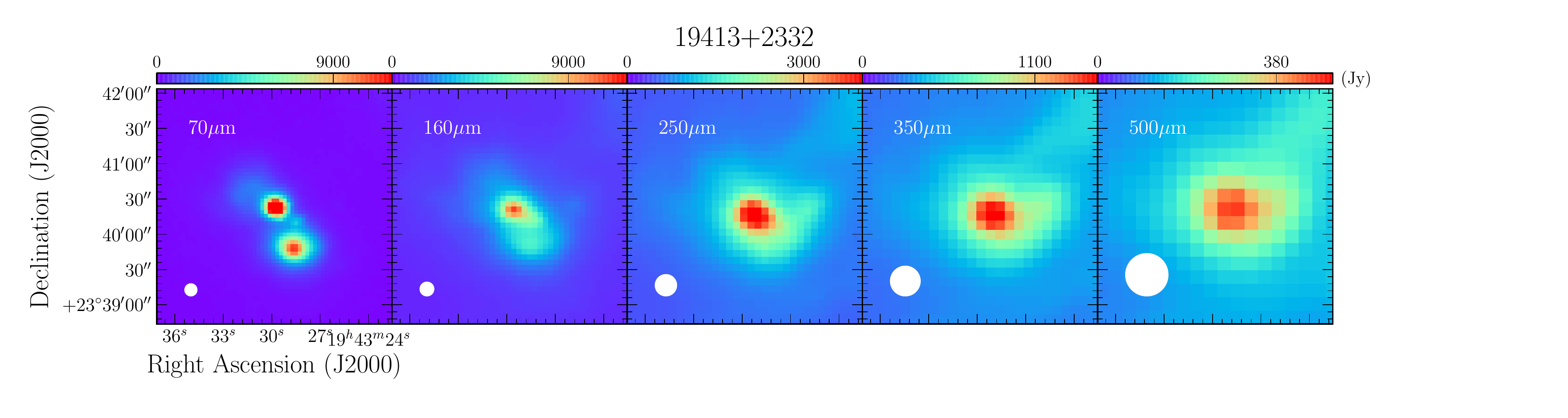}\hspace{-0.8cm} &
  \includegraphics[width=0.52\textwidth,  trim = 20 20 20 15, clip, angle = 0]{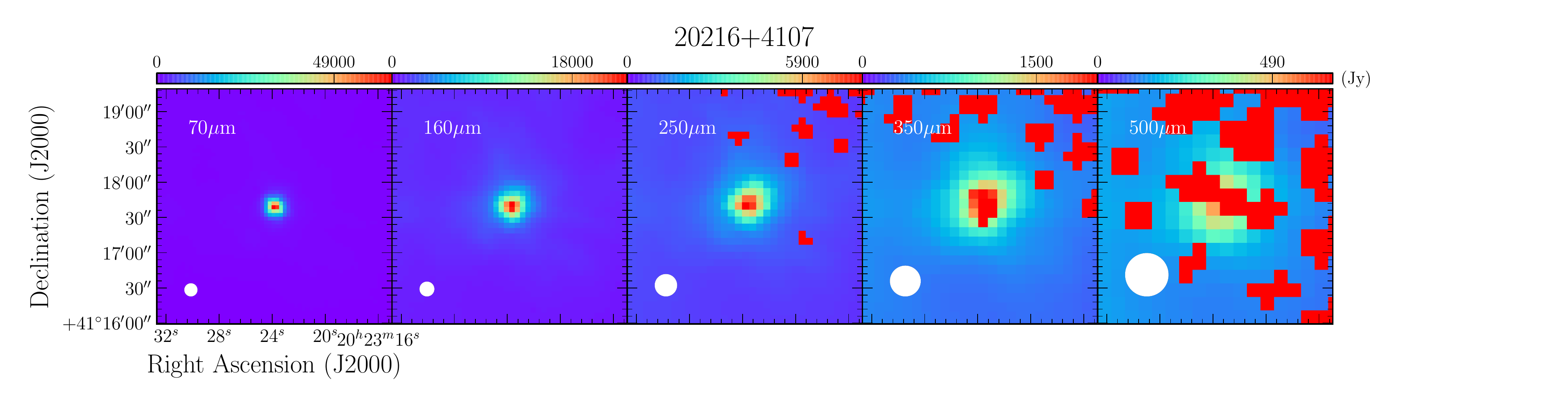}  \\
 \includegraphics[width=0.52\textwidth,  trim = 20 20 20 15, clip, angle = 0]{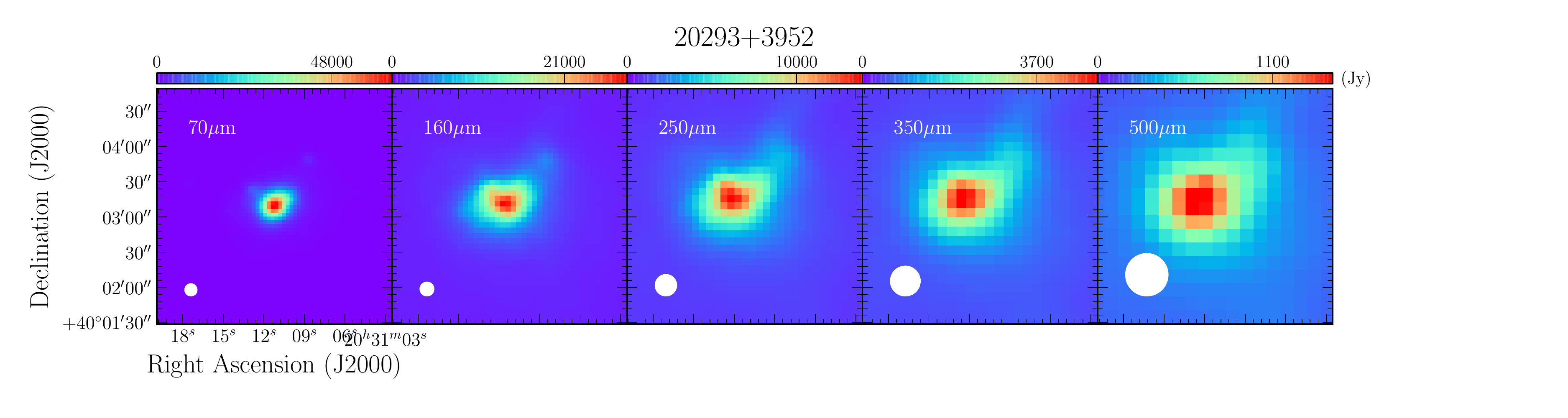}\hspace{-0.8cm} &
  \includegraphics[width=0.52\textwidth,  trim = 20 20 20 15, clip, angle = 0]{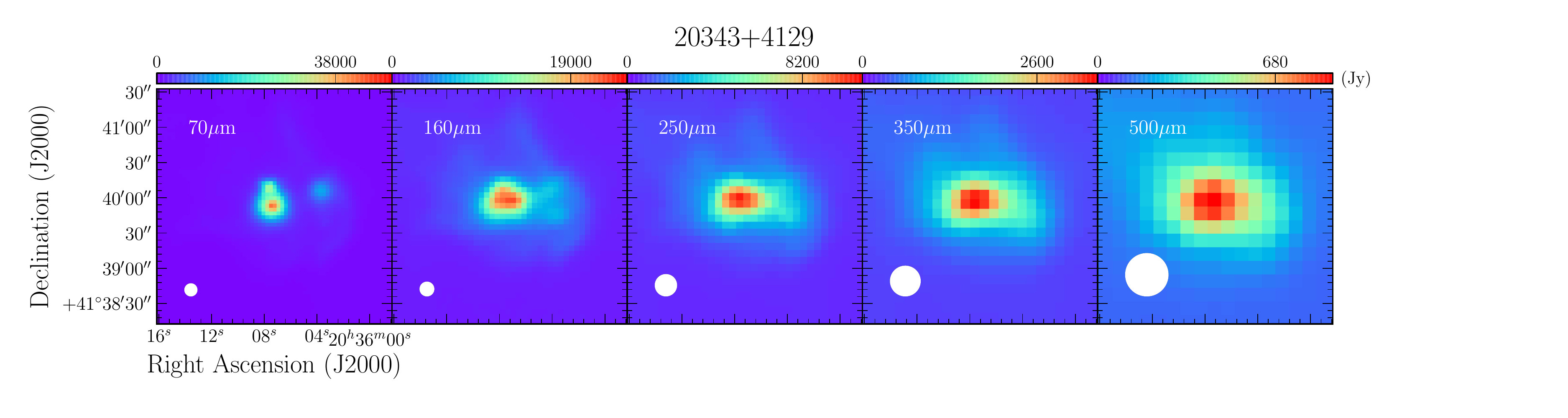} 
\end{tabular}
\caption{\small{Hi--GAL images for 52 of the regions reported in \citetalias{2016ApJS..227...25R}. For 20216$+$4107 the images at 250, 350 and 500 $\mu$m contain bad pixels (seen as bright red color).
}}
 \label{fig:Higal_images}
\end{figure}

\begin{figure}[htbp]
\centering
\begin{tabular}{cc}
  \ContinuedFloat
\hspace*{\fill}%
 \includegraphics[width=0.52\textwidth,  trim = 20 20 20 15, clip, angle = 0]{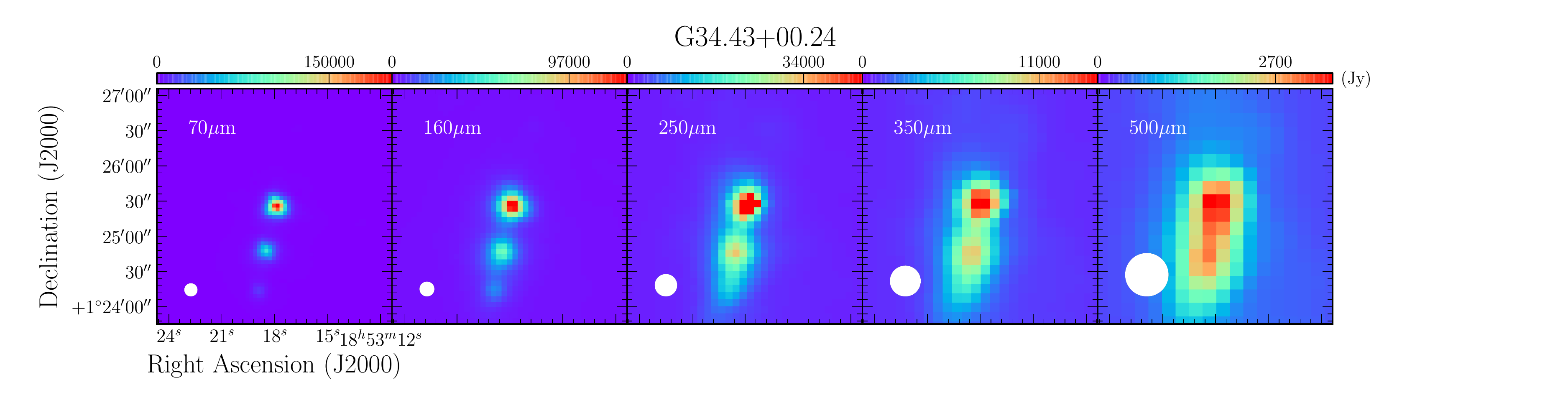}\hspace{-0.8cm} &
 \includegraphics[width=0.52\textwidth,  trim = 20 20 20 15, clip, angle = 0]{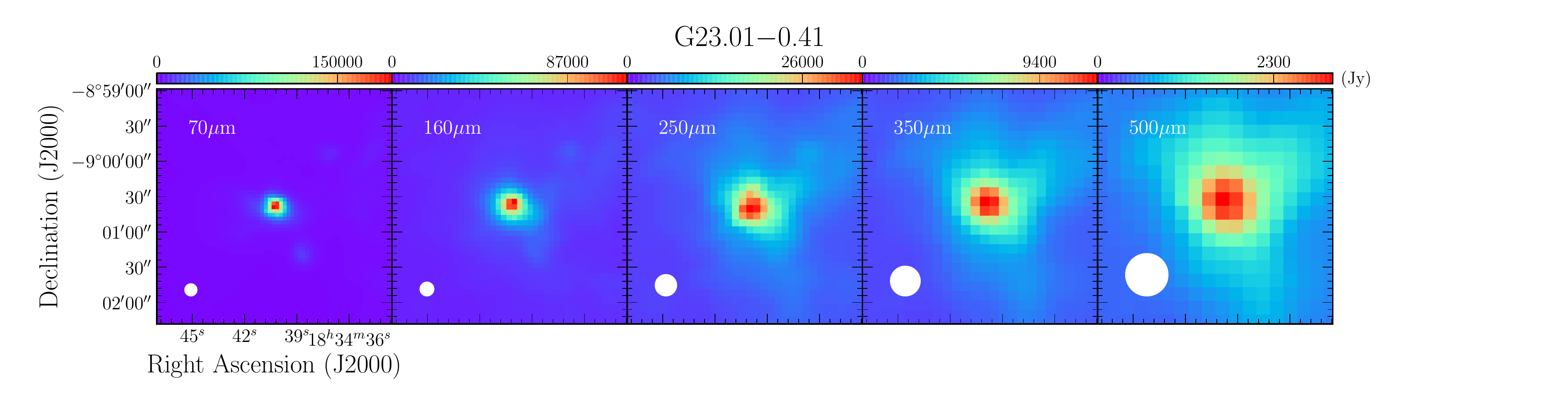} \\
 \includegraphics[width=0.52\textwidth,  trim = 20 20 20 15, clip, angle = 0]{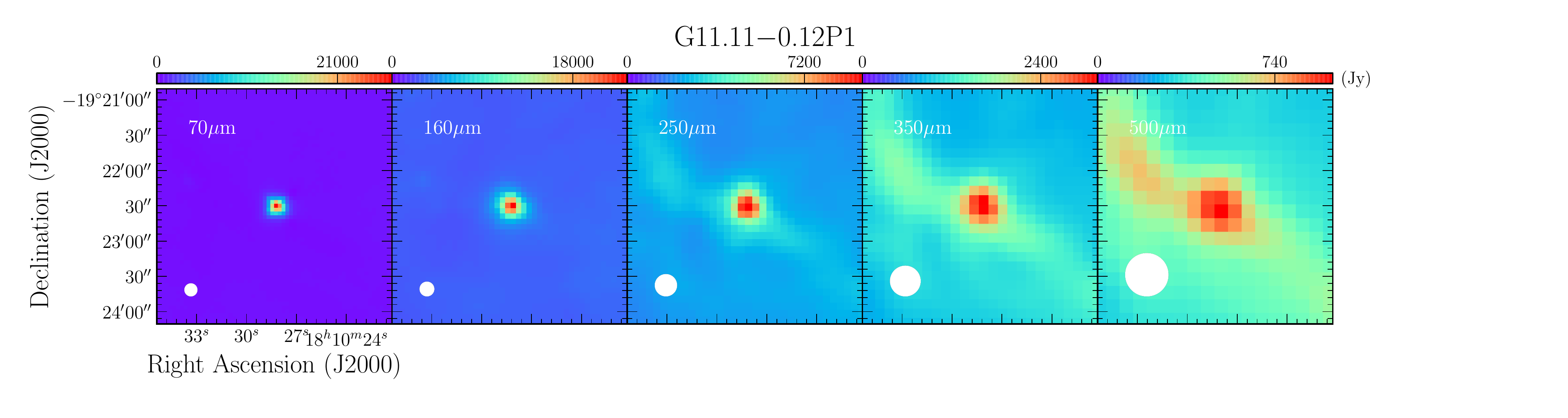}\hspace{-0.8cm} &
 \includegraphics[width=0.52\textwidth,  trim = 20 20 20 15, clip, angle = 0]{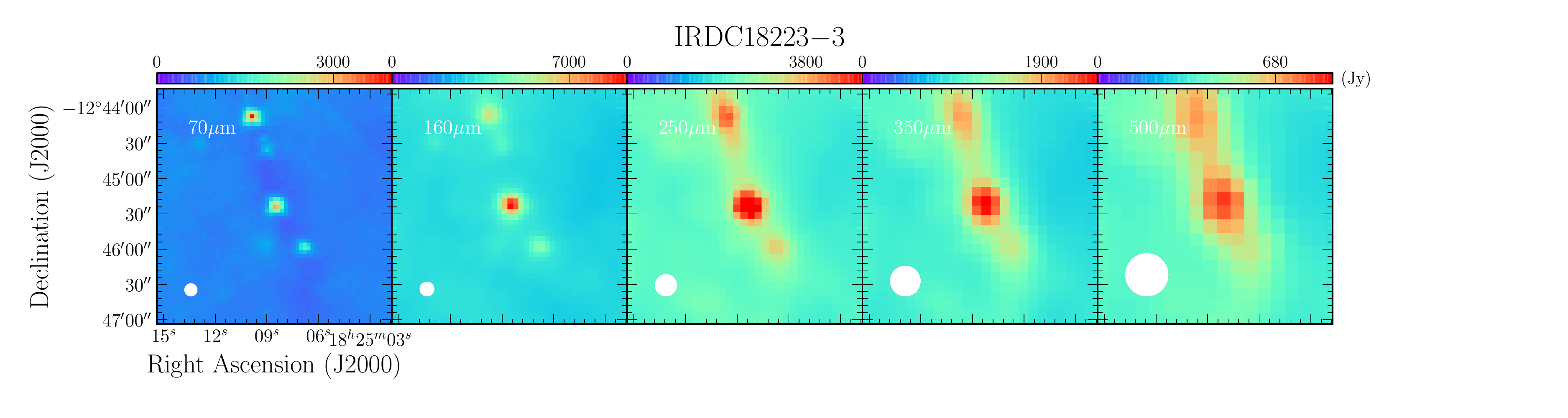} \\
 \includegraphics[width=0.52\textwidth,  trim = 20 20 20 15, clip, angle = 0]{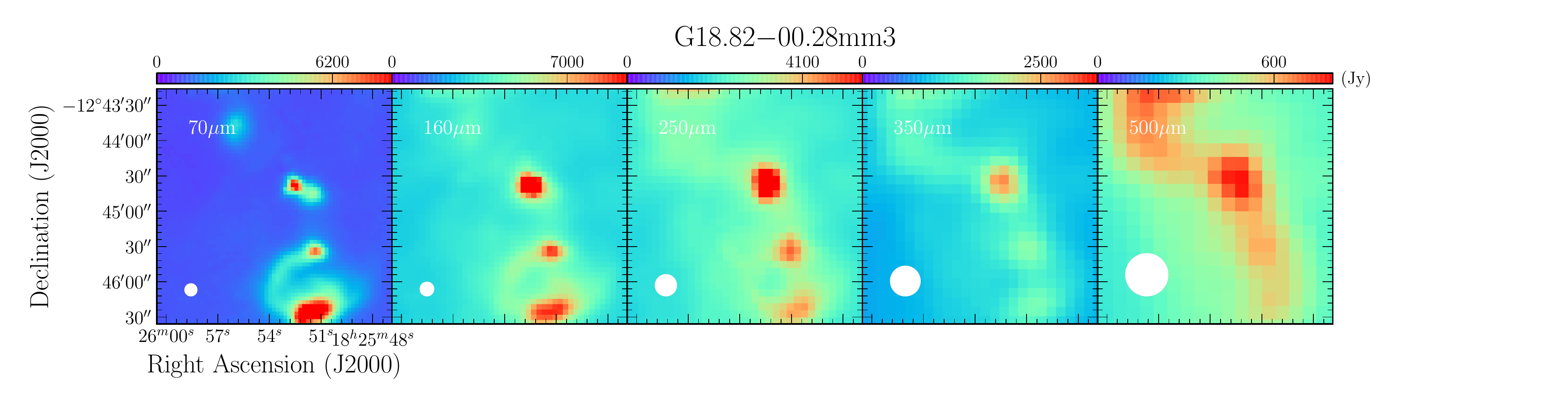}\hspace{-0.8cm} &
 \includegraphics[width=0.52\textwidth,  trim = 20 20 20 15, clip, angle = 0]{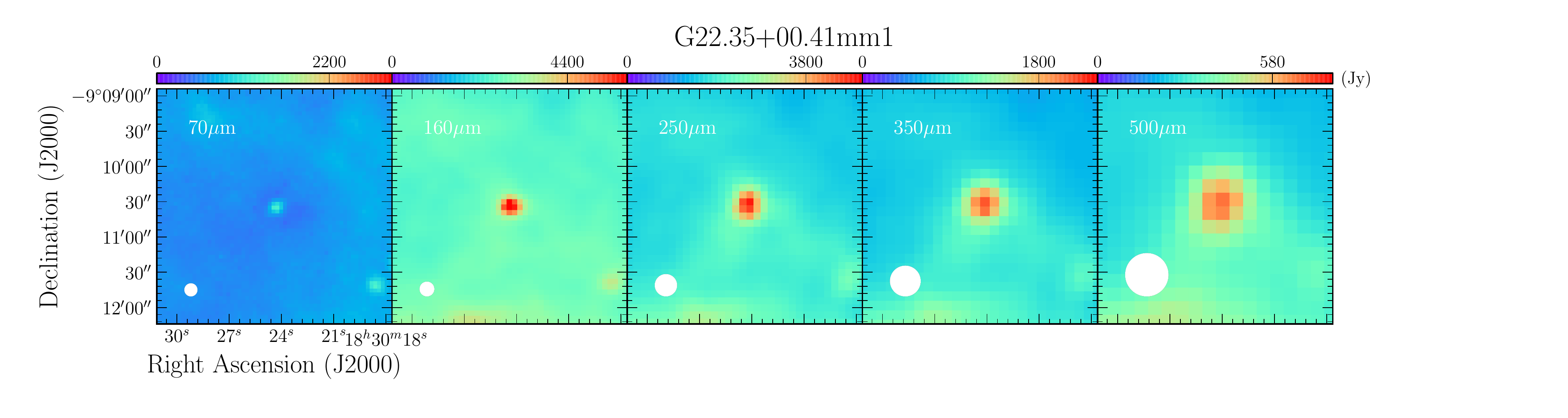} \\
 \includegraphics[width=0.52\textwidth,  trim = 20 20 20 15, clip, angle = 0]{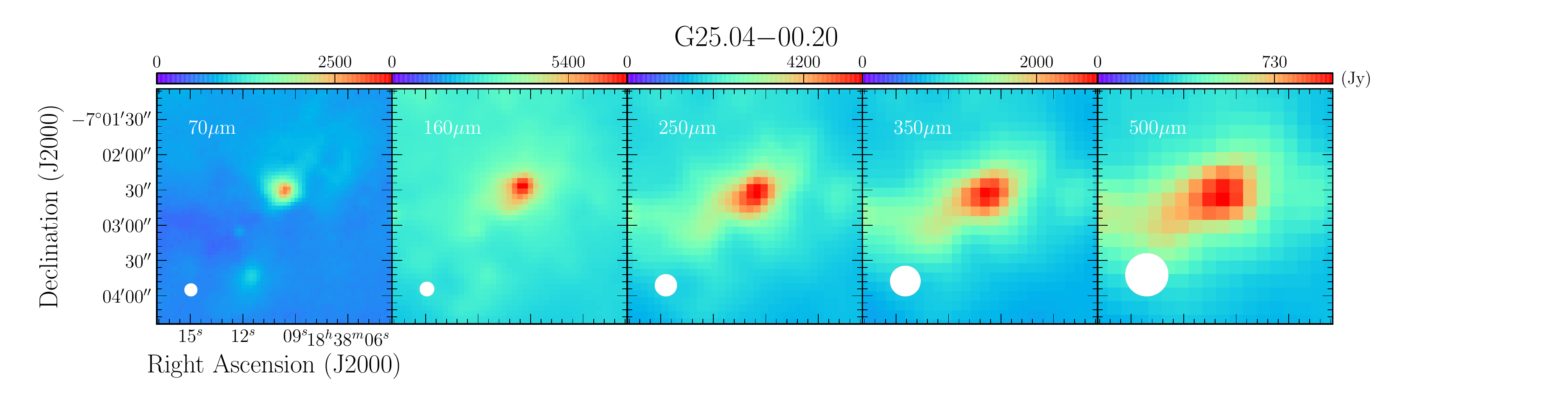}\hspace{-0.8cm}  &
 \includegraphics[width=0.52\textwidth,  trim = 20 20 20 15, clip, angle = 0]{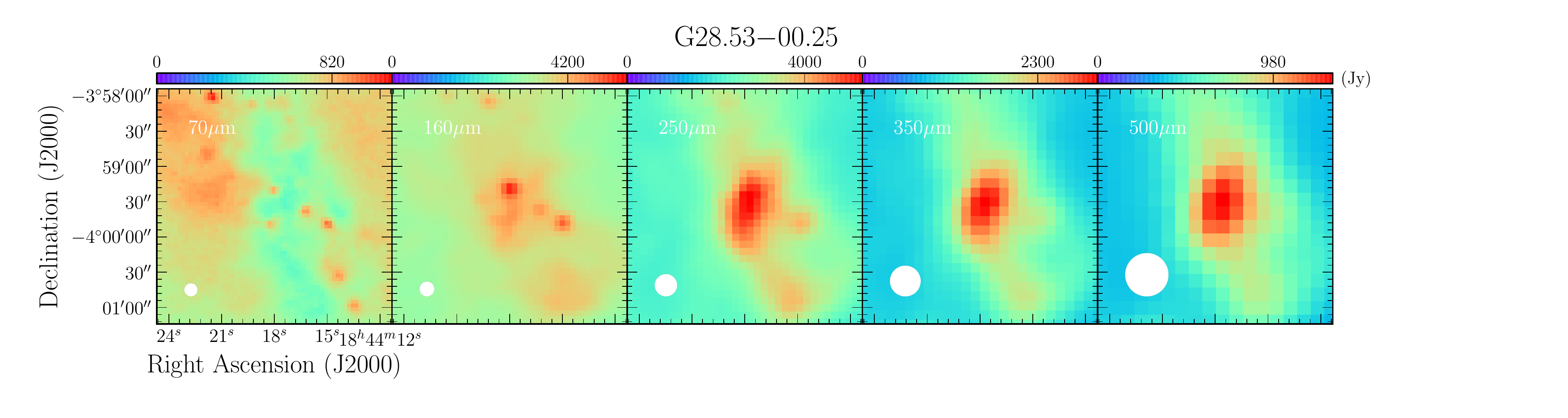} \\
 \includegraphics[width=0.52\textwidth,  trim = 20 20 20 15, clip, angle = 0]{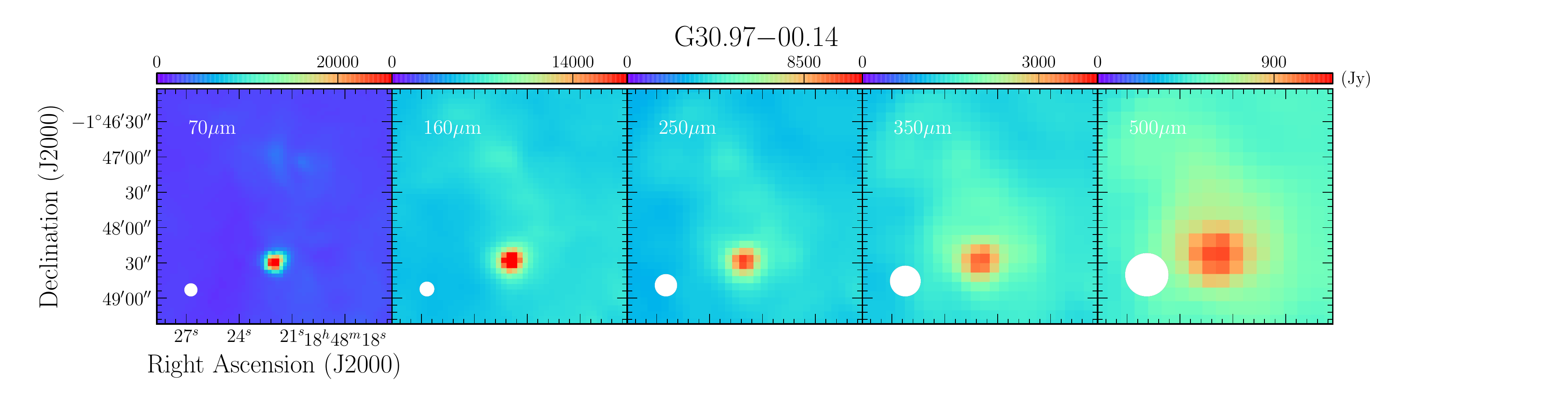}\hspace{-0.8cm}  &
 \includegraphics[width=0.52\textwidth,  trim = 20 20 20 15, clip, angle = 0]{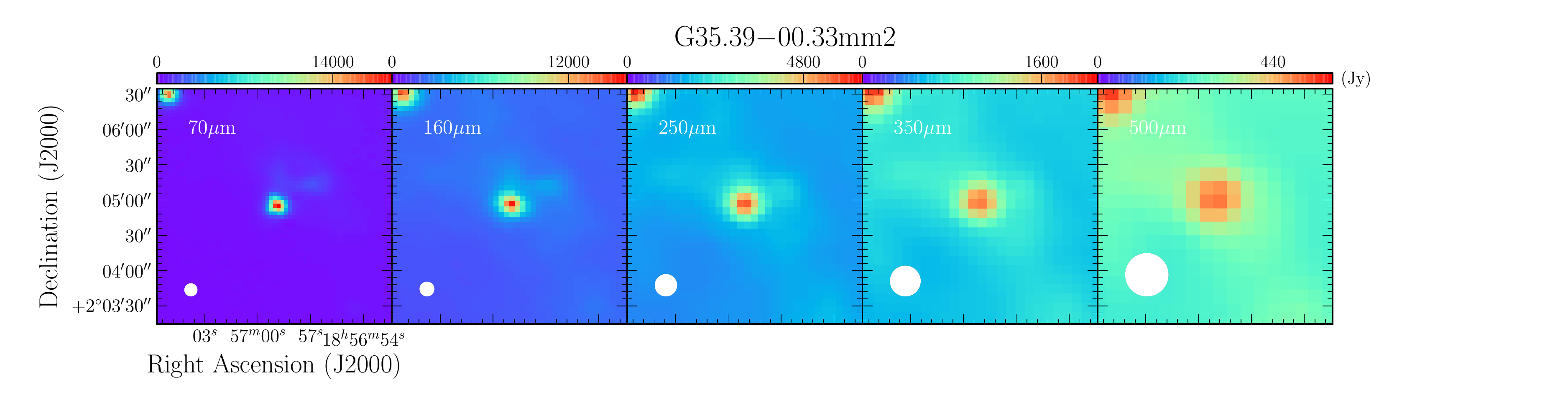} \\
 \includegraphics[width=0.52\textwidth,  trim = 20 20 20 15, clip, angle = 0]{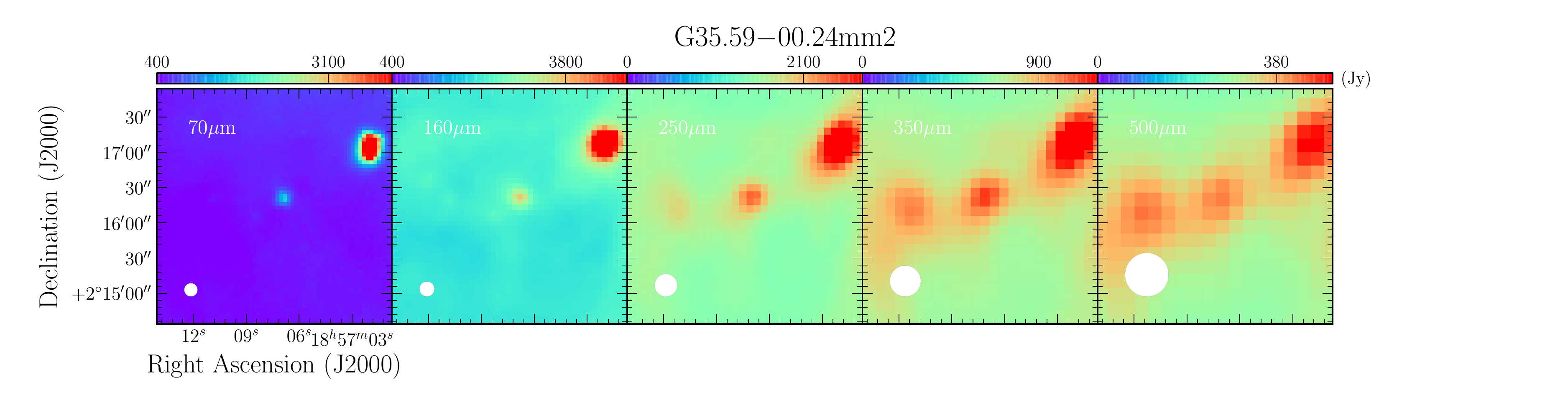}\hspace{-0.8cm}  &
 \includegraphics[width=0.52\textwidth,  trim = 20 20 20 15, clip, angle = 0]{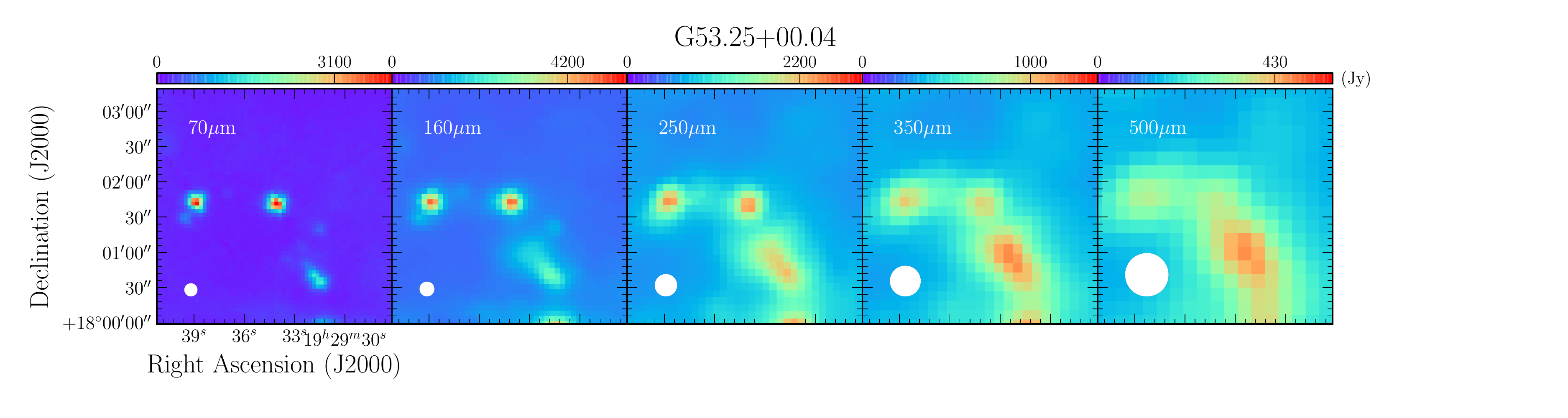}  \\
 \includegraphics[width=0.52\textwidth,  trim = 20 20 20 15, clip, angle = 0]{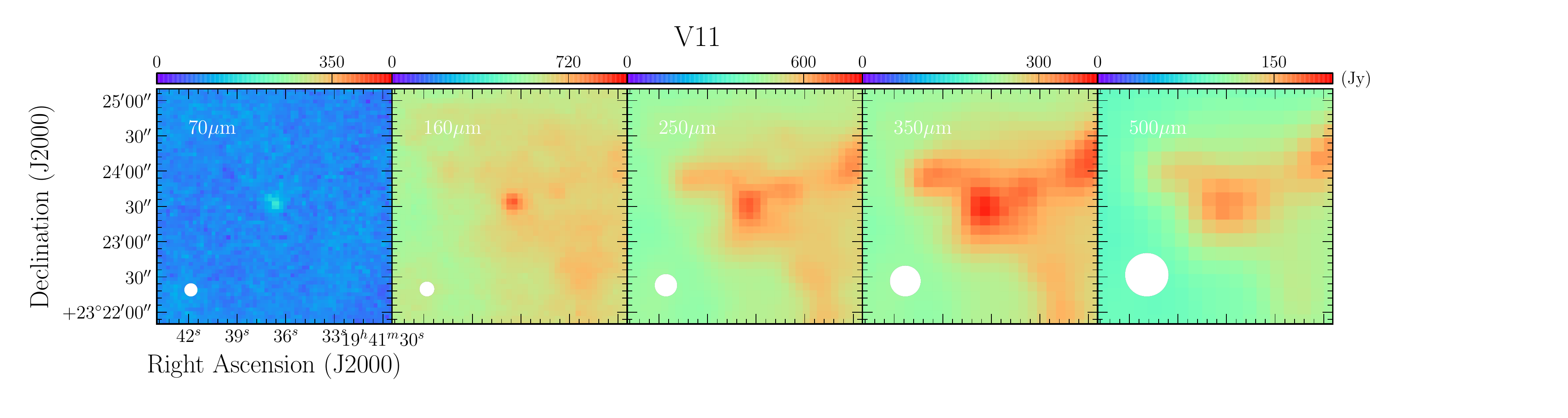}\hspace{-0.8cm}  &
 \includegraphics[width=0.52\textwidth,  trim = 20 20 20 15, clip, angle = 0]{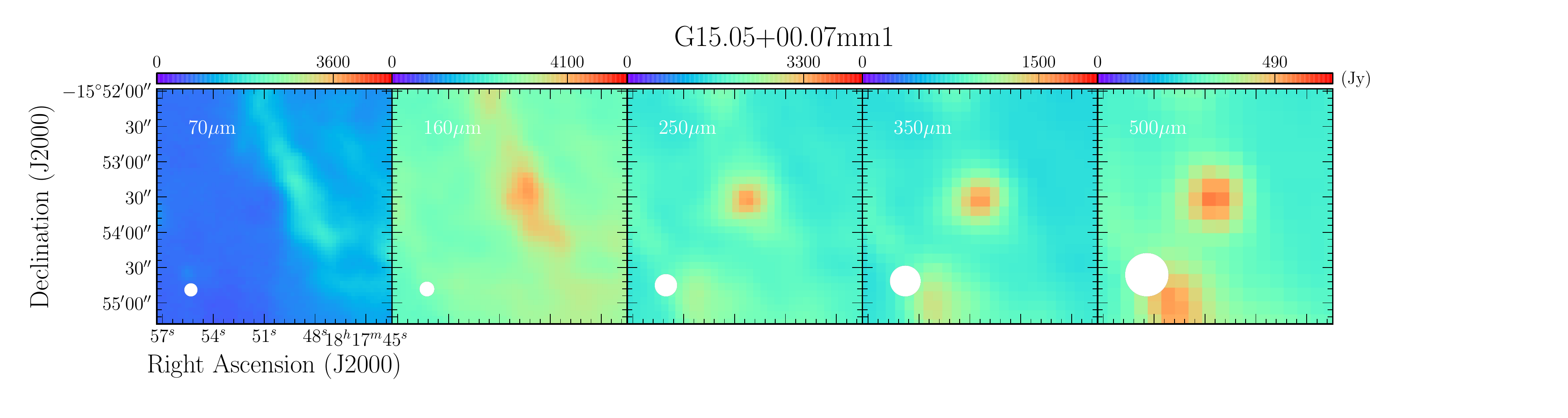}  \\
  \includegraphics[width=0.52\textwidth,  trim = 20 20 20 15, clip, angle = 0]{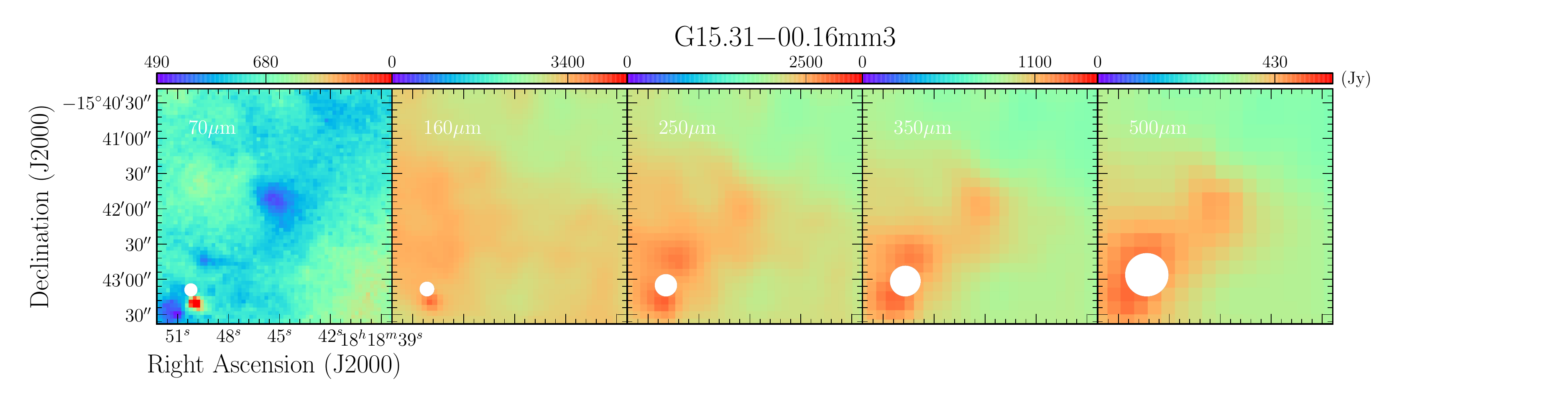}\hspace{-0.8cm}  &
 \includegraphics[width=0.52\textwidth,  trim = 20 20 20 15, clip, angle = 0]{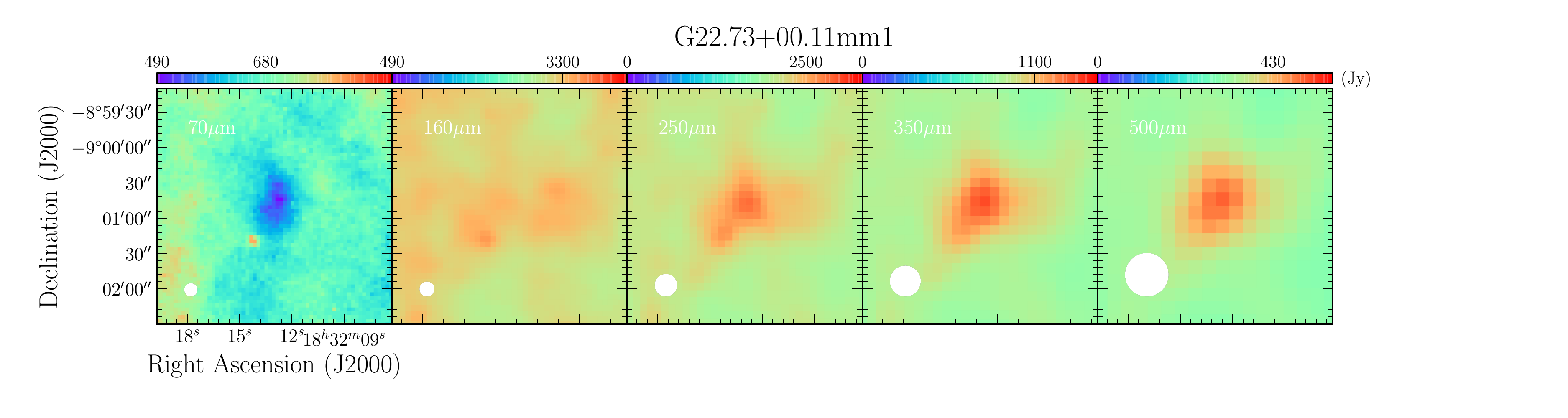} \\
 \includegraphics[width=0.52\textwidth,  trim = 20 20 20 15, clip, angle = 0]{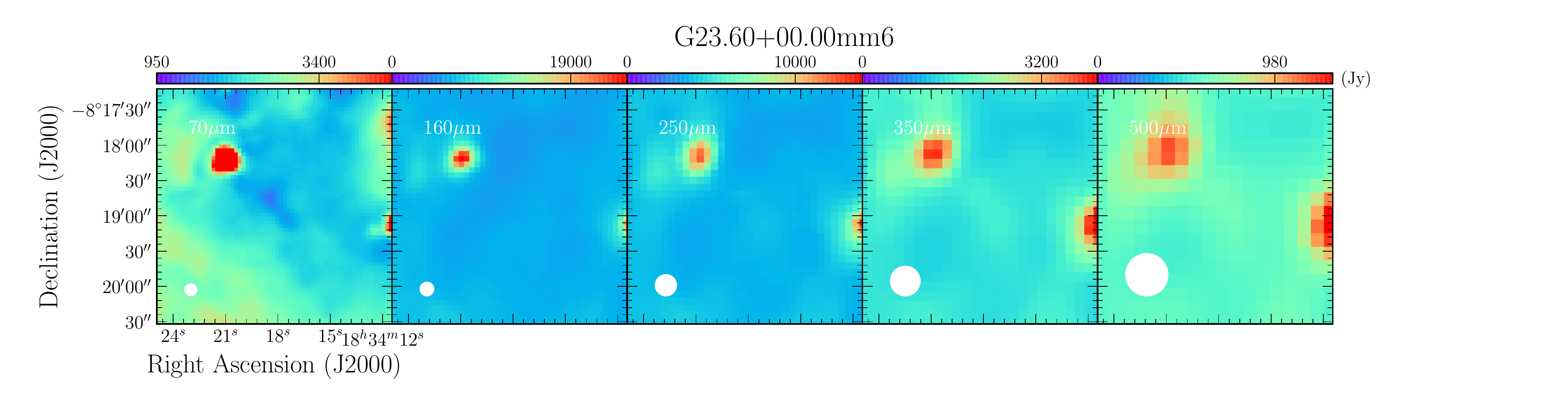}\hspace{-0.8cm}  &
 \includegraphics[width=0.52\textwidth,  trim = 20 20 20 15, clip, angle = 0]{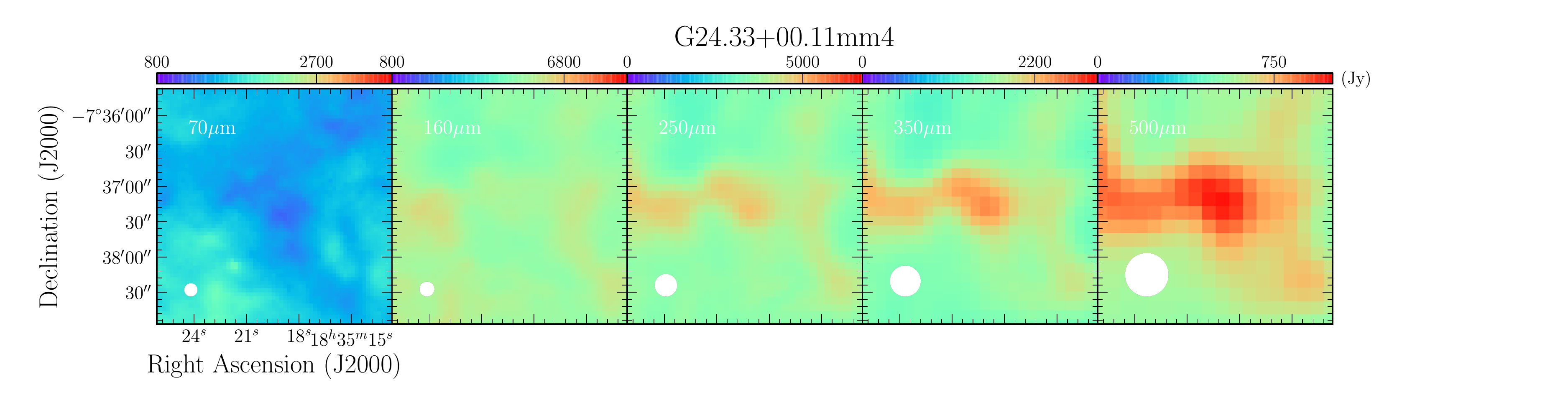}  

\end{tabular}
 \hspace*{\fill}%
\caption{\small{Continued.}}
 \label{fig:Higal_images}
\end{figure}

\begin{figure}[htbp]
\centering
\begin{tabular}{cc}
  \ContinuedFloat
\hspace*{\fill}%
 \includegraphics[width=0.52\textwidth,  trim = 20 20 20 15, clip, angle = 0]{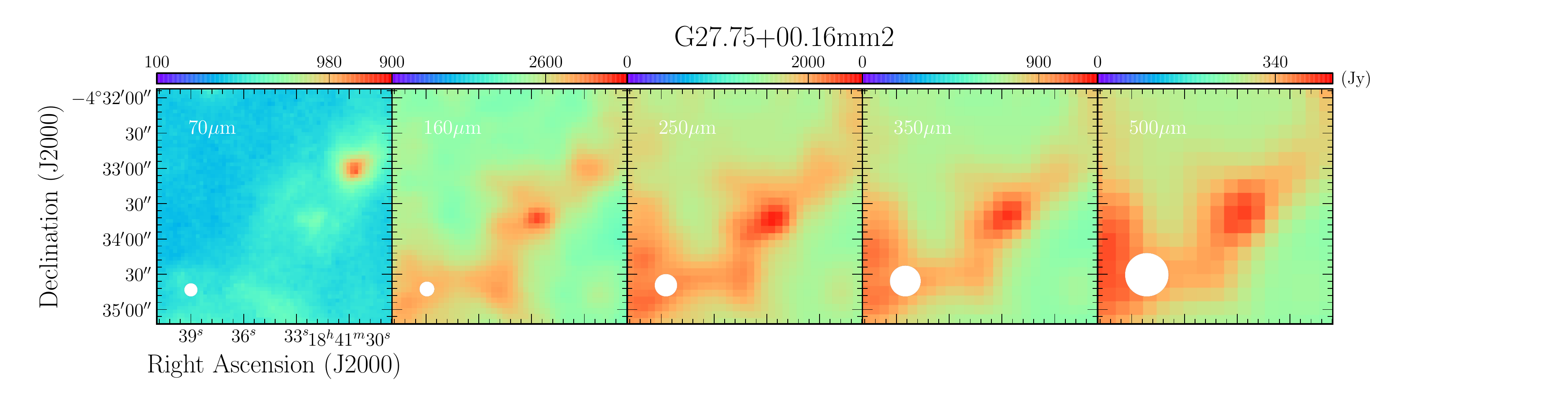}\hspace{-0.8cm}  &
 \includegraphics[width=0.52\textwidth,  trim = 20 20 20 15, clip, angle = 0]{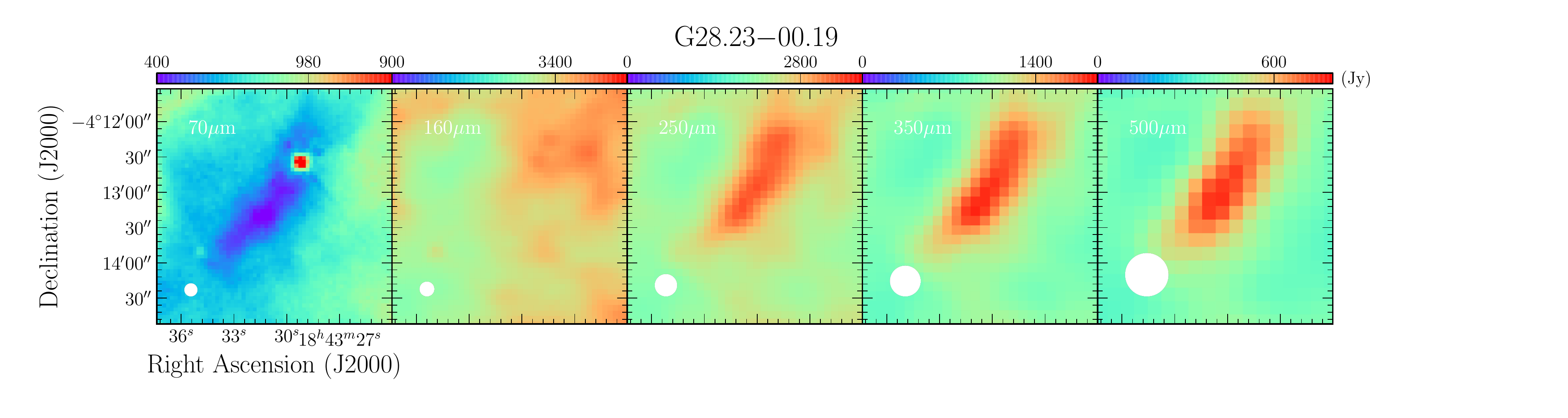} \\
 \includegraphics[width=0.52\textwidth,  trim = 20 20 20 15, clip, angle = 0]{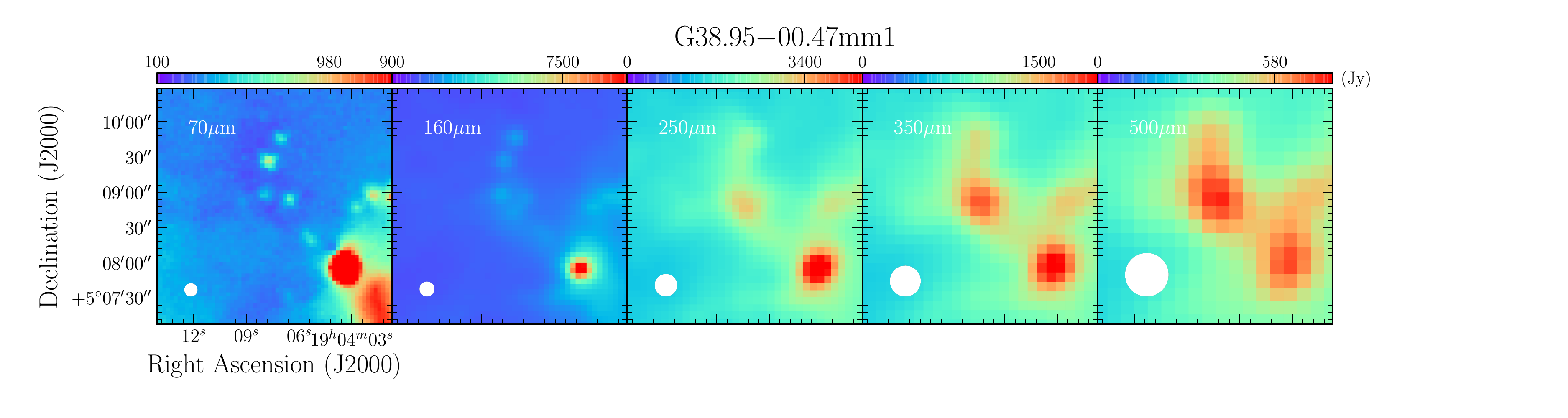}\hspace{-0.8cm}  &
 \includegraphics[width=0.52\textwidth,  trim = 20 20 20 15, clip, angle = 0]{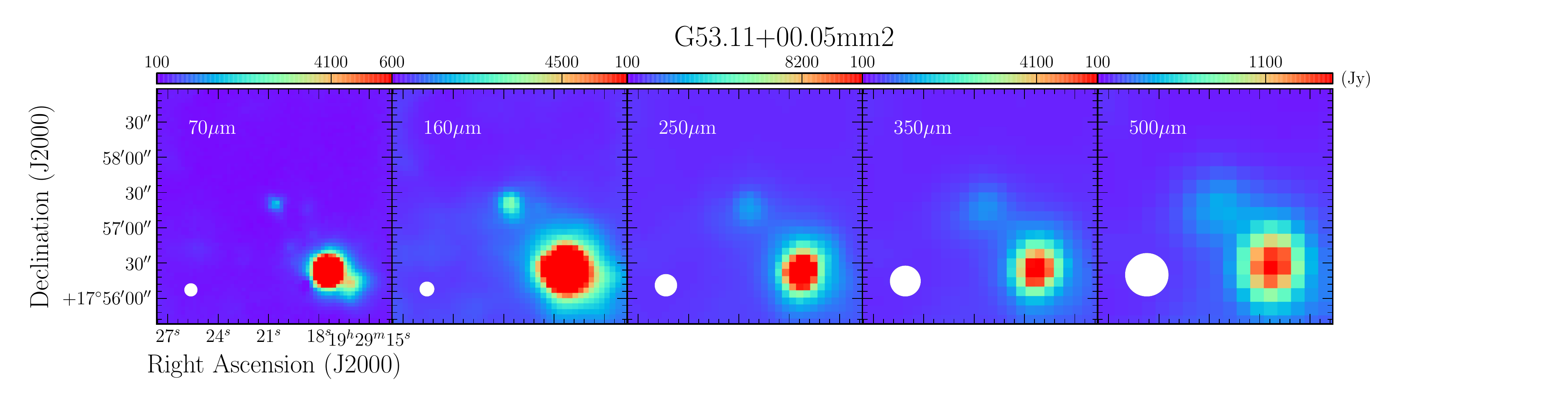} \\
 \includegraphics[width=0.52\textwidth,  trim = 20 20 20 15, clip, angle = 0]{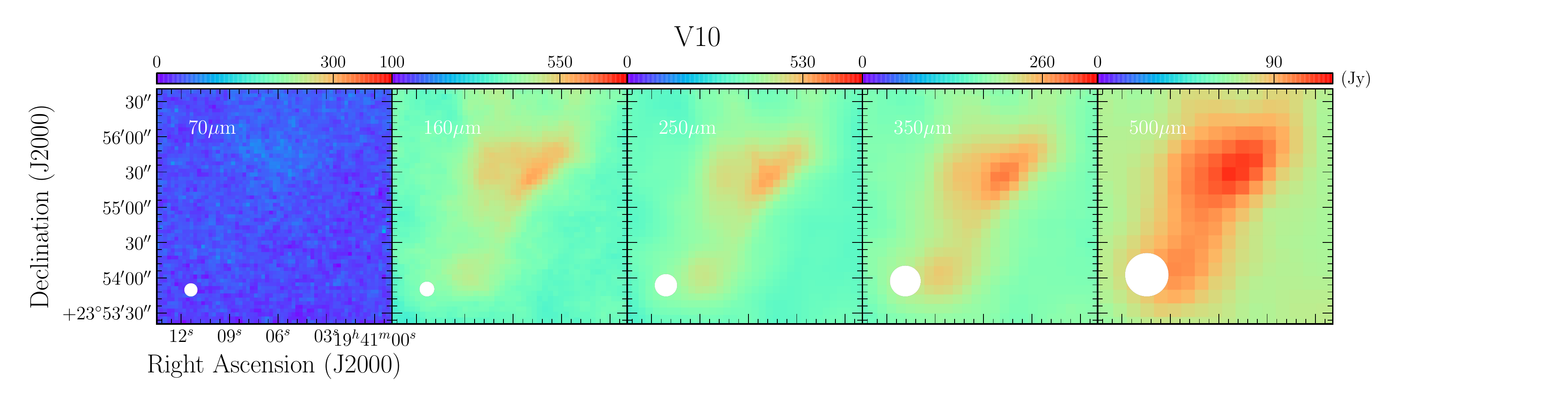}\hspace{-0.8cm}  &
 \includegraphics[width=0.52\textwidth,  trim = 20 20 20 15, clip, angle = 0]{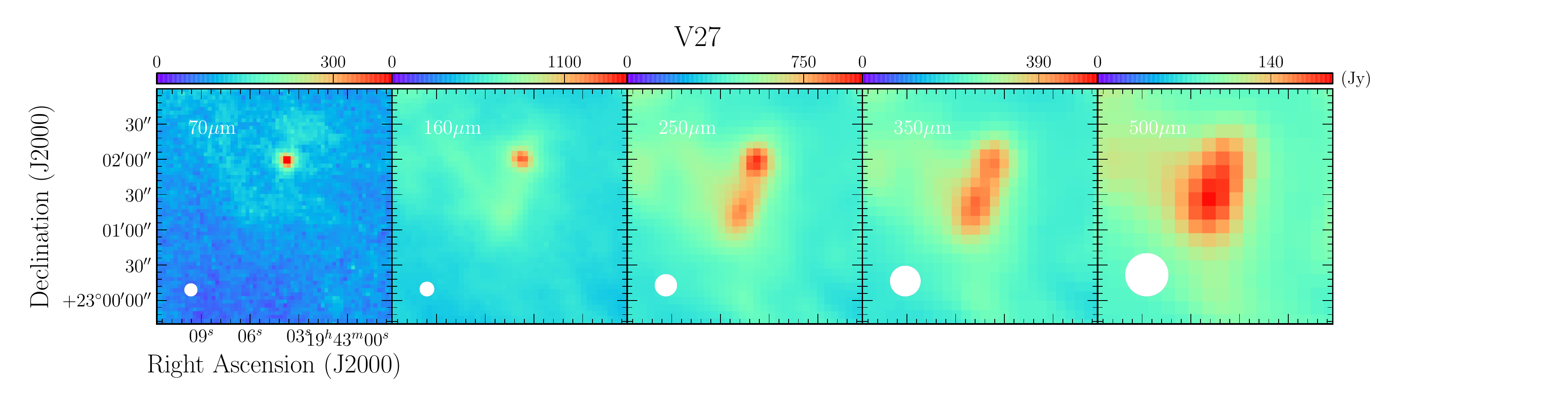} \\
 \includegraphics[width=0.52\textwidth,  trim = 20 20 20 15, clip, angle = 0]{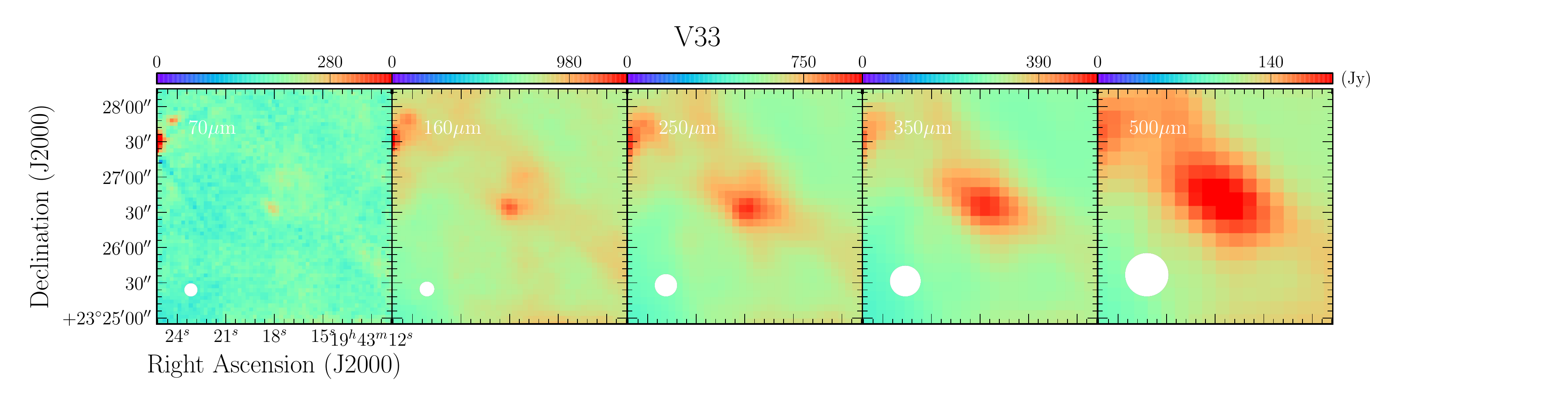}\hspace{-0.8cm}  
\end{tabular}

 \hspace*{\fill}%
\caption{\small{Continued.}}
 \label{fig:Higal_images}
\end{figure}

\begin{deluxetable}{l c c c c c c c c c c c }
\tabletypesize{\scriptsize}
\tablecaption{SED Parameters \label{SED_Parameters}}
\tablewidth{0pt}

\tablehead{
\colhead{Region}                  & 
\colhead{S$_{24\,\mu m}$}  &
\colhead{S$_{70\,\mu m}$}   &
\colhead{S$_{160\,\mu m}$ }               & 
\colhead{S$_{250\,\mu m}$ }  & 
\colhead{S$_{350\,\mu m}$ }             & 
\colhead{S$_{500\,\mu m}$ } & 
\colhead{S$_{870\,\mu m}$ } & 
\colhead{S$_{1100\,\mu m}$ }& 
\colhead{Distance\tablenotemark{$\star$}} & 
\colhead{Log$_{10}${\it L}\tablenotemark{$\dagger$}} & 
\colhead{Notes} \\[2pt]
\colhead{}                            & 
\colhead{(Jy)}                      & 
\colhead{(Jy)}                      & 
\colhead{(Jy)}                      & 
\colhead{(Jy)}                      & 
\colhead{(Jy)}                      &  
\colhead{(Jy)}                      & 
\colhead{(Jy)}                     & 
\colhead{(Jy)}                     &  
\colhead{(kpc)}                   & 
\colhead{(Log$_{10}${\it L$_{\odot}$})}   &   
\colhead{}                                \\[-20pt]\\}
\colnumbers
\startdata
LDN1657A$-$3        & \nodata          & \nodata          & \nodata          & \nodata          & \nodata          & \nodata          & \nodata          & \nodata          & 1.0              & \nodata          & \nodata         \\[3pt]
UYSO1               & \nodata          & \nodata          & \nodata          & \nodata          & \nodata          & \nodata          & \nodata          & \nodata          & 1.0              & \nodata          & \nodata         \\[3pt]
G11.11$-$0.12P1     & 1.75             & 79.88            & 133.78           & 110.59           & 68.39            & 42.79            & 15.00            & 8.56             & 3.6              & 3.31             & \nodata         \\[3pt]
18089$-$1732        & 12.31            & 1270.00          & 1580.00          & 685.30           & 452.33           & 212.59           & 45.63            & 15.25            & 2.34             & 4.01             & \nodata         \\[3pt]
G15.05+00.07mm1     & 0.10\tablenotemark{a} & 2.59\tablenotemark{a} & 100.49           & 71.69            & 42.13            & 19.21            & 2.32             & 0.84             & 2.5              & 2.23             & \nodata         \\[3pt]
18151$-$1208        & \nodata          & \nodata          & \nodata          & \nodata          & \nodata          & \nodata          & \nodata          & \nodata          & 2.8              & \nodata          & \nodata         \\[3pt]
G15.31$-$00.16mm3   & 0.91             & 16.27            & \nodata          & 34.79            & 23.98            & 12.13            & \nodata          & \nodata          & 3.0              & 2.52             & \nodata\tablenotemark{b}~ \tablenotemark{c} \\[3pt]
18182$-$1433        & 13.94            & 521.25           & 911.62           & 557.09           & 264.79           & 116.26           & 14.82            & \nodata          & 3.58             & 4.10             & \nodata         \\[3pt]
IRDC18223$-$3       & 0.31\tablenotemark{a} & 6.29             & 48.75            & 82.11            & 66.91            & 32.95            & 9.48             & 2.00             & 3.7              & 2.74             & \nodata         \\[3pt]
G18.82$-$00.28mm3   & 0.39             & 30.19            & 107.83           & 128.05           & 70.79            & 24.91            & 10.12            & 1.60             & 3.7              & 3.11             & \nodata         \\[3pt]
18264$-$1152        & 5.60             & 373.79           & 855.74           & 614.30           & 298.49           & 111.63           & 27.37            & 8.38             & 3.3              & 3.94             & \nodata         \\[3pt]
G22.35+00.41mm1     & 0.02             & 1.26             & 37.35            & 54.43            & 41.79            & 33.20            & 7.92             & 2.46             & 3.2              & 2.43             & \nodata         \\[3pt]
G22.73+00.11mm1     & 0.82             & 15.96            & 26.71            & 31.27            & 29.34            & 17.01            & 3.33             & \nodata          & 4.2              & 2.82             & (i)\tablenotemark{b}~ \tablenotemark{c} \\[3pt]
G23.60+00.00mm6     & 1.29             & 35.18            & 49.79            & 37.99            & 21.88            & 9.49             & \nodata          & \nodata          & 3.7              & 2.95             & (ii)\tablenotemark{b}~ \tablenotemark{c} \\[3pt]
G23.01$-$0.41       & 6.68             & 996.69           & 1560.00          & 1090.00          & 585.02           & 270.58           & 71.94            & 23.99            & 4.59             & 4.55             & \nodata         \\[3pt]
G24.33+00.11mm4     & 0.92\tablenotemark{a} & 22.17\tablenotemark{a} & 60.62            & 90.51            & 74.59            & 43.74            & 11.90            & 4.81             & 3.7              & 2.63             & (iii)\tablenotemark{b} \\[3pt]
18337$-$0743        & 19.46            & 407.39           & 342.56           & 182.35           & 87.86            & 37.94            & 4.82             & \nodata          & 3.8              & 3.95             & \nodata         \\[3pt]
18345$-$0641        & 9.69             & 139.83           & 199.04           & 137.33           & 67.91            & 39.19            & 4.17             & 1.13             & 5.2              & 3.87             & \nodata         \\[3pt]
G25.04$-$00.20mm1   & 0.53             & 21.95            & 155.96           & 253.55           & 151.45           & 92.72            & 24.55            & 6.25             & 4.3              & 3.39             & \nodata\tablenotemark{d} \\[3pt]
G25.04$-$00.20mm3   & 0.15\tablenotemark{a} & 3.27\tablenotemark{a} & 155.96           & 253.55           & 151.45           & 92.72            & 24.55            & 6.25             & 3.5              & 2.98             & \nodata\tablenotemark{d} \\[3pt]
G27.75+00.16mm2     & 0.15\tablenotemark{a} & 3.75\tablenotemark{a} & 16.78            & 18.89            & 19.28            & 7.28             & 1.32             & 1.02             & 3.5              & 1.94             & \nodata         \\[3pt]
G28.23$-$00.19mm1   & 1.14             & 25.14            & \nodata          & 86.22            & 71.08            & 43.17            & 15.87            & 5.11             & 4.1              & 3.09             & (i)\tablenotemark{b}~ \tablenotemark{c}~ \tablenotemark{e} \\[3pt]
G28.23$-$00.19mm3   & \nodata          & \nodata          & \nodata          & 131.62           & 95.99            & 46.39            & \nodata          & \nodata          & 5.1              & 2.67             & \nodata\tablenotemark{e} \\[3pt]
G28.53$-$00.25mm1   & 2.10\tablenotemark{a} & 43.91            & 107.09           & 245.98           & 149.55           & 79.56            & 45.41            & 26.09            & 4.4              & 3.37             & (iv)\tablenotemark{f}~ \tablenotemark{b}~ \tablenotemark{c} \\[3pt]
G28.53$-$00.25mm2   & 0.12\tablenotemark{a} & 2.22             & 36.00            & 53.64            & 30.80            & 15.36            & \nodata          & \nodata          & 4.4              & 2.65             & (v)(iv)\tablenotemark{f} \\[3pt]
G28.53$-$00.25mm4   & 0.31\tablenotemark{a} & 10.13\tablenotemark{a} & 49.30            & 75.11            & 59.75            & 27.65            & 45.41            & \nodata          & 5.4              & 2.88             & (vi)(iv)\tablenotemark{f}~ \tablenotemark{g} \\[3pt]
G28.53$-$00.25mm6   & 0.31\tablenotemark{a} & 10.13\tablenotemark{a} & 49.30            & 75.11            & 59.75            & 27.65            & 45.41            & \nodata          & 5.5              & 2.90             & (vi)(iv)\tablenotemark{f}~ \tablenotemark{g} \\[3pt]
18437$-$0216        & 10.01            & 221.98           & 260.35           & 197.03           & 103.44           & 55.81            & \nodata          & \nodata          & 7.3              & 4.31             & \nodata         \\[3pt]
18440$-$0148        & 9.82             & 333.87           & 314.08           & 160.75           & 69.37            & 23.55            & 2.74             & 1.08             & 8.3              & 4.53             & \nodata         \\[3pt]
G30.97$-$00.14mm1   & 6.78             & 130.15           & 155.99           & 136.68           & 104.85           & 76.50            & 17.45            & 6.52             & 5.0              & 3.79             & \nodata\tablenotemark{h} \\[3pt]
G30.97$-$00.14mm2   & 0.53\tablenotemark{a} & 24.59\tablenotemark{a} & 68.63            & 45.99            & 26.99            & 21.40            & \nodata          & \nodata          & 4.8              & 2.62             & (vii)\tablenotemark{h} \\[3pt]
18470$-$0044        & 11.09            & 466.09           & 625.48           & 345.01           & 148.20           & 65.22            & 14.15            & 3.49             & 8.2              & 4.72             & \nodata         \\[3pt]
G34.43+00.24mm1     & 6.40             & 824.25           & 1090.00          & 545.50           & 354.29           & 200.23           & 34.74            & 34.30            & 1.56/3.7\tablenotemark{i} & 3.49/4.24               & \nodata         \\[3pt]
G34.43+00.24mm2     & 10.81            & 308.29           & 626.98           & 575.98           & 321.41           & 165.57           & 40.67            & \nodata          & 3.7              & 3.98             & \nodata         \\[3pt]
18517+0437          & \nodata          & \nodata          & \nodata          & \nodata          & \nodata          & \nodata          & 18.90            & \nodata          & 1.88             & \nodata          & \nodata         \\[3pt]
18521+0134          & 8.65             & 191.91           & 273.66           & 162.42           & 75.65            & 31.05            & 4.00             & 2.50             & 9.1              & 4.46             & \nodata         \\[3pt]
G35.39$-$00.33mm2   & 0.89             & 75.12            & 114.36           & 75.19            & 34.74            & 17.82            & 2.27             & 0.60             & 2.3              & 2.83             & \nodata         \\[3pt]
G35.59$-$00.24mm2   & 0.22             & 3.56             & 10.12            & 16.39            & 18.94            & 10.03            & 2.54             & 0.53             & 2.3              & 1.84             & \nodata\tablenotemark{b} \\[3pt]
18553+0414          & 13.27            & 266.24           & 289.30           & 203.20           & 69.52            & 32.75            & 5.17             & 2.02             & 12.3             & 4.83             & \nodata         \\[3pt]
18566+0408          & 20.27            & 722.30           & 860.88           & 561.42           & 295.14           & 156.81           & 17.65            & 5.46             & 6.7              & 4.73             & \nodata         \\[3pt]
19012+0536          & 6.57             & 424.28           & 402.52           & 227.30           & 93.36            & 39.49            & 5.51             & 1.97             & 4.2              & 4.02             & \nodata         \\[3pt]
G38.95$-$00.47mm1   & 0.28\tablenotemark{a} & 2.33\tablenotemark{a} & 36.69            & 62.33            & 57.17            & 38.94            & 14.71            & 5.60             & 2.1              & 2.00             & (iii)           \\[3pt]
19035+0641          & 18.94            & 790.82           & 727.19           & 456.96           & 212.83           & 92.89            & 11.17            & 4.63             & 2.3              & 3.78             & \nodata         \\[3pt]
19266+1745          & 8.89             & 147.10           & 255.57           & 180.76           & 86.89            & 34.73            & 4.95             & \nodata          & 9.5              & 4.45             & \nodata         \\[3pt]
G53.11+00.05mm2     & 0.37             & 4.59             & 19.98            & 44.69            & 33.55            & 12.99            & 2.97             & \nodata          & 1.9              & 1.93             & \nodata         \\[3pt]
G53.25+00.04mm2     & 0.90             & 6.34             & 30.83            & 41.93            & 31.77            & 18.09            & 7.57             & 3.98             & 2.0              & 2.10             & \nodata         \\[3pt]
G53.25+00.04mm4     & 0.14             & 13.48            & 27.23            & 26.58            & 19.02            & 11.13            & \nodata          & \nodata          & 2.0              & 2.06             & \nodata         \\[3pt]
V10                 & 0.11\tablenotemark{a} & 1.63\tablenotemark{a} & 21.49            & 25.63            & 12.49            & 6.79             & \nodata          & \nodata          & 2.3              & 1.61             & (iii)           \\[3pt]
V11                 & 0.09             & 0.53             & 4.96             & 14.89            & 12.76            & 9.86             & \nodata          & \nodata          & 2.3              & 1.51             & \nodata         \\[3pt]
V27                 & 0.23\tablenotemark{a} & 5.03\tablenotemark{a} & 19.82            & 31.28            & 25.19            & 12.36            & \nodata          & \nodata          & 2.3              & 1.71             & (iv)            \\[3pt]
19411+2306          & 15.76            & 139.80           & 188.78           & 136.78           & 68.84            & 30.06            & 8.98             & \nodata          & 2.9/5.8\tablenotemark{i} & 3.39/3.99                & \nodata         \\[3pt]
V33                 & 0.09\tablenotemark{a} & 2.41\tablenotemark{a} & 18.62            & 22.13            & 14.94            & 10.76            & \nodata          & \nodata          & 2.3              & 1.59             & (iii)           \\[3pt]
19413+2332          & 11.57            & 120.89           & 179.25           & 157.97           & 78.16            & 38.28            & 7.47             & \nodata          & 1.8/6.8\tablenotemark{i} & 2.93/4.08             & \nodata         \\[3pt]
20126+4104          & \nodata          & \nodata          & \nodata          & \nodata          & \nodata          & \nodata          & \nodata          & \nodata          & 1.64             & \nodata          & \nodata         \\[3pt]
20216+4107          & \nodata          & 248.13           & 280.84           & 157.06           & 49.64            & 15.19            & \nodata          & \nodata          & 1.7              & 2.87             & \nodata         \\[3pt]
20293+3952          & \nodata          & 678.50           & 879.92           & 673.65           & 331.51           & 137.16           & \nodata          & \nodata          & 1.3/2.0\tablenotemark{i} & 3.14/3.51             & \nodata         \\[3pt]
20343+4129          & \nodata          & 468.34           & 629.64           & 452.00           & 200.76           & 77.17            & \nodata          & \nodata          & 1.4              & 3.04             & \nodata         \\[3pt]

\enddata
\tablenotetext{\text{$\star$}}{See \citetalias{2016ApJS..227...25R} for references.}
\tablenotetext{\text{$\dagger$}}{ This luminosity corresponds to  L$_{This\,work}$ plotted in Figure \ref{IRAS_Herschel}.}
\tablenotetext{\text{a}}{Upper limit at 24$\,\mu$m and 70$\,\mu$m.}
\tablenotetext{\text{b}}{Absorption in 24$\,\mu$m map.}
\tablenotetext{\text{c}}{Absorption in 70$\,\mu$m map.}
\tablenotetext{\text{d}}{Regions G25.04--00.20mm1 and G25.04--00.20mm3 are blended at  wavelengths $>$160$\,\mu$m. Listed luminosities are upper limits.}
\tablenotetext{\text{e}}{Regions G28.23-00.19mm1 and G28.23-00.19mm3 are blended.}
\tablenotetext{\text{f}}{Regions G28.53--00.25mm1, G28.53--00.25mm2, G28.53--00.25mm4 and G28.53--00.25mm6 are blended at  wavelengths $>$160$\,\mu$m. Listed luminosities are upper limits. Region G28.53--00.25mm1 is the dominant one in the map.}
\tablenotetext{\text{g}}{We used the same estimated fluxes for   G28.53--00.25mm4 and G28.53--00.25mm6.}
\tablenotetext{\text{h}}{Regions G30.97--00.14mm1 and G30.97--00.14mm2 are blended at  wavelengths $>$160$\,\mu$m. Listed luminosities are upper limits.  Region G30.97--00.14mm1 is the dominant one in the map.}
\tablenotetext{\text{i}}{Distance ambiguity.}

\tablecomments{(i) Polygon at 70$\,\mu$m used to calculate flux at 24$\,\mu$m; (ii) Polygon at 350$\,\mu$m used to calculate flux at 500$\,\mu$m;
(iii) Polygon at 250$\,\mu$m was used to estimate fluxes at the shortest wavelengths; (iv) Polygon at 160$\,\mu$m was used to estimate fluxes at the shortest wavelengths;
(v) Polygon at 250$\,\mu$m was used to estimate fluxes at the longest wavelengths; (vi) Polygon at 160$\,\mu$m was used to estimate fluxes at the largest wavelengths;
(vii) Polygon at 350$\,\mu$m was used to estimate fluxes at the shortest wavelengths.\\
In this paper, we have excluded 19282$+$1814 from the original survey sample. See  \citetalias{2016ApJS..227...25R} for details.}

\end{deluxetable}



\begin{figure}[!h]%
    \centering
    \includegraphics[width=0.8\linewidth, clip]{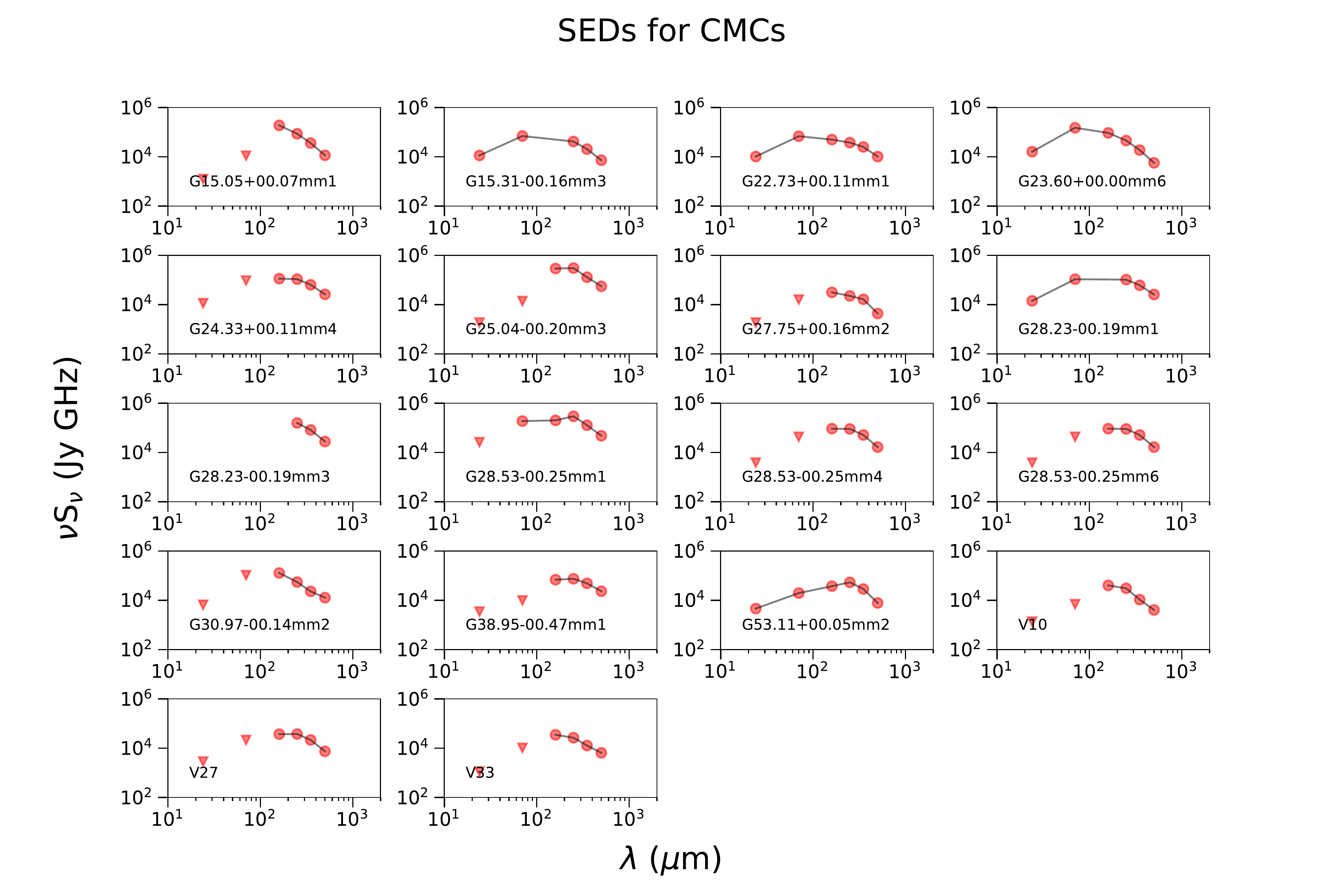}%
     \caption{Spectral energy distributions of the 18 CMC in our VLA survey with Hi--GAL counterparts. Additionally, we have included {\it Herschel}/MIPS 24$\, \mu$m, ATLASGAL 870\,$\mu$m and BGPS v2 1.1 mm flux densities. Triangles represent upper  limits of the flux densities. The solid black line represents the linear interpolation of the flux densities in the SED used to estimate the luminosity.}%
    \label{SED_CMC}%
\end{figure}

\begin{figure}[!h]%
    \centering
    \includegraphics[width=0.8\linewidth, clip]{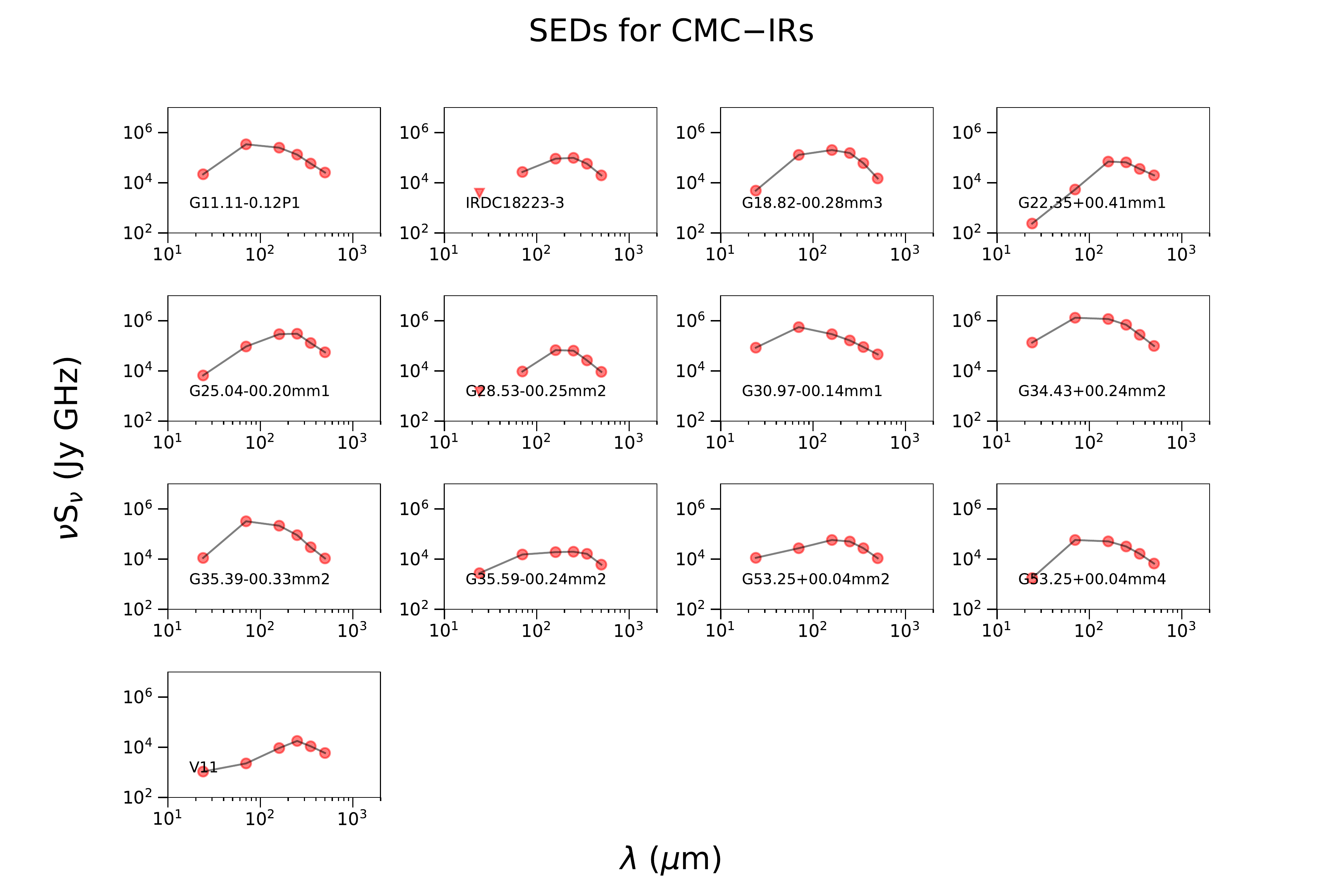}%
     \caption{Spectral energy distributions of  13 CMC---IR from our VLA survey with Hi--GAL counterparts. Additionally, we have included {\it Herschel}/MIPS 24$\, \mu$m, ATLASGAL 870\,$\mu$m and BGPS v2 1.1 mm flux densities. Triangles represent upper  limits of the flux densities. The solid black line represents the linear interpolation of the flux densities in the SED used to estimate the luminosity.}%
    \label{SED_CMC_IR}%
\end{figure}

\begin{figure}[!h]%
    \centering
    \includegraphics[width=0.8\linewidth, clip]{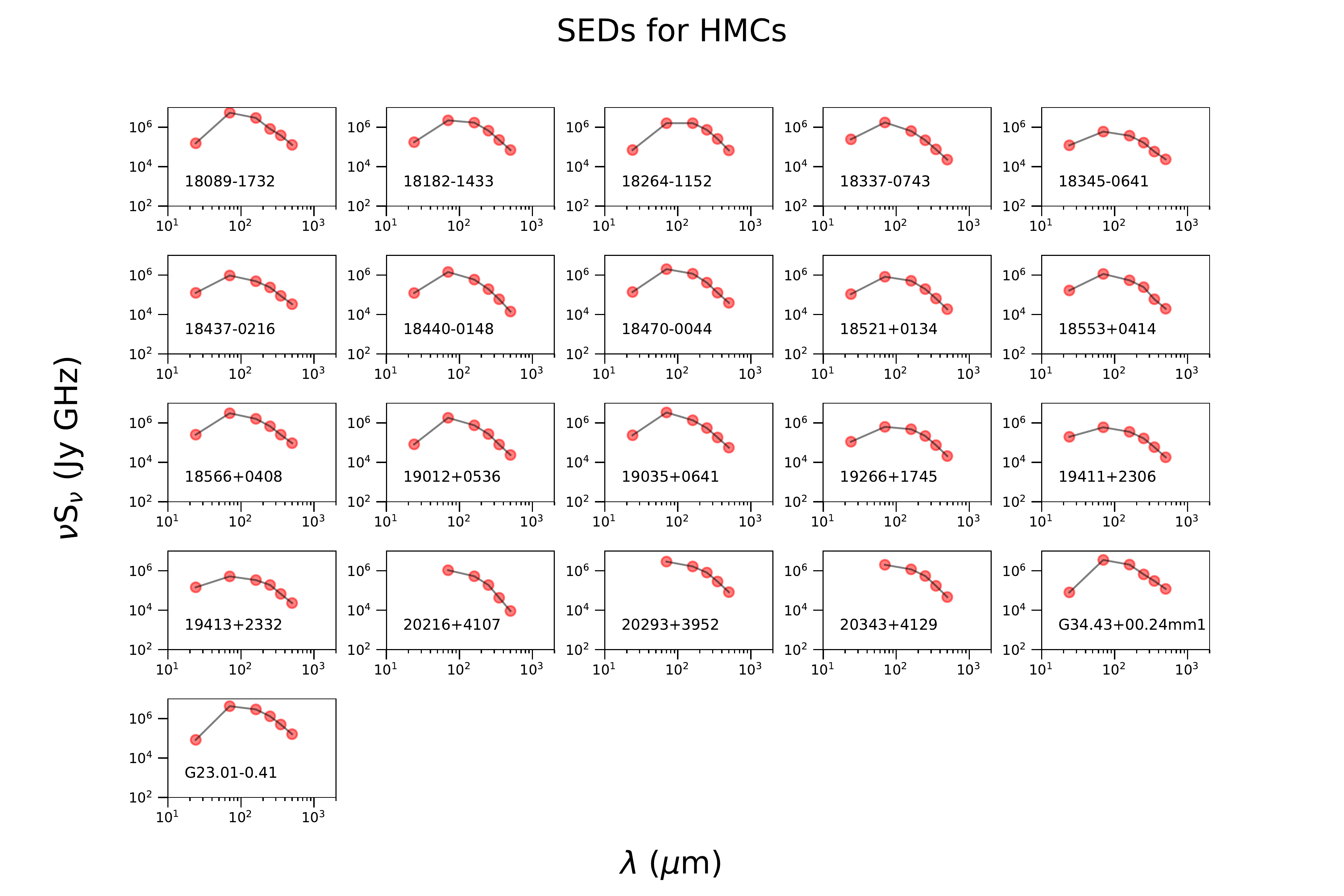}%
     \caption{Spectral energy distributions of 23 HMC from our VLA survey with Hi--GAL counterparts. Additionally, we have included {\it Herschel}/MIPS 24$\, \mu$m, ATLASGAL 870\,$\mu$m and BGPS v2 1.1 mm flux densities. Triangles represent upper limits of the flux densities. The solid black line represents the linear interpolation of the flux densities in the SED used to estimate the luminosity.}%
    \label{SED_HMC}%
\end{figure}

\begin{figure}[!h]%
    \centering
    \includegraphics[width=0.5\linewidth, clip]{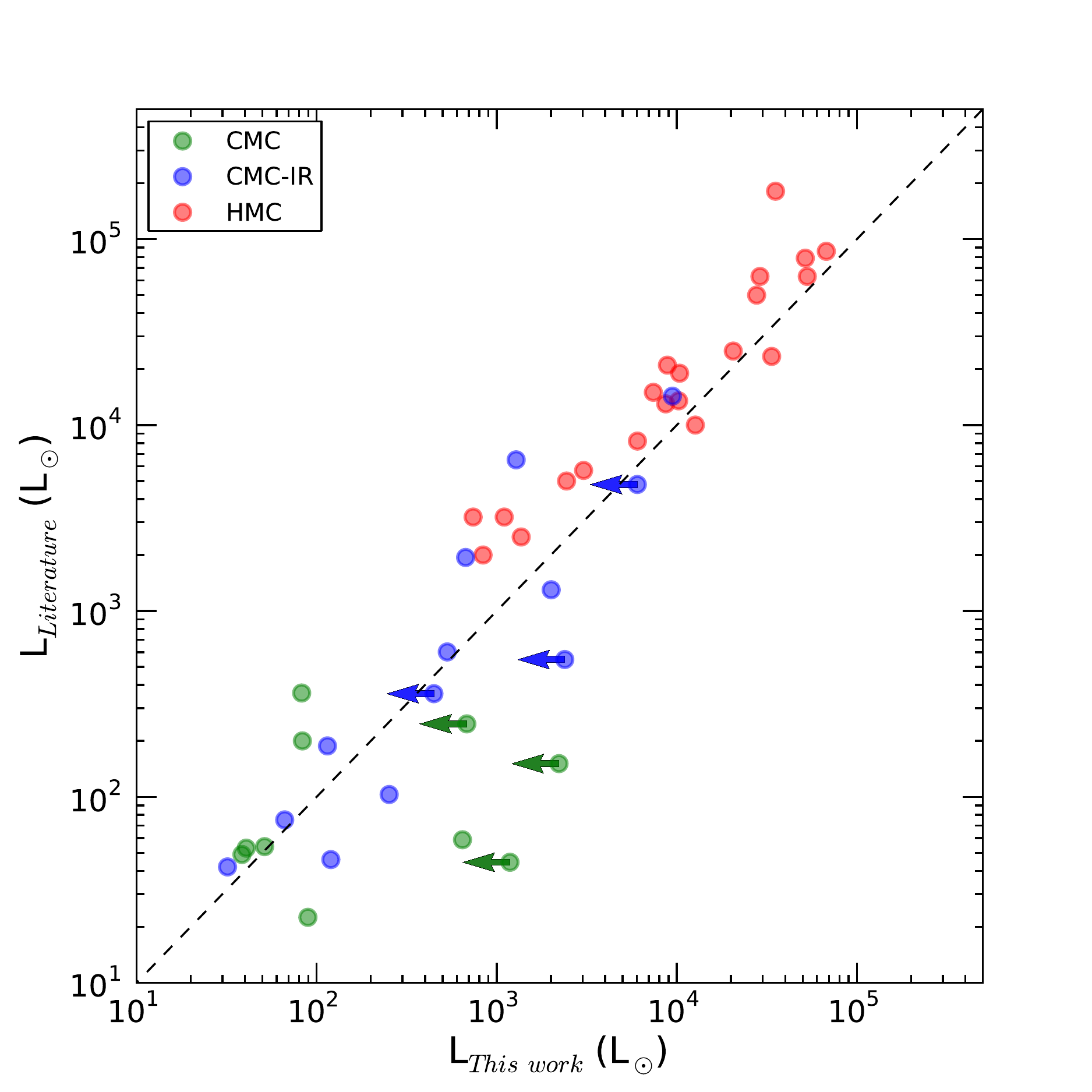}%
     \caption{Comparison between our luminosity estimates with respect to luminosities from the literature. Most of the previous luminosities estimated for the HMCs in our sample are IRAS luminosities. The dashed line corresponds to L$_{Lit}=$ L$_{This\,work}$. The arrows represent upper limits of our luminosity calculations for the given region. We are using the near distance for sources with distance ambiguity.  The values for  L$_{This\,work}$ and L$_{Lit}$ are presented in Table \ref{SED_Parameters} and \citetalias{2016ApJS..227...25R} Table 1, respectively.}%
    \label{IRAS_Herschel}%
\end{figure}


\clearpage
\section{Understanding the Momentum Rate of the Ionized Jet}\label{mom_rate_appe}

If we assume that our radio sources with rising spectrum have an ionized jet nature, we can use the models of \citet*{1987RMxAA..14..595C} and  \citet{1986ApJ...304..713R} to derive some interesting parameters of the jet in order to understand its behavior. From the shock- induced ionization model assuming a plane-parallel shock, \citet*{1987RMxAA..14..595C} and \citet{1989ApL&C..27..299C} derived a relationship between the momentum rate of the molecular outflow ($\dot{P}_{Curiel+}$) and the centimeter emission from the ionized jet given by \citep{1989ApL&C..27..299C,1998AJ....116.2953A, 2013A&A...551A..43J}:

\begin{equation}
\begin{aligned}
\left( \frac{\dot{P}_{Curiel+}}{M_{\odot}\,\text{yr}^{-1}\,\text{km}\,\text{s}^{-1}} \right) = \frac{3.13\times10^{-4}}
{\eta}\left(\frac{S_{\nu}d^{2}}{\text{mJy kpc}^{2}}\right)\left(\frac{v_{\star}} {200\,\text{km s}^{-1}}\right)^{0.32} 
\left(\frac{T} {10^{4}\,\text{K}}\right)^{-0.45} 
 \left(\frac{\nu} {5\,\text{GHz}}\right)^{0.1} \left(\frac{\tau} {1-e^{-\tau}}\right), 
\end{aligned}
\end{equation}
 
 \noindent
where $\eta$ is  the shock efficiency fraction\footnote{The shock efficiency $\eta=\frac{\Omega}{4\pi}$ is the fraction of the solid angle that is shocked \citep{1996ASPC...93....3A}.} observationally found to be only partial with  $\eta \sim$0.1 for low-mass stars \citep[e.g.,][]{1995RMxAC...1...67A, 1998AJ....116.2953A}. 
The initial velocity of the jet is ${\it v}_{\star}$ which for high-mass stars is $\sim$ 700 km s$^{-1}$ \citep[e.g.,][]{1994ApJ...430L..65R}, $T$ is the temperature of the ionized gas which is  assumed to be 10$^{4}$ K, $S_{\nu}$ is the observed flux density at a frequency $\nu$  and $\tau$ is the optical depth of the  ionized gas. If the flux is measured also at another frequency, $\nu^{\prime}$, one can measure the spectral index $\alpha$ and obtain $\tau$ from the expression:


\begin{equation}
\alpha = 2 + \frac{1}{\text{ln}\left( \frac{\nu}{\nu^{\prime}} \right)} \text{ln}\left[\frac{1- \frac{2}{\tau^{2}}\{ 1-(\tau + 1)e^{-\tau} \}}{1- \frac{2}{\left[\tau\left(\frac{\nu}{\nu^{\prime}}\right)^{2.1}\right]^{2}}\{ 1-(\tau\left(\frac{\nu}{\nu^{\prime}}\right)^{2.1} + 1)e^{-\tau\left(\frac{\nu}{\nu^{\prime}}\right)^{2.1}}\}}\right] .
\end{equation}

This relationship is equivalent to the one  derived by \citet{1998AJ....116.2953A} for a cylindrical distribution.

Similarly,  the momentum rate ($\dot{P}_{Reynolds}$) from a partially optically thick jet can be estimated using the model from  \citet{1986ApJ...304..713R}: 

\begin{equation}
\begin{aligned}
\left( \frac{\dot{P}_{Reynolds}}{M_{\odot}\,\text{yr}^{-1}\,\text{km}\,\text{s}^{-1}} \right) =  9.38\times 10^{-6} \left(\frac{\it{v_{jet}}} {100\,\text{km s}^{-1}}\right)^{2} \left(\frac{1} {x_{0}}\right) \left(\frac{\mu} {m_{p}}\right)  \left[\left(\frac{S_{\nu}} {\text{mJy}}\right) \left(\frac{\nu} {10\,\text{GHz}}\right)^{-\alpha}\right]^{0.75}  \left(\frac{d} {\text{kpc}}\right)^{1.5}  \left(\frac{\nu_{m}} {10\,\text{GHz}}\right)^{-0.45+0.75\alpha} \\
\times \theta_{0}^{0.75} \left(\frac{T} {10^{4}\,\text{K}}\right)^{-0.075} (\text{sin}\,i)^{-0.25}\,F^{-0.75}. 
\end{aligned}
\end{equation}

Most of these parameters are not known except for $\alpha$ and $S_{\nu}$ measured at $\nu$= 4.9 GHz, thus for the rest of parameters we assume  values that are typical of jets associated with luminous objects. The expanding velocity of the jet $v_{jet}$= 700 km s$^{-1}$, the opening angle $\theta_{0}=1\, $rad, the ionized gas temperature $T=10^{4}$ K, the inclination with the line of sight $i=45^{\circ}$, the hydrogen ionization fraction $x_{0}=0.1$  and  the ratio of the mean particle mass per hydrogen atom and the proton mass $\mu/m_{p}=1/(1+x_{0})$ \citep{1994ApJ...430L..65R}. The turn-over frequency value that we adopted for our calculations is $\nu_{m}=$ 50 GHz, which is the highest frequency of the VLA Q-band receiver. The parameter $F$ which is an index for the jet optical depth is estimated from Eq. (17) of \citet{1986ApJ...304..713R}:

\begin{equation}
F = \frac{(2.1)^{2}}{q_{\tau}(\alpha-2)(\alpha+0.1)} ,
\end{equation}

\noindent
and assuming the parameters for the standard case (isothermal, constant velocity and fully ionized jet) with $q_{\tau}=-3\epsilon$ and $\epsilon= \frac{0.7}{1.3-\alpha}$ (see \S \ref{sec:intro}).

\begin{figure}[htbp]
\centering
\includegraphics[width=.5\textwidth,clip=true, angle = 0]{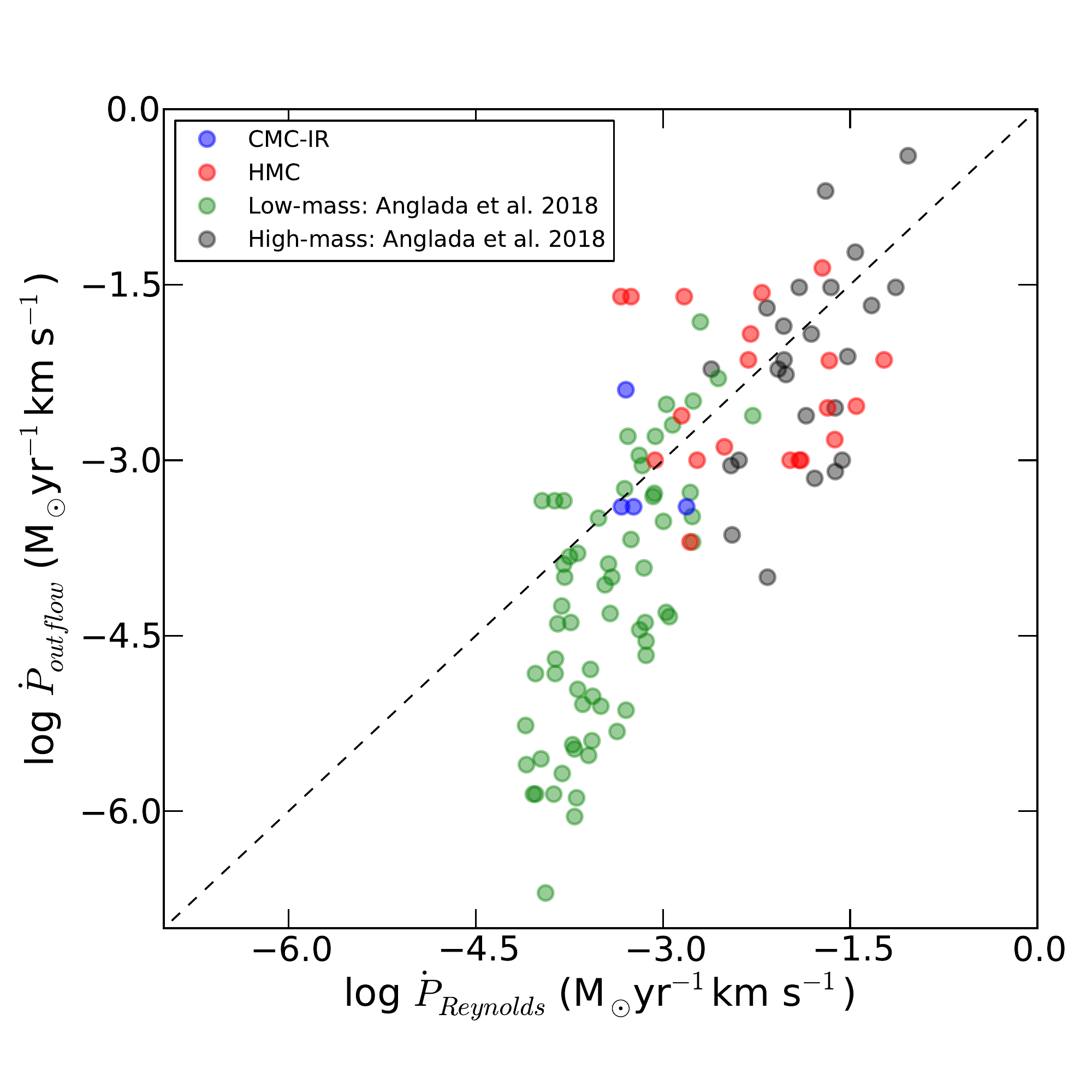}
\caption{\small{Momentum rate of the molecular outflow vs the momentum rate of an associated partially optically thick ionized jet as modeled by \citet{1986ApJ...304..713R} at $\nu=$ 4.9~GHz.  The green and gray circles represent ionized jets associated with low-mass   ($1\,L_{\odot} \leq L_{bol} \leq 1000\,L_{\odot}$)  and high-mass  ( $L_{bol} > 1000\,L_{\odot}$) protostars  from \citet{2018A&ARv..26....3A}, respectively. The momentum rate of the molecular outflows is data from the literature, usually from single dish observations.  The dashed line represents the case where the  momentum rate of the molecular outflow and the momentum rate of the ionized jet are the same. The momentum rate of the molecular outflow and the momentum rate of the associated ionized jet as estimated using the model of  \citet{1986ApJ...304..713R} is significantly similar for sources with $L_{bol} \gtrsim 100\,L_{\odot}$.}}
 \label{fig:mom_rates_v3}
\end{figure}

Figure \ref{fig:mom_rates_v3} shows the momentum rate of the molecular outflow ($\dot{P}_{outflow}$) vs the momentum rate of an associated partially optically thick ionized jet ($\dot{P}_{Reynolds}$) as modeled by \citet{1986ApJ...304..713R} at $\nu=$ 4.9~GHz assuming $\alpha = 0.6$ and the same parameters listed above. The green and gray circles represent ionized jets associated with low-mass   ($1\,L_{\odot} \leq L_{bol} \leq 1000\,L_{\odot}$)  and high-mass  ( $L_{bol} > 1000\,L_{\odot}$) protostars  from \citet{2018A&ARv..26....3A}, respectively, scaled to a frequency of $\nu=$ 4.9~GHz. The momentum rate of the molecular outflows is taken from the literature, usually from single dish observations.  The red and blue circles are our radio detections toward our HMCs and CMC--IRs, respectively, with $0.2 \leq \alpha \leq 1.2$ and for those sources that $\dot{P}_{outflow}$ is available in the literature.  The dashed line represents the case where the  momentum rate of the molecular outflow and the momentum rate of the ionized jet are the same. We noticed that the momentum rate of the molecular outflow and the momentum rate of the associated ionized jet as estimated using the model of  \citet{1986ApJ...304..713R} is significantly similar for sources with $L_{bol} \gtrsim 100\,L_{\odot}$.

\begin{figure}[htbp]
\centering
\includegraphics[width=.5\textwidth,clip=true, angle = 0]{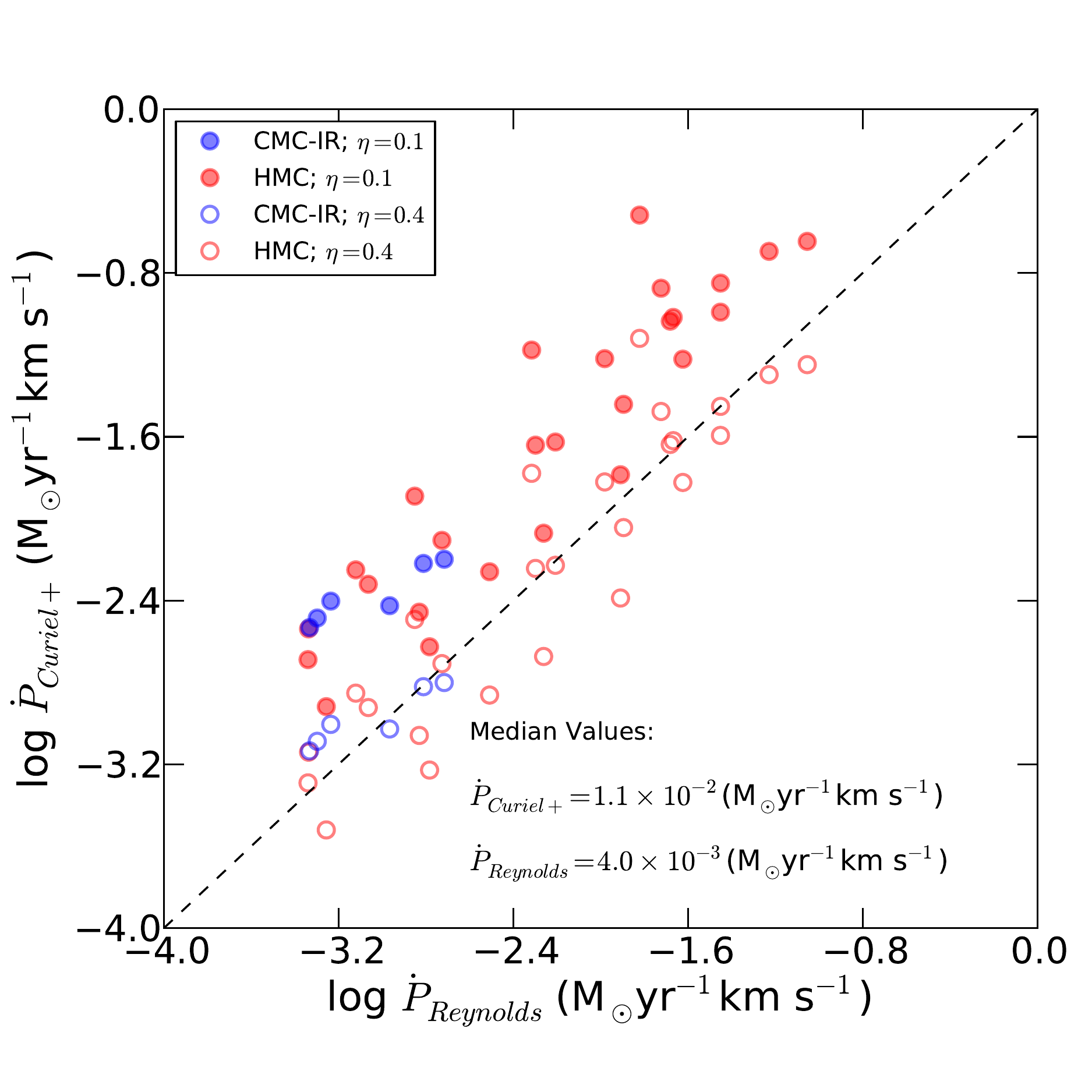}
\caption{\small{Momentum rate of an ionized jet  from a shock of a neutral wind against the surrounding high density envelope as modeled \citet*{1987RMxAA..14..595C} vs the momentum rate of a partially optically thick ionized jet as modeled by \citet{1986ApJ...304..713R} at $\nu=$ 4.9~GHz. The colored dots and the empty circles represent the momentum rate calculated from shock ionization assuming a shock efficiency of $\eta=$0.1 and 0.4, respectively. The dashed line represents the case where the  momentum rate of the jet is  sufficient to ionize itself. }}
 \label{fig:mom_rates_v2}
\end{figure}

Figure  \ref{fig:mom_rates_v2} shows the momentum rate of the ionized jet  as estimated from shock induced ionization \citep*{1987RMxAA..14..595C}  assuming that it is the same as the momentum rate of the molecular outflow  vs the momentum rate of a partially optically thick ionized jet as estimated using the model from \citet{1986ApJ...304..713R}. On average we found that the required momentum rate in the jet as estimated from shock ionization is larger than the momentum rate calculated from a collimated jet as derived by \citet{1986ApJ...304..713R} assuming a shock efficiency $\eta=$0.1 (colored dots in Figure  \ref{fig:mom_rates_v2}).  However, if the shock efficiency $\eta$ is increased to $\gtrsim$ 30$\%$  or if the jet velocities are $\sim$1600 km s$^{-1}$ the momentum rate of the jet would be sufficient to ionize itself (as an example,  the empty circles in Figure \ref{fig:mom_rates_v2} represent the momentum rate from ionized jets assuming $\eta=$0.4). For the very extreme cases where the momentum rate calculated from shock ionization is larger than the momentum rate calculated from Reynolds' model by a factor $\gtrsim$10 and the velocities are $\sim$ 700 km s$^{-1}$, the required shock efficiency would be required to be near  unity.


\section{Additional Tables}

\subsection{Table for Figure 3: Initial Str\"{o}mgren sphere radius as a function of the Lyman continuum for compact sources with rising spectra}

\begin{deluxetable}{l c c c c c}
\tabletypesize{\scriptsize}
 \renewcommand*{\arraystretch}{1.5}
\tablecaption{Table for Figure 3 \label{tab:fig3}}
\tablewidth{0pt}
\tablehead{
\colhead{Region}                  & 
\colhead{Radio Source}        &
\colhead{N$_{Ly}$}   &
 \colhead{EM}      &
 \colhead{n$_{e}$} &
  \colhead{R$_{s}$} \\
\colhead{}                  & 
\colhead{}        &
\colhead{($10^{43}$ s$^{-1}$)}   &
 \colhead{($10^{8}$ pc cm$^{-6}$)}      &
 \colhead{($10^{6}$ cm$^{-3}$)} &
  \colhead{(au)} \\  }
\startdata
\setcounter{iso}{0}	
18151-1208      & A  & 8.9 &  85.1&  10.3&  8.3   \\ 
18151-1208      & B  & 50.6&  11.9&   1.8&  36.3  \\ 
18182-1433      & D  & 13.6&  16.8&   3.2&  17.3 \\  
18182-1433      & G  & 8.3 &  51.9&  7.8 &  8.9   \\
18264-1152      & A  & 8.9 &  3.6 & 1.2  &  24.9  \\ 
18264-1152      & H  & 8.1&  1.9 & 0.8  &  29.9  \\ 
18264-1152      & E  & 13.8&  16.5&  3.3 & 15.3   \\
18264-1152      & D  & 10.0&  8.1&  2.3 & 15.7   \\
18264-1152      & F  & 82.4&  4.3 & 0.8 & 64.9   \\
18345-0641      & A  & 32.3&  4.1 & 0.9 & 46.9  \\ 
18440-0148      & A  & 302.0&  14.1&  1.4&  77.7  \\ 
18470-0044      & B  & 106.1 & 7.7  & 1.2 & 58.6   \\
18517+0437      & A  & 8.4&  10.7&  2.7&  14.9  \\ 
18521+0134      & A  & 216.2 & 13.8&  1.4&  68.3  \\ 
18553+0414      & A  & 581.0 & 1.4 & 0.2 & 301.6 \\  
18566+0408      & A  & 104.6 & 1.0  & 0.3 & 157.9  \\ 
18566+0408      & B  & 153.4 & 13.4&  1.5&  59.8  \\ 
19012+0536      & A  & 168.2 & 8.8  & 1.1 & 73.2   \\
19035+0641      & A  & 41.4&  8.0 & 1.5 & 37.7   \\
19413+2332      & A  & 9.2&  1.1 &  0.5&  44.3  \\ 
20293+3952      & A  & 2.8&  7.3&  2.7 & 10.2   \\ 
20293+3952      & B  & 4.2&  1.5 & 0.7 & 27.4  \\ 
20293+3952      & E  & 2.1&  12.4&  4.2&   7.3 \\       
20293+3952      & D  & 1.3&  5.5 & 2.7 & 7.8    \\ 
20343+4129      & B  & 2.8&  7.2 & 2.7 & 10.4   \\
G34.43+00.24mm1&  A  & 37.5&  5.6 & 1.2 & 40.1   \\ 
G23.01-0.41    &  A  & 188.2 &  4.4 &  0.7&  103.8\\   
UYSO1          &  A  & 3.1&  23.5&  6.0&  6.8  \\
UYSO1          &  B  & 1.0& 34.1&  11.3&  2.7   \\
G11.11-0.12P1  &  A  & 6.7&  3.8 & 1.4 & 21.1   \\
G11.11-0.12P1  &  C  & 6.1&  1.6 & 0.7&  33.3  \\ 
IRDC18223-3    &  A  & 5.4&  1.5 & 0.7 & 32.9   \\ 
IRDC18223-3    &  B  & 6.2& 32.2 & 6.2 & 8.8   \\ 
G35.39-00.33mm2&  A  & 7.8&  4.7 &  1.5&  21.7  \\  
G53.25+00.04mm2&  A  & 6.0&  3.1 & 1.2 & 21.8   \\
G53.11+00.05mm2&  A  & 1.5&  17.0&  5.7&  5.5   \\

\enddata
\tablecomments{For the 36 compact sources with rising spectrum shown in Figure 3}

\end{deluxetable}

\subsection{Table for Figure 4: Lyman continuum measured at 25.5 GHz as a function of the bolometric luminosity for all detected sources with flat or rising spectra in our sample.}

\begin{deluxetable}{l c c c c c}
\tabletypesize{\scriptsize}
 \renewcommand*{\arraystretch}{1.5}
\tablecaption{Table for Figure 4 \label{tab:fig4}}
\tablewidth{0pt}
\tablehead{
\colhead{Region}                  & 
\colhead{Radio Source}        &
 \colhead{S$_\nu$}      &
\colhead{d}   &
 \colhead{L$_{bol}$}      &
 \colhead{N$_{Ly}$} \\
\colhead{}                  & 
\colhead{}        &
\colhead{($\mu$Jy)}   &
 \colhead{(kpc)}      &
 \colhead{($10^{3}$ L$_{\odot}$)} &
  \colhead{($10^{43}$ s$^{-1}$)} \\  }
\startdata
\setcounter{iso}{0}	
18089-1732      & A& 1637 & 2.34& 10.2& 94.0 \\   
18151-1208      & A& 108  & 2.8 & 17.4\tablenotemark{a}& 8.9  \\  
18151-1208      & B& 616  & 2.8 & 17.4\tablenotemark{a}& 50.6 \\
18182-1433      & C& 686  & 3.58& 12.7& 92.3 \\
18182-1433      & E& 166  & 3.58& 12.7& 22.3 \\
18182-1433      & D& 101  & 3.58& 12.7& 13.6 \\
18182-1433      & G& 62   & 3.58& 12.7& 8.3  \\ 
18182-1433      & F& 105  & 3.58& 12.7& 14.1 \\
18264-1152      & A& 78   & 3.3 & 8.7 & 8.9 \\
18264-1152      & E& 121  & 3.3 & 8.7 & 13.8 \\
18264-1152      & D& 87   & 3.3 & 8.7 & 10.0 \\
18264-1152      & G& 102  & 3.3 & 8.7 & 11.6 \\
18264-1152      & F& 721  & 3.3 & 8.7 & 82.4 \\
18264-1152      & H& 71   & 3.3 & 8.7 & 8.1 \\
18345-0641      & A& 114  & 5.2 & 7.4 & 32.3 \\
18440-0148      & A& 418  & 8.3 & 33.6& 302.0\\ 
18470-0044      & C& 3494 & 8.2 & 51.7& 2464.5\\ 
18470-0044      & B& 150  & 8.2 & 51.7& 106.1 \\
18517+0437      & A& 226  & 1.88& 7.2\tablenotemark{a} & 8.4 \\
18521+0134      & A& 249  & 9.1 & 28.9& 216.2\\ 
18521+0134      & B& 378  & 9.1 & 28.9& 327.9\\ 
18553+0414      & A& 366  & 12.3& 67.6& 581.0\\ 
18566+0408      & A& 222  & 6.7 & 53.0& 104.6\\ 
18566+0408      & B& 326  & 6.7 & 53.0& 153.4\\ 
19012+0536      & A& 909  & 4.2 & 10.4& 168.2\\ 
19035+0641      & A& 746  & 2.3 & 6.0 & 41.4 \\
19035+0641      & B& 2269 & 2.3 & 6.0 & 125.9\\ 
19266+1745      & B& 82   & 9.5 & 27.7& 77.2 \\
19411+2306      & A& 247  & 2.9 & 2.4 & 21.8 \\
19413+2332      & A& 270  & 1.8 & 0.8 & 9.2 \\  
20126+4104      & A& 97   & 1.64& 10.0\tablenotemark{a}& 2.7 \\
20126+4104      & B& 942  & 1.64& 10.0\tablenotemark{a}& 26.6 \\
20216+4107      & A& 199  & 1.7 & 0.7 & 6.0 \\
20293+3952      & A& 155  & 1.3 & 1.4 & 2.8 \\
20293+3952      & C& 1556 & 1.3 & 1.4 & 27.6 \\
20293+3952      & B& 237  & 1.3 & 1.4 & 4.2 \\
20293+3952      & E& 120  & 1.3 & 1.4 & 2.1 \\
20293+3952      & D& 76   & 1.3 & 1.4 & 1.3 \\
20343+4129      & A& 881  & 1.4 & 1.1 & 18.1 \\
20343+4129      & B& 135  & 1.4 & 1.1 & 2.8 \\
G34.43+00.24mm1 & A& 1467 & 1.56& 3.0 & 37.5 \\
G23.01-0.41     & A& 852  & 4.59& 35.3& 188.2 \\
UYSO1           & A& 297  & 1.0 & 0.24\tablenotemark{a}& 3.1 \\ 
UYSO1           & B& 92   & 1.0 & 0.24\tablenotemark{a}& 1.0  \\
G11.11-0.12P1   & A& 49   & 3.6 & 2.0 & 6.7 \\
G11.11-0.12P1   & C& 45   & 3.6 & 2.0 & 6.1 \\
G11.11-0.12P1   & B& 78   & 3.6 & 2.0 & 10.6 \\
IRDC18223-3     & A& 38   & 3.7 & 0.5 & 5.4\\ 
IRDC18223-3     & B& 43   & 3.7 & 0.5 & 6.2\\ 
G35.39-00.33mm2 & A& 140  & 2.3 & 0.7 & 7.8\\ 
G53.25+00.04mm2 & A& 143  & 2.0 & 0.1 & 6.0\\ 
G53.25+00.04mm4 & A& 67   & 2.0 & 0.1 & 2.8\\ 
G53.11+00.05mm2 & A& 41   & 1.9 & 0.1 & 1.5\\ 

\enddata
\tablenotetext{\text{a}}{Bolometric luminosity is from the literature.}
\tablecomments{For the 54 sources with rising spectrum shown in Figure 4. All the fluxes are at 25.5 GHz. 18566 C does not have flux at 25.5 GHz. }

\end{deluxetable}

\subsection{Table for Figures 8 and 9: Radio luminosity at 4.9 GHz as a function of the bolometric luminosity and Momentum rate of the molecular outflow as a function of the radio luminosity at 4.9 GHz for the ionized jets and ionized jet/wind candidates in Figures \ref{fig:rad_bol_lum} and \ref{fig:Curiel_plot}, respectively.}

\begin{deluxetable}{l c c c  }
\tabletypesize{\scriptsize}
 \renewcommand*{\arraystretch}{1.5}
\tablecaption{Table for Figure 8 and 9 \label{c}}
\tablewidth{0pt}
\tablehead{
\colhead{Region}                  & 
 \colhead{L$_{bol}$}      &
 \colhead{S$_\nu$d$^{2}$}      &
\colhead{$\dot{P\,}$ }    \\
\colhead{}                  & 
 \colhead{($10^{3}$ L$_{\odot}$)} &
\colhead{($\mu$Jy kpc$^{2}$)}   &
 \colhead{(10$^{-3}$ M$_{\odot}$yr$^{-1}$km s$^{-1}$)}       \\  }
\startdata
\setcounter{iso}{0}	
18089-1732 & 10.2 & 2.27	& \nodata  \\
18151-1208 & 17.4\tablenotemark{a} & 1.34		& 1.5 \\
18182-1433 & 12.7 & 3.02	& 2.8 \\
18264-1152 & 8.7 & 2.95 	& 1.0 \\
18345-0641 & 7.4 & 1.12	& 12.0 \\
18440-0148 & 33.6 & 4.23	& \nodata \\
18470-0044 & 51.7 & 3.15	& 7.1 \\
18517+0437 & 7.2\tablenotemark{a} & 0.19		& 0.4 \\
18521+0134 & 28.9 & 4.13	& \nodata \\
18553+0414 & 67.6 & 34.84	& \nodata \\
18566+0408 & 53.0 & 12.13\tablenotemark{b}	& 7.2 \\
19012+0536 & 10.4 & 3.82	& 2.9 \\
19035+0641 & 6.0 & 0.98	& 1.0 \\
19411+2306 & 2.4 & 1.48	& 2.4 \\
19413+2332 & 0.8 & 0.64	& \nodata \\
20126+4104 & 10.0\tablenotemark{a} & 0.34		& 24.0 \\
20216+4107 & 0.7 & 0.15	& 1.3 \\
20343+4129 & 1.1 & 0.06	& 0.2 \\
G23.01-0.41 & 35.3 & 6.52	& 44.0 \\
UYSO1 & 0.24\tablenotemark{a} & 0.04 			& 11.0 \\
G11.11-0.12P1 & 2.0 &  1.15	& 0.4 \\
IRDC18223-3 & 0.5 & 0.38	& 4.0 \\
G35.39-00.33mm2 & 0.7 & 0.31	& \nodata \\
G53.25+00.04mm2 & 0.1 & 0.23	& \nodata \\

\enddata
\tablenotetext{\text{a}}{Bolometric luminosity is from the literature.}
\tablenotetext{\text{b}}{Radio sources C and D have  upper limit spectral indices that are consistent with being negative and their fluxes were not included in Figures \ref{fig:rad_bol_lum} and \ref{fig:Tanaka_tracks} (right panel). }
\tablecomments{ The fluxes are at 4.9 GHz. Radio source 20293+3952 E has an upper limit value for the flux at 4.9 GHz and its value is not included in these correlations.}

\end{deluxetable}

\section{Spectra of the compact radio sources with rising spectral index}\label{all_HII_fit}

\begin{figure}[htbp]
\centering
\vspace{.2cm} 
\begin{tabular}{ccc}
 \hspace{-0.6cm} 
    \includegraphics[width=0.34\linewidth, clip]{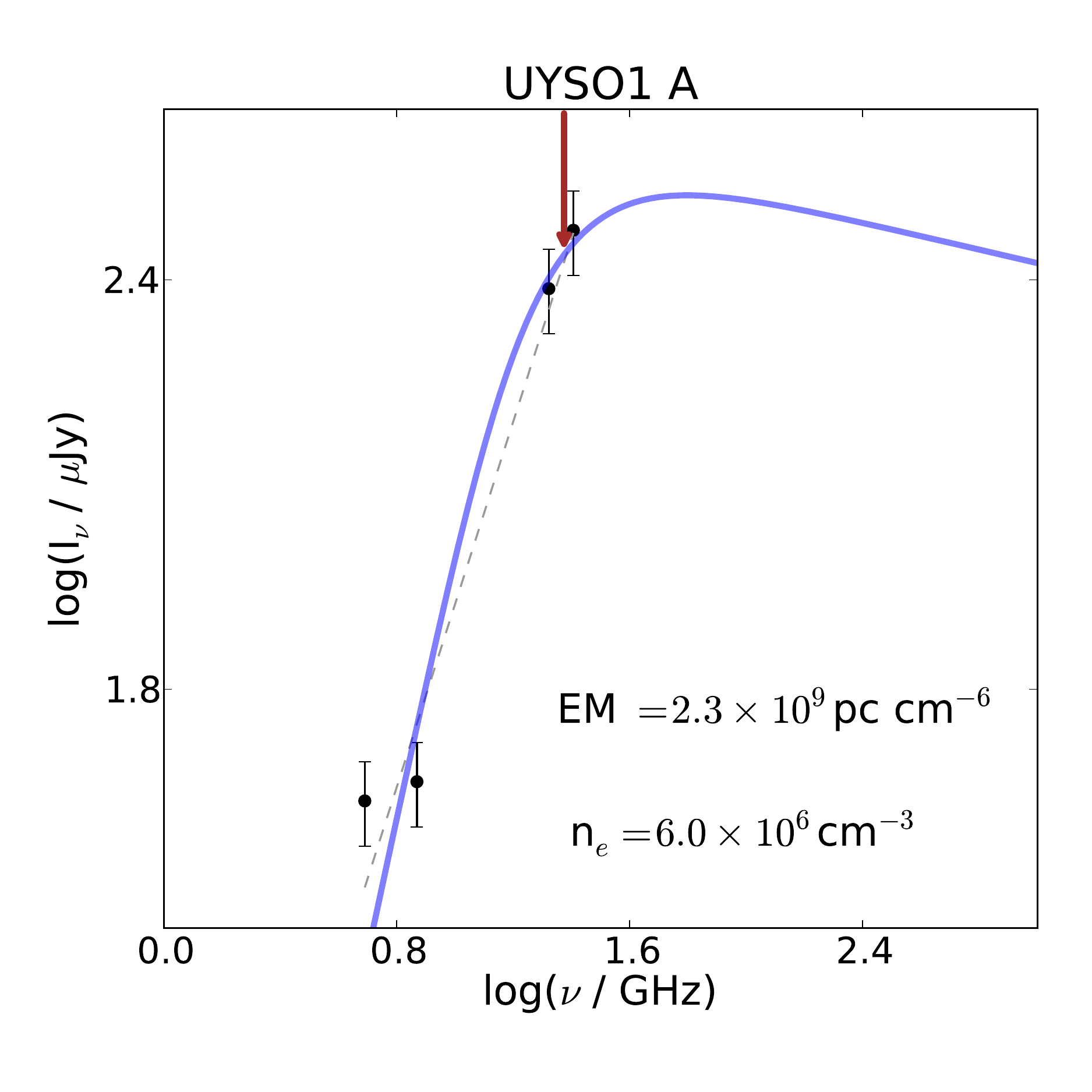} \vspace{-.2cm}  &
    \hspace{-0.8cm} 
    \includegraphics[width=0.34\linewidth, clip]{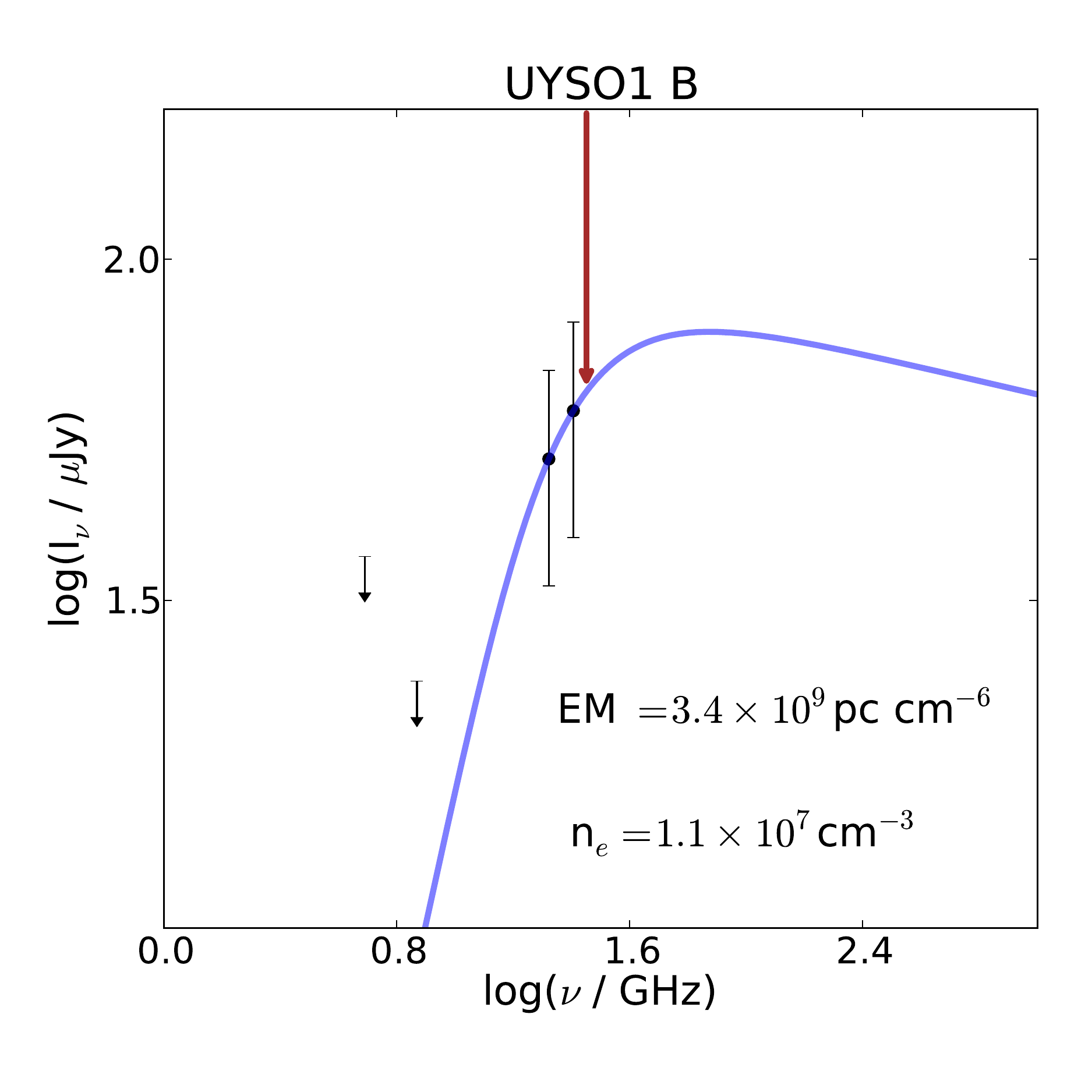}\vspace{-.2cm}  &
    \hspace{-0.8cm} 
    \includegraphics[width=0.34\linewidth, clip]{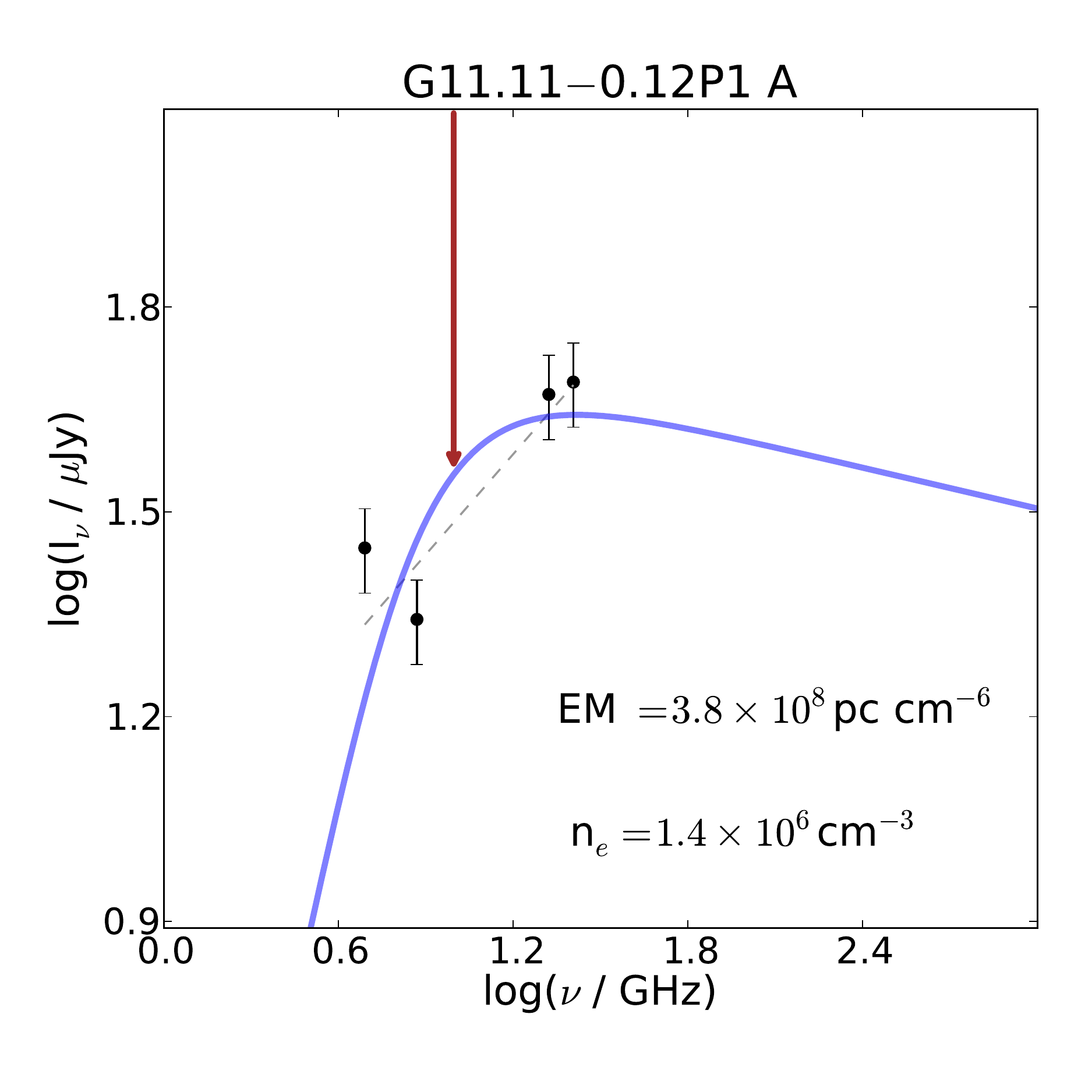}\vspace{-.2cm}\\
  \hspace{-0.6cm} 
        \includegraphics[width=0.34\linewidth, clip]{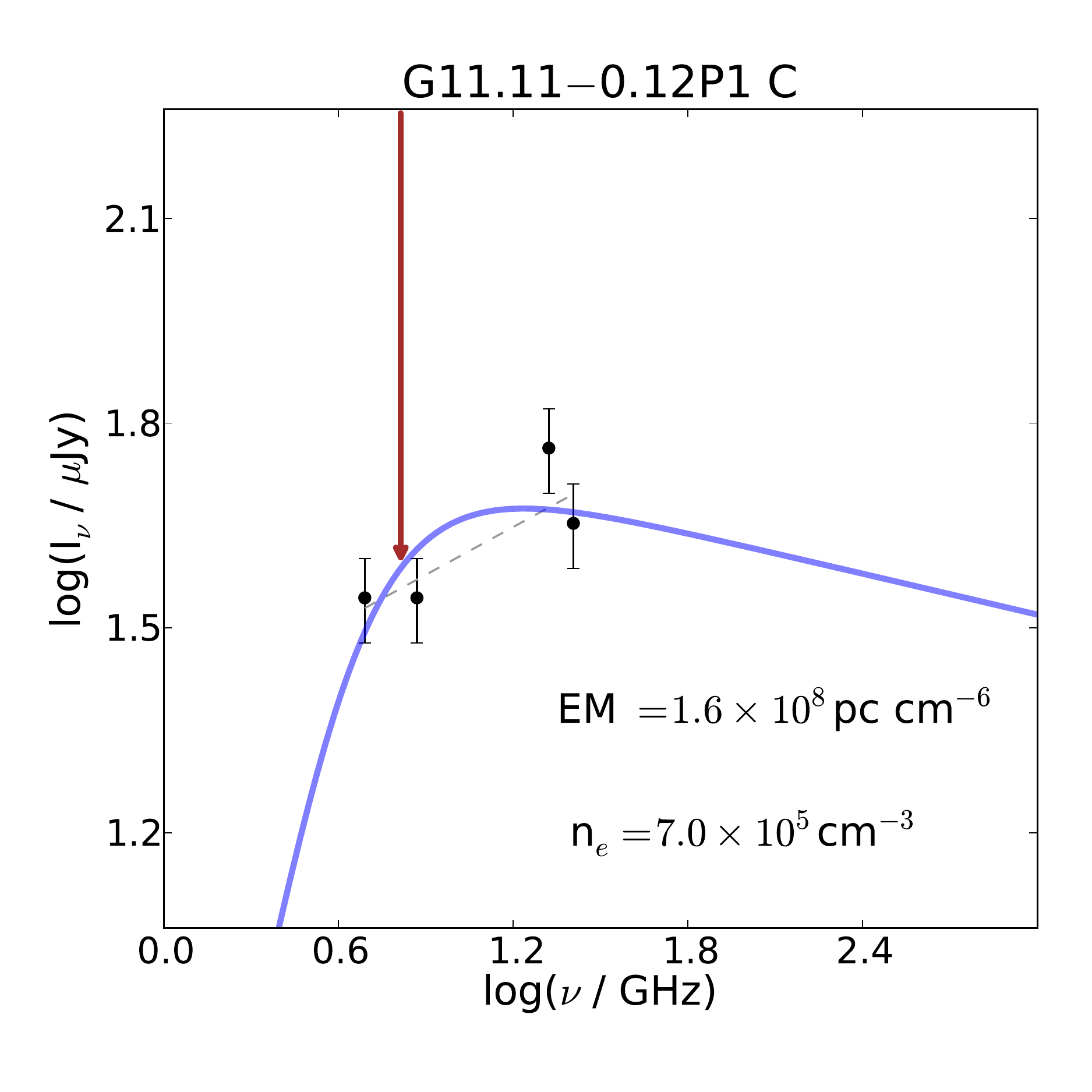}\vspace{-.2cm}  &
        \hspace{-0.8cm} 
    \includegraphics[width=0.34\linewidth, clip]{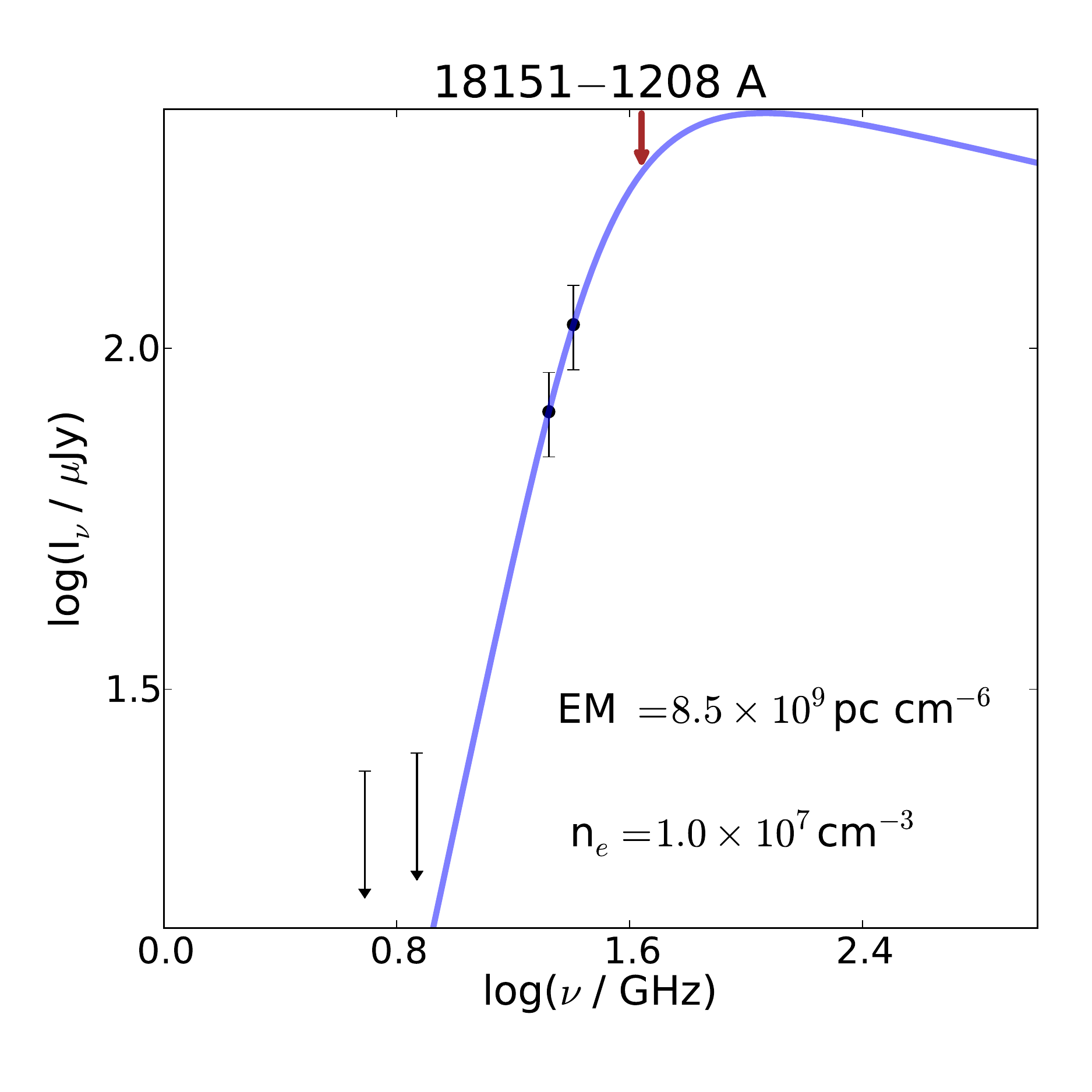}\vspace{-.2cm}  &
    \hspace{-0.8cm} 
        \includegraphics[width=0.34\linewidth, clip]{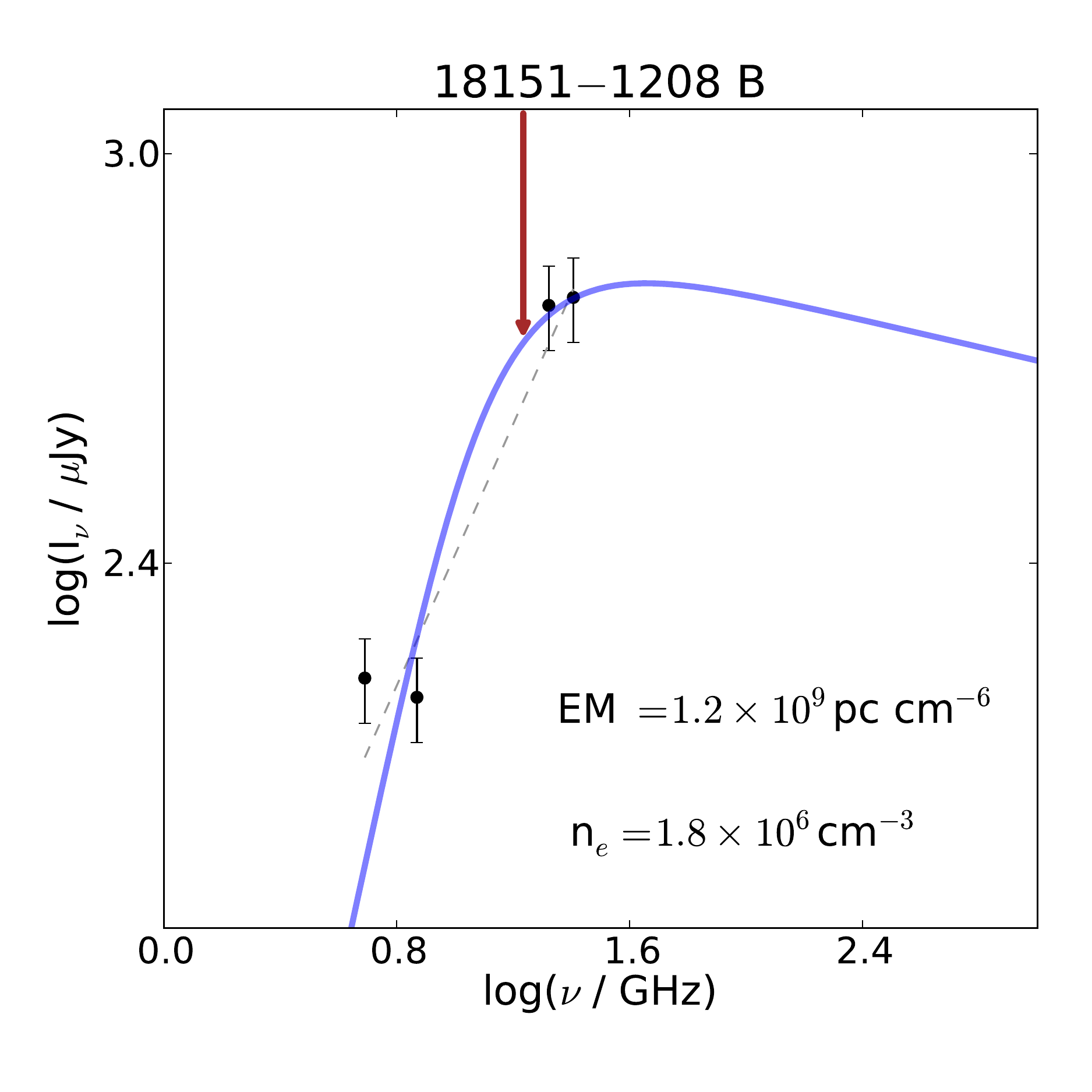}\vspace{-.2cm}\\ 
         \hspace{-.6cm} 
    \includegraphics[width=0.34\linewidth, clip]{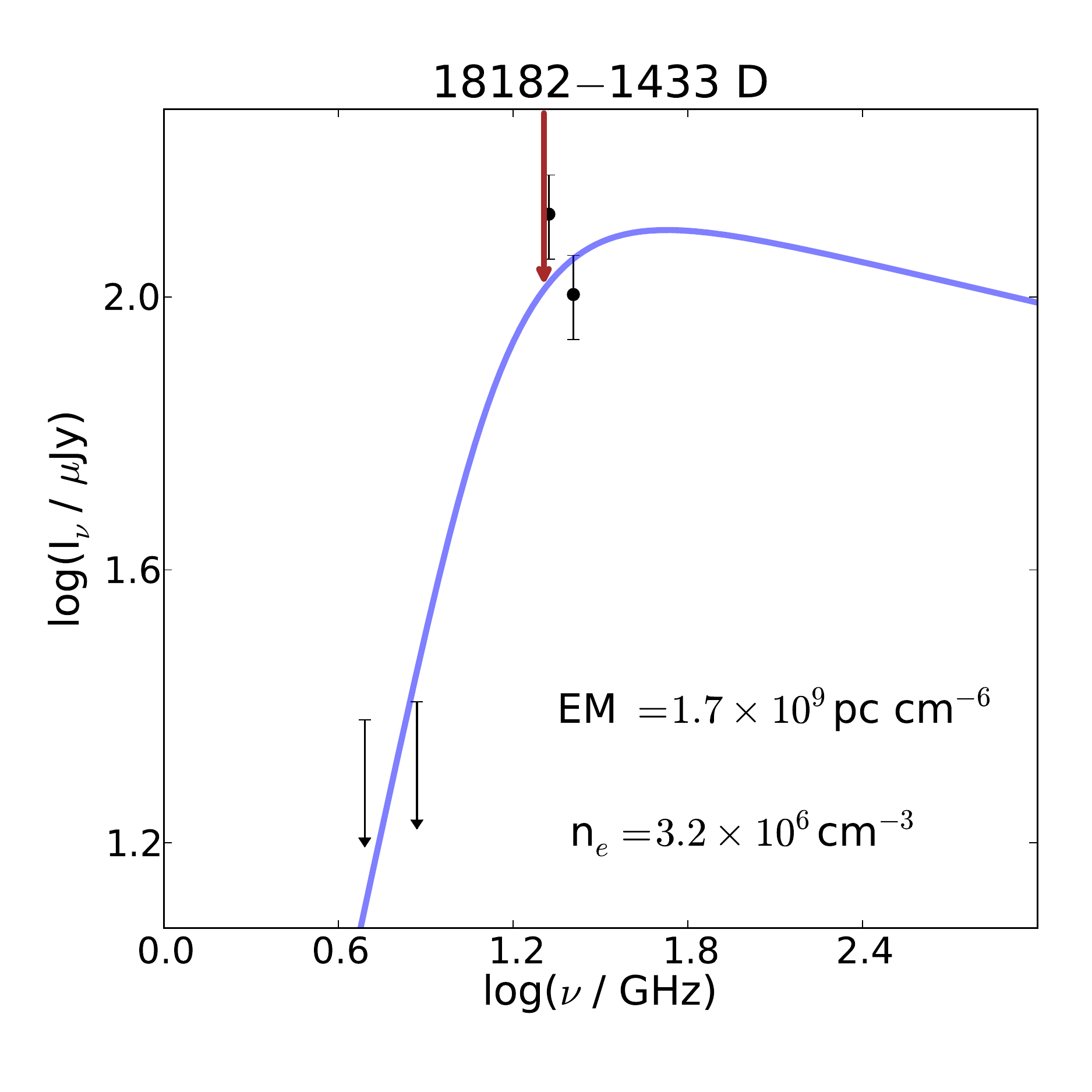}\vspace{-.2cm}    &
    \hspace{-0.8cm} 
    \includegraphics[width=0.34\linewidth, clip]{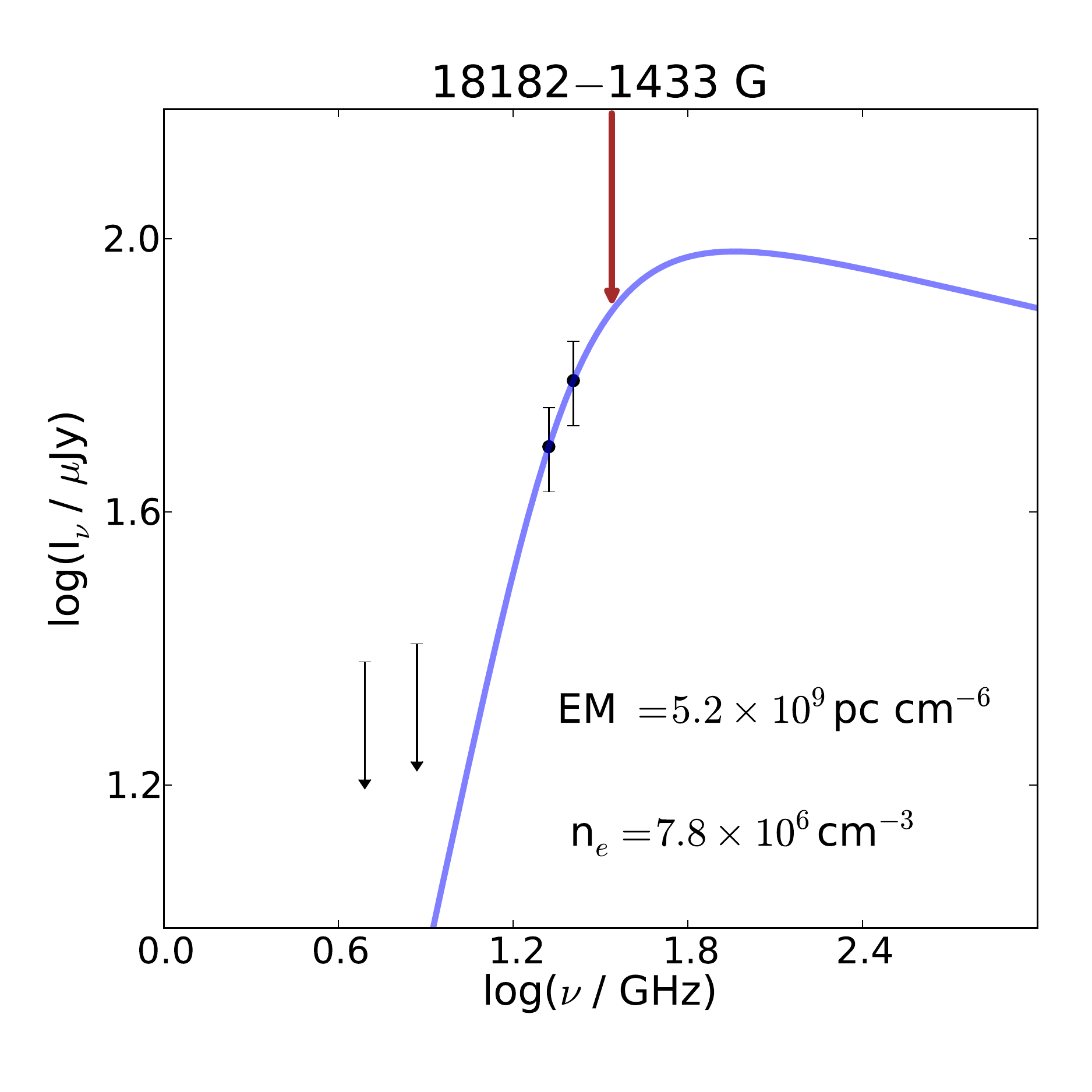}\vspace{-.2cm}    &
    \hspace{-0.8cm} 
        \includegraphics[width=0.34\linewidth, clip]{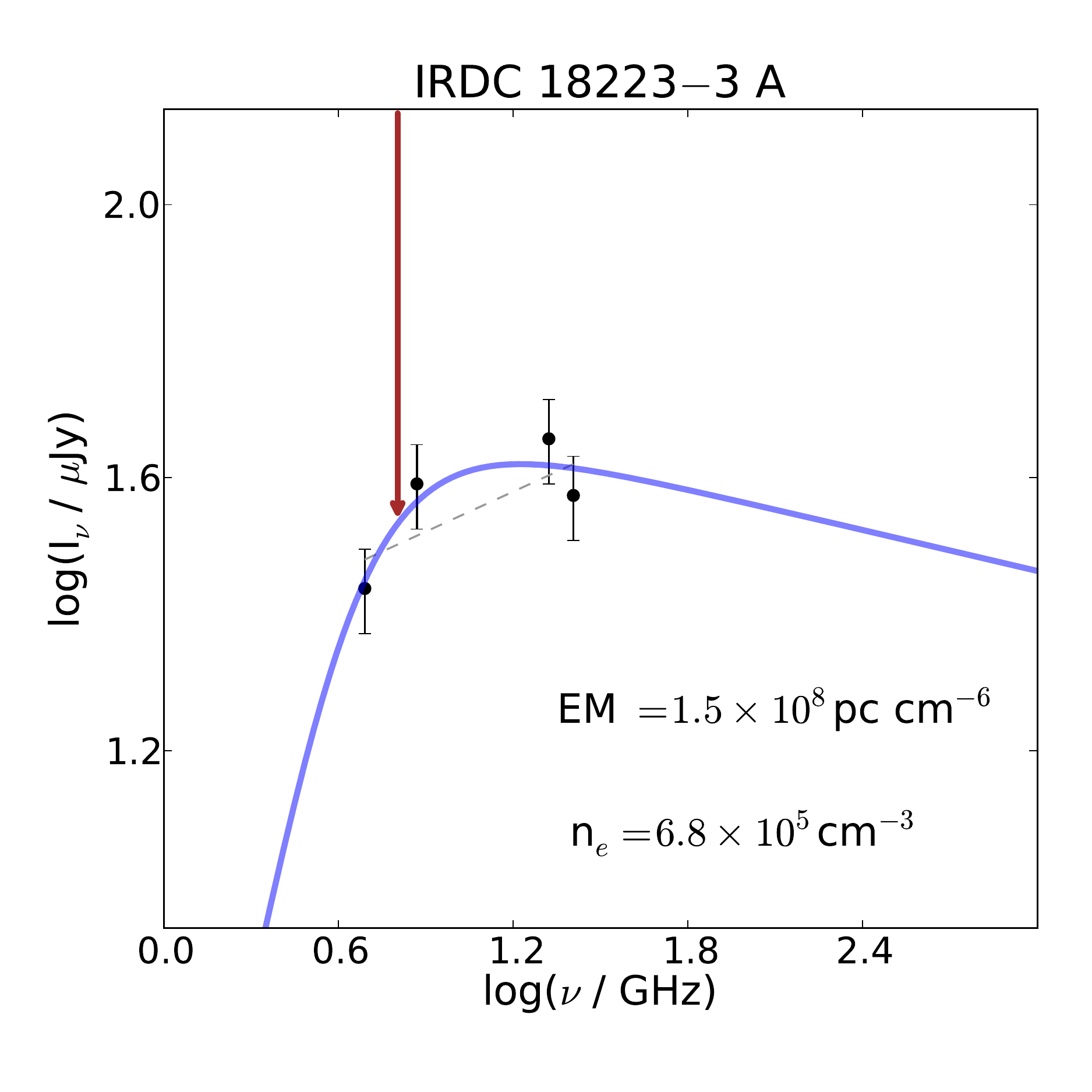}\vspace{-.2cm}\\
        \hspace{-.6cm} 
    \includegraphics[width=0.34\linewidth, clip]{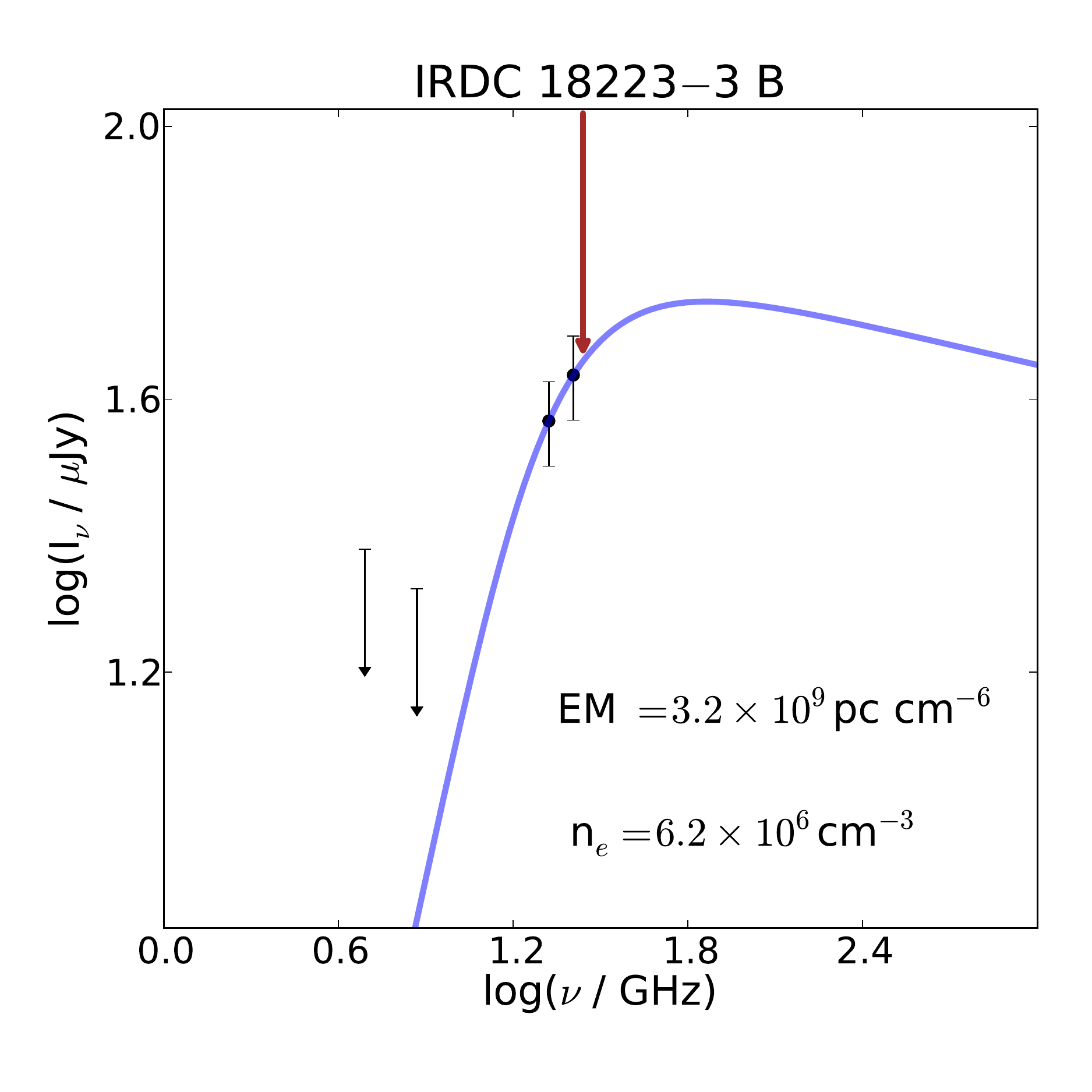} \vspace{-.2cm}&
  \hspace{-0.8cm} 
     \includegraphics[width=0.34\linewidth, clip]{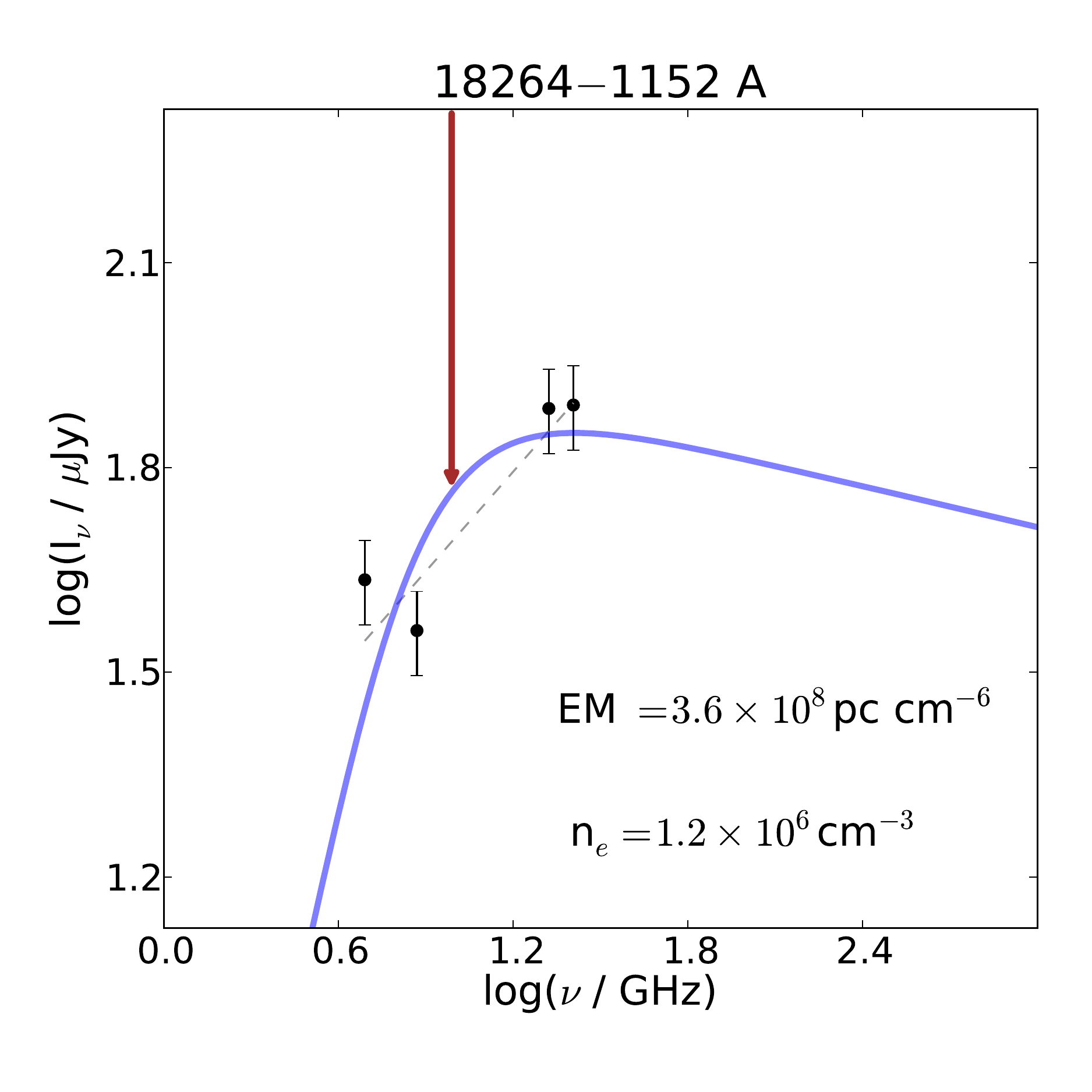} \vspace{-.2cm}&
   \hspace{-0.8cm} 
         \includegraphics[width=0.34\linewidth, clip]{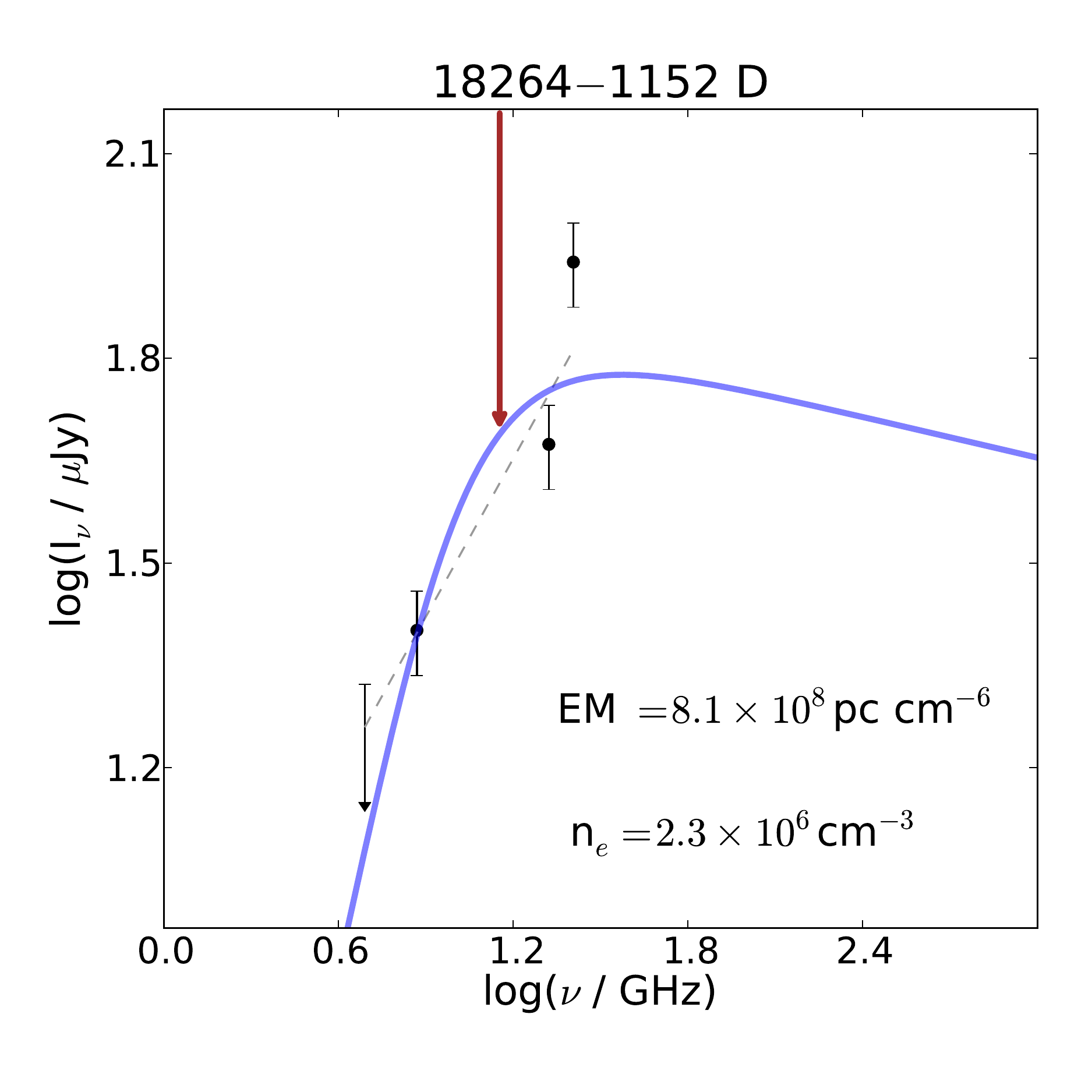} \vspace{-.2cm}\\
\end{tabular}
     \caption[Spectra of the compact radio sources with rising spectral index]{\small{Spectra of the compact radio sources with rising spectral index. Error bars are an assumed uncertainty of 10$\%$ from the flux densities added in quadrature with an assumed 10$\%$ error in calibration. The  continuous blue line is the HII region fit using a spherical distribution. The dashed line is the best fit to the data from a power law of the form $S_{\nu}$ $\propto$ $\nu^{\alpha}$.}}%
    \label{HII_fit_app}%
\end{figure}

\begin{figure}
\centering
  \ContinuedFloat
\begin{tabular}{ccc}

\hspace{-.6cm}
    \includegraphics[width=0.34\linewidth, clip]{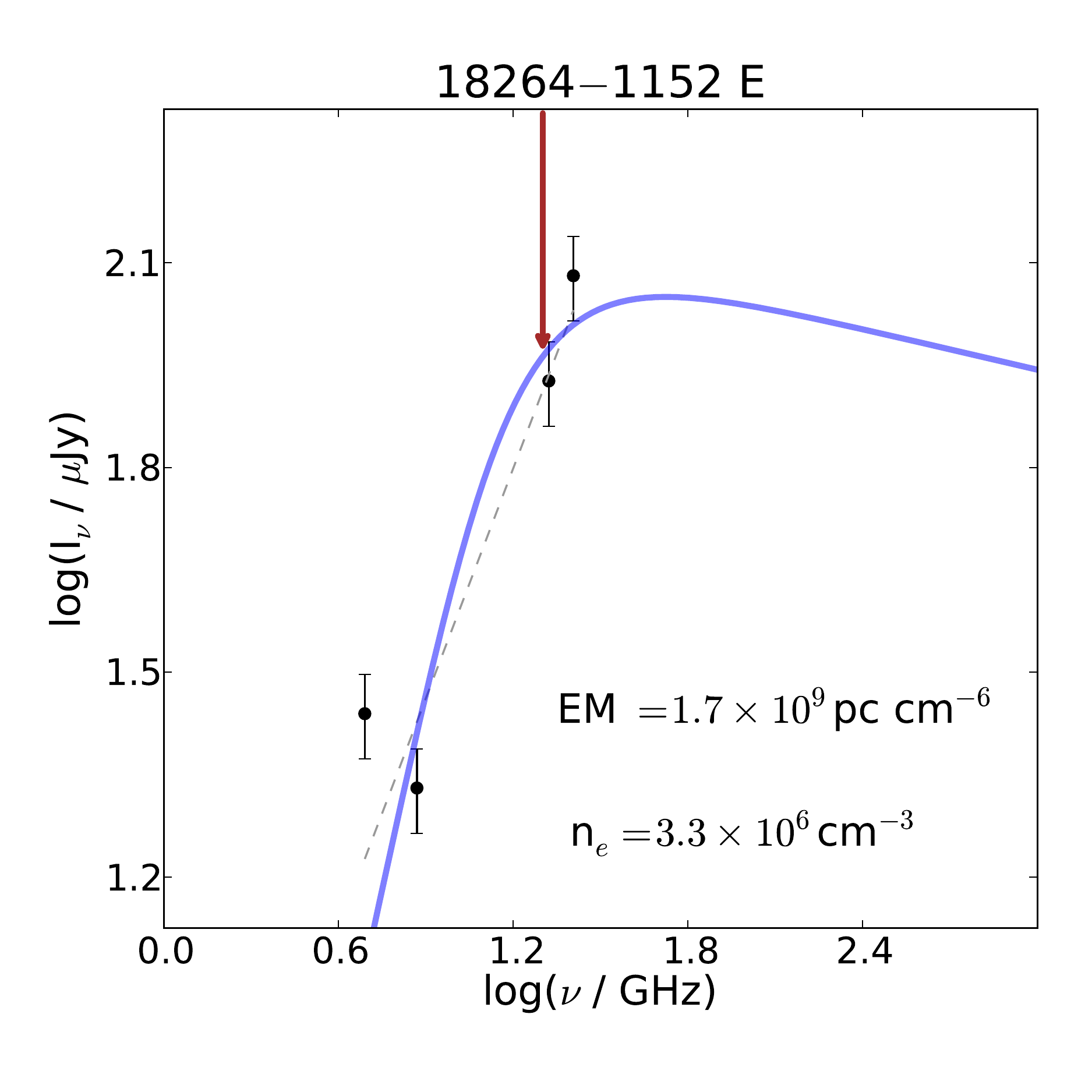} \vspace{-.2cm}  &
    \hspace{-0.8cm} 
    \includegraphics[width=0.34\linewidth, clip]{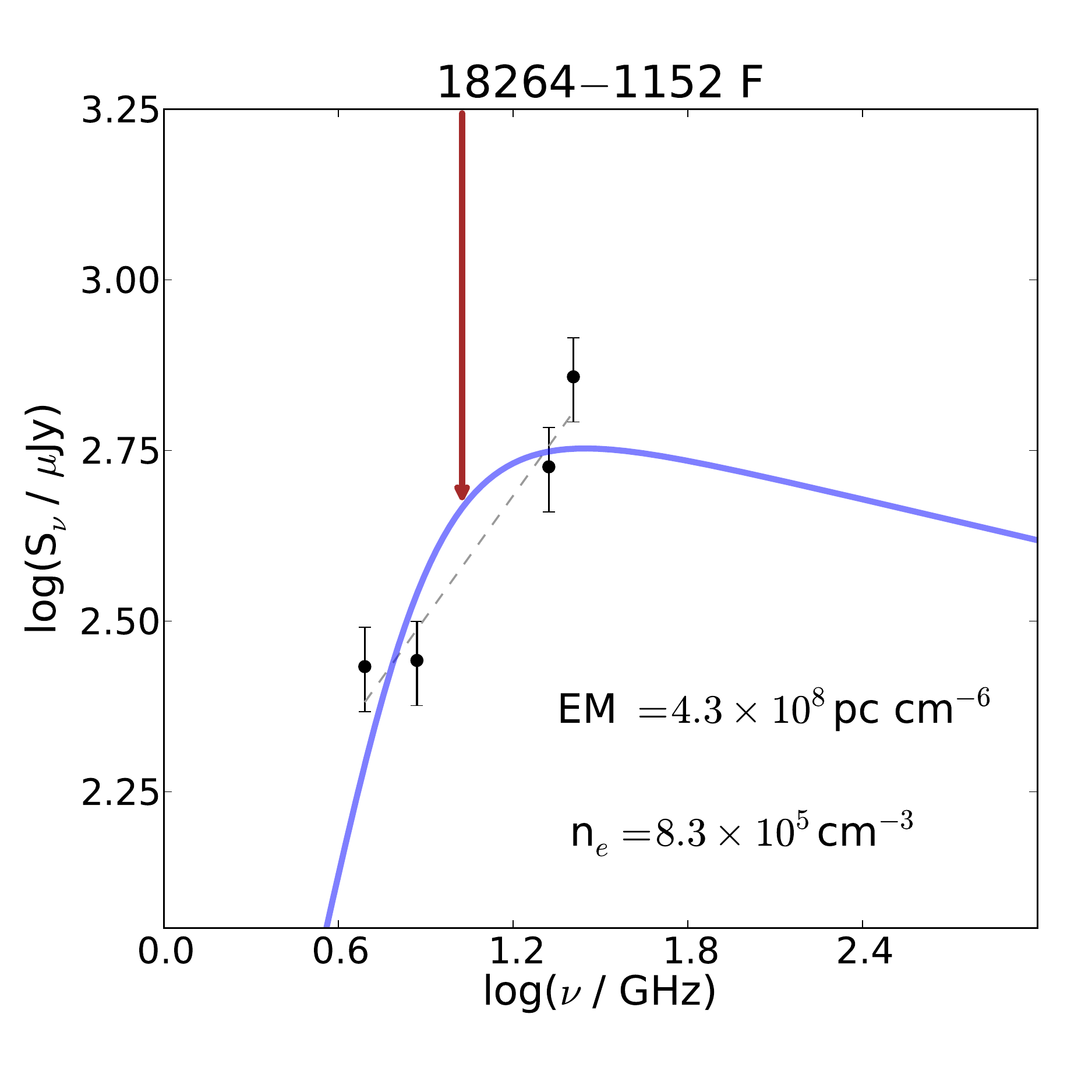} \vspace{-.2cm}  &
    \hspace{-0.8cm} 
    \includegraphics[width=0.34\linewidth, clip]{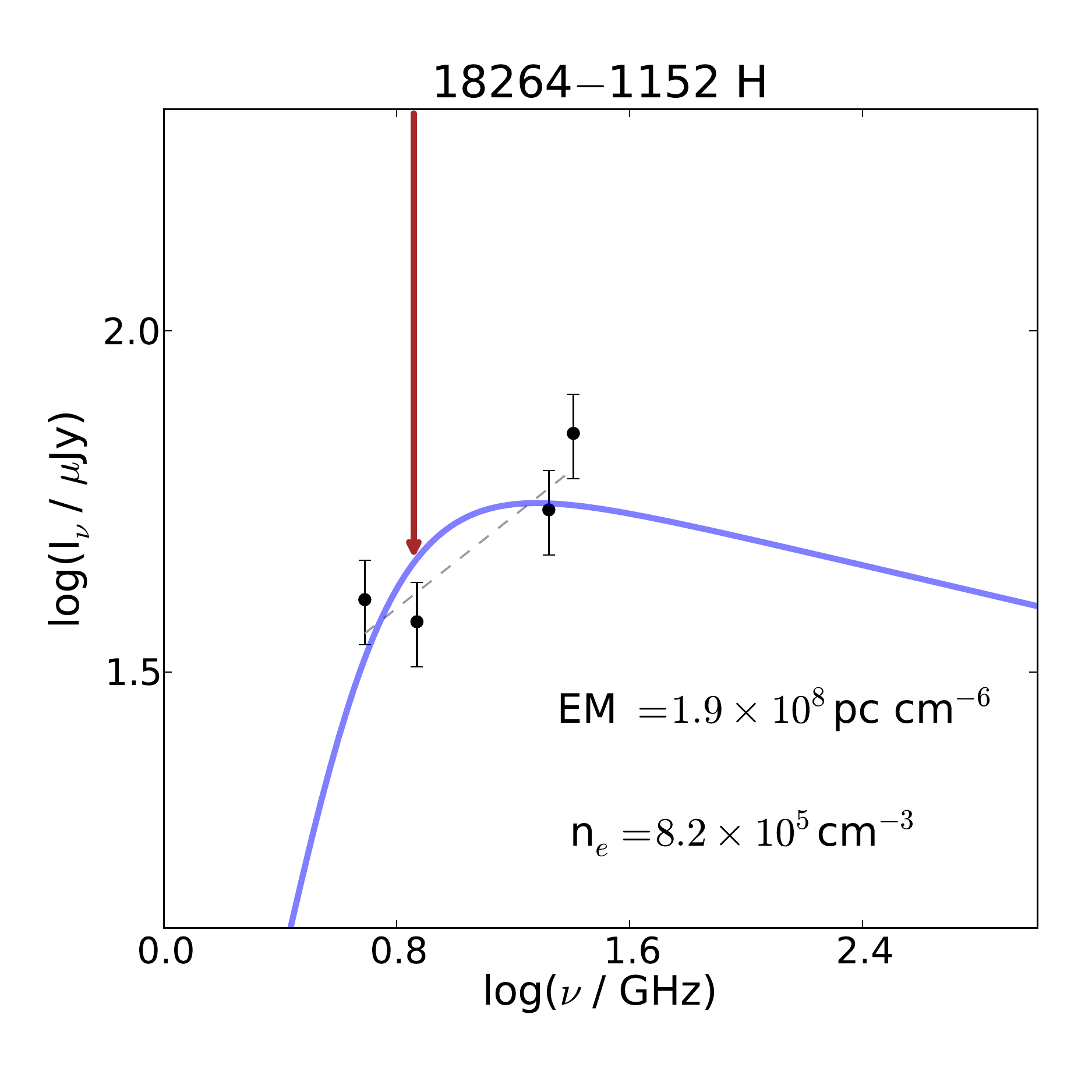}\vspace{-.2cm}\\
        \hspace{-.6cm}
    \includegraphics[width=0.34\linewidth, clip]{g23_01_alpha_HII_A} \vspace{-.2cm}  &
    \hspace{-0.8cm} 
    \includegraphics[width=0.34\linewidth, clip]{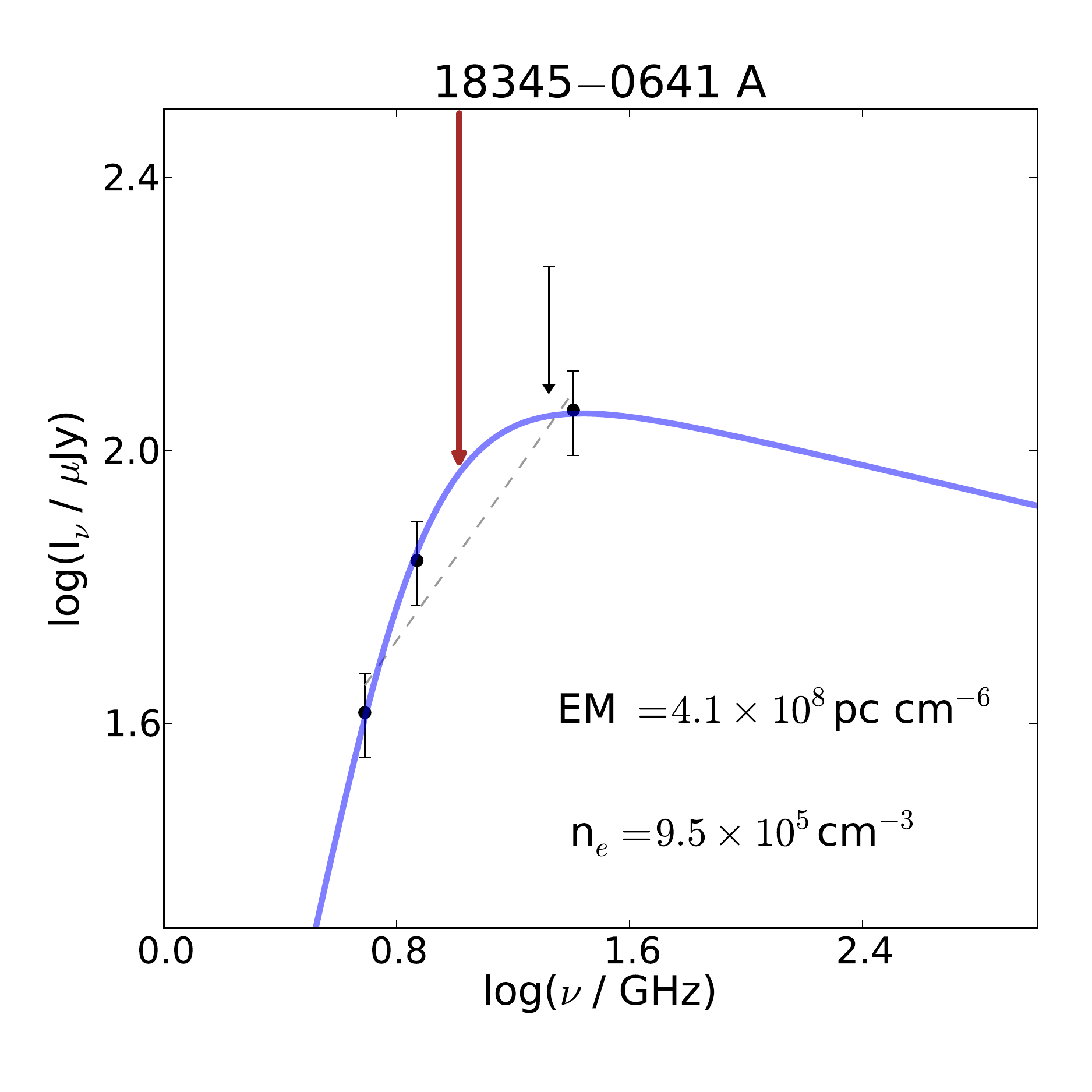} \vspace{-.2cm}  &
    \hspace{-0.8cm} 
    \includegraphics[width=0.34\linewidth, clip]{18440_alpha_HII_A}\vspace{-.2cm}\\
        \hspace{-.6cm}
    \includegraphics[width=0.34\linewidth, clip]{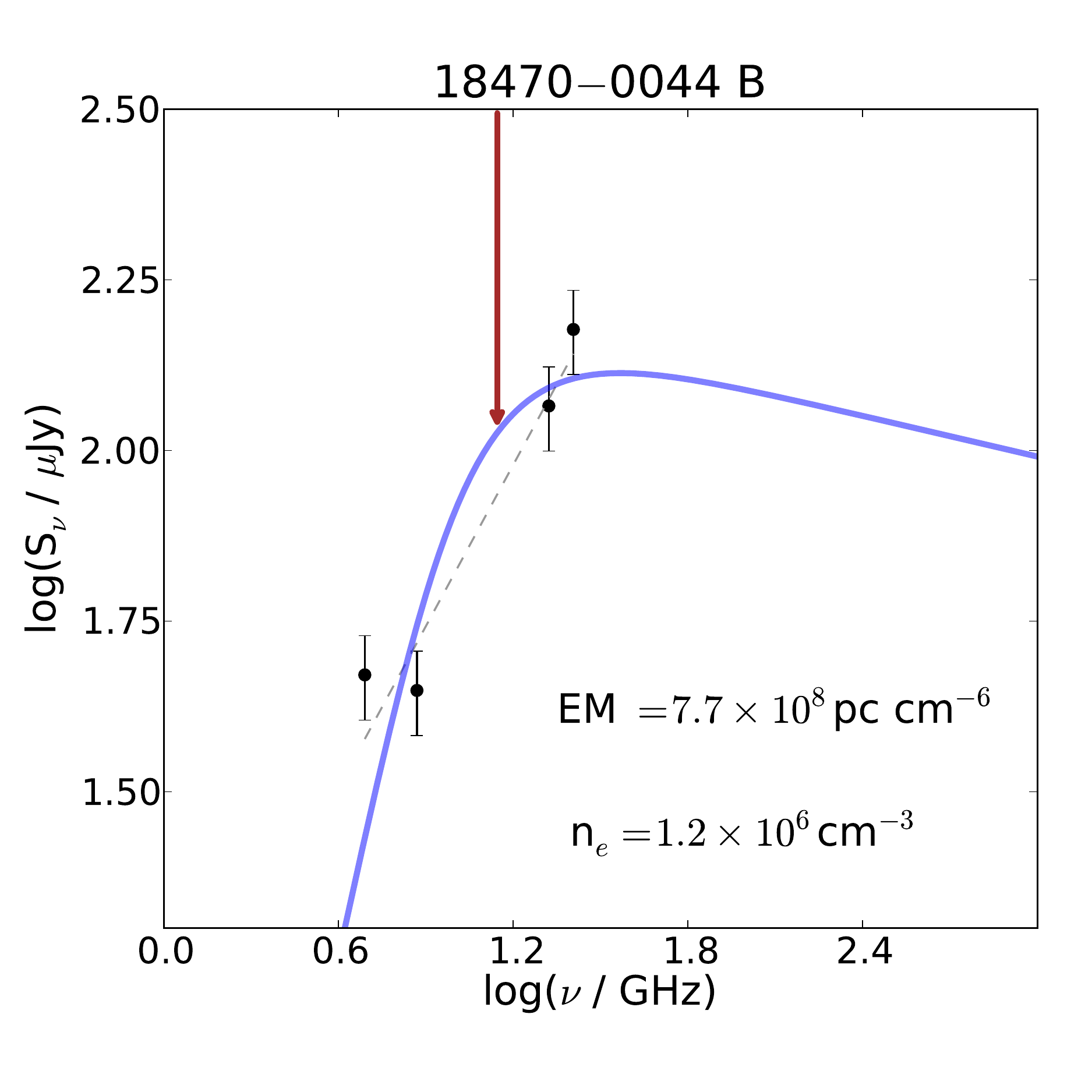} \vspace{-.2cm}  &
    \hspace{-0.8cm} 
    \includegraphics[width=0.34\linewidth, clip]{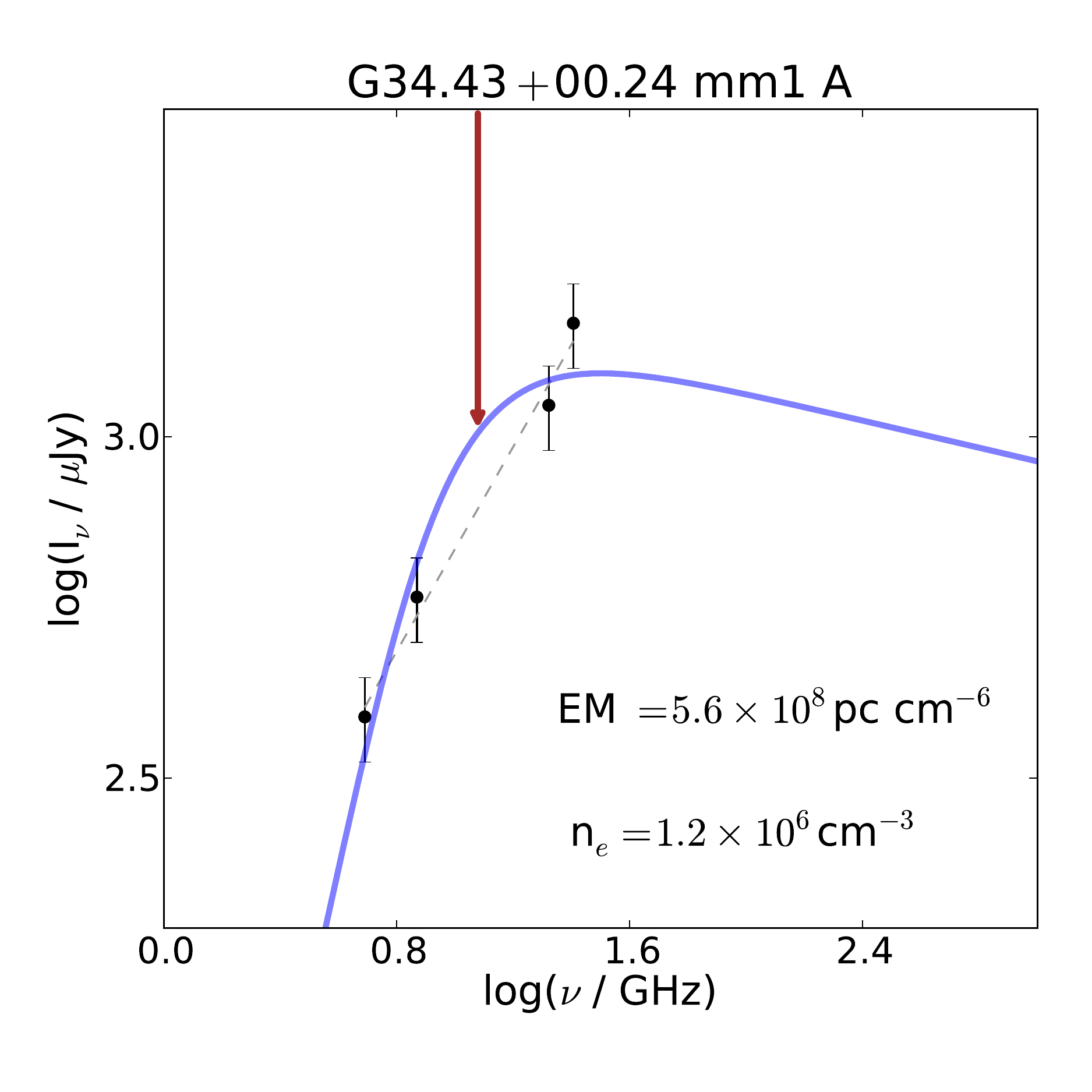} \vspace{-.2cm}  &
    \hspace{-0.8cm}   
    \includegraphics[width=0.34\linewidth, clip]{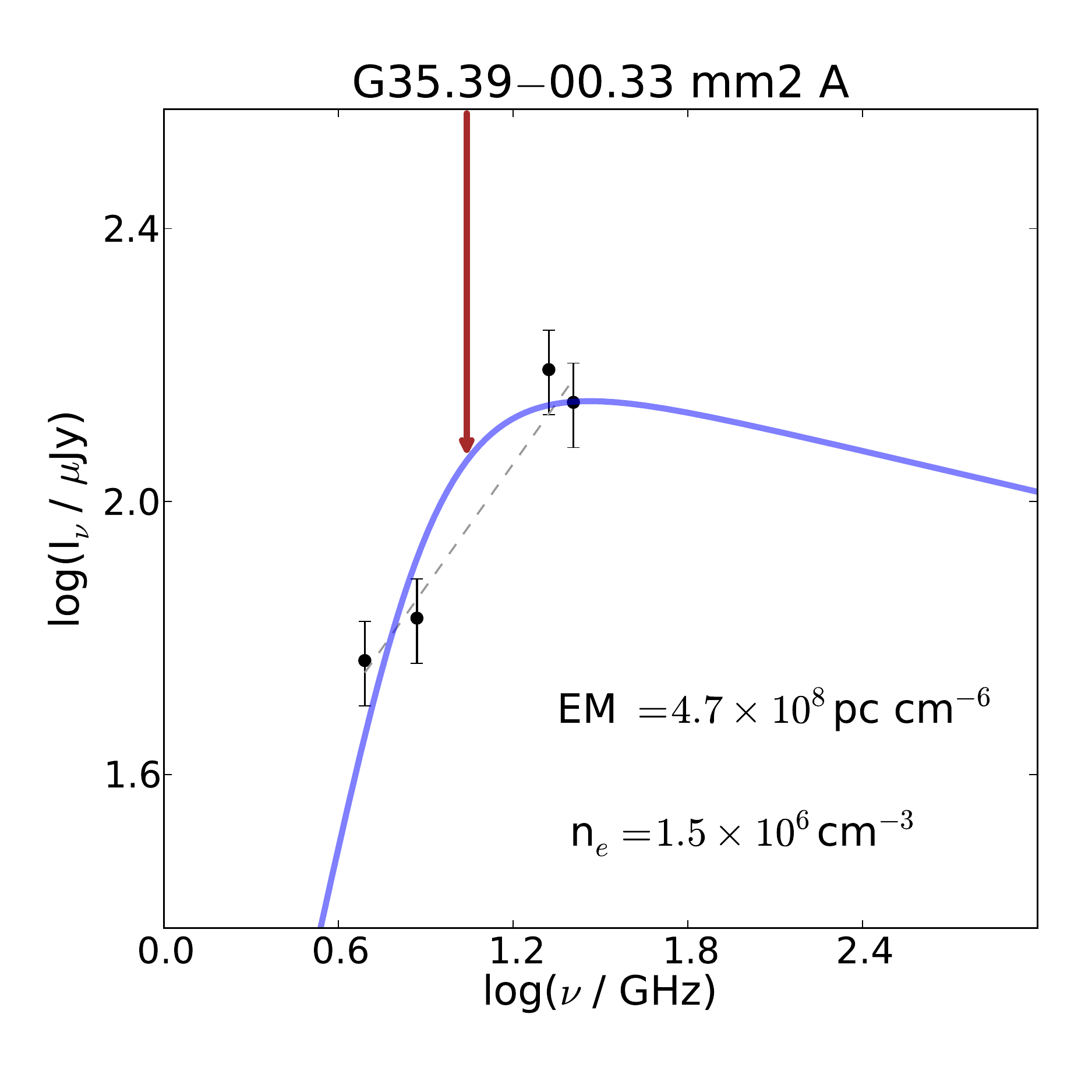}\vspace{-.2cm}\\
        \hspace{-.6cm}
    \includegraphics[width=0.34\linewidth, clip]{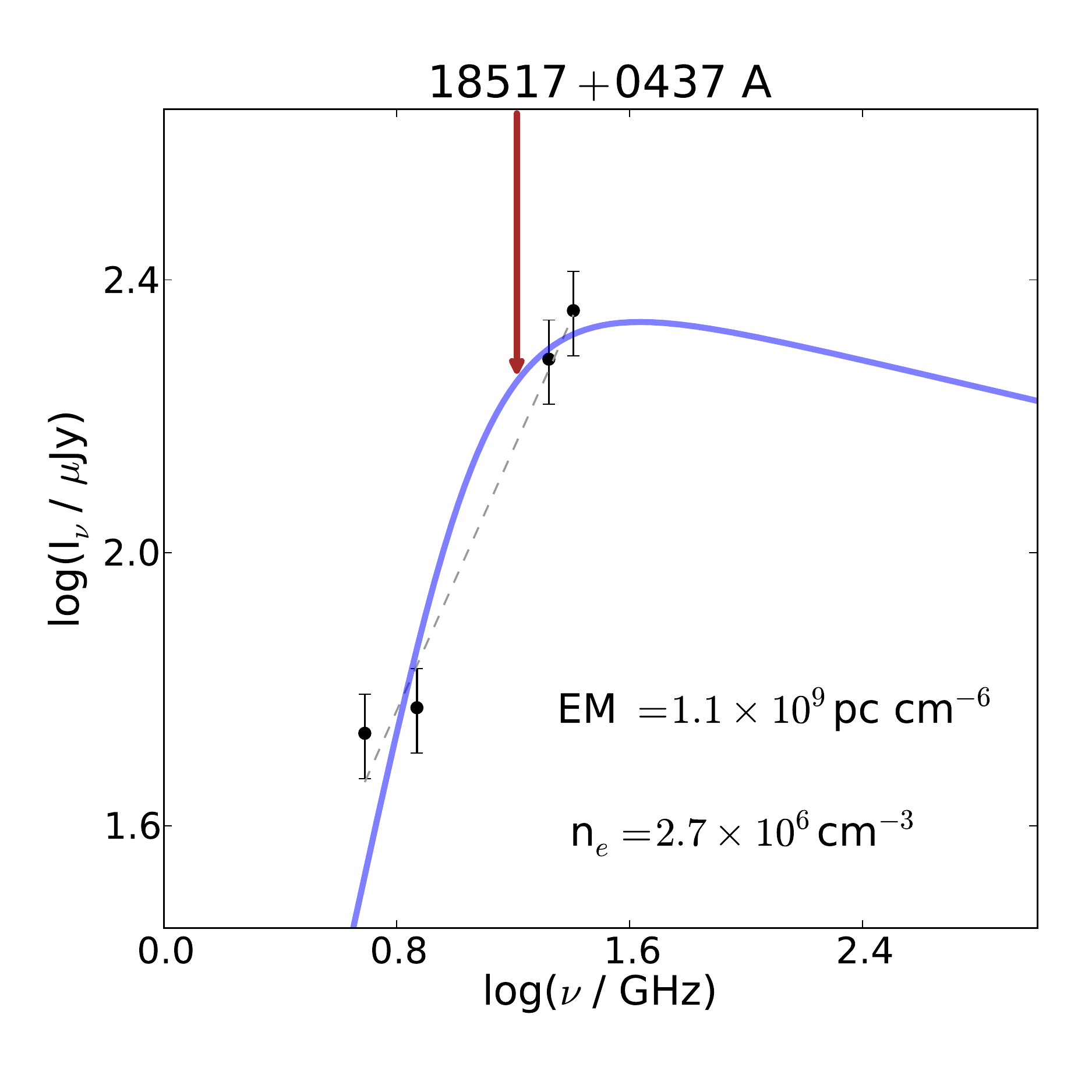} \vspace{-.2cm}  &
    \hspace{-0.8cm} 
    \includegraphics[width=0.34\linewidth, clip]{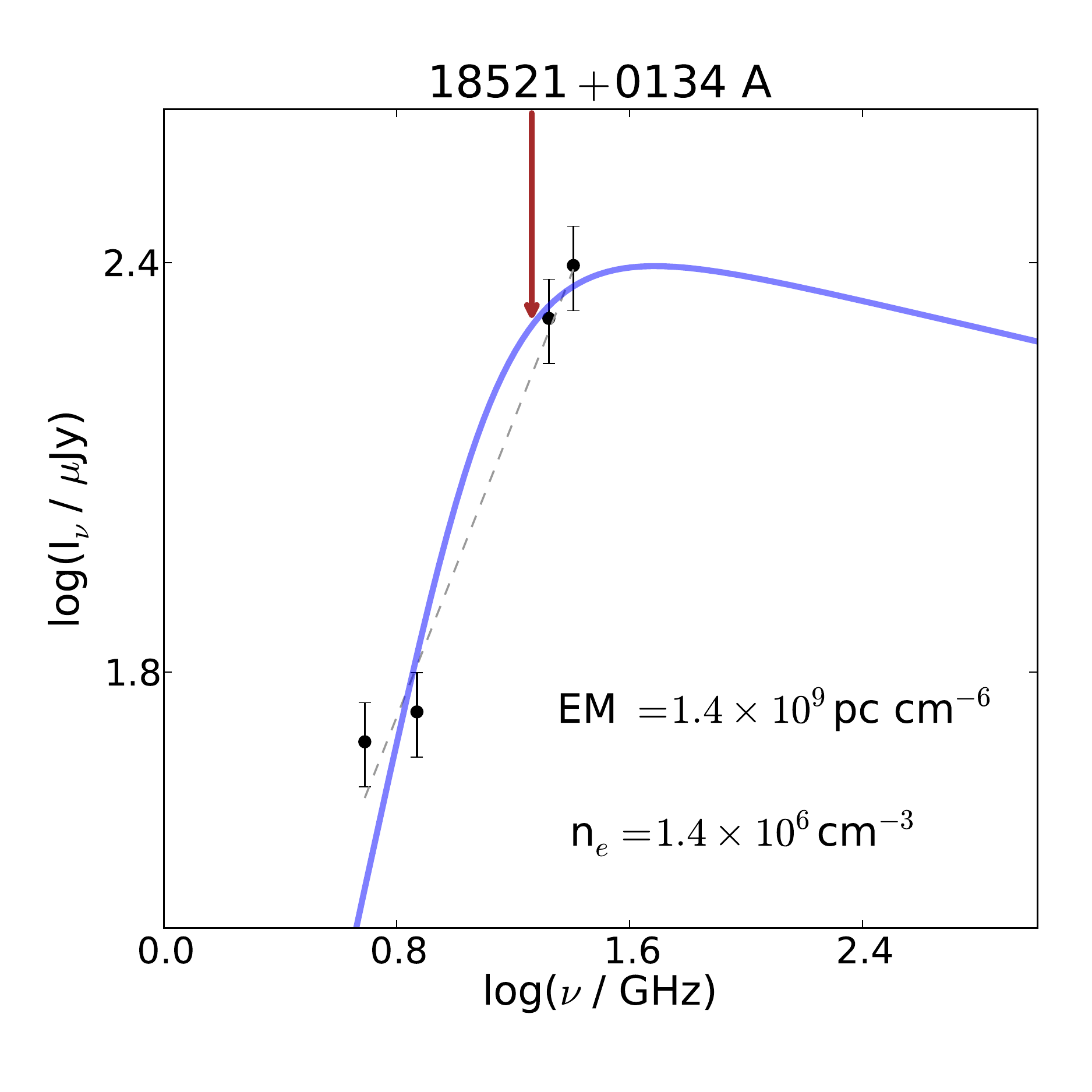} \vspace{-.2cm}  &
    \hspace{-0.8cm} 
        \includegraphics[width=0.34\linewidth, clip]{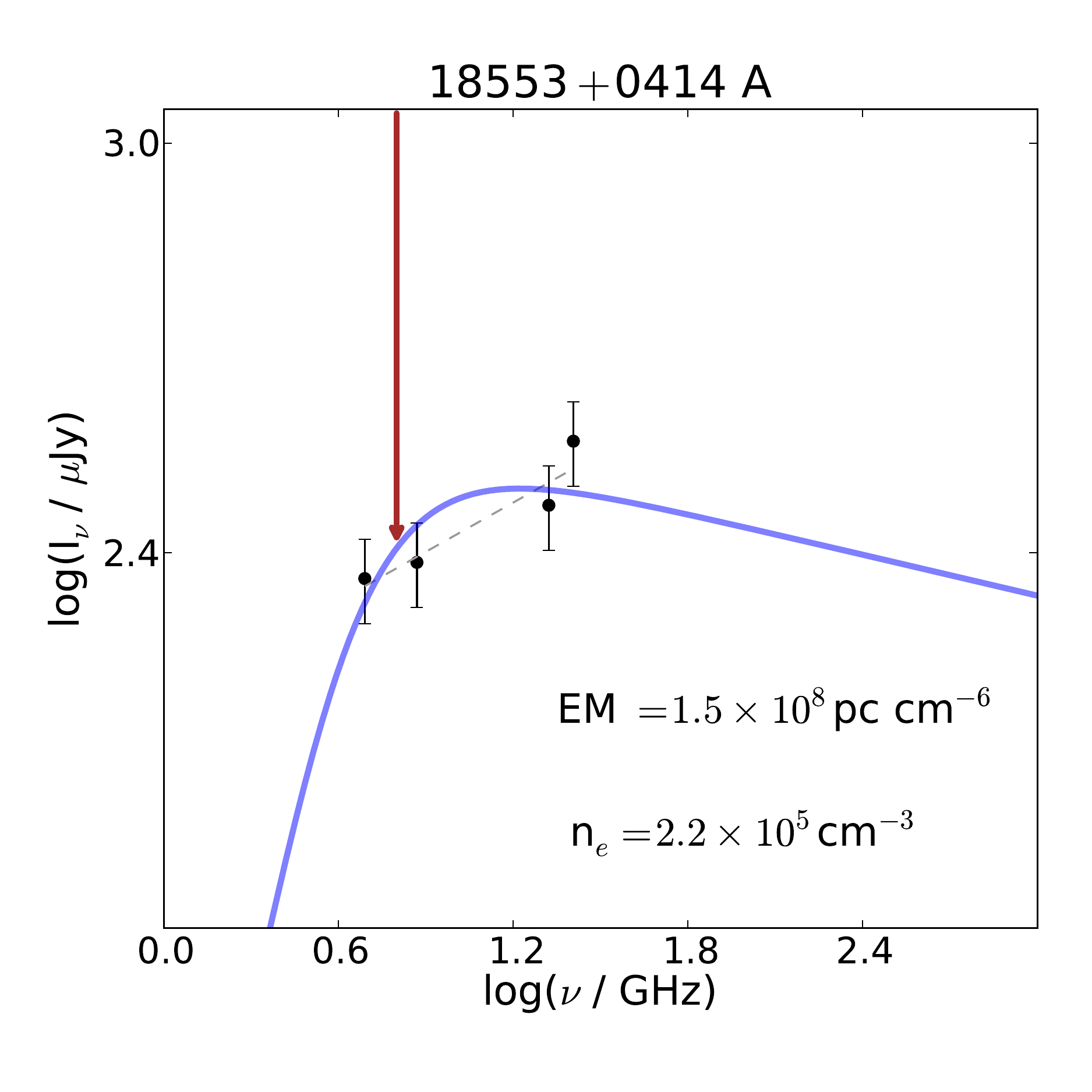}\vspace{-.2cm}\\

\end{tabular}
 \hspace*{\fill}%
\caption[]{Continued.}
    \label{HII_fit_app}%
\end{figure}

\begin{figure}
\centering
  \ContinuedFloat
\begin{tabular}{ccc}

\hspace{-.6cm}
    \includegraphics[width=0.34\linewidth, clip]{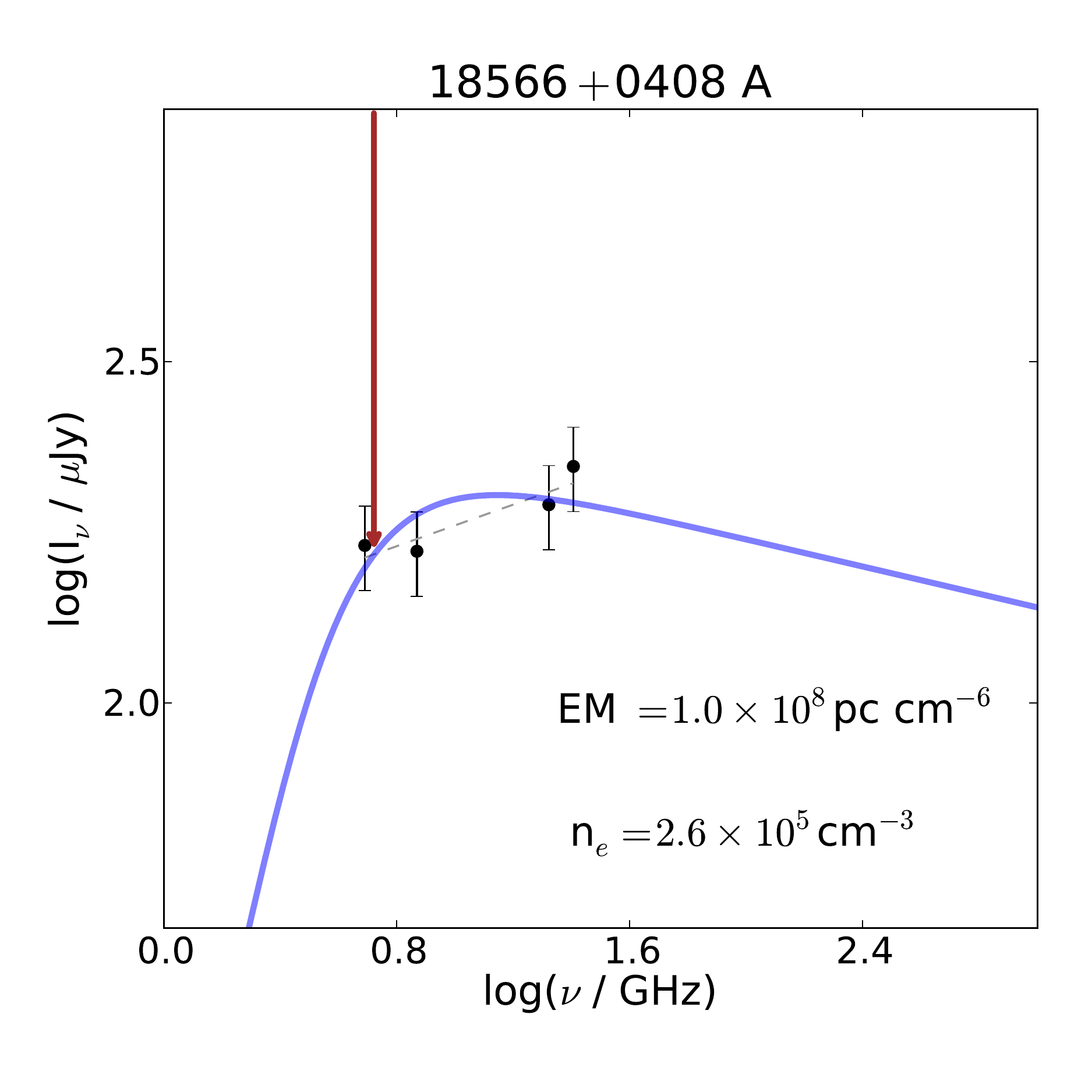} \vspace{-.2cm}  &
    \hspace{-0.8cm} 
    \includegraphics[width=0.34\linewidth, clip]{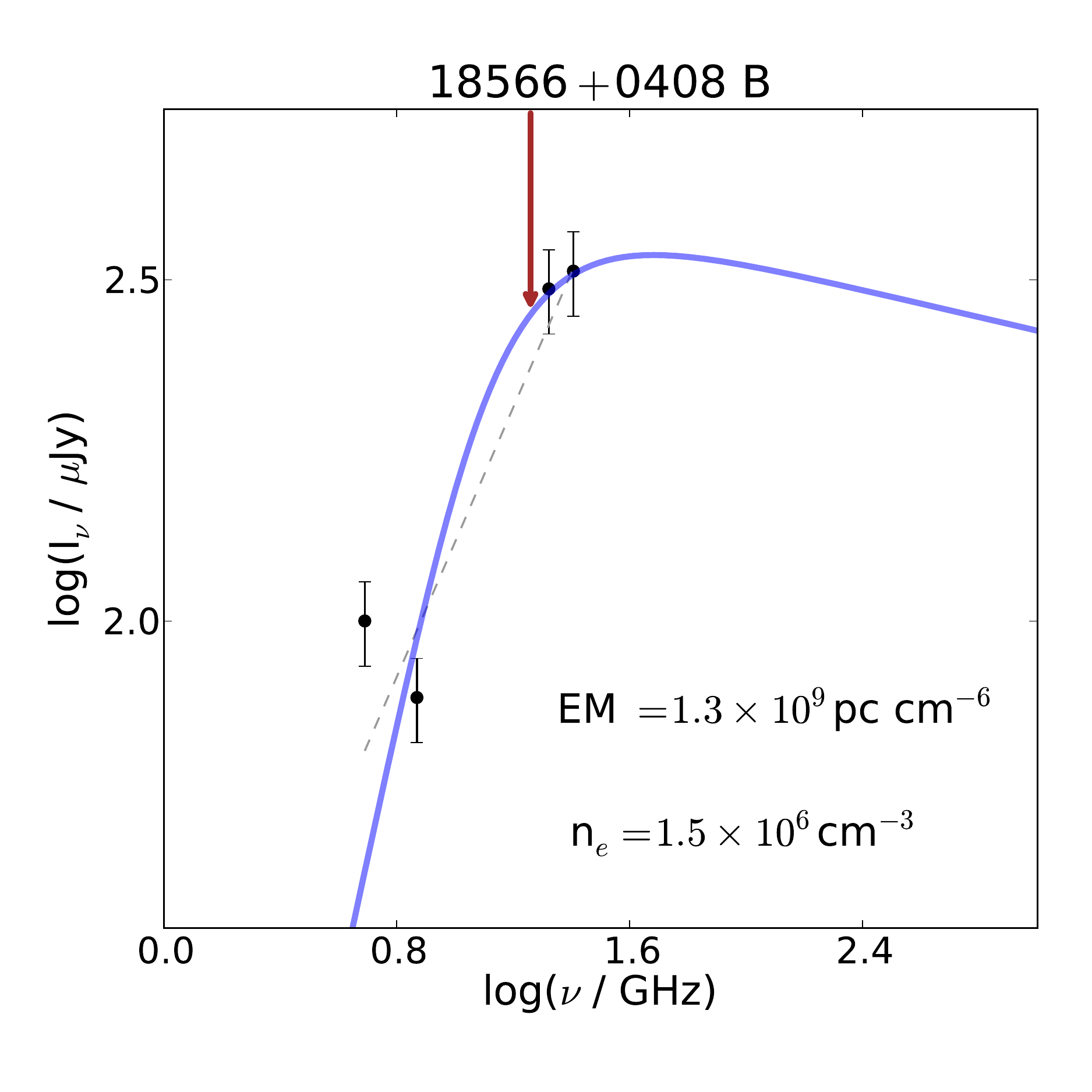} \vspace{-.2cm}  &
    \hspace{-0.8cm} 
 \includegraphics[width=0.34\linewidth, clip]{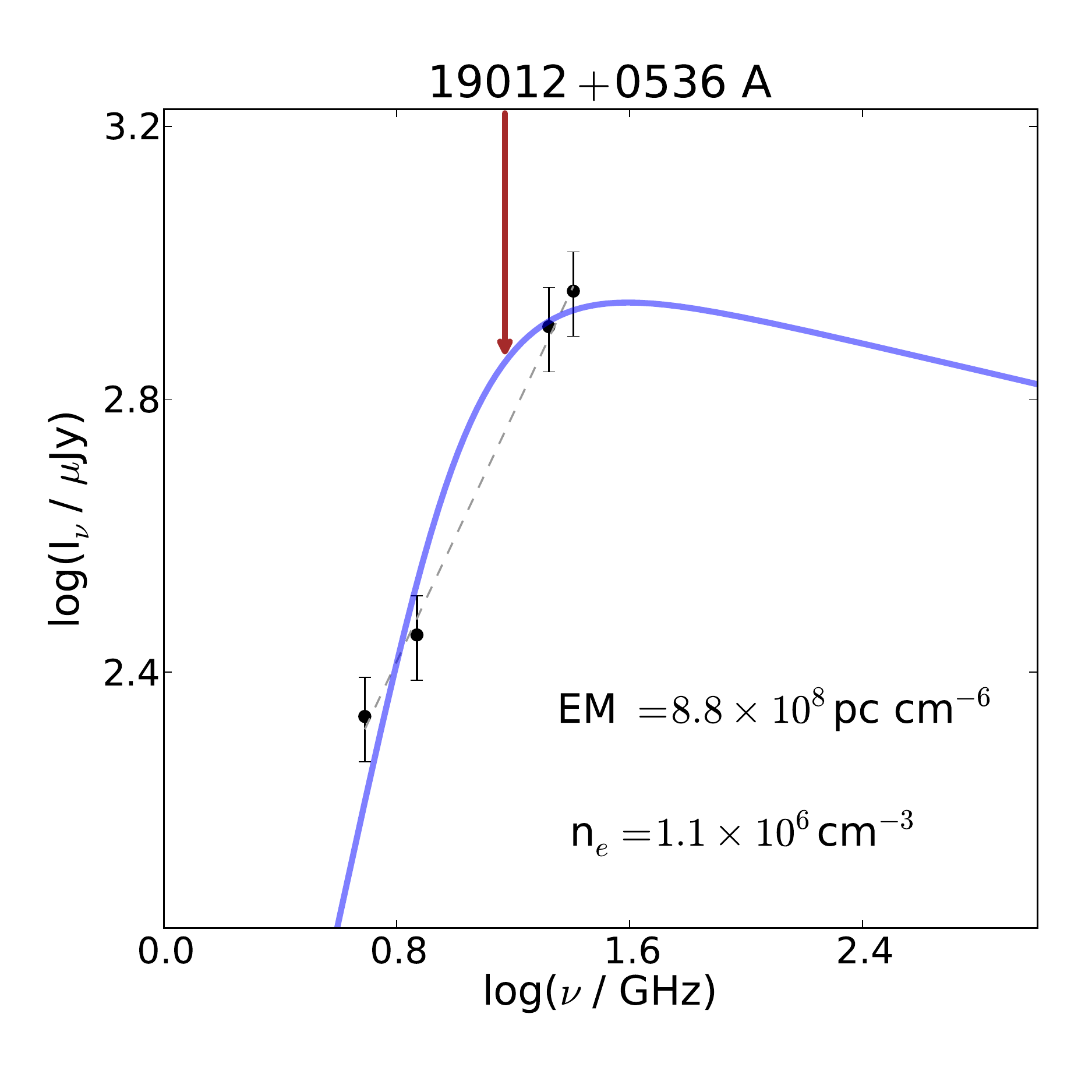}\vspace{-.2cm}\\
        \hspace{-.6cm}
                 \includegraphics[width=0.34\linewidth, clip]{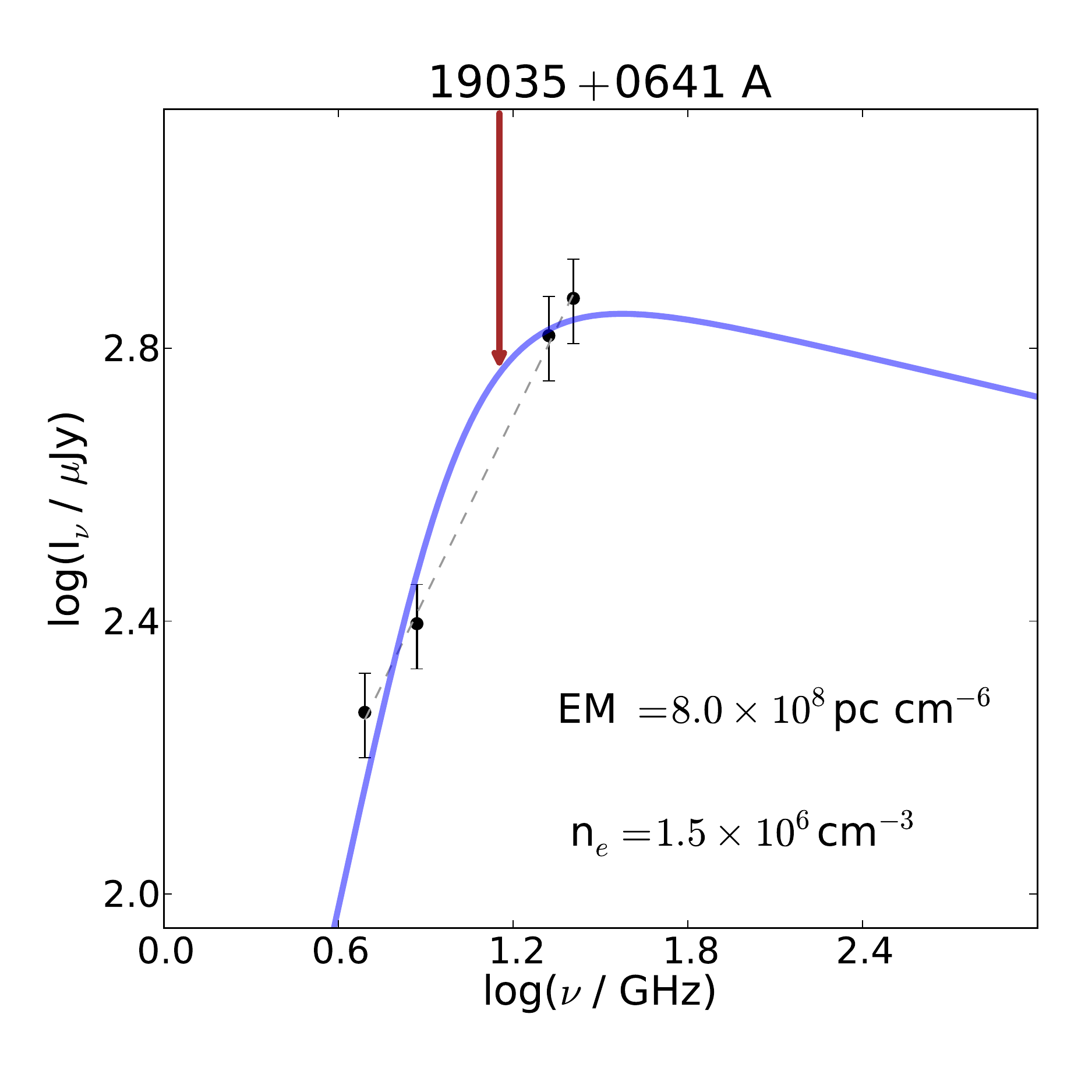}\vspace{-.2cm}  &
    \hspace{-0.6cm} 
    \includegraphics[width=0.34\linewidth, clip]{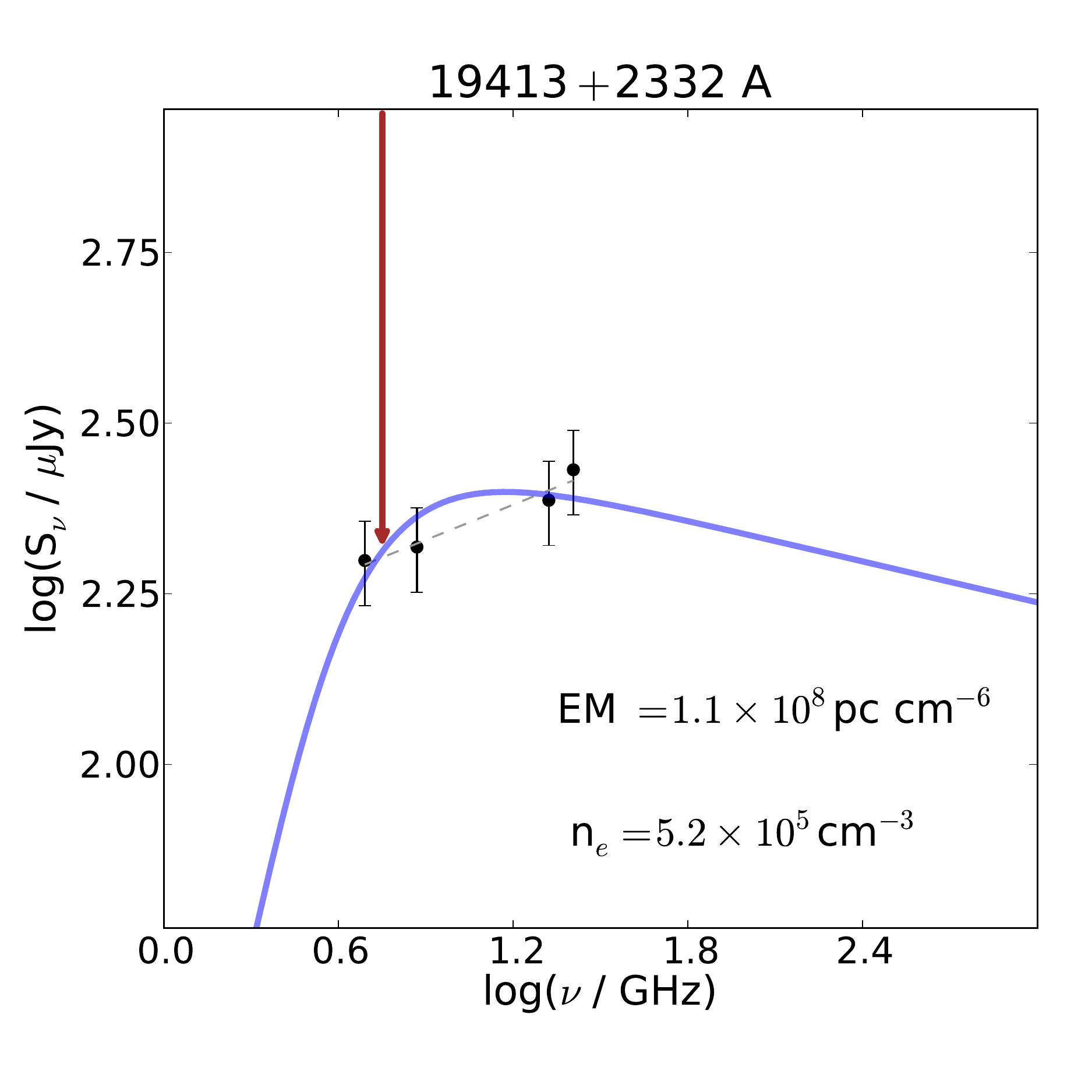}\vspace{-.2cm}  &
    \hspace{-0.6cm} 
    \includegraphics[width=0.34\linewidth, clip]{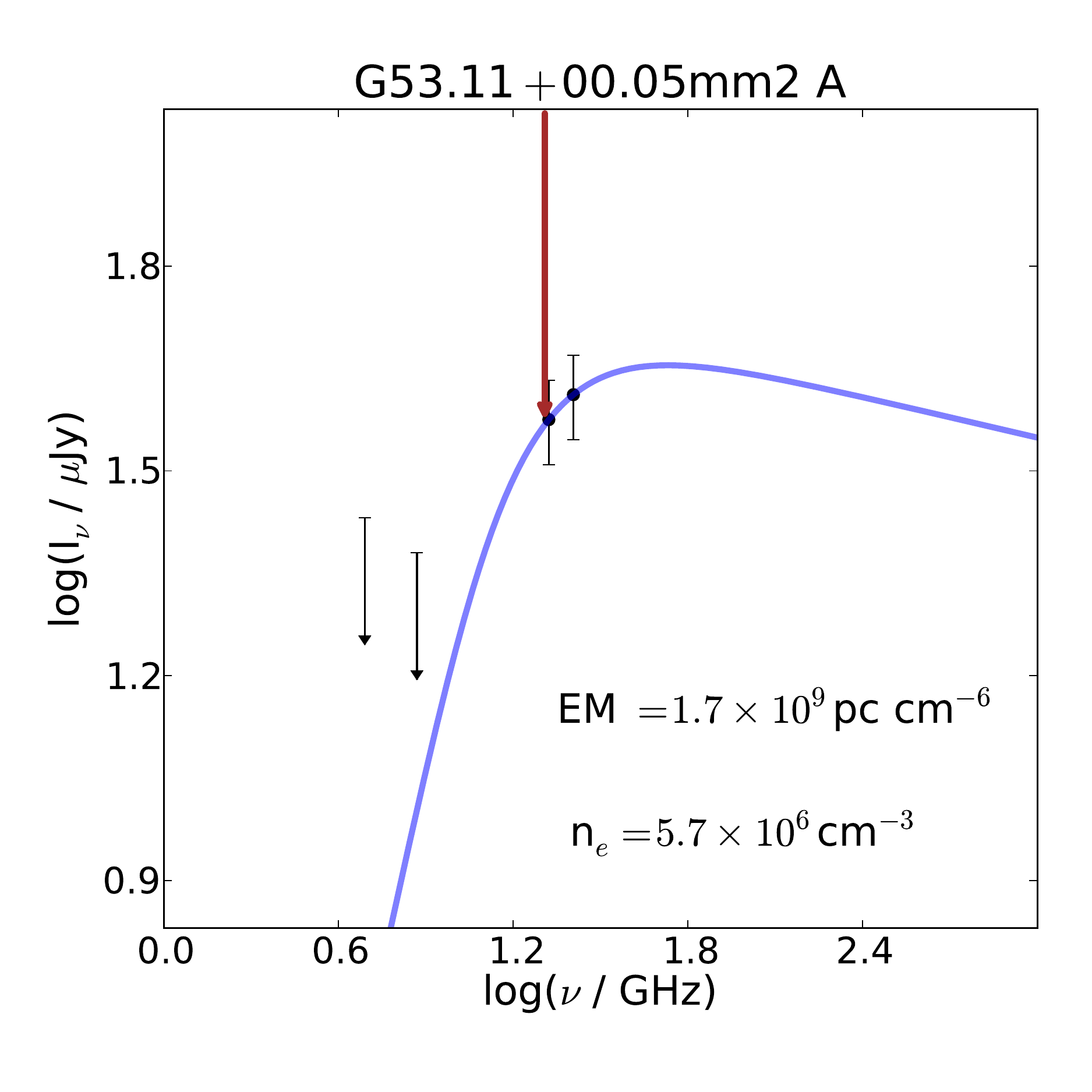}\vspace{-.2cm}\\
        \hspace{-.6cm}
    \includegraphics[width=0.34\linewidth, clip]{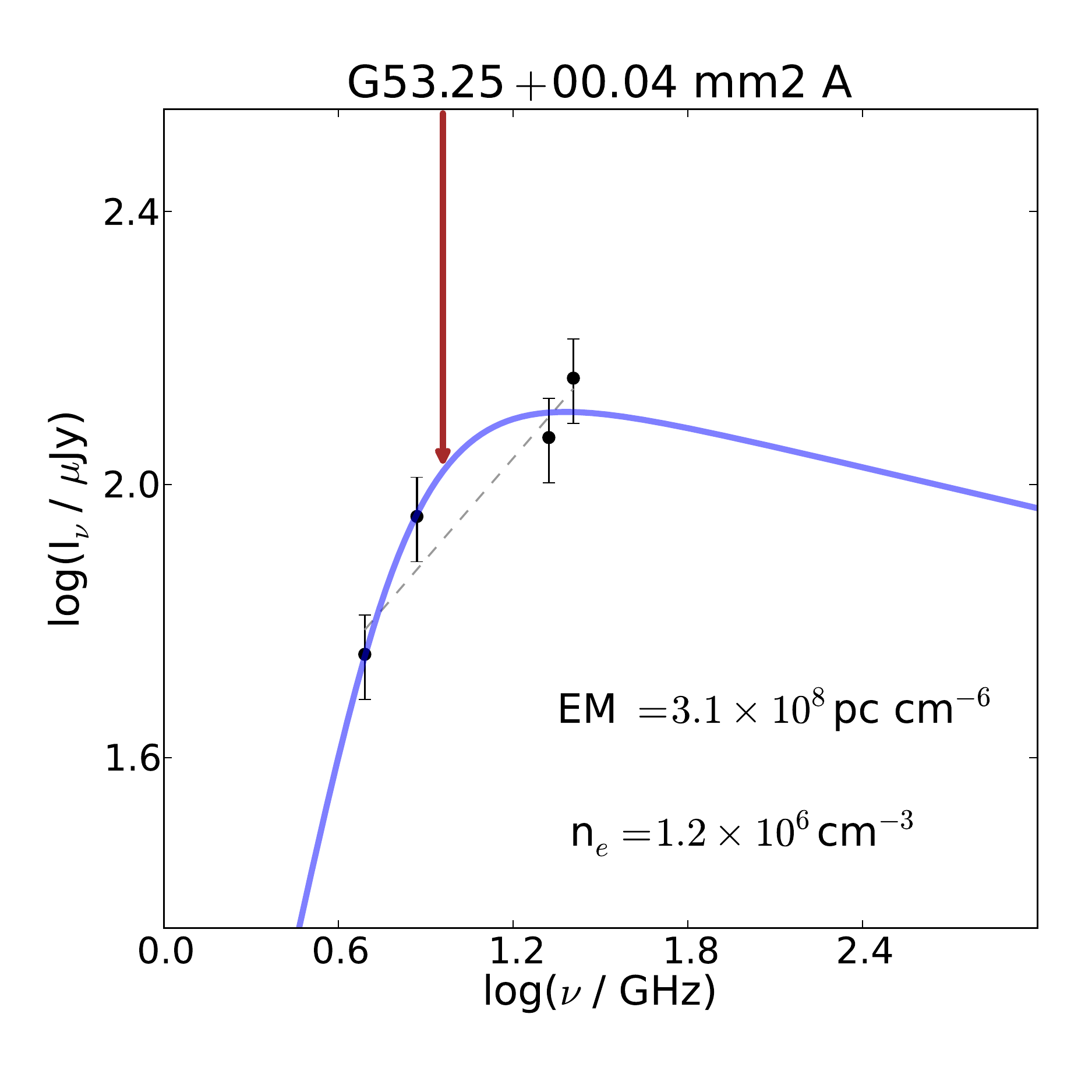}\vspace{-.2cm}  &
    \hspace{-0.8cm} 
    \includegraphics[width=0.34\linewidth, clip]{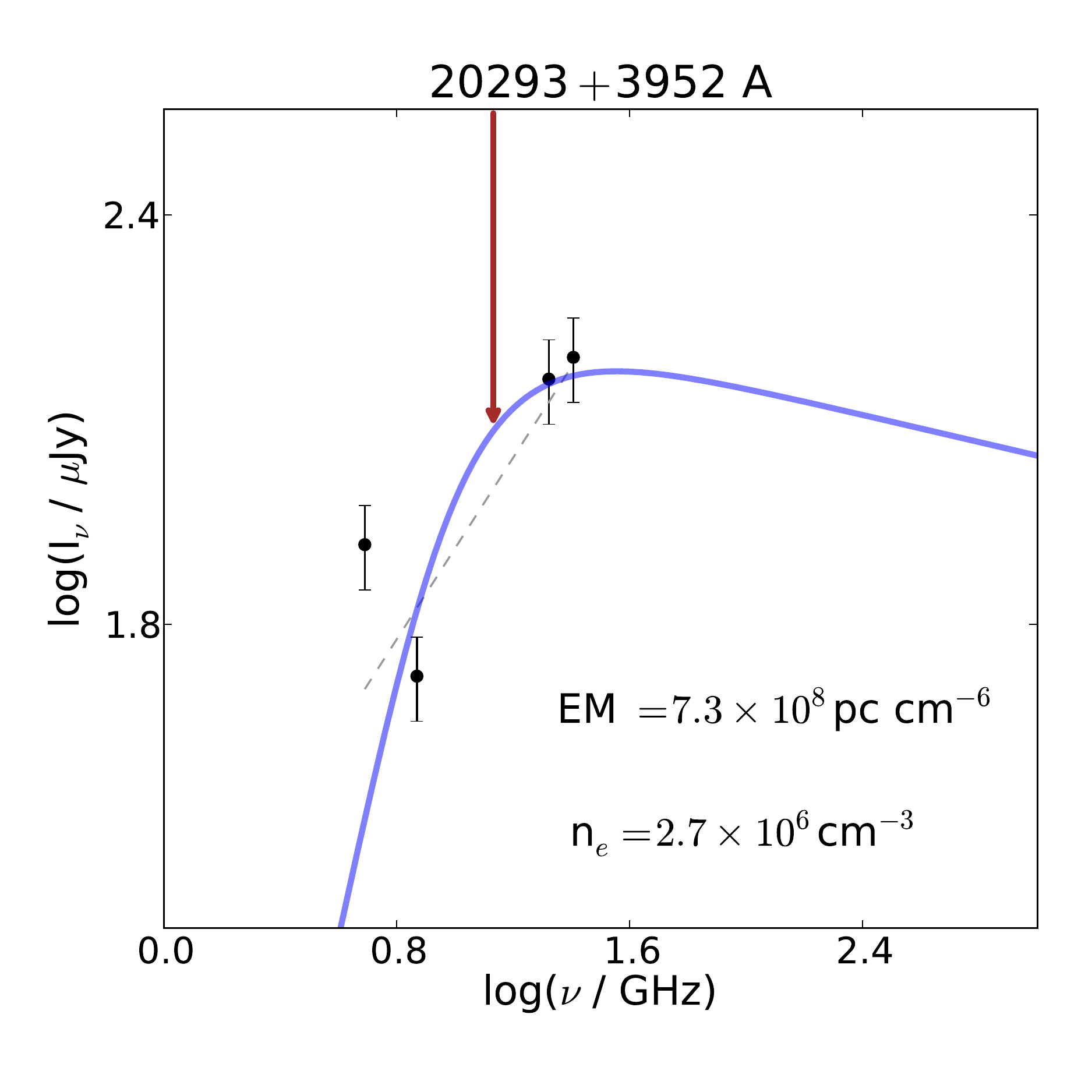}\vspace{-.2cm}  &
    \hspace{-0.8cm} 
        \includegraphics[width=0.34\linewidth, clip]{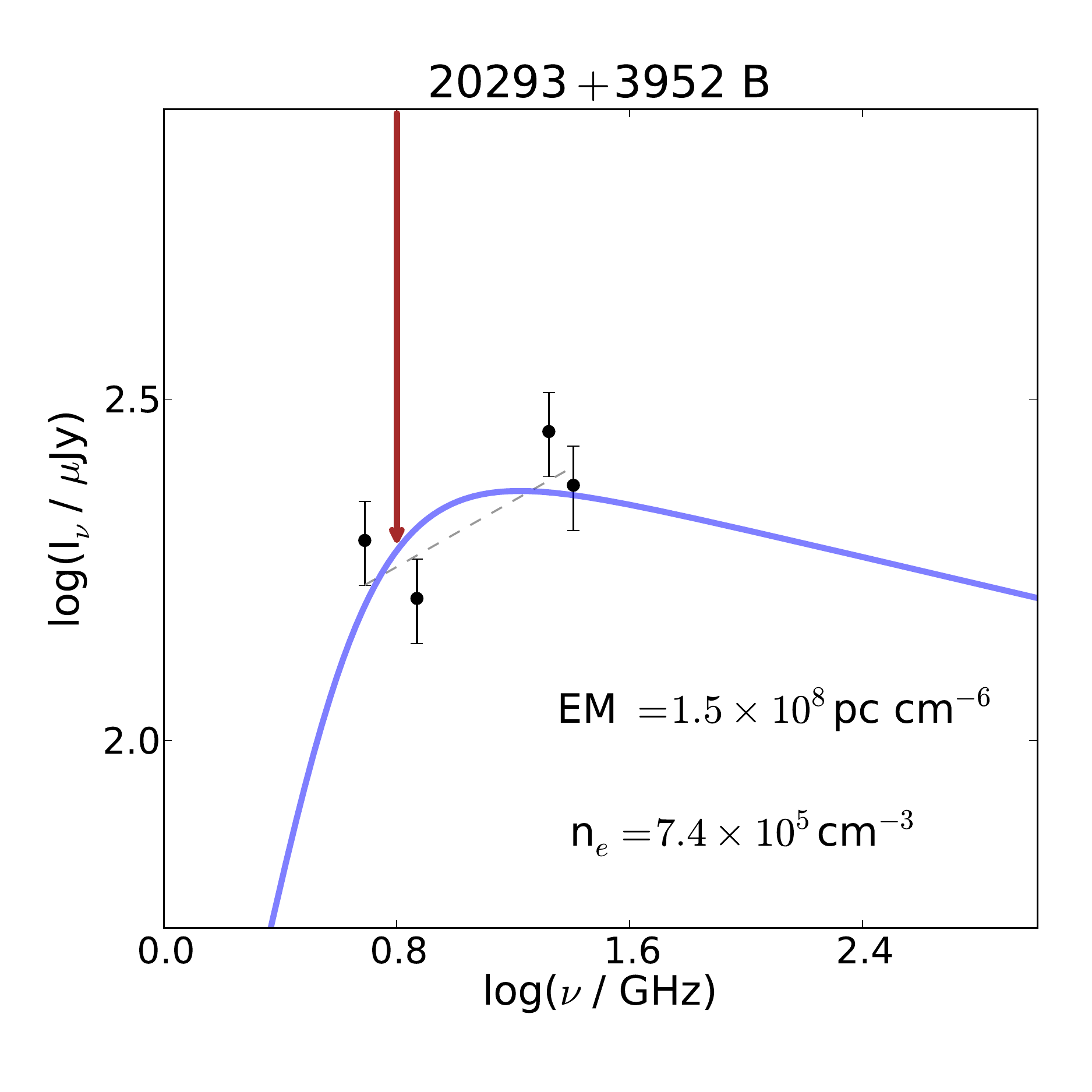}\vspace{-.2cm}\\
        \hspace{-.6cm}
     \includegraphics[width=0.34\linewidth, clip]{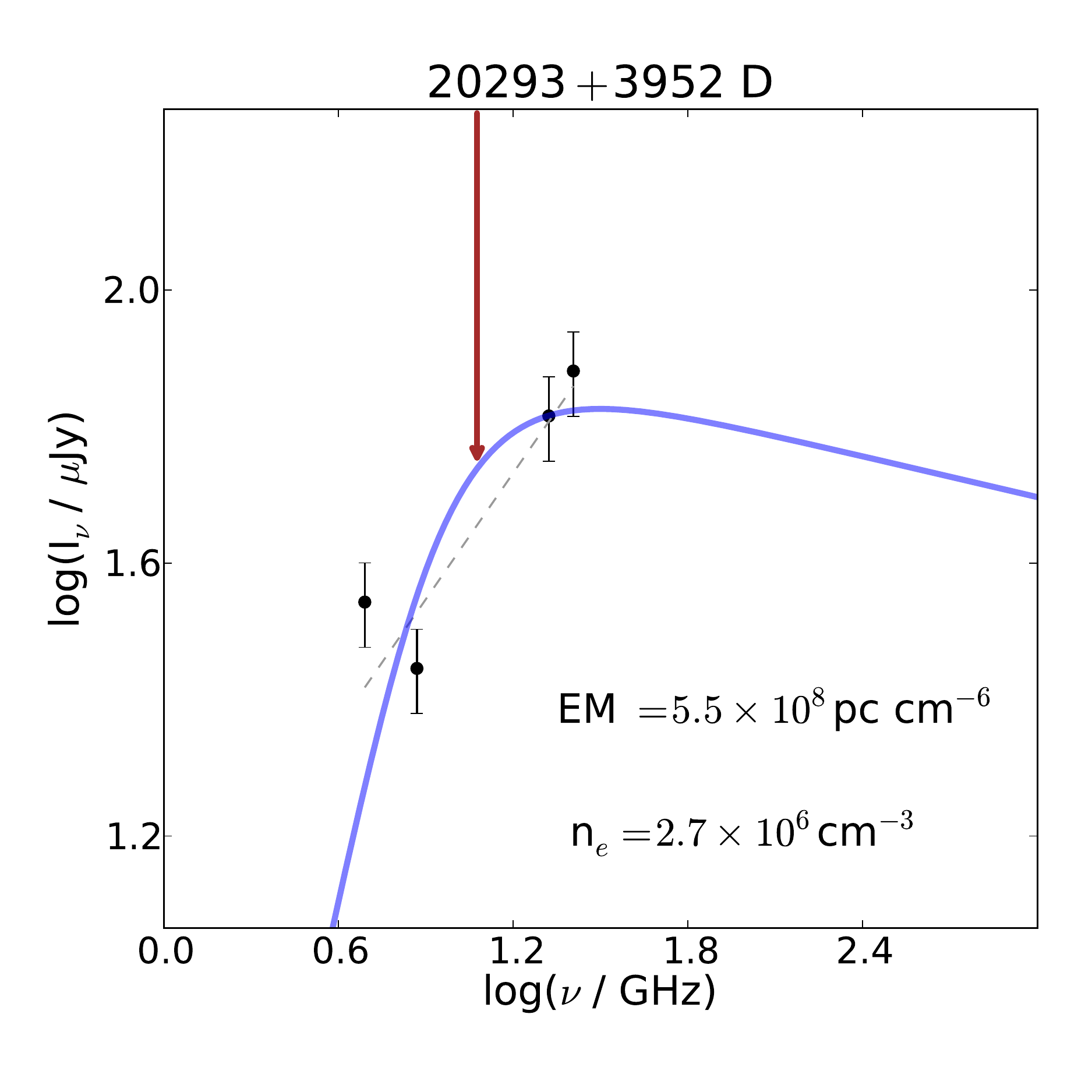}\vspace{-.2cm}  &
    \hspace{-0.8cm} 
    \includegraphics[width=0.34\linewidth, clip]{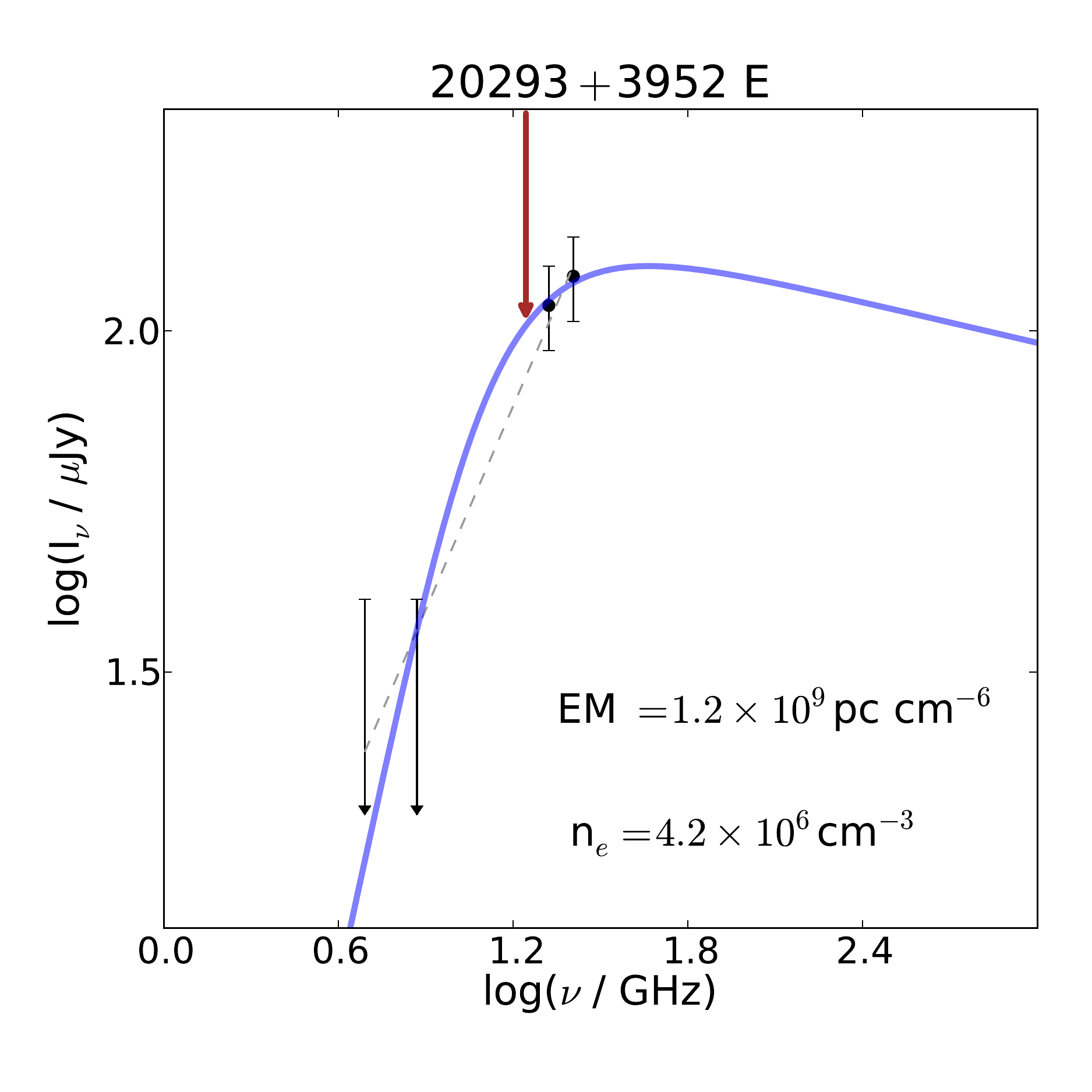}\vspace{-.2cm}  &
    \hspace{-0.8cm} 
    \includegraphics[width=0.34\linewidth, clip]{20343_alpha_HII_B}\vspace{-.2cm}\\   
    
\end{tabular}
 \hspace*{\fill}%
\caption[]{Continued.}
    \label{HII_fit_app}%
\end{figure}


\begin{thebibliography}{}
\expandafter\ifx\csname natexlab\endcsname\relax\def\natexlab#1{#1}\fi

\bibitem[{{AMI Consortium} {et~al.}(2011){AMI Consortium}, {Scaife},
  {Hatchell}, {Davies}, {Franzen}, {Grainge}, {Hobson}, {Hurley-Walker},
  {Lasenby}, {Olamaie}, {Perrott}, {Pooley}, {Rodr{\'{\i}}guez-Gonz{\'a}lvez},
  {Saunders}, {Schammel}, {Scott}, {Shimwell}, {Titterington}, \&
  {Waldram}}]{2011MNRAS.415..893A}
{AMI Consortium}, {Scaife}, A.~M.~M., {Hatchell}, J., {et~al.} 2011, \mnras,
  415, 893

\bibitem[{{AMI Consortium} {et~al.}(2012){AMI Consortium}, {Scaife},
  {Hatchell}, {Davies}, {Franzen}, {Grainge}, {Hobson}, {Hurley-Walker},
  {Lasenby}, {Olamaie}, {Perrott}, {Pooley}, {Rodr{\'{\i}}guez-Gonz{\'a}lvez},
  {Saunders}, {Schammel}, {Scott}, {Shimwell}, {Titterington}, \&
  {Waldram}}]{2012MNRAS.420.1019A}
---. 2012, \mnras, 420, 1019

\bibitem[{{Anglada}(1995)}]{1995RMxAC...1...67A}
{Anglada}, G. 1995, in Revista Mexicana de Astronomia y Astrofisica, vol.~27,
  Vol.~1, Revista Mexicana de Astronomia y Astrofisica Conference Series, ed.
  S.~{Lizano} \& J.~M. {Torrelles}, 67

\bibitem[{{Anglada}(1996)}]{1996ASPC...93....3A}
{Anglada}, G. 1996, in Astronomical Society of the Pacific Conference Series,
  Vol.~93, Radio Emission from the Stars and the Sun, ed. A.~R. {Taylor} \&
  J.~M. {Paredes}, 3--14

\bibitem[{{Anglada} {et~al.}(2015){Anglada}, {Rodr{\'{\i}}guez}, \&
  {Carrasco-Gonzalez}}]{2015aska.confE.121A}
{Anglada}, G., {Rodr{\'{\i}}guez}, L.~F., \& {Carrasco-Gonzalez}, C. 2015,
  Advancing Astrophysics with the Square Kilometre Array (AASKA14), 121

\bibitem[{{Anglada} {et~al.}(2018){Anglada}, {Rodr{\'{\i}}guez}, \&
  {Carrasco-Gonz{\'a}lez}}]{2018A&ARv..26....3A}
{Anglada}, G., {Rodr{\'{\i}}guez}, L.~F., \& {Carrasco-Gonz{\'a}lez}, C. 2018,
  \aapr, 26, 3

\bibitem[{{Anglada} {et~al.}(1998){Anglada}, {Villuendas}, {Estalella},
  {Beltr{\'a}n}, {Rodr{\'{\i}}guez}, {Torrelles}, \&
  {Curiel}}]{1998AJ....116.2953A}
{Anglada}, G., {Villuendas}, E., {Estalella}, R., {et~al.} 1998, \aj, 116, 2953

\bibitem[{{Araya} {et~al.}(2007){Araya}, {Hofner}, {Sewi{\l}o}, {Goss}, {Linz},
  {Kurtz}, {Olmi}, {Churchwell}, {Rodr{\'{\i}}guez}, \&
  {Garay}}]{2007ApJ...669.1050A}
{Araya}, E., {Hofner}, P., {Sewi{\l}o}, M., {et~al.} 2007, \apj, 669, 1050

\bibitem[{{Araya} {et~al.}(2008){Araya}, {Hofner}, {Goss}, {Linz}, {Kurtz}, \&
  {Olmi}}]{2008ApJS..178..330A}
{Araya}, E.~D., {Hofner}, P., {Goss}, W.~M., {et~al.} 2008, \apjs, 178, 330

\bibitem[{{Bally} \& {Zinnecker}(2005)}]{2005AJ....129.2281B}
{Bally}, J., \& {Zinnecker}, H. 2005, \aj, 129, 2281

\bibitem[{{Beltr{\'a}n} \& {de Wit}(2016)}]{2016A&ARv..24....6B}
{Beltr{\'a}n}, M.~T., \& {de Wit}, W.~J. 2016, \aapr, 24, 6

\bibitem[{{Beuther} {et~al.}(2004{\natexlab{a}}){Beuther}, {Schilke}, \&
  {Gueth}}]{2004ApJ...608..330B}
{Beuther}, H., {Schilke}, P., \& {Gueth}, F. 2004{\natexlab{a}}, \apj, 608, 330

\bibitem[{{Beuther} {et~al.}(2002{\natexlab{a}}){Beuther}, {Schilke}, {Menten},
  {Motte}, {Sridharan}, \& {Wyrowski}}]{2002ApJ...566..945B}
{Beuther}, H., {Schilke}, P., {Menten}, K.~M., {et~al.} 2002{\natexlab{a}},
  \apj, 566, 945

\bibitem[{{Beuther} {et~al.}(2002{\natexlab{b}}){Beuther}, {Schilke},
  {Sridharan}, {Menten}, {Walmsley}, \& {Wyrowski}}]{2002A&A...383..892B}
{Beuther}, H., {Schilke}, P., {Sridharan}, T.~K., {et~al.} 2002{\natexlab{b}},
  \aap, 383, 892

\bibitem[{{Beuther} {et~al.}(2005){Beuther}, {Sridharan}, \&
  {Saito}}]{2005ApJ...634L.185B}
{Beuther}, H., {Sridharan}, T.~K., \& {Saito}, M. 2005, \apjl, 634, L185

\bibitem[{{Beuther} {et~al.}(2010){Beuther}, {Vlemmings}, {Rao}, \& {van der
  Tak}}]{2010ApJ...724L.113B}
{Beuther}, H., {Vlemmings}, W.~H.~T., {Rao}, R., \& {van der Tak}, F.~F.~S.
  2010, \apjl, 724, L113

\bibitem[{{Beuther} {et~al.}(2002{\natexlab{c}}){Beuther}, {Walsh}, {Schilke},
  {Sridharan}, {Menten}, \& {Wyrowski}}]{2002A&A...390..289B}
{Beuther}, H., {Walsh}, A., {Schilke}, P., {et~al.} 2002{\natexlab{c}}, \aap,
  390, 289

\bibitem[{{Beuther} {et~al.}(2006){Beuther}, {Zhang}, {Sridharan}, {Lee}, \&
  {Zapata}}]{2006A&A...454..221B}
{Beuther}, H., {Zhang}, Q., {Sridharan}, T.~K., {Lee}, C.-F., \& {Zapata},
  L.~A. 2006, \aap, 454, 221

\bibitem[{{Beuther} {et~al.}(2004{\natexlab{b}}){Beuther}, {Hunter}, {Zhang},
  {Sridharan}, {Zhao}, {Sollins}, {Ho}, {Ohashi}, {Su}, {Lim}, \&
  {Liu}}]{2004ApJ...616L..23B}
{Beuther}, H., {Hunter}, T.~R., {Zhang}, Q., {et~al.} 2004{\natexlab{b}},
  \apjl, 616, L23

\bibitem[{{Bonnell} {et~al.}(1998){Bonnell}, {Bate}, \&
  {Zinnecker}}]{1998MNRAS.298...93B}
{Bonnell}, I.~A., {Bate}, M.~R., \& {Zinnecker}, H. 1998, \mnras, 298, 93

\bibitem[{{Caratti o Garatti} {et~al.}(2015){Caratti o Garatti}, {Stecklum},
  {Linz}, {Garcia Lopez}, \& {Sanna}}]{2015A&A...573A..82C}
{Caratti o Garatti}, A., {Stecklum}, B., {Linz}, H., {Garcia Lopez}, R., \&
  {Sanna}, A. 2015, \aap, 573, A82

\bibitem[{{Carey} {et~al.}(2005){Carey}, {Noriega-Crespo}, {Price}, {Padgett},
  {Kraemer}, {Indebetouw}, {Mizuno}, {Ali}, {Berriman}, {Boulanger}, {Cutri},
  {Ingalls}, {Kuchar}, {Latter}, {Marleau}, {Miville-Deschenes}, {Molinari},
  {Rebull}, \& {Testi}}]{2005AAS...207.6333C}
{Carey}, S.~J., {Noriega-Crespo}, A., {Price}, S.~D., {et~al.} 2005, in
  Bulletin of the American Astronomical Society, Vol.~37, American Astronomical
  Society Meeting Abstracts, 1252

\bibitem[{{Carral} {et~al.}(1999){Carral}, {Kurtz}, {Rodr{\'{\i}}guez},
  {Mart{\'{\i}}}, {Lizano}, \& {Osorio}}]{1999RMxAA..35...97C}
{Carral}, P., {Kurtz}, S., {Rodr{\'{\i}}guez}, L.~F., {et~al.} 1999, \rmxaa,
  35, 97

\bibitem[{{Carrasco-Gonz{\'a}lez} {et~al.}(2010){Carrasco-Gonz{\'a}lez},
  {Rodr{\'{\i}}guez}, {Anglada}, {Mart{\'{\i}}}, {Torrelles}, \&
  {Osorio}}]{2010Sci...330.1209C}
{Carrasco-Gonz{\'a}lez}, C., {Rodr{\'{\i}}guez}, L.~F., {Anglada}, G., {et~al.}
  2010, Science, 330, 1209

\bibitem[{{Cesaroni} {et~al.}(1999){Cesaroni}, {Felli}, {Jenness}, {Neri},
  {Olmi}, {Robberto}, {Testi}, \& {Walmsley}}]{1999A&A...345..949C}
{Cesaroni}, R., {Felli}, M., {Jenness}, T., {et~al.} 1999, \aap, 345, 949

\bibitem[{{Cesaroni} {et~al.}(2005){Cesaroni}, {Neri}, {Olmi}, {Testi},
  {Walmsley}, \& {Hofner}}]{2005A&A...434.1039C}
{Cesaroni}, R., {Neri}, R., {Olmi}, L., {et~al.} 2005, \aap, 434, 1039

\bibitem[{{Cesaroni} {et~al.}(2016){Cesaroni}, {S{\'a}nchez-Monge},
  {Beltr{\'a}n}, {Molinari}, {Olmi}, \&
  {Trevi{\~n}o-Morales}}]{2016A&A...588L...5C}
{Cesaroni}, R., {S{\'a}nchez-Monge}, {\'A}., {Beltr{\'a}n}, M.~T., {et~al.}
  2016, \aap, 588, L5

\bibitem[{{Cesaroni} {et~al.}(2013){Cesaroni}, {Massi}, {Arcidiacono},
  {Beltr{\'a}n}, {McCarthy}, {Kulesa}, {Boutsia}, {Paris},
  {Quir{\'o}s-Pacheco}, \& {Xompero}}]{2013A&A...549A.146C}
{Cesaroni}, R., {Massi}, F., {Arcidiacono}, C., {et~al.} 2013, \aap, 549, A146

\bibitem[{{Cesaroni} {et~al.}(2015){Cesaroni}, {Pestalozzi}, {Beltr{\'a}n},
  {Hoare}, {Molinari}, {Olmi}, {Smith}, {Stringfellow}, {Testi}, \&
  {Thompson}}]{2015A&A...579A..71C}
{Cesaroni}, R., {Pestalozzi}, M., {Beltr{\'a}n}, M.~T., {et~al.} 2015, \aap,
  579, A71

\bibitem[{{Cesaroni} {et~al.}(2018){Cesaroni}, {Moscadelli}, {Neri}, {Sanna},
  {Caratti o Garatti}, {Eisloffel}, {Stecklum}, {Ray}, \&
  {Walmsley}}]{2018A&A...612A.103C}
{Cesaroni}, R., {Moscadelli}, L., {Neri}, R., {et~al.} 2018, \aap, 612, A103

\bibitem[{{Cooper} {et~al.}(2013){Cooper}, {Lumsden}, {Oudmaijer}, {Hoare},
  {Clarke}, {Urquhart}, {Mottram}, {Moore}, \& {Davies}}]{2013MNRAS.430.1125C}
{Cooper}, H.~D.~B., {Lumsden}, S.~L., {Oudmaijer}, R.~D., {et~al.} 2013,
  \mnras, 430, 1125

\bibitem[{{Curiel} {et~al.}(1987){Curiel}, {Canto}, \&
  {Rodriguez}}]{1987RMxAA..14..595C}
{Curiel}, S., {Canto}, J., \& {Rodriguez}, L.~F. 1987, \rmxaa, 14, 595

\bibitem[{{Curiel} {et~al.}(1989){Curiel}, {Rodriguez}, {Bohigas}, {Roth},
  {Canto}, \& {Torrelles}}]{1989ApL&C..27..299C}
{Curiel}, S., {Rodriguez}, L.~F., {Bohigas}, J., {et~al.} 1989, Astrophysical
  Letters and Communications, 27, 299

\bibitem[{{Curiel} {et~al.}(2006){Curiel}, {Ho}, {Patel}, {Torrelles},
  {Rodr{\'{\i}}guez}, {Trinidad}, {Cant{\'o}}, {Hern{\'a}ndez}, {G{\'o}mez},
  {Garay}, \& {Anglada}}]{2006ApJ...638..878C}
{Curiel}, S., {Ho}, P.~T.~P., {Patel}, N.~A., {et~al.} 2006, \apj, 638, 878

\bibitem[{{Cyganowski} {et~al.}(2011){Cyganowski}, {Brogan}, {Hunter},
  {Churchwell}, \& {Zhang}}]{2011ApJ...729..124C}
{Cyganowski}, C.~J., {Brogan}, C.~L., {Hunter}, T.~R., {Churchwell}, E., \&
  {Zhang}, Q. 2011, \apj, 729, 124

\bibitem[{{Davis} {et~al.}(2004){Davis}, {Varricatt}, {Todd}, \& {Ramsay
  Howat}}]{2004A&A...425..981D}
{Davis}, C.~J., {Varricatt}, W.~P., {Todd}, S.~P., \& {Ramsay Howat}, S.~K.
  2004, \aap, 425, 981

\bibitem[{{de Pree} {et~al.}(1995){de Pree}, {Rodriguez}, \&
  {Goss}}]{1995RMxAA..31...39D}
{de Pree}, C.~G., {Rodriguez}, L.~F., \& {Goss}, W.~M. 1995, \rmxaa, 31, 39

\bibitem[{{Dyson} \& {Williams}(1980)}]{1980pim..book.....D}
{Dyson}, J.~E., \& {Williams}, D.~A. 1980, {Physics of the interstellar medium}

\bibitem[{{Dzib} {et~al.}(2013){Dzib}, {Loinard}, {Mioduszewski},
  {Rodr{\'{\i}}guez}, {Ortiz-Le{\'o}n}, {Pech}, {Rivera}, {Torres}, {Boden},
  {Hartmann}, {Evans}, {Brice{\~n}o}, \& {Tobin}}]{2013ApJ...775...63D}
{Dzib}, S.~A., {Loinard}, L., {Mioduszewski}, A.~J., {et~al.} 2013, \apj, 775,
  63

\bibitem[{{Dzib} {et~al.}(2015){Dzib}, {Loinard}, {Rodr{\'{\i}}guez},
  {Mioduszewski}, {Ortiz-Le{\'o}n}, {Kounkel}, {Pech}, {Rivera}, {Torres},
  {Boden}, {Hartmann}, {Evans}, {Brice{\~n}o}, \&
  {Tobin}}]{2015ApJ...801...91D}
{Dzib}, S.~A., {Loinard}, L., {Rodr{\'{\i}}guez}, L.~F., {et~al.} 2015, \apj,
  801, 91

\bibitem[{{Ellsworth-Bowers} {et~al.}(2015){Ellsworth-Bowers}, {Rosolowsky},
  {Glenn}, {Ginsburg}, {Evans}, {Battersby}, {Shirley}, \&
  {Svoboda}}]{2015ApJ...799...29E}
{Ellsworth-Bowers}, T.~P., {Rosolowsky}, E., {Glenn}, J., {et~al.} 2015, \apj,
  799, 29

\bibitem[{{Fallscheer} {et~al.}(2011){Fallscheer}, {Beuther}, {Sauter}, {Wolf},
  \& {Zhang}}]{2011ApJ...729...66F}
{Fallscheer}, C., {Beuther}, H., {Sauter}, J., {Wolf}, S., \& {Zhang}, Q. 2011,
  \apj, 729, 66

\bibitem[{{Fallscheer} {et~al.}(2009){Fallscheer}, {Beuther}, {Zhang}, {Keto},
  \& {Sridharan}}]{2009A&A...504..127F}
{Fallscheer}, C., {Beuther}, H., {Zhang}, Q., {Keto}, E., \& {Sridharan}, T.~K.
  2009, \aap, 504, 127

\bibitem[{{Forbrich} {et~al.}(2004){Forbrich}, {Schreyer}, {Posselt}, {Klein},
  \& {Henning}}]{2004ApJ...602..843F}
{Forbrich}, J., {Schreyer}, K., {Posselt}, B., {Klein}, R., \& {Henning}, T.
  2004, \apj, 602, 843

\bibitem[{{Franco} {et~al.}(2007){Franco}, {Garc{\'{\i}}a-Segura}, {Kurtz}, \&
  {Arthur}}]{2007ApJ...660.1296F}
{Franco}, J., {Garc{\'{\i}}a-Segura}, G., {Kurtz}, S.~E., \& {Arthur}, S.~J.
  2007, \apj, 660, 1296

\bibitem[{{Garay} {et~al.}(2003){Garay}, {Brooks}, {Mardones}, \&
  {Norris}}]{2003ApJ...587..739G}
{Garay}, G., {Brooks}, K.~J., {Mardones}, D., \& {Norris}, R.~P. 2003, \apj,
  587, 739

\bibitem[{{Garay} \& {Rodriguez}(1990)}]{1990ApJ...362..191G}
{Garay}, G., \& {Rodriguez}, L.~F. 1990, \apj, 362, 191

\bibitem[{{Goddi} {et~al.}(2015){Goddi}, {Zhang}, \&
  {Moscadelli}}]{2015A&A...573A.108G}
{Goddi}, C., {Zhang}, Q., \& {Moscadelli}, L. 2015, \aap, 573, A108

\bibitem[{{Guzm{\'a}n} {et~al.}(2012){Guzm{\'a}n}, {Garay}, {Brooks}, \&
  {Voronkov}}]{2012ApJ...753...51G}
{Guzm{\'a}n}, A.~E., {Garay}, G., {Brooks}, K.~J., \& {Voronkov}, M.~A. 2012,
  \apj, 753, 51

\bibitem[{{Hoare} {et~al.}(2007){Hoare}, {Kurtz}, {Lizano}, {Keto}, \&
  {Hofner}}]{2007prpl.conf..181H}
{Hoare}, M.~G., {Kurtz}, S.~E., {Lizano}, S., {Keto}, E., \& {Hofner}, P. 2007,
  Protostars and Planets V, 181

\bibitem[{{Hofner} {et~al.}(2017){Hofner}, {Cesaroni}, {Kurtz}, {Rosero},
  {Anderson}, {Furuya}, {Araya}, \& {Molinari}}]{2017ApJ...843...99H}
{Hofner}, P., {Cesaroni}, R., {Kurtz}, S., {et~al.} 2017, \apj, 843, 99

\bibitem[{{Hofner} {et~al.}(2007){Hofner}, {Cesaroni}, {Olmi},
  {Rodr{\'{\i}}guez}, {Mart{\'{\i}}}, \& {Araya}}]{2007A&A...465..197H}
{Hofner}, P., {Cesaroni}, R., {Olmi}, L., {et~al.} 2007, \aap, 465, 197

\bibitem[{{Hofner} {et~al.}(2000){Hofner}, {Wyrowski}, {Walmsley}, \&
  {Churchwell}}]{2000ApJ...536..393H}
{Hofner}, P., {Wyrowski}, F., {Walmsley}, C.~M., \& {Churchwell}, E. 2000,
  \apj, 536, 393

\bibitem[{{Hofner} {et~al.}(2011){Hofner}, {Kurtz}, {Ellingsen}, {Menten},
  {Wyrowski}, {Araya}, {Loinard}, {Rodr{\'{\i}}guez}, \&
  {Cesaroni}}]{2011ApJ...739L..17H}
{Hofner}, P., {Kurtz}, S., {Ellingsen}, S.~P., {et~al.} 2011, \apjl, 739, L17

\bibitem[{{Johnston} {et~al.}(2013){Johnston}, {Shepherd}, {Robitaille}, \&
  {Wood}}]{2013A&A...551A..43J}
{Johnston}, K.~G., {Shepherd}, D.~S., {Robitaille}, T.~P., \& {Wood}, K. 2013,
  \aap, 551, A43

\bibitem[{{Keto}(2002)}]{2002ApJ...568..754K}
{Keto}, E. 2002, \apj, 568, 754

\bibitem[{{Keto}(2003)}]{2003ApJ...599.1196K}
---. 2003, \apj, 599, 1196

\bibitem[{{Keto}(2007)}]{2007ApJ...666..976K}
---. 2007, \apj, 666, 976

\bibitem[{{Konigl} \& {Pudritz}(2000)}]{2000prpl.conf..759K}
{Konigl}, A., \& {Pudritz}, R.~E. 2000, Protostars and Planets IV, 759

\bibitem[{{Kounkel} {et~al.}(2014){Kounkel}, {Hartmann}, {Loinard},
  {Mioduszewski}, {Dzib}, {Ortiz-Le{\'o}n}, {Rodr{\'{\i}}guez}, {Pech},
  {Rivera}, {Torres}, {Boden}, {Evans}, {Brice{\~n}o}, \&
  {Tobin}}]{2014ApJ...790...49K}
{Kounkel}, M., {Hartmann}, L., {Loinard}, L., {et~al.} 2014, \apj, 790, 49

\bibitem[{{Kurtz}(2005)}]{2005IAUS..227..111K}
{Kurtz}, S. 2005, in IAU Symposium, Vol. 227, Massive Star Birth: A Crossroads
  of Astrophysics, ed. R.~{Cesaroni}, M.~{Felli}, E.~{Churchwell}, \&
  M.~{Walmsley}, 111--119

\bibitem[{{Kurtz} {et~al.}(1994){Kurtz}, {Churchwell}, \&
  {Wood}}]{1994ApJS...91..659K}
{Kurtz}, S., {Churchwell}, E., \& {Wood}, D.~O.~S. 1994, \apjs, 91, 659

\bibitem[{{Lawrence} {et~al.}(2007){Lawrence}, {Warren}, {Almaini}, {Edge},
  {Hambly}, {Jameson}, {Lucas}, {Casali}, {Adamson}, {Dye}, {Emerson},
  {Foucaud}, {Hewett}, {Hirst}, {Hodgkin}, {Irwin}, {Lodieu}, {McMahon},
  {Simpson}, {Smail}, {Mortlock}, \& {Folger}}]{2007MNRAS.379.1599L}
{Lawrence}, A., {Warren}, S.~J., {Almaini}, O., {et~al.} 2007, \mnras, 379,
  1599

\bibitem[{{Lee} {et~al.}(2012){Lee}, {Takami}, {Duan}, {Karr}, {Su}, {Liu},
  {Froebrich}, \& {Yeh}}]{2012ApJS..200....2L}
{Lee}, H.-T., {Takami}, M., {Duan}, H.-Y., {et~al.} 2012, \apjs, 200, 2

\bibitem[{{Lee} {et~al.}(2013){Lee}, {Liao}, {Froebrich}, {Karr}, {Ioannidis},
  {Lee}, {Su}, {Liu}, {Duan}, \& {Takami}}]{2013ApJS..208...23L}
{Lee}, H.-T., {Liao}, W.-T., {Froebrich}, D., {et~al.} 2013, \apjs, 208, 23

\bibitem[{{L{\'o}pez-Sepulcre} {et~al.}(2010){L{\'o}pez-Sepulcre}, {Cesaroni},
  \& {Walmsley}}]{2010A&A...517A..66L}
{L{\'o}pez-Sepulcre}, A., {Cesaroni}, R., \& {Walmsley}, C.~M. 2010, \aap, 517,
  A66

\bibitem[{{Marti} {et~al.}(1995){Marti}, {Rodriguez}, \&
  {Reipurth}}]{1995ApJ...449..184M}
{Marti}, J., {Rodriguez}, L.~F., \& {Reipurth}, B. 1995, \apj, 449, 184

\bibitem[{{Mart{\'{\i}}} {et~al.}(1998){Mart{\'{\i}}}, {Rodr{\'{\i}}guez}, \&
  {Reipurth}}]{1998ApJ...502..337M}
{Mart{\'{\i}}}, J., {Rodr{\'{\i}}guez}, L.~F., \& {Reipurth}, B. 1998, \apj,
  502, 337

\bibitem[{{McMullin} {et~al.}(2007){McMullin}, {Waters}, {Schiebel}, {Young},
  \& {Golap}}]{2007ASPC..376..127M}
{McMullin}, J.~P., {Waters}, B., {Schiebel}, D., {Young}, W., \& {Golap}, K.
  2007, in Astronomical Society of the Pacific Conference Series, Vol. 376,
  Astronomical Data Analysis Software and Systems XVI, ed. R.~A. {Shaw},
  F.~{Hill}, \& D.~J. {Bell}, 127

\bibitem[{{Molinari} {et~al.}(1998){Molinari}, {Brand}, {Cesaroni}, {Palla}, \&
  {Palumbo}}]{1998A&A...336..339M}
{Molinari}, S., {Brand}, J., {Cesaroni}, R., {Palla}, F., \& {Palumbo},
  G.~G.~C. 1998, \aap, 336, 339

\bibitem[{{Molinari} {et~al.}(2010){Molinari}, {Swinyard}, {Bally}, {Barlow},
  {Bernard}, {Martin}, {Moore}, {Noriega-Crespo}, {Plume}, {Testi}, {Zavagno},
  {Abergel}, {Ali}, {Andr{\'e}}, {Baluteau}, {Benedettini}, {Bern{\'e}},
  {Billot}, {Blommaert}, {Bontemps}, {Boulanger}, {Brand}, {Brunt}, {Burton},
  {Campeggio}, {Carey}, {Caselli}, {Cesaroni}, {Cernicharo}, {Chakrabarti},
  {Chrysostomou}, {Codella}, {Cohen}, {Compiegne}, {Davis}, {de Bernardis}, {de
  Gasperis}, {Di Francesco}, {di Giorgio}, {Elia}, {Faustini}, {Fischera},
  {Fukui}, {Fuller}, {Ganga}, {Garcia-Lario}, {Giard}, {Giardino}, {Glenn},
  {Goldsmith}, {Griffin}, {Hoare}, {Huang}, {Jiang}, {Joblin}, {Joncas},
  {Juvela}, {Kirk}, {Lagache}, {Li}, {Lim}, {Lord}, {Lucas}, {Maiolo},
  {Marengo}, {Marshall}, {Masi}, {Massi}, {Matsuura}, {Meny}, {Minier},
  {Miville-Desch{\^e}nes}, {Montier}, {Motte}, {M{\"u}ller}, {Natoli}, {Neves},
  {Olmi}, {Paladini}, {Paradis}, {Pestalozzi}, {Pezzuto}, {Piacentini},
  {Pomar{\`e}s}, {Popescu}, {Reach}, {Richer}, {Ristorcelli}, {Roy}, {Royer},
  {Russeil}, {Saraceno}, {Sauvage}, {Schilke}, {Schneider-Bontemps},
  {Schuller}, {Schultz}, {Shepherd}, {Sibthorpe}, {Smith}, {Smith},
  {Spinoglio}, {Stamatellos}, {Strafella}, {Stringfellow}, {Sturm}, {Taylor},
  {Thompson}, {Tuffs}, {Umana}, {Valenziano}, {Vavrek}, {Viti}, {Waelkens},
  {Ward-Thompson}, {White}, {Wyrowski}, {Yorke}, \&
  {Zhang}}]{2010PASP..122..314M}
{Molinari}, S., {Swinyard}, B., {Bally}, J., {et~al.} 2010, \pasp, 122, 314

\bibitem[{{Moscadelli} {et~al.}(2013){Moscadelli}, {Cesaroni},
  {S{\'a}nchez-Monge}, {Goddi}, {Furuya}, {Sanna}, \&
  {Pestalozzi}}]{2013A&A...558A.145M}
{Moscadelli}, L., {Cesaroni}, R., {S{\'a}nchez-Monge}, {\'A}., {et~al.} 2013,
  \aap, 558, A145

\bibitem[{{Moscadelli} {et~al.}(2016){Moscadelli}, {S{\'a}nchez-Monge},
  {Goddi}, {Li}, {Sanna}, {Cesaroni}, {Pestalozzi}, {Molinari}, \&
  {Reid}}]{2016A&A...585A..71M}
{Moscadelli}, L., {S{\'a}nchez-Monge}, {\'A}., {Goddi}, C., {et~al.} 2016,
  \aap, 585, A71

\bibitem[{{Motte} {et~al.}(2018){Motte}, {Bontemps}, \&
  {Louvet}}]{2018ARA&A..56...41M}
{Motte}, F., {Bontemps}, S., \& {Louvet}, F. 2018, \araa, 56, 41

\bibitem[{{Navarete} {et~al.}(2015){Navarete}, {Damineli}, {Barbosa}, \&
  {Blum}}]{2015MNRAS.450.4364N}
{Navarete}, F., {Damineli}, A., {Barbosa}, C.~L., \& {Blum}, R.~D. 2015,
  \mnras, 450, 4364

\bibitem[{{Olnon}(1975)}]{1975A&A....39..217O}
{Olnon}, F.~M. 1975, \aap, 39, 217

\bibitem[{{Ortiz-Le{\'o}n} {et~al.}(2015){Ortiz-Le{\'o}n}, {Loinard},
  {Mioduszewski}, {Dzib}, {Rodr{\'{\i}}guez}, {Pech}, {Rivera}, {Torres},
  {Boden}, {Hartmann}, {Evans}, {Brice{\~n}o}, {Tobin}, {Kounkel}, \&
  {Gonz{\'a}lez-L{\'o}pezlira}}]{2015ApJ...805....9O}
{Ortiz-Le{\'o}n}, G.~N., {Loinard}, L., {Mioduszewski}, A.~J., {et~al.} 2015,
  \apj, 805, 9

\bibitem[{{Palau} {et~al.}(2007{\natexlab{a}}){Palau}, {Estalella}, {Girart},
  {Ho}, {Zhang}, \& {Beuther}}]{2007A&A...465..219P}
{Palau}, A., {Estalella}, R., {Girart}, J.~M., {et~al.} 2007{\natexlab{a}},
  \aap, 465, 219

\bibitem[{{Palau} {et~al.}(2007{\natexlab{b}}){Palau}, {Estalella}, {Ho},
  {Beuther}, \& {Beltr{\'a}n}}]{2007A&A...474..911P}
{Palau}, A., {Estalella}, R., {Ho}, P.~T.~P., {Beuther}, H., \& {Beltr{\'a}n},
  M.~T. 2007{\natexlab{b}}, \aap, 474, 911

\bibitem[{{Panagia}(1973)}]{1973AJ.....78..929P}
{Panagia}, N. 1973, \aj, 78, 929

\bibitem[{{Panagia} \& {Felli}(1975)}]{1975A&A....39....1P}
{Panagia}, N., \& {Felli}, M. 1975, \aap, 39, 1

\bibitem[{{Pech} {et~al.}(2016){Pech}, {Loinard}, {Dzib}, {Mioduszewski},
  {Rodr{\'{\i}}guez}, {Ortiz-Le{\'o}n}, {Rivera}, {Torres}, {Boden},
  {Hartmann}, {Kounkel}, {Evans}, {Brice{\~n}o}, {Tobin}, \&
  {Zapata}}]{2016ApJ...818..116P}
{Pech}, G., {Loinard}, L., {Dzib}, S.~A., {et~al.} 2016, \apj, 818, 116

\bibitem[{{Pilbratt} {et~al.}(2010){Pilbratt}, {Riedinger}, {Passvogel},
  {Crone}, {Doyle}, {Gageur}, {Heras}, {Jewell}, {Metcalfe}, {Ott}, \&
  {Schmidt}}]{2010A&A...518L...1P}
{Pilbratt}, G.~L., {Riedinger}, J.~R., {Passvogel}, T., {et~al.} 2010, \aap,
  518, L1

\bibitem[{{Purcell} {et~al.}(2013){Purcell}, {Hoare}, {Cotton}, {Lumsden},
  {Urquhart}, {Chandler}, {Churchwell}, {Diamond}, {Dougherty}, {Fender},
  {Fuller}, {Garrington}, {Gledhill}, {Goldsmith}, {Hindson}, {Jackson},
  {Kurtz}, {Mart{\'{\i}}}, {Moore}, {Mundy}, {Muxlow}, {Oudmaijer}, {Pandian},
  {Paredes}, {Shepherd}, {Smethurst}, {Spencer}, {Thompson}, {Umana}, \&
  {Zijlstra}}]{2013ApJS..205....1P}
{Purcell}, C.~R., {Hoare}, M.~G., {Cotton}, W.~D., {et~al.} 2013, \apjs, 205, 1

\bibitem[{{Purser} {et~al.}(2016){Purser}, {Lumsden}, {Hoare}, {Urquhart},
  {Cunningham}, {Purcell}, {Brooks}, {Garay}, {G{\'u}zman}, \&
  {Voronkov}}]{2016MNRAS.460.1039P}
{Purser}, S.~J.~D., {Lumsden}, S.~L., {Hoare}, M.~G., {et~al.} 2016, \mnras,
  460, 1039

\bibitem[{{Rathborne} {et~al.}(2010){Rathborne}, {Jackson}, {Chambers},
  {Stojimirovic}, {Simon}, {Shipman}, \& {Frieswijk}}]{2010ApJ...715..310R}
{Rathborne}, J.~M., {Jackson}, J.~M., {Chambers}, E.~T., {et~al.} 2010, \apj,
  715, 310

\bibitem[{{Rathborne} {et~al.}(2006){Rathborne}, {Jackson}, \&
  {Simon}}]{2006ApJ...641..389R}
{Rathborne}, J.~M., {Jackson}, J.~M., \& {Simon}, R. 2006, \apj, 641, 389

\bibitem[{{Reynolds}(1986)}]{1986ApJ...304..713R}
{Reynolds}, S.~P. 1986, \apj, 304, 713

\bibitem[{{Rieke} {et~al.}(2004){Rieke}, {Young}, {Engelbracht}, {Kelly},
  {Low}, {Haller}, {Beeman}, {Gordon}, {Stansberry}, {Misselt}, {Cadien},
  {Morrison}, {Rivlis}, {Latter}, {Noriega-Crespo}, {Padgett}, {Stapelfeldt},
  {Hines}, {Egami}, {Muzerolle}, {Alonso-Herrero}, {Blaylock}, {Dole}, {Hinz},
  {Le Floc'h}, {Papovich}, {P{\'e}rez-Gonz{\'a}lez}, {Smith}, {Su}, {Bennett},
  {Frayer}, {Henderson}, {Lu}, {Masci}, {Pesenson}, {Rebull}, {Rho}, {Keene},
  {Stolovy}, {Wachter}, {Wheaton}, {Werner}, \&
  {Richards}}]{2004ApJS..154...25R}
{Rieke}, G.~H., {Young}, E.~T., {Engelbracht}, C.~W., {et~al.} 2004, \apjs,
  154, 25

\bibitem[{{Rivilla} {et~al.}(2013){Rivilla}, {Mart{\'{\i}}n-Pintado},
  {Jim{\'e}nez-Serra}, \& {Rodr{\'{\i}}guez-Franco}}]{2013A&A...554A..48R}
{Rivilla}, V.~M., {Mart{\'{\i}}n-Pintado}, J., {Jim{\'e}nez-Serra}, I., \&
  {Rodr{\'{\i}}guez-Franco}, A. 2013, \aap, 554, A48

\bibitem[{{Robitaille} \& {Bressert}(2012)}]{2012ascl.soft08017R}
{Robitaille}, T., \& {Bressert}, E. 2012, {APLpy: Astronomical Plotting Library
  in Python}, Astrophysics Source Code Library, , , ascl:1208.017

\bibitem[{{Rodriguez} {et~al.}(1994){Rodriguez}, {Garay}, {Curiel}, {Ramirez},
  {Torrelles}, {Gomez}, \& {Velazquez}}]{1994ApJ...430L..65R}
{Rodriguez}, L.~F., {Garay}, G., {Curiel}, S., {et~al.} 1994, \apjl, 430, L65

\bibitem[{{Rodr{\'{\i}}guez} {et~al.}(2012){Rodr{\'{\i}}guez}, {Gonz{\'a}lez},
  {Montes}, {Asiri}, {Raga}, \& {Cant{\'o}}}]{2012ApJ...755..152R}
{Rodr{\'{\i}}guez}, L.~F., {Gonz{\'a}lez}, R.~F., {Montes}, G., {et~al.} 2012,
  \apj, 755, 152

\bibitem[{{Rodr{\'{\i}}guez} {et~al.}(2008){Rodr{\'{\i}}guez}, {Moran},
  {Franco-Hern{\'a}ndez}, {Garay}, {Brooks}, \&
  {Mardones}}]{2008AJ....135.2370R}
{Rodr{\'{\i}}guez}, L.~F., {Moran}, J.~M., {Franco-Hern{\'a}ndez}, R., {et~al.}
  2008, \aj, 135, 2370

\bibitem[{{Rosero} {et~al.}(2014){Rosero}, {Hofner}, {McCoy}, {Kurtz},
  {Menten}, {Wyrowski}, {Araya}, {Loinard}, {Carrasco-Gonz{\'a}lez},
  {Rodr{\'{\i}}guez}, {Cesaroni}, \& {Ellingsen}}]{2014ApJ...796..130R}
{Rosero}, V., {Hofner}, P., {McCoy}, M., {et~al.} 2014, \apj, 796, 130

\bibitem[{{Rosero} {et~al.}(2016){Rosero}, {Hofner}, {Claussen}, {Kurtz},
  {Cesaroni}, {Araya}, {Carrasco-Gonz{\'a}lez}, {Rodr{\'{\i}}guez}, {Menten},
  {Wyrowski}, {Loinard}, \& {Ellingsen}}]{2016ApJS..227...25R}
{Rosero}, V., {Hofner}, P., {Claussen}, M., {et~al.} 2016, \apjs, 227, 25

\bibitem[{{S{\'a}nchez-Monge} {et~al.}(2013{\natexlab{a}}){S{\'a}nchez-Monge},
  {Beltr{\'a}n}, {Cesaroni}, {Fontani}, {Brand}, {Molinari}, {Testi}, \&
  {Burton}}]{2013A&A...550A..21S}
{S{\'a}nchez-Monge}, {\'A}., {Beltr{\'a}n}, M.~T., {Cesaroni}, R., {et~al.}
  2013{\natexlab{a}}, \aap, 550, A21

\bibitem[{{S{\'a}nchez-Monge} {et~al.}(2013{\natexlab{b}}){S{\'a}nchez-Monge},
  {Kurtz}, {Palau}, {Estalella}, {Shepherd}, {Lizano}, {Franco}, \&
  {Garay}}]{2013ApJ...766..114S}
{S{\'a}nchez-Monge}, {\'A}., {Kurtz}, S., {Palau}, A., {et~al.}
  2013{\natexlab{b}}, \apj, 766, 114

\bibitem[{{S{\'a}nchez-Monge} {et~al.}(2013{\natexlab{c}}){S{\'a}nchez-Monge},
  {L{\'o}pez-Sepulcre}, {Cesaroni}, {Walmsley}, {Codella}, {Beltr{\'a}n},
  {Pestalozzi}, \& {Molinari}}]{2013A&A...557A..94S}
{S{\'a}nchez-Monge}, {\'A}., {L{\'o}pez-Sepulcre}, A., {Cesaroni}, R., {et~al.}
  2013{\natexlab{c}}, \aap, 557, A94

\bibitem[{{Sanna} {et~al.}(2016){Sanna}, {Moscadelli}, {Cesaroni}, {Caratti o
  Garatti}, {Goddi}, \& {Carrasco-Gonz{\'a}lez}}]{2016A&A...596L...2S}
{Sanna}, A., {Moscadelli}, L., {Cesaroni}, R., {et~al.} 2016, \aap, 596, L2

\bibitem[{{Sanna} {et~al.}(2018){Sanna}, {Moscadelli}, {Goddi}, {Krishnan}, \&
  {Massi}}]{2018A&A...619A.107S}
{Sanna}, A., {Moscadelli}, L., {Goddi}, C., {Krishnan}, V., \& {Massi}, F.
  2018, \aap, 619, A107

\bibitem[{{Sanna} {et~al.}(2019{\natexlab{a}}){Sanna}, {K{\"o}lligan},
  {Moscadelli}, {Kuiper}, {Cesaroni}, {Pillai}, {Menten}, {Zhang}, {Caratti o
  Garatti}, {Goddi}, {Leurini}, \&
  {Carrasco-Gonz{\'a}lez}}]{2019A&A...623A..77S}
{Sanna}, A., {K{\"o}lligan}, A., {Moscadelli}, L., {et~al.} 2019{\natexlab{a}},
  \aap, 623, A77

\bibitem[{{Sanna} {et~al.}(2019{\natexlab{b}}){Sanna}, {Moscadelli}, {Goddi},
  {Beltr{\'a}n}, {Brogan}, {Caratti o Garatti}, {Carrasco-Gonz{\'a}lez},
  {Hunter}, {Massi}, \& {Padovani}}]{2019A&A...623L...3S}
{Sanna}, A., {Moscadelli}, L., {Goddi}, C., {et~al.} 2019{\natexlab{b}}, \aap,
  623, L3

\bibitem[{{Shepherd} \& {Churchwell}(1996)}]{1996ApJ...457..267S}
{Shepherd}, D.~S., \& {Churchwell}, E. 1996, \apj, 457, 267

\bibitem[{{Shepherd} {et~al.}(2004){Shepherd}, {N{\"u}rnberger}, \&
  {Bronfman}}]{2004ApJ...602..850S}
{Shepherd}, D.~S., {N{\"u}rnberger}, D.~E.~A., \& {Bronfman}, L. 2004, \apj,
  602, 850

\bibitem[{{Shepherd} {et~al.}(2000){Shepherd}, {Yu}, {Bally}, \&
  {Testi}}]{2000ApJ...535..833S}
{Shepherd}, D.~S., {Yu}, K.~C., {Bally}, J., \& {Testi}, L. 2000, \apj, 535,
  833

\bibitem[{{Shepherd} {et~al.}(2007){Shepherd}, {Povich}, {Whitney},
  {Robitaille}, {N{\"u}rnberger}, {Bronfman}, {Stark}, {Indebetouw}, {Meade},
  \& {Babler}}]{2007ApJ...669..464S}
{Shepherd}, D.~S., {Povich}, M.~S., {Whitney}, B.~A., {et~al.} 2007, \apj, 669,
  464

\bibitem[{{Shu} {et~al.}(1987){Shu}, {Adams}, \&
  {Lizano}}]{1987ARA&A..25...23S}
{Shu}, F.~H., {Adams}, F.~C., \& {Lizano}, S. 1987, \araa, 25, 23

\bibitem[{{Shu} {et~al.}(1988){Shu}, {Lizano}, {Ruden}, \&
  {Najita}}]{1988ApJ...328L..19S}
{Shu}, F.~H., {Lizano}, S., {Ruden}, S.~P., \& {Najita}, J. 1988, \apjl, 328,
  L19

\bibitem[{{Sicilia-Aguilar} {et~al.}(2005){Sicilia-Aguilar}, {Hartmann},
  {Szentgyorgyi}, {Fabricant}, {F{\H u}r{\'e}sz}, {Roll}, {Conroy}, {Calvet},
  {Tokarz}, \& {Hern{\'a}ndez}}]{2005AJ....129..363S}
{Sicilia-Aguilar}, A., {Hartmann}, L.~W., {Szentgyorgyi}, A.~H., {et~al.} 2005,
  \aj, 129, 363

\bibitem[{{Sridharan} {et~al.}(2002){Sridharan}, {Beuther}, {Schilke},
  {Menten}, \& {Wyrowski}}]{2002ApJ...566..931S}
{Sridharan}, T.~K., {Beuther}, H., {Schilke}, P., {Menten}, K.~M., \&
  {Wyrowski}, F. 2002, \apj, 566, 931

\bibitem[{{Su} {et~al.}(2007){Su}, {Liu}, {Chen}, {Zhang}, \&
  {Cesaroni}}]{2007ApJ...671..571S}
{Su}, Y.-N., {Liu}, S.-Y., {Chen}, H.-R., {Zhang}, Q., \& {Cesaroni}, R. 2007,
  \apj, 671, 571

\bibitem[{{Tan} {et~al.}(2014){Tan}, {Beltr{\'a}n}, {Caselli}, {Fontani},
  {Fuente}, {Krumholz}, {McKee}, \& {Stolte}}]{2014prpl.conf..149T}
{Tan}, J.~C., {Beltr{\'a}n}, M.~T., {Caselli}, P., {et~al.} 2014, Protostars
  and Planets VI, 149

\bibitem[{{Tan} {et~al.}(2016){Tan}, {Kong}, {Zhang}, {Fontani}, {Caselli}, \&
  {Butler}}]{2016ApJ...821L...3T}
{Tan}, J.~C., {Kong}, S., {Zhang}, Y., {et~al.} 2016, \apjl, 821, L3

\bibitem[{{Tanaka} {et~al.}(2016){Tanaka}, {Tan}, \&
  {Zhang}}]{2016ApJ...818...52T}
{Tanaka}, K.~E.~I., {Tan}, J.~C., \& {Zhang}, Y. 2016, \apj, 818, 52

\bibitem[{{Tobin} {et~al.}(2009){Tobin}, {Hartmann}, {Furesz}, {Mateo}, \&
  {Megeath}}]{2009ApJ...697.1103T}
{Tobin}, J.~J., {Hartmann}, L., {Furesz}, G., {Mateo}, M., \& {Megeath}, S.~T.
  2009, \apj, 697, 1103

\bibitem[{{Urquhart} {et~al.}(2014){Urquhart}, {Csengeri}, {Wyrowski},
  {Schuller}, {Bontemps}, {Bronfman}, {Menten}, {Walmsley}, {Contreras},
  {Beuther}, {Wienen}, \& {Linz}}]{2014A&A...568A..41U}
{Urquhart}, J.~S., {Csengeri}, T., {Wyrowski}, F., {et~al.} 2014, \aap, 568,
  A41

\bibitem[{{van Buren} {et~al.}(1990){van Buren}, {Mac Low}, {Wood}, \&
  {Churchwell}}]{1990ApJ...353..570V}
{van Buren}, D., {Mac Low}, M.-M., {Wood}, D.~O.~S., \& {Churchwell}, E. 1990,
  \apj, 353, 570

\bibitem[{{Varricatt} {et~al.}(2010){Varricatt}, {Davis}, {Ramsay}, \&
  {Todd}}]{2010MNRAS.404..661V}
{Varricatt}, W.~P., {Davis}, C.~J., {Ramsay}, S., \& {Todd}, S.~P. 2010,
  \mnras, 404, 661

\bibitem[{{Varricatt} {et~al.}(2013){Varricatt}, {Thomas}, {Davis}, {Ramsay},
  \& {Currie}}]{2013A&A...554A...9V}
{Varricatt}, W.~P., {Thomas}, H.~S., {Davis}, C.~J., {Ramsay}, S., \& {Currie},
  M.~J. 2013, \aap, 554, A9

\bibitem[{{Walmsley}(1995)}]{1995RMxAC...1..137W}
{Walmsley}, M. 1995, in Revista Mexicana de Astronomia y Astrofisica Conference
  Series, Vol.~1, Revista Mexicana de Astronomia y Astrofisica Conference
  Series, ed. S.~{Lizano} \& J.~M. {Torrelles}, 137

\bibitem[{{Wang} {et~al.}(2014){Wang}, {Zhang}, {Testi}, {van der Tak}, {Wu},
  {Zhang}, {Pillai}, {Wyrowski}, {Carey}, {Ragan}, \&
  {Henning}}]{2014MNRAS.439.3275W}
{Wang}, K., {Zhang}, Q., {Testi}, L., {et~al.} 2014, \mnras, 439, 3275

\bibitem[{{Wilner} {et~al.}(2001){Wilner}, {De Pree}, {Welch}, \&
  {Goss}}]{2001ApJ...550L..81W}
{Wilner}, D.~J., {De Pree}, C.~G., {Welch}, W.~J., \& {Goss}, W.~M. 2001,
  \apjl, 550, L81

\bibitem[{{Wolf-Chase} {et~al.}(2017){Wolf-Chase}, {Arvidsson}, \&
  {Smutko}}]{2017ApJ...844...38W}
{Wolf-Chase}, G., {Arvidsson}, K., \& {Smutko}, M. 2017, \apj, 844, 38

\bibitem[{{Wood} \& {Churchwell}(1989{\natexlab{a}})}]{1989ApJ...340..265W}
{Wood}, D.~O.~S., \& {Churchwell}, E. 1989{\natexlab{a}}, \apj, 340, 265

\bibitem[{{Wood} \& {Churchwell}(1989{\natexlab{b}})}]{1989ApJS...69..831W}
---. 1989{\natexlab{b}}, \apjs, 69, 831

\bibitem[{{Xie} {et~al.}(1996){Xie}, {Mundy}, {Vogel}, \&
  {Hofner}}]{1996ApJ...473L.131X}
{Xie}, T., {Mundy}, L.~G., {Vogel}, S.~N., \& {Hofner}, P. 1996, \apjl, 473,
  L131

\bibitem[{{Zapata} {et~al.}(2006){Zapata}, {Rodr{\'{\i}}guez}, {Ho}, {Beuther},
  \& {Zhang}}]{2006AJ....131..939Z}
{Zapata}, L.~A., {Rodr{\'{\i}}guez}, L.~F., {Ho}, P.~T.~P., {Beuther}, H., \&
  {Zhang}, Q. 2006, \aj, 131, 939

\bibitem[{{Zhang} {et~al.}(2007){Zhang}, {Sridharan}, {Hunter}, {Chen},
  {Beuther}, \& {Wyrowski}}]{2007A&A...470..269Z}
{Zhang}, Q., {Sridharan}, T.~K., {Hunter}, T.~R., {et~al.} 2007, \aap, 470, 269

\bibitem[{{Zhang} {et~al.}(2014){Zhang}, {Tan}, \&
  {Hosokawa}}]{2014ApJ...788..166Z}
{Zhang}, Y., {Tan}, J.~C., \& {Hosokawa}, T. 2014, \apj, 788, 166

\end{thebibliography}

\end{document}